\documentclass[traditabstract]{aa} 
\usepackage{graphicx}
\usepackage{longtable,lscape}   
\usepackage{ulem}
\usepackage{latexsym}
\usepackage{color}
\usepackage{natbib}
\bibpunct{(}{)}{;}{a}{}{,} 
\usepackage{multirow}

\newcommand{\teff}{$T_{\rm eff}$}
\newcommand{\logg}{$\log\,g$}
\newcommand{\vsini}{$v$\,sin\,$i$}

\newcommand{\rotfit}{{\sf ROTFIT}}
\newcommand{\ROTFIT}{{\sf ROTFIT}}
\newcommand{\COMPO}{{\sf COMPO2}}
\newcommand{\kms}{km\,s$^{-1}$}
 
\newcommand{\gaia}{{\it Gaia}}
\newcommand{\tess}{{\it TESS}}
\newcommand{\eagles}{{\sf EAGLES}}

\definecolor{blu}{rgb}{0,0,1}
\definecolor{mag}{rgb}{1,0,1}
\definecolor{dgreen}{rgb}{0.1, 0.53, 0.22}
\definecolor{orange}{RGB}{255,127,0}

\begin{document}

\title{The Radcliffe Wave as traced by young open clusters\thanks{Based on observations made with the Italian Telescopio Nazionale Galileo (TNG)
operated on the island of La Palma by the Fundaci\'on Galileo Galilei of the INAF (Istituto Nazionale di Astrofisica) at the Observatorio
del Roque de los Muchachos.
}}
\subtitle{Stellar parameters, activity indicators, and abundances of solar-type members of eight young clusters} 

\author{J. Alonso-Santiago\inst{1}\and
        A. Frasca\inst{1}\and 
        A. Bragaglia\inst{2} \and
        G. Catanzaro\inst{1}\and
        X. Fu\inst{3}\and
        G. Andreuzzi\inst{4,5}\and
        L. Magrini\inst{6}\and
         S. Lucatello\inst{7}\and
         A. Vallenari\inst{7}\and
         M. Jian\inst{8}
}

\offprints{J. Alonso-Santiago\\ \email{javier.alonso@inaf.it}}

\institute{
INAF--Osservatorio Astrofisico di Catania, via S. Sofia 78, 95123 Catania, Italy
\and
INAF--Osservatorio di Astrofisica e Scienza dello Spazio, via P. Gobetti 93/3, 40129 Bologna, Italy
\and
Purple Mountain Observatory, Chinese Academy of Sciences, Nanjing 210023, China
\and
Fundaci\'on Galileo Galilei - INAF, Rambla Jos\'e Ana Fern\'andez P\'erez 7, 38712 Bre\~na Baja, Tenerife - Spain
\and
INAF--Osservatorio Astronomico di Roma, Via Frascati 33, 00078 Monte Porzio Catone, Italy
\and
INAF--Osservatorio Astrofisico di Arcetri,  Largo  E.  Fermi  5, 50125 Firenze, Italy
\and
INAF--Osservatorio Astronomico di Padova, vicolo dell'Osservatorio 5, 35122 Padova, Italy
\and
Department of Astronomy, Stockholm University, AlbaNova University centre, Roslagstullsbacken 21, 114 21 Stockholm, Sweden
}

\date{Received  / accepted}

 
\abstract{The Radcliffe Wave has only recently been recognised as a $\approx$ 3\,kpc long coherent gas structure encompassing most of the
star forming regions in the solar vicinity. 
Since its discovery, it has been mainly studied from the perspective of dynamics, 
but a detailed chemical study is necessary to understand its nature and the composition of the natal clouds that gave rise to it. In this paper we used some of the connected young open clusters (age $\lesssim$ 100 Myr) as tracers of the molecular clouds. 
We performed high-resolution spectroscopy with GIARPS at the TNG of 53 stars that are bona fide members of seven clusters located at different positions along the Radcliffe Wave. We provide radial velocities and atmospheric parameters for all of them. For a subsample consisting of 41 FGK stars we also studied the chromospheric activity and the content of Li, from which we inferred the age of the parent clusters. These values agree with the evolutionary ages reported in the literature. For these FGK stars we determined the chemical abundances for 25 species. Pleiades, ASCC\,16 and NGC\,7058 exhibit a solar metallicity while Melotte\,20, ASCC\,19, NGC\,2232, and Roslund\,6 show a slightly subsolar value ($\approx$\,$-$0.1\,dex). On average, the clusters show a chemical composition compatible with that of the Sun, especially for $\alpha$- and Fe-peak elements. Neutron-capture elements, on the other hand, present a slight overabundance of about 0.2\,dex, specially barium. Finally, considering also ASCC\,123, studied by our group in a previous research, we infer a correlation between the chemical composition and the age or position of the clusters along the Wave, demonstrating their physical connection within an inhomogeneous mixing scenario.
}

   \titlerunning{Young open clusters in the Radcliffe Wave}

      \authorrunning{J. Alonso-Santiago et al.}

\maketitle

\section{Introduction}
\label{Sec:intro}

We are currently living in a golden age for Galactic Archaeology. The ESA $Gaia$ mission \citep{Gaia16,GaiaDR3}, in combination with other large photometric and spectroscopic surveys, is revolutionizing our knowledge of the Milky Way \citep[for a review, see e.g.][]{Deason24}. Never before have we had at our disposal a set of astrometric, photometric and spectroscopic data of such high quality and for such a large sample of hundreds of thousands of stars. Thanks to this huge amount of information we considerably improved our understanding of the history \citep[see][]{Antoja18,Helmi18} and components of the Galaxy. It is worth mentioning the discovery of new open clusters (OCs) and new members of already known clusters \citep[e.g.,][]{Cantat2020,Castro-Ginard2020,Castro-Ginard2022,HR23,Cavallo24}, as well as the discovery of new associations and comoving groups in the vicinity (1--3\,kpc) of the Sun \citep[e.g.,][]{KounkelCovey2019,Kounkel2020}.

Recently, \citet{Alves2020} discovered a spatially coherent filamentary structure of dense gas in the solar neighbourhood when studying the 3D distribution of the local cloud complexes. This discovery was possible thanks to the accurate astrometry provided by $Gaia$, which allowed us to significantly improve the 3D mapping of the Galactic interstellar dust \citep{Green2019,Lallement2019} and determine the distances to the local molecular clouds with great precision \citep{Zucker20}. This structure, the so-called Radcliffe Wave (RW), is a narrow (aspect ratio 1:20) and 2.7-kpc long band with a wave-like shape that, at its closest part, is 300 pc from the Sun \citep{Alves2020}. Its origin can be due to a disc instability \citep{Fleck2020} or to a perturbation generated by the passage of a satellite \citep{Li22,Thulasidharan22}.

Many of the most important star forming regions (SFRs) in the solar vicinity, such as Canis Major, Monoceros R2, Orion, Taurus, Perseus, Cepheus, North America nebula and Cygnus are aligned in the XY Galactic plane along the RW, which could be associated to the Orion spiral arm \citep{Swiggum2022}. It should be noted that four of these complexes (Orion, Taurus, Perseus and Cepheus) have classically been part of the Gould Belt \citep{Gould74,Palous17}. It was assumed to be a Sun-centred ring-shaped structure with a size of about 500 pc involving the nearest molecular clouds and OB associations as well as a sparse population of X-ray point-like sources associated with young stellar objects (YSOs) \citep[e.g.,][]{Guillout1998}. However, in light of $Gaia$'s very precise astrometry, our view of the Gould Belt has changed, becoming in fact just the apparent projection on the sky of two larger physical structures: the RW and the `split' \citep{Alves2020,Swiggum2022}. The latter is also a linear arrangement of cloud complexes 1\,kpc long and contains, among others, the Ophiucus SFR, the only one of the five large Gould clouds not associated to the RW.

Since its discovery, the kinematics of the RW has been studied to investigate the possible existence of an oscillation pattern. Different RW tracers such as YSOs, OB stars, young OCs or even masers and radio sources were used, which, however, did not lead to conclusive results \citep{DonadaFigueras2021,Bobylev22,Li22,Thulasidharan22,Tu22,Alfaro22,Bobylev24}. Finally, \citet{Konietzka24}, by making use of line-of-sight velocity measurements of $^{12}$CO and 3D velocities of young OCs, demonstrated that the RW is not only sinusoidal in shape but  actually oscillates across the Galactic plane and moves radially outward from the Galactic centre. It implies that all the SFRs composing the RW 
share a common history. In fact, it is possible that the star cluster whose supernovae 
originated the Local Bubble was also part of the RW in the past \citep{Zucker22}.

In this context, the RW and its individual components constitute an unparalleled laboratory for the study of star formation in the solar neighbourhood.
Indeed, metal mixing in giant molecular clouds is key to understanding the star formation history and the chemical evolution of the Galaxy. Traditionally, the nearby molecular clouds were assumed to be well mixed. However, \citet{DeCia2021} observed nearby hot, young stars and found differences of a factor of ten in metallicity, which was ascribed to
inhomogeneous metal mixing in the natal cloud(s), on a scale of a few parsec to tens of parsec at most. Whether inhomogeneous mixing is a general feature should be ascertained by looking at molecular clouds systems, such as those defining the RW. 
In this respect, \citet{Fu2022}, claimed the possible detection of metallicty variations along the Wave in a work based on low-resolution LAMOST spectra of OCs stars.

To date, the 
RW has always been studied from a kinematical perspective. This work represents the first attempt to characterise its chemical pattern. The use of high-resolution spectroscopy to derive precise chemical abundances of some young OCs associated to the Wave can shed light on the gas mixing and star formation processes inside the structure. Given the young age of the Wave, we expect that stellar evolution and migration across the Galactic disc have not significantly affected the clusters yet and, thus, their chemical composition should be representative of the composition of the local gas. The analysis of elements belonging to all main nucleosynthesis channels allows us to establish the chemical coherence of OCs belonging to the RW, and to estimate local enrichment effects due, for instance, to recent supernovae explosions. We might expect that star formation had propagated in nearby clusters and that the formation of stars in one subgroup might trigger or affect star formation in the other subgroups \citep[e.g.,][]{Armillotta2018}. 

The paper is structured as follows. In Sect.~\ref{Sec:Observations} we present our observations and the criteria used to select our targets. The spectral analysis that we followed is described in Sect.~\ref{Sec:analysis} along with the stellar parameters and chemical abundances we obtained. The chromospheric activity and the lithium content are evaluated in Sect.~\ref{Sec:chrom_lithium}. Then, we compare our results with the literature and place them in the Galactic context in Sect.~\ref{Sec:disc}, while the discussion and interpretation of our findings relative to the Radcliffe Wave are presented in Sect.~\ref{sec:rw}. Finally, in Sect.~\ref{Sec:summary} we summarize our results and present our conclusions. 

\section{Observations and data reduction}
\label{Sec:Observations}

To study the RW, instead of relying on field stars (for which the age is always difficult to measure with a sufficient precision) or on molecular gas (due to the uncertainties in determining its chemical and kinematical properties), it is convenient to use young OCs as tracers. For those, one can obtain a robust age determination and precise kinematics and element abundances. 
Many candidate clusters have been proposed to be associated to the Wave \citep[e.g.,][]{DonadaFigueras2021,Fu2022}. As stars in a cluster share essentially the same age,  initial metallicity and chemical composition, observations of a few members is sufficient to characterise the cluster, as already shown in some papers based on data of the Stellar Populations Astrophysics (SPA) large program conducted at the TNG. In those papers, we have chemically characterised a sample of dwarfs and giants in nearby OCs using high-resolution spectra \citep[e.g.,][]{Frasca2019,Casali20,DOrazi2020,Alonso2021,zhang21,zhang22}. 

\subsection{Sample selection}
\label{Subsec:sample}

We have initially selected five nearby and low-extinction OCs associated with the RW, 
namely Melotte\,20 ($\alpha$ Per), Pleiades (Melotte\,22), ASCC\,16, NGC\,2232, and Roslund\,6.
To measure the metallicity and abundance ratios with a precision sufficient to resolve cluster-to-cluster variations, we need to use high-resolution spectra of slow rotators (\vsini\,$\leq$10--15\,\kms; higher rotation would blur and blend lines). In these young OCs, this means observing main-sequence (MS) stars of mid-G spectral type.
To this aim, we have selected four to six members of the above clusters from the lists of \citet{Cantat2020} with a color index $(G_{\rm BP}-G_{\rm RP})$ in the range 0.75--0.95\,mag, corresponding approximately to G1--G9 spectral types, which have no visual companion with a comparable magnitude within 10\arcsec.  
To reject fast-rotating stars, which are frequent in young ($age<$300\,Myr) clusters \citep[e.g.,][]{Barnes2003,Fritzewski2020}, we have initially relied on the spectral line broadening parameter $V_{\rm broad}$, which is reported in the \gaia~DR3 catalogue \citep{GaiaDR3} and is based on the medium-resolution \gaia~RVS spectra \citep{Fremat2023}. This parameter is not available for all the pre-selected targets. Moreover, it provides only a rough indication of line broadening, due to the moderate resolution (R=11,500) of RVS and the presence of broad \ion{Ca}{ii} IRT lines in its spectral domain. Therefore, we have used an inclusive threshold of $V_{\rm broad}\leq 40$\,\kms\ and resorted to additional criteria to select the slowly-rotating stars.
For the pre-selected sources, we have retrieved from the MAST\footnote{\url{https://mast.stsci.edu/portal/Mashup/Clients/Mast/Portal.html}} archive the light curves produced by the NASA's Transiting Exoplanet Survey Satellite ({\it TESS}; \citealt{Ricker2015})
and searched for the rotation period, $P_{\rm rot}$ by means of the periodogram analysis \citep{Scargle1982} and the CLEAN deconvolution 
algorithm \citep{Roberts1987}, similar to what we did in \citet{Frasca2023a}.
We have then chosen the targets with the longest period ($P_{\rm rot}\geq 3$\,d) for which an equatorial velocity $v_{\rm eq}\leq 16$\,\kms\ is expected. 
For the Pleiades we used the rotational periods derived by \citet{Rebull2016} for a sample of members observed with {\it Kepler}-K2. These selection criteria do not prevent some binary system from being included. Indeed, four double-lined spectroscopic binaries (SB2s) were observed and their spectra were also used, as discussed in Sect.\,\ref{Sec:analysis}.

In addition to the solar-type stars, a few brighter and faster rotating F5--G0 type stars have been observed as well. For them, a value of \vsini\,$\leq$\,60\,\kms\ is a safe threshold for determining the atmospheric parameters, \vsini, and RV with a sufficient precision with \ROTFIT, as we did in \citet{Frasca2019} and \citet{Alonso2021}, where we were also able to derive parameters for binary systems. 

We have increased our target sample with archive spectra of two members of NGC\,2232 and three members of Melotte\,20, as well as several stars in three additional OCs associated with the RW, namely ASCC~19, NGC~7058, and ASCC\,123. Spectra of these sources were taken in the framework of the SPA large program before this structure was discovered by Alves et al. (2020). However, the selection criteria for the candidate members of these three clusters are different, with the result of including more hotter and faster rotating stars. The analysis of ASCC\,123 was previously published in a dedicated paper and will not be further discussed throughout the present work. The reader can find all the information in \citet{Frasca2019,Frasca2023a}.

\begin{figure*}[ht]
\begin{center}
\includegraphics[width=15cm]{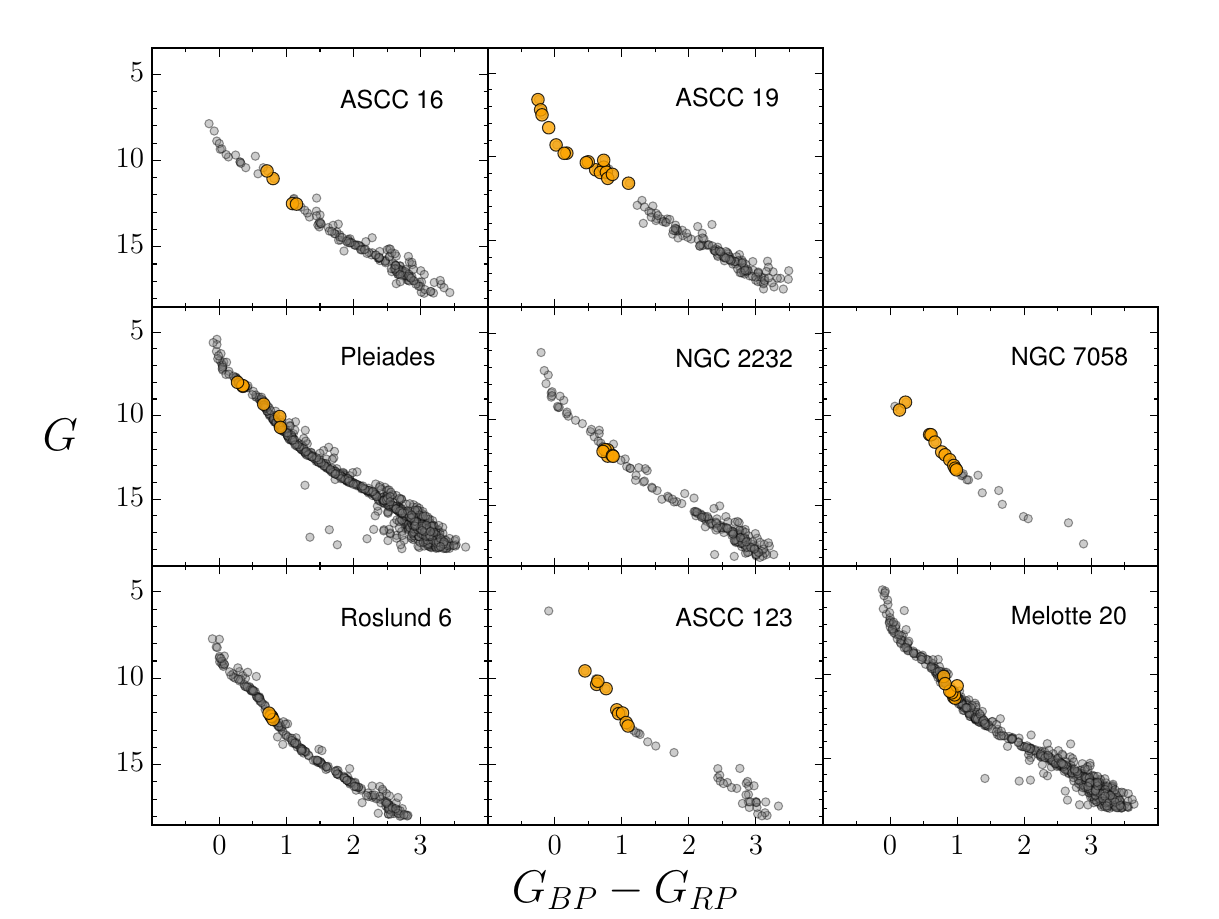}
\caption{CMDs of the clusters studied in this work. Small grey circles represent the likely members identified in \citet{Cantat2020} while the big orange ones mark the stars composing our sample for which we have GIARPS spectra.}

\label{Fig:CMD_radcliffe}
\end{center}
\end{figure*}

The main properties of the clusters targeted in this work, according to \citet{Cantat2020}, are reported in Table~\ref{Tab:radcliffe_clusters}. Additionally, Fig.~\ref{Fig:CMD_radcliffe} shows the position on the color-magnitude diagram (CMD) of the likely members observed in each cluster. Note that  five of these clusters (ASCC\,16, ASCC\,19, NGC\,2232, NGC\,7058, and Roslund\,6) have not been studied with high signal-to-noise ratio (S/N), high resolution before, while Melotte~20  and Pleiades have been included in this study to compare with literature and keep systematics under control. ASCC\,123 is also a very little-studied cluster whose only chemical study to date was carried out, as we have already mentioned, by our group.

In total, we observed 59 stars that are listed in Table~\ref{Tab:obs_log}. We followed an internal numbering that consists of adding a sequential number to the (short)name of the cluster to which they belong. For Melotte\,20 and NGC\,2232, we distinguished between the stars observed during the main run (letter S) and those taken from the SPA archive (A). Not all the targets in our preliminary list were observed and that is why some numbers are missing. In any case, to facilitate their identification we also provide the $Gaia$ source name in Table~\ref{Tab:obs_log} and the equatorial coordinates (J2000.0) in Table~\ref{Tab:astrom}.

\begin{table}
\caption{Main properties of the clusters investigated in this work according to \citet{Cantat2020}: radius containing half of the members ($\it{r50}$), number of high-probability members ($N$), age (log\,$\tau$), extinction ($A_V$) and average distance ($d$).}
\begin{center}
\begin{tabular}{lccccc}
\hline
\hline
\noalign{\smallskip}
Cluster    & $\it{r50}$ ($^{\circ}$) & $N$  & log\,$\tau$ & $A_V$  & $d$ (pc) \\
\noalign{\smallskip}
\hline
\noalign{\smallskip}
{\scriptsize Melotte~20 / $\alpha$~Per} &2.027   & 747  &  7.71    &  0.30  &  171    \\
{\scriptsize Pleiades / Melotte~22}   &  1.274    & 952  &  7.89   &  0.18  &  128     \\
ASCC~16$^*$&  0.376      & 175  &  7.13       &  0.20  &  344     \\
ASCC~19    &  0.605      & 149  &  7.02       &  0.13  &  346     \\
NGC~2232   &  0.516      & 188  &  7.25       &  0.01  &  315     \\
Roslund~6  &  1.004      & 247  &  7.95       &  0.00  &  375     \\
NGC~7058   &  0.109      &  26  &  7.61       &  0.18  &  350     \\
ASCC~123   &  1.294      &  49  &  7.65       &  0.00  &  229     \\
\hline
\end{tabular}
\end{center}
\label{Tab:radcliffe_clusters}
{\bf Note.} $^*$ Identified as Briceno\,1 in \citet{HR23} and \citet{Cavallo24}.
\end{table}

\subsection{Spectroscopic observations}

The observations were conducted from 26 to 28 October 2022 and in the fall of 2019 and 2020 for those collected in the framework of the SPA large program (see Table~\ref{Tab:obs_log} for a logbook of the observations). We used the GIARPS 
(GIANO-B \& HARPS-N; \citealt{Claudi2017}) facility at the 3.6m TNG telescope located at El Roque de los Muchachos Observatory, in La Palma (Spain). With GIARPS we are able to simultaneously collect both optical 
(HARPS-N, $R\simeq 115\,000$, range\,=\,0.39--0.68\,$\mu$m, \citealt{Cosentino2014}) and near-infrared (NIR) high-resolution spectra (GIANO-B, $R\simeq 50\,000$, range\,=\,0.97--2.45\,$\mu$m; \citealt{Oliva2012a,Origlia2014}).
The HARPS-N spectra were acquired with the second fiber on-sky.
The total exposure times ranged from 300 to 7200 sec, depending on the star brightness and sky conditions, and were chosen so as to reach a signal-to-noise ratio per pixel S/N$\geq$\,30 at red wavelengths. Exposures longer than 1800 sec were usually split in two or three to reduce the contamination of cosmic rays.
The coordinates, TIC identifier, as well as the main kinematical properties of the stars observed at the TNG are listed in Table~\ref{Tab:astrom}. In the present paper we only  make  use of HARPS-N spectra, which are best suited for the determination of stellar parameters and abundances of solar-like stars.

The HARPS-N spectra were reduced by dedicated personnel using the instrument Data Reduction Software pipeline. 
Radial velocities (RVs) were derived by this pipeline using the weighted cross-correlation function (CCF) method \citep{Baranne1996, Pepe2002}. 
However, we preferred to perform an ad-hoc CCF analysis using synthetic template spectra and a broader RV window because some of our targets turned out to be spectroscopic binaries or rapidly rotating stars whose CCF peaks were not entirely covered by the RV range of the online CCF procedure. 
The telluric H$_2$O lines at the H$\alpha$ and Na\,{\sc i}\,D$_2$ wavelengths, as well as those of O$_2$ at 6300 \AA, were removed from the
extracted HARPS-N spectra using an interactive procedure described in \citet{Frasca2000} and adopting telluric templates (spectra of hot, fast-rotating stars) acquired during the observing run. 
For the following analysis we have normalized the spectra to the local continuum by fitting a low-order polynomial.

\setlength{\tabcolsep}{10pt}

\begin{table*}[h!]
\caption{Observation log.}
\tiny
\begin{center}
\begin{tabular}{lcccrr}   
\hline\hline
\noalign{\smallskip}
\multirow{2}{*}{Star$^{a}$}     & \multirow{2}{*}{$Gaia$DR3 source} & Date\_obs                & UT\_mid             & $t_{\rm exp}$  & \multirow{2}{*}{S/N$^b$} \\             
              & & {\scriptsize yyyy-mm-dd} & {\scriptsize hh:mm} &    (s)~        &         \\
\hline
\noalign{\smallskip}
Mel20-A1      &  435977186717605504  & 2019-08-11  &  05:34 & 1380 &  54 \\
"  "          &  435977186717605504  & 2019-08-14  &  05:51 & 1380 &  69 \\
Mel20-A2      &  436477017829810944  & 2019-08-12  &  05:28 & 3000 &  47 \\
Mel20-A3      &  441585726810505472  & 2019-08-15  &  04:54 & 3000 & 106 \\
Mel20-S1      &  247787692782638976  & 2022-10-27  &  00:27 & 3600 &  74 \\
Mel20-S2      &  441466189279599104  & 2022-10-27  &  01:21 & 2400 &  67 \\
Mel20-S3      &  461768327888957440  & 2022-10-28  &  00:20 & 1800 &  66 \\
Mel20-S8      &  442036938892925824  & 2022-10-28  &  00:53 & 1500 &  63 \\
Mel20-S11     &  244833446777247488  & 2022-10-28  &  01:16 &  900 &  76 \\
\noalign{\smallskip}
Plei-S2       &  65199978672758272   & 2022-10-26  &  23:42 & 1200 &  80 \\
Plei-S3       &  64808204638390912   & 2022-10-27  &  23:20 &  700 &  84 \\
Plei-S5       &  68129764842889600   & 2022-10-27  &  23:45 & 1500 &  75 \\
Plei-S7       &  66715273197982848   & 2022-10-26  &  23:13 &  300 &  78 \\
Plei-S8       &  68293729514855040   & 2022-10-26  &  23:27 &  300 &  86 \\
Plei-S9       &  68306099020642432   & 2022-10-29  &  03:23 &  900 & 153 \\
\noalign{\smallskip}
ASCC16-S1     &  3222084843416897664 & 2022-10-28  &  02:16 & 5400 &  57 \\
ASCC16-S2     &  3222099274507005056 & 2022-10-27  &  02:55 & 3600 &  37 \\
ASCC16-S6     &  3222160885813451776 & 2022-10-27  &  02:07 & 1800 &  53 \\
ASCC16-S12    &  3221874669193382016 & 2022-10-28  &  05:17 & 1200 &  64 \\
\noalign{\smallskip} 
ASCC19-S1     &  3217109248360315904 & 2019-11-09  &  06:11 & 2800 &  66 \\
ASCC19-S2     &  3220091879809599616 & 2019-11-11  &  05:11 & 2800 &  60 \\
ASCC19-S3     &  3214057416398183296 & 2019-11-11  &  06:04 & 2800 &  52 \\
ASCC19-S4     &  3220302676802812928 & 2019-11-12  &  06:09 & 2800 &  12 \\
ASCC19-S5     &  3220244986802246400 & 2020-09-29  &  04:58 & 4260 &  96 \\
ASCC19-S6     &  3220215162549378176 & 2020-10-01  &  05:28 & 2840 &  82 \\	
ASCC19-S7     &  3220189976861927808 & 2020-10-01  &  06:10 & 4260 &  65 \\
ASCC19-S14    &  3220092532644062720 & 2019-12-08  &  01:55 & 4500 &  70 \\
ASCC19-S15    &  3217211812178947072 & 2019-12-08  &  04:00 & 4500 &  71 \\ 
ASCC19-S16    &  3217189753227270144 & 2020-09-29  &  06:06 & 4420 & 126 \\
ASCC19-S17    &  3217072174202699264 & 2019-12-08  &  22:23 &  690 &  81 \\
ASCC19-S18    &  3220290376016482304 & 2019-12-08  &  22:38 &  690 &  77 \\
ASCC19-S19    &  3217081898008606464 & 2019-12-08  &  23:22 &  690 &  56 \\
ASCC19-S20    &  3214070301300036992 & 2020-10-01  &  04:48 & 1400 & 166 \\
ASCC19-S21    &  3220707468880621568 & 2020-10-02  &  03:27 & 1400 &  74 \\
ASCC19-S22    &  3220110983823622400 & 2019-12-09  &  00:22 & 4200 &  23 \\
ASCC19-S23    &  3210985999385616000 & 2020-10-02  &  04:19 & 2800 &  91 \\ 
\noalign{\smallskip} 
NGC2232-A2    &  3104532761057366016 & 2020-11-25  &  02:11 & 6300 &  68 \\
NGC2232-A3    &  3104339964265249536 & 2020-11-25  &  04:29 & 6300 &  83 \\
NGC2232-S3    &  3104457096612037504 & 2022-10-27  &  04:17 & 5400 &  47 \\
NGC2232-S4    &  3104591653648488192 & 2022-10-27  &  05:20 & 1800 &  41 \\
"  "          &  "  "                & 2022-10-28  &  04:47 & 1800 &  43 \\ 
"  "          &  "  "                & 2022-10-29  &  04:53 & 1800 &  33 \\
NGC2232-S9    &  3104453905456953216 & 2022-10-28  &  03:27 & 3600 &  54 \\
NGC2232-S11   &  3104153120306292096 & 2022-10-29  &  04:53 & 2100 &  46 \\
\noalign{\smallskip}
Ros6-S1       &  2058141112700761472 & 2022-10-26  &  20:33 & 7200 &  62 \\
Ros6-S2       &  2061151889077202688 & 2022-10-26  &  22:14 & 3600 &  46 \\
Ros6-S5       &  2064092361123030912 & 2022-10-27  &  20:52 & 7200 &  49 \\
Ros6-S9       &  2067418413856205056 & 2022-10-27  &  22:33 & 3600 &  58 \\
\noalign{\smallskip}
NGC7058-S1    &  2171977885162801920 & 2019-08-11  &  03:25 & 1380 &  75 \\
NGC7058-S2    &  2171977782083589888 & 2019-08-11  &  03:58 & 2070 &  78 \\ 
NGC7058-S3    &  2172020903547164416 & 2019-08-11  &  21:46 & 2760 &  48 \\
NGC7058-S4    &  2171979843667676416 & 2019-08-11  &  22:38 & 2760 &  47 \\
NGC7058-S5    &  2171976510773081728 & 2019-08-12  &  02:11 & 2760 &  32 \\
NGC7058-S6    &  2172072370130551936 & 2019-08-12  &  03:07 & 3000 &  25 \\
NGC7058-S7    &  2171979427039771648 & 2019-08-15  &  00:42 & 3600 &  29 \\
NGC7058-S9    &  2171974827145934464 & 2019-08-16  &  01:19 & 3600 &  16 \\
NGC7058-S10   &  2171977782083591680 & 2019-08-14  &  01:52 & 5550 &  29 \\
NGC7058-S11   &  2171929747167359488 & 2019-08-14  &  00:13 & 5550 &  38 \\
NGC7058-S12   &  2171982145770310528 & 2019-08-12  &  00:50 & 5520 &  19 \\
"  "          &  "  "                & 2019-08-14  &  05:15 & 1980 &  20 \\
NGC7058-S13   &  2171970497818874368 & 2019-08-15  &  23:45 & 6300 &  22 \\
\noalign{\smallskip}
\hline  
\end{tabular}
\label{Tab:obs_log}
\end{center}
{\bf Notes.} $^a$ ``Internal" identifier. $^b$ Signal-to-noise ratio per pixel at 6500\,\AA.
\end{table*}

\section{Stellar parameters and abundances}
\label{Sec:analysis}

\subsection{Radial velocity}
\label{subsec:RV}

We begin our spectroscopic analysis by measuring the RV, which is obtained by cross-correlating  the HARPS-N spectra of our targets with synthetic BTSettl templates \citep{Allard2012}. 
The cross-correlation is carried out separately for 14 spectral segments of 200\,\AA\ each, from 4000\,\AA\ to 6800\,\AA, after masking emission lines and regions strongly affected by telluric absorption lines. The model spectrum producing the highest correlation peak is degraded to the spectral resolution of HARPS-N and resampled to the spectral points of the target spectrum. To measure the centroid of the cross-correlation peak, we fitted it with a Gaussian  by the procedure \textsc{curvefit} \citep{Bevington}, taking the CCF noise, $\sigma_{\rm CCF}$, into account.  The latter was evaluated as the standard deviation of the CCF values in two windows at the two sides of the peak.
The RV error per each segment, $\sigma_{\rm RV}$, was estimated as the error of the centre of the Gaussian fitted to the CCF \citep[see also][]{Frasca2017}.  The RV  and associated uncertainty were computed as the weighted mean and weighted standard deviation of the RVs obtained from individual segments, adopting $w=1/\sigma_{\rm RV}^2$ as the weight.
For the components of a SB2 we evaluated the RVs  by means of a two-Gaussian fit.
As a check, we used also the IRAF\footnote{IRAF is distributed by the National Optical Astronomy Observatory, which is operated by the Association of the Universities for Research in Astronomy, Inc. (AURA), under cooperative agreement with the National Science Foundation.}  task  {\sc fxcor} for a few targets, obtaining values and errors in close agreement with those determined  with our procedure.
The RV values are listed in Table~\ref{Tab:Spectra_param} for the single (or single-lined) stars and in Table~\ref{Tab:Spectra_param_SB2} for the components of SB2 systems. For the latter we also estimated the systemic velocity, $\gamma$, following the guidelines of \citet[][Eq. 1]{Frasca2019}.

Stars belonging to the same cluster share a similar RV and this is what we see for most of the stars in Table~\ref{Tab:Spectra_param}. However, stars A1 and S3 in Melotte\,20 and S1, S2, and S9 in NGC\,7058 show somewhat different RVs with respect to the other cluster members. We note that Mel20-A1 was not selected on the basis of \gaia\ data and came out as a non-member on the basis of our chemical analysis. The remaining four stars are bona fide members according to both \citet{Cantat2020} and \citet{HR23}. Additionally, Mel20-S3 and NGC7058-S9, cool stars for which we derived the abundances (Sect.~\ref{subsec:abundance}), show a chemical composition fully compatible with that of their respective clusters. On the other hand, the warm star Plei-S8, despite having an RV compatible with that of the cluster, is not included in the list of \citet{Cantat2020} and has a low membership probability, Prob=0.2, according to \citet{HR23}.

\subsection{Atmospheric parameters}
\label{subsec:APS}

\setlength{\tabcolsep}{9pt}
\begin{table*}[htb]
\begin{center}
\caption{Stellar parameters of the single-lined members of Radcliffe Wave clusters derived in the present work with the code \ROTFIT.}
\tiny
\begin{tabular}{lccccccccc}
\hline
\hline
\noalign{\smallskip}
\multirow{2}{*}{Star}  & \teff  	&  \logg   &   [Fe/H]	& \multirow{2}{*}{SpT} & \vsini	 &  $RV$  & $P_{\rm rot}^a$ & \multirow{2}{*}{Memb?}\\  
                     &   (K)    & (dex)    &   (dex)    &                      & (\kms)  & (\kms) &        (d)      &                       \\   
\hline
\noalign{\smallskip}
Mel20-A1   &  5752\,$\pm$\,69  &  4.35\,$\pm$\,0.12 &    0.19\,$\pm$\,0.07 &  G2.5V &   1.9\,$\pm$\,1.4  &  $-$5.98\,$\pm$\,0.07 & \dots & N \\ 
Mel20-A2   &  5843\,$\pm$\,71  &  4.44\,$\pm$\,0.11 & $-$0.03\,$\pm$\,0.10 &  G1V   &  11.6\,$\pm$\,0.4  &  $-$1.62\,$\pm$\,0.12 &  4.08 & Y \\  
Mel20-A3   &  6101\,$\pm$\,70  &  4.20\,$\pm$\,0.10 &    0.09\,$\pm$\,0.08 &  F8V   &   4.4\,$\pm$\,0.8  &  $-$0.68\,$\pm$\,0.06 &  2.64 & Y \\  
Mel20-S1   &  5774\,$\pm$\,71  &  4.37\,$\pm$\,0.12 &    0.06\,$\pm$\,0.08 &  G2V   &   8.2\,$\pm$\,0.5  &     1.71\,$\pm$\,0.08 &  3.98 & Y \\ 
Mel20-S2   &  5781\,$\pm$\,74  &  4.41\,$\pm$\,0.12 &    0.10\,$\pm$\,0.09 &  G2V   &  11.9\,$\pm$\,0.3  &     0.41\,$\pm$\,0.18 &  3.26 & Y \\ 
Mel20-S3   &  5888\,$\pm$\,110 &  4.44\,$\pm$\,0.11 & $-$0.03\,$\pm$\,0.10 &  G1V   &  14.2\,$\pm$\,0.3  &  $-$4.34\,$\pm$\,0.16 &  3.90 & Y? \\    
Mel20-S8   &  5828\,$\pm$\,72  &  4.44\,$\pm$\,0.11 & $-$0.02\,$\pm$\,0.10 &  G2V   &  13.5\,$\pm$\,0.5  &  $-$0.88\,$\pm$\,0.15 &  3.05 & Y \\    
\noalign{\smallskip}
Plei-S3    &  6601\,$\pm$\,119 &  4.14\,$\pm$\,0.15 & $-$0.00\,$\pm$\,0.07 &  F6V   &  34.4\,$\pm$\,1.2  &     5.96\,$\pm$\,0.59 &  6.16$^b$ & Y  \\  
Plei-S5    &  5752\,$\pm$\,67  &  4.39\,$\pm$\,0.12 &    0.18\,$\pm$\,0.08 &  G2V   &   9.3\,$\pm$\,0.3  &     5.54\,$\pm$\,0.12 &  6.84$^b$ & Y \\ Plei-S7$^*$&  7800\,$\pm$\,170 &  3.90\,$\pm$\,0.20 & $-$0.20\,$\pm$\,0.11 &  A7III &  18.4\,$\pm$\,6.8  &     4.99\,$\pm$\,0.23 & \dots & Y \\	
Plei-S8$^*$&  7900\,$\pm$\,160 &  3.90\,$\pm$\,0.20 & $-$0.16\,$\pm$\,0.11 &  A7III &  31.8\,$\pm$\,6.2  &     4.31\,$\pm$\,0.43 & \dots & N \\	
Plei-S9$^*$&  8200\,$\pm$\,100 &  4.10\,$\pm$\,0.20 & $-$0.20\,$\pm$\,0.15 &  A4IV  &  40.7\,$\pm$\,6.2  &     4.35\,$\pm$\,1.02 & \dots & Y \\	
\noalign{\smallskip}
ASCC16-S1  &  5271\,$\pm$\,89  &  4.53\,$\pm$\,0.13 & $-$0.00\,$\pm$\,0.08 &  K0V   &  17.9\,$\pm$\,0.4  &    21.27\,$\pm$\,0.37 &  4.15 & Y \\    
ASCC16-S2  &  5184\,$\pm$\,84  &  4.54\,$\pm$\,0.10 & $-$0.03\,$\pm$\,0.08 &  K0V   &  18.3\,$\pm$\,0.4  &    20.57\,$\pm$\,0.34 &  4.06 & Y \\    
ASCC16-S6  &  5947\,$\pm$\,114 &  4.34\,$\pm$\,0.13 &    0.07\,$\pm$\,0.08 &  G2V   &  32.5\,$\pm$\,0.4  &    21.37\,$\pm$\,0.43 &  2.34 & Y \\ 	
ASCC16-S12 &  6233\,$\pm$\,107 &  3.87\,$\pm$\,0.14 & $-$0.02\,$\pm$\,0.09 &  F8IV  &  32.8\,$\pm$\,0.6  &    20.58\,$\pm$\,0.48 &  2.82 & Y \\ 
\noalign{\smallskip}
ASCC19-S1  &  6724\,$\pm$\,135 &  4.14\,$\pm$\,0.14 & $-$0.02\,$\pm$\,0.08 &  F4V   &  53.5\,$\pm$\,1.4  &    22.53\,$\pm$\,0.52 &  3.77 & Y \\   
ASCC19-S2  &  6549\,$\pm$\,170 &  4.09\,$\pm$\,0.16 & $-$0.06\,$\pm$\,0.12 &  F5V   & 171.3\,$\pm$\,7.2  &    25.16\,$\pm$\,3.62 &  0.642 & Y \\   
ASCC19-S3  &  6657\,$\pm$\,119 &  4.14\,$\pm$\,0.15 & $-$0.01\,$\pm$\,0.08 &  F4V   & 110.8\,$\pm$\,4.4  &    21.16\,$\pm$\,1.62 & \dots& Y \\   
ASCC19-S4  &  6062\,$\pm$\,119 &  4.15\,$\pm$\,0.14 & $-$0.03\,$\pm$\,0.12 & F9IV-V &  13.8\,$\pm$\,0.6  &    21.76\,$\pm$\,0.28 &  4.26& Y \\   
ASCC19-S5  &  6084\,$\pm$\,73  &  4.20\,$\pm$\,0.10 &    0.07\,$\pm$\,0.10 & F9IV-V &  18.2\,$\pm$\,1.0  &    23.89\,$\pm$\,0.32 &  3.42 & Y \\    
ASCC19-S6  &  5895\,$\pm$\,96  &  4.35\,$\pm$\,0.14 &    0.11\,$\pm$\,0.09 &  G0V   &  45.5\,$\pm$\,1.1  &    22.71\,$\pm$\,1.20 &  2.09& Y \\    
ASCC19-S14 &  6513\,$\pm$\,96  &  3.96\,$\pm$\,0.12 & $-$0.03\,$\pm$\,0.09 &  F4V   &  17.0\,$\pm$\,0.7  &    23.28\,$\pm$\,0.18 &  0.665 & Y\\    
ASCC19-S15 &  6988\,$\pm$\,121 &  4.13\,$\pm$\,0.12 & $-$0.04\,$\pm$\,0.08 &  F1V   &  62.6\,$\pm$\,1.9  &    22.70\,$\pm$\,1.01 &  0.622 & Y\\    
ASCC19-S16 &  6750\,$\pm$\,220 &  4.21\,$\pm$\,0.13 & $-$0.35\,$\pm$\,0.15 &  F2V   &  64.8\,$\pm$\,4.4  &    21.29\,$\pm$\,0.84 & \dots & Y \\	
ASCC19-S17$^*$ &16\,800\,$\pm$\,900 &  4.00\,$\pm$\,0.15 &   0.00          &  B4V   & 250.0\,$\pm$\,10.0 &	  22.06\,$\pm$\,2.10 & \dots & Y \\    
ASCC19-S18$^*$ &14\,700\,$\pm$\,600 &  4.10\,$\pm$\,0.10 &   0.00          &  B6V   &  38.0\,$\pm$\,5.0  &	  21.48\,$\pm$\,0.70 & \dots & Y \\    
ASCC19-S19$^*$ &14\,600\,$\pm$\,800 &  4.00\,$\pm$\,0.20 &   0.00          &  B6V   & 190.0\,$\pm$\,10.0 &	  18.85\,$\pm$\,1.85 & \dots & Y \\    
ASCC19-S20$^*$ &12\,800\,$\pm$\,300 &  4.00\,$\pm$\,0.15 &   0.00          &  B8V   & 120.0\,$\pm$\,13.2 &	  23.60\,$\pm$\,0.75 & \dots & Y \\    
ASCC19-S21$^*$ &  9300\,$\pm$\,110  &  4.10\,$\pm$\,0.15 &   0.00          &  A1V   &  36.9\,$\pm$\,2.9  &	  20.81\,$\pm$\,0.27 & \dots & Y\\    
ASCC19-S22$^*$ &  9400\,$\pm$\,190  &  4.10\,$\pm$\,0.20 &   0.00          &  A1V   &  49.5\,$\pm$\,2.9  &	  22.35\,$\pm$\,0.46 & \dots& Y \\    
ASCC19-S23$^*$ &  9200\,$\pm$\,360  &  4.20\,$\pm$\,0.25 &   0.00          &  A1V   & 137.4\,$\pm$\,15.7 &	  23.08\,$\pm$\,1.34 & \dots& Y \\    
\noalign{\smallskip} 
NGC2232-A2 &  5778\,$\pm$\,73  &  4.39\,$\pm$\,0.12 &    0.10\,$\pm$\,0.09 &  G2V   &  17.9\,$\pm$\,0.7  &    26.12\,$\pm$\,0.26 &  2.79 & Y \\   
NGC2232-A3 &  5803\,$\pm$\,69  &  4.40\,$\pm$\,0.12 &    0.05\,$\pm$\,0.07 &  G2V   &  13.1\,$\pm$\,0.8  &    26.73\,$\pm$\,0.24 &  3.97 & Y \\   
NGC2232-S3 &  5895\,$\pm$\,111 &  4.45\,$\pm$\,0.10 & $-$0.04\,$\pm$\,0.10 &  G1V   &  17.5\,$\pm$\,0.4  &    26.05\,$\pm$\,0.35 &  3.11 & Y \\  
NGC2232-S9 &  5849\,$\pm$\,70  &  4.44\,$\pm$\,0.11 & $-$0.04\,$\pm$\,0.10 &  G1V   &  17.6\,$\pm$\,0.4  &    25.96\,$\pm$\,0.32 &  2.61 & Y \\    
NGC2232-S11&  6099\,$\pm$\,75  &  4.17\,$\pm$\,0.12 &    0.06\,$\pm$\,0.09 &  F8V   &  12.1\,$\pm$\,0.7  &    26.94\,$\pm$\,0.14 &  3.06 & Y \\  
\noalign{\smallskip}
Ros6-S1    &  5942\,$\pm$\,133 &  4.45\,$\pm$\,0.10 & $-$0.03\,$\pm$\,0.10 &  G1V   &  13.7\,$\pm$\,0.4  &  $-$8.13\,$\pm$\,0.21 &  3.13 & Y \\ 
Ros6-S2    &  6082\,$\pm$\,71  &  4.30\,$\pm$\,0.14 &    0.05\,$\pm$\,0.09 &  G1V   &  17.3\,$\pm$\,0.5  &  $-$8.20\,$\pm$\,0.28 &  2.78 & Y \\  
Ros6-S5    &  5844\,$\pm$\,68  &  4.43\,$\pm$\,0.11 & $-$0.03\,$\pm$\,0.10 &  G1V   &  14.1\,$\pm$\,0.3  &  $-$7.47\,$\pm$\,0.22 &  3.24 & Y \\  
Ros6-S9    &  6088\,$\pm$\,72  &  4.20\,$\pm$\,0.10 &    0.07\,$\pm$\,0.09 &  F8V   &  13.6\,$\pm$\,0.4  &  $-$8.24\,$\pm$\,0.12 &  3.27 & Y \\  
\noalign{\smallskip}
NGC7058-S1$^*$ &  9200\,$\pm$\,200  &  4.10\,$\pm$\,0.10  &    0.00        &  A1    &  95.7\,$\pm$\,2.6  &   $-$5.51\,$\pm$\,2.02 & \dots & Y? \\   
NGC7058-S2$^*$ &  9500\,$\pm$\,200  &  4.30\,$\pm$\,0.10  &    0.00        &  A0    &  42.3\,$\pm$\,1.2  &   $-$8.05\,$\pm$\,0.47 & \dots & Y? \\  
NGC7058-S3 &  6700\,$\pm$\,148 &  4.15\,$\pm$\,0.16 & $-$0.02\,$\pm$\,0.09 &  F5V   & 116.4\,$\pm$\,5.4  &  $-$17.25\,$\pm$\,3.18 & \dots & Y \\   
NGC7058-S4 &  6654\,$\pm$\,122 &  4.14\,$\pm$\,0.15 & $-$0.01\,$\pm$\,0.09 &  F3V   & 119.3\,$\pm$\,4.3  &  $-$18.21\,$\pm$\,2.85 & \dots & Y \\   
NGC7058-S5 &  6324\,$\pm$\,147 &  4.10\,$\pm$\,0.17 &    0.00\,$\pm$\,0.08 &  F6IV  &  41.7\,$\pm$\,2.2  &  $-$19.25\,$\pm$\,1.01 &  1.48 & Y \\   
NGC7058-S6 &  6087\,$\pm$\,72  &  4.21\,$\pm$\,0.10 &    0.07\,$\pm$\,0.10 &  F8V   &  24.7\,$\pm$\,0.8  &  $-$19.13\,$\pm$\,0.40 & 13.65 & Y \\   
NGC7058-S7 &  5901\,$\pm$\,120 &  4.44\,$\pm$\,0.11 &    0.05\,$\pm$\,0.07 &  G1V   &  14.8\,$\pm$\,0.9  &  $-$19.03\,$\pm$\,0.22 &  2.87 & Y \\  
NGC7058-S9 &  5785\,$\pm$\,88  &  4.42\,$\pm$\,0.12 & $-$0.00\,$\pm$\,0.10 &  G2V   &   6.6\,$\pm$\,0.5  &  $-$33.25\,$\pm$\,0.08 &  7.30 & Y? \\   
NGC7058-S10&  5744\,$\pm$\,83  &  4.38\,$\pm$\,0.12 &    0.18\,$\pm$\,0.08 &  G2V   &   4.0\,$\pm$\,0.7  &  $-$19.41\,$\pm$\,0.07 &  6.77 & Y \\   
NGC7058-S11&  5682\,$\pm$\,114 &  4.45\,$\pm$\,0.15 &    0.19\,$\pm$\,0.10 &  G2.5V &   5.8\,$\pm$\,0.4  &  $-$19.43\,$\pm$\,0.07 &  4.63 & Y \\   
NGC7058-S12&  5491\,$\pm$\,157 &  4.53\,$\pm$\,0.11 &    0.11\,$\pm$\,0.11 &  K0V   &   7.5\,$\pm$\,0.5  &  $-$19.08\,$\pm$\,0.09 &  4.82 & Y \\  
NGC7058-S13&  5285\,$\pm$\,90  &  4.55\,$\pm$\,0.11 & $-$0.00\,$\pm$\,0.09 &  K0V   &   3.3\,$\pm$\,0.8  &  $-$18.96\,$\pm$\,0.06 &  5.69 & Y \\ 

\noalign{\smallskip}
\hline
\end{tabular}
\begin{list}{}{}
\item[$^a$] Rotation periods derived in the present paper analyzing the \tess\ light curves.  
\item[$^b$] Rotation periods derived by \citet{Rebull2016} with {\it Kepler}-K2. 
\item[$^*$] Warm stars. Atmospheric parameters derived with synthetic templates with solar metallicity as described in the text.
\end{list}
\label{Tab:Spectra_param}
\end{center}
\end{table*}

\begin{figure}
\begin{center}
\hspace{-0.5cm}
\includegraphics[width=9.3cm]{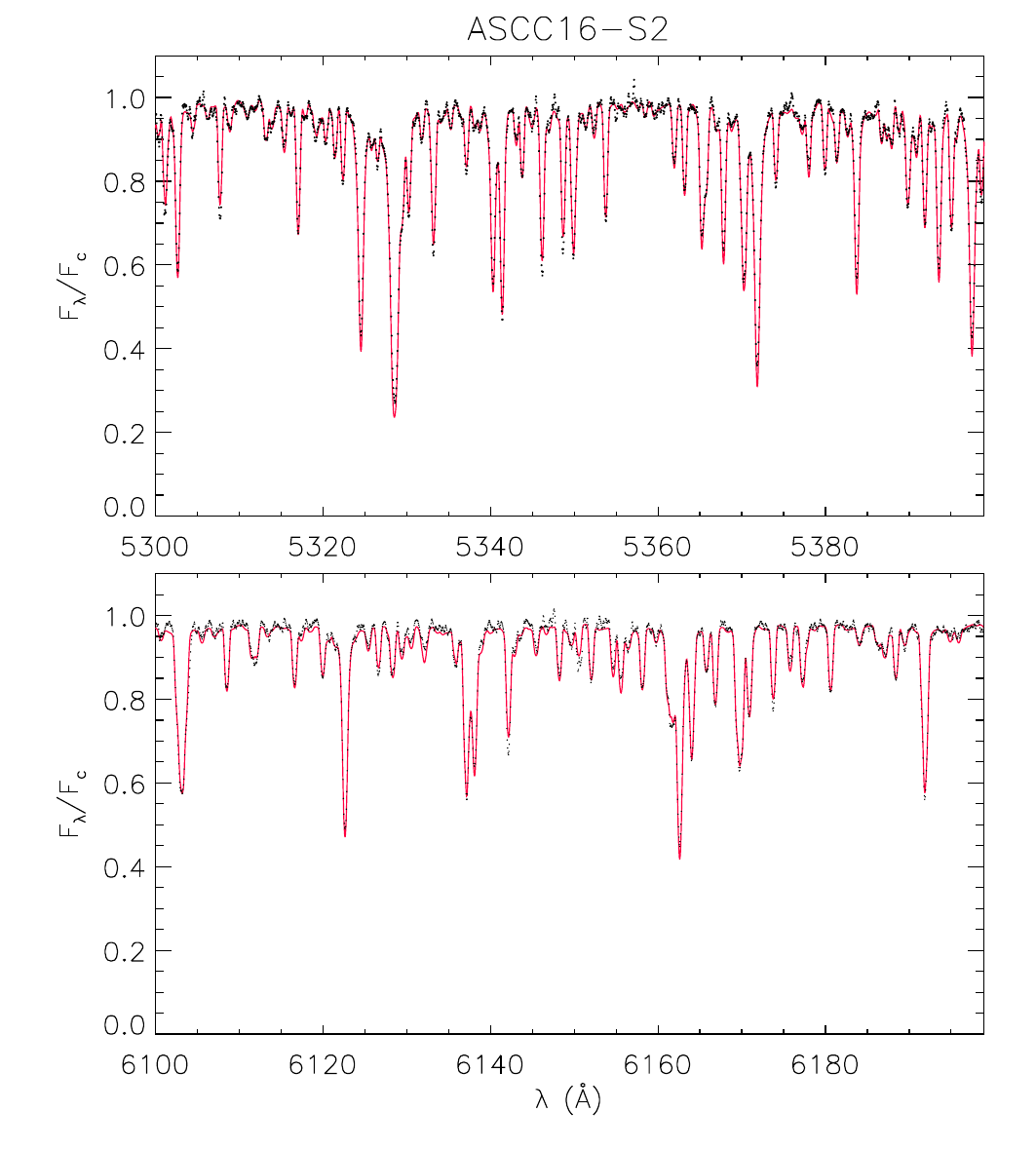}
\caption{Observed HARPS-N spectrum of ASCC16-S2 (black dots) in the $\lambda$\,5300\,\AA\ ({\it upper panel}) and $\lambda$\,6100\,\AA\
({\it lower panel}) spectral regions. In each panel the template spectrum broadened at the \vsini\ of the target is overlaid with a full red line. }
\label{fig:spe_ASCC16}
\end{center}
\end{figure}

The second step of our analysis is the determination of the stellar atmospheric parameters (APs) namely, effective temperature ($T_{\rm{eff}}$), surface gravity (log\,$g$) and metallicity (using [Fe/H] as a proxy).
For the stars with single-lined HARPS-N
spectra, APs were derived with the code \ROTFIT\ \citep{Frasca2006}, which also allows us to measure the projected rotational velocity (\vsini) and to perform an MK spectral classification (SpT). 
\ROTFIT\  uses a grid of template spectra and is based on a $\chi^2$ minimization of the difference \textit{observed--template} in 28  selected spectral regions of 100\,\AA\ each in which we split our HARPS-N spectra. 
As templates, we adopted high-resolution ELODIE spectra of real stars, whose APs are well-known.
This grid is the same used by the Catania (OACT) node for the analysis of FGK-type spectra in the \gaia-ESO Survey 
\citep[GES; see, e.g.,][]{Smiljanic2014,Frasca2015}. 
The spectra of our targets are degraded to the resolution of ELODIE ($R=42\,000$) and resampled on the ELODIE spectral points ($\Delta\lambda=0.05$\,\AA). 
This also has the advantage of improving the S/N of the spectra to be analyzed. For a detailed description of the application of \ROTFIT\ to the HARPS-N spectra the reader is referred to \citet{Frasca2019}. An example of the application of the code \ROTFIT\ to two spectral segments of ASCC16-S2 is shown in Fig.~\ref{fig:spe_ASCC16}. 

For the hottest stars, those near or beyond the early-type edge of the ELODIE template grid ($\approx$ F0), we used a different approach, as done in previous works \citep[see for example][]{M39}. For these objects, we made use of synthetic templates (see Sect.~\ref{subsec:abundance}) and focused on the wings and cores of the Balmer lines, which are efficient diagnostics of \logg\ and \teff\ respectively, as well as additional spectral regions with weaker lines to constrain the \vsini, mainly the \ion{Mg}{I} $\lambda$\,4481 {\AA}, clearly visible at these effective temperatures.

\setlength{\tabcolsep}{4pt}

As mentioned above, the observed targets include four SB2 systems.
One of these, Plei-S2 (= HII~1117 = HD~282975), was already discovered as a double-lined spectroscopic binary by \citet{Mermilliod1992}, but we decided to keep it in our list and observe it anyway for comparison purposes.
Another SB2, NGC2232-S4, was discovered by us as an eclipsing binary by examination of \tess\ light curves and was observed purposely to obtain high-precision stellar parameters which were the subject of a recent work \citep{Frasca2023b}. 

\begin{figure}
\begin{center}
\hspace{-0.5cm}
\includegraphics[width=9.3cm]{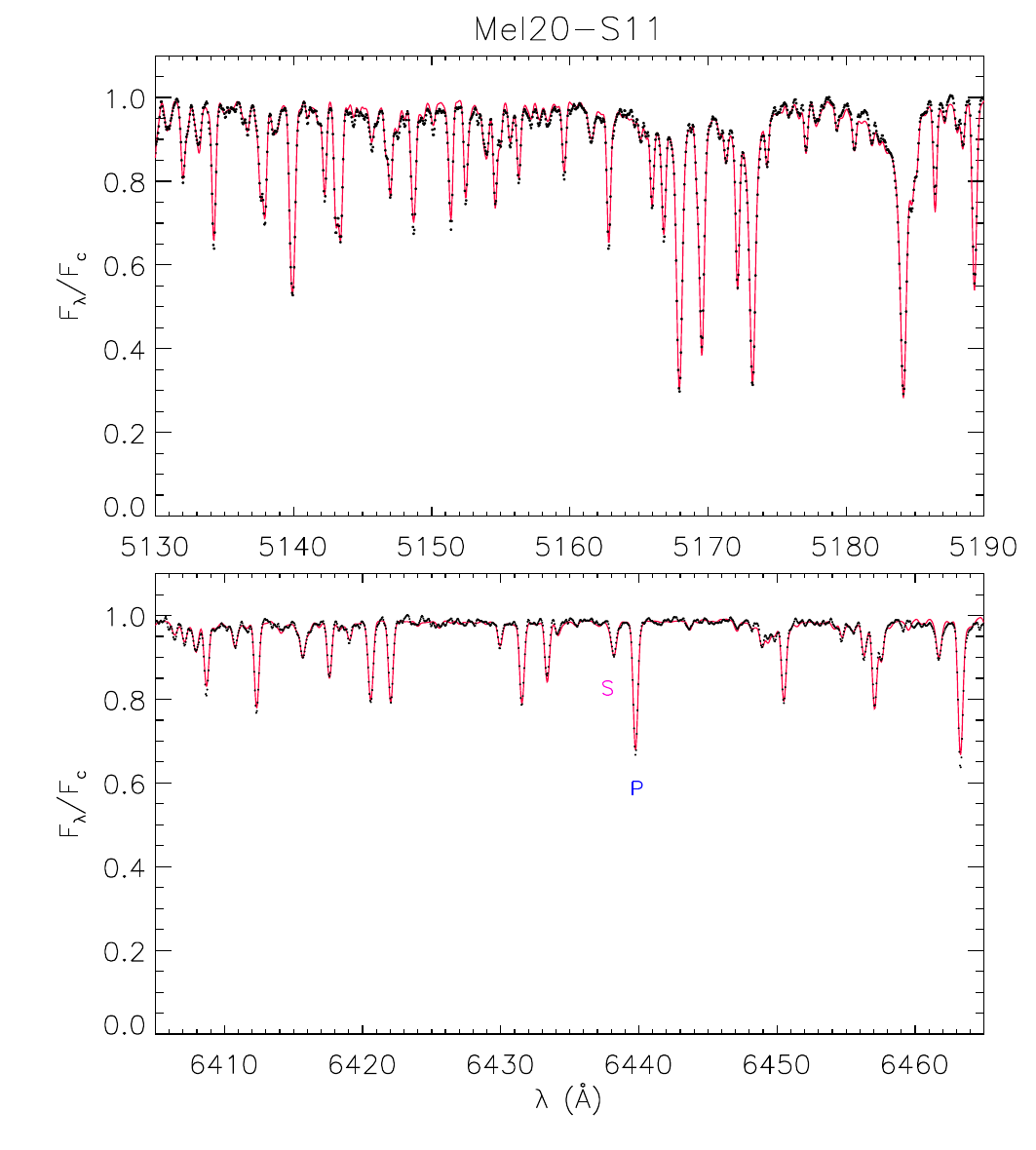}
\caption{Observed HARPS-N spectrum of the SB2 system Mel20-S11 (black dots) around the \ion{Mg}{i}\,b lines ($\lambda$\,5160\,\AA, {\it upper panel}) and around 
$\lambda$\,6440\,\AA\ ({\it lower panel}). In each panel the synthetic spectrum, which is the weighted sum of two standard star spectra mimicking
the primary and secondary component of Mel20-S115, is overplotted with a full red line. 
}
\label{fig:spe_SB2}
\end{center}
\end{figure}

The spectral lines of the components of these SB2s are sufficiently separated in wavelength to allow us to infer their stellar parameters from a proper analysis of their spectra. For the SB2 in NGC2232 (TIC~43152097), which is also a totally eclipsing binary, we have already provided the parameters of the two components in \citet{Frasca2023b} by simultaneously solving the RV and light curves. 
To get the stellar parameters for SB2 systems, we used \COMPO, a code developed in {\sf IDL}\footnote{IDL (Interactive Data Language) is a registered trademark of  Harris Corporation.} 
environment by \citet{Frasca2006}, which, like \ROTFIT, was adapted to the HARPS-N spectra. A description of this version of \COMPO\ can also
be found in \citet{Frasca2019} to which paper the reader is referred to for further details.
The continuum flux ratio (i.e., the flux contribution of the primary component in units of the continuum, $w^{\rm P}$) is  an adjustable parameter, which is also an output of the code. An example of the application of \COMPO\ to the binary Mel20-S11 is shown in Fig.~\ref{fig:spe_SB2} for two spectral segments, one around the \ion{Mg}{i}\,b triplet and the other centred at 6440\,\AA, in which the lines of the two components are clearly distinguishable. The flux contribution was evaluated at three wavelengths (4400\,\AA, 5500\,\AA, and 6400\,\AA) using only the spectral segments around these wavelengths. In this case, the SpTs of the components are taken as the mode of the spectral-type distributions, while the the APs are estimated as the weighted averages of the best 100 combinations per each analyzed spectral segment, considering the $\chi^2$ and the line intensity in the weight. The APs obtained for the single stars are reported in Table~\ref{Tab:Spectra_param}, in which we also list the rotational periods measured by analyzing the \tess\ light curves, while those for SB2s systems are listed in Table~\ref{Tab:Spectra_param_SB2}.

\setlength{\tabcolsep}{3pt}

\begin{table*}[htb]
\caption{Stellar parameters of the components of SB2 systems derived with the code \COMPO.}
\begin{tabular}{cccccccccccccc}
\hline
\hline
\noalign{\smallskip}
Star		    & \teff & err & \logg &  err  & $RV$ & err & $\gamma$  & $w^{\rm P}_{4400}$ &  $w^{\rm P}_{5500}$ & $w^{\rm P}_{6400}$ & SpT & \vsini  \\ 
		    &  (K)  & (K) & (dex) & (dex) &  \multicolumn{2}{c}{(\kms)} &  (\kms) &		     &  	     &  		  &	    & (\kms)  \\  
		    & [P/S] &[P/S]& [P/S] & [P/S] &  [P/S] & [P/S] &   & &		  &		     &  [P/S]  & [P/S]  \\  
\hline
\noalign{\smallskip}
Mel20-S11 & {\scriptsize 6119/5172} & {\scriptsize 75/240} & {\scriptsize 4.02/3.92} & {\scriptsize 0.14/0.62} &{\scriptsize 31.74/-41.46} & {\scriptsize 0.15/2.39} &  {\scriptsize 4.2$\pm$1.2} &  {\scriptsize 0.91$\pm$0.02} & {\scriptsize 0.89$\pm$0.02} & {\scriptsize 0.90$\pm$0.03} & {\scriptsize F7IV/K1IV}   & {\scriptsize 8/8 }   \\
Plei-S2 & {\scriptsize 5674/5371} & {\scriptsize  139/172} & {\scriptsize  4.36/4.39} & {\scriptsize 0.13/0.14} & {\scriptsize 14.32/0.20} & {\scriptsize 0.22/0.30} & {\scriptsize 7.4$\pm$0.2}$^a$  & {\scriptsize 0.56$\pm$0.03} & {\scriptsize 0.55$\pm$0.03} & {\scriptsize 0.53$\pm$0.03} & {\scriptsize G2V/G2V}    & {\scriptsize 4/3 }   \\
ASCC19-S7 & {\scriptsize 5376/4894} & {\scriptsize  139/226} & {\scriptsize  4.41/4.36} & {\scriptsize 0.11/0.40} & {\scriptsize 35.49/9.11} & {\scriptsize 0.25/1.16} & {\scriptsize 24.6$\pm$0.5}  & {\scriptsize 0.87$\pm$0.02} & {\scriptsize 0.83$\pm$0.02} & {\scriptsize 0.79$\pm$0.04} & {\scriptsize G8V/K4V}    & {\scriptsize 7/7 }  \\
\hline
\end{tabular}
\begin{list}{}{}
\item[$^a$] In very good agreement with the value of 7.434\,\kms\  reported by \citet{Torres2021}.
\end{list}
\label{Tab:Spectra_param_SB2}
\end{table*}

\subsection{Elemental abundances}
\label{subsec:abundance}
Once the APs are known, we can determine the chemical abundances.
We followed the same strategy implemented in our previous works  \citep{Alonso2021,M39,Catanzaro24}, which is based on the spectral synthesis technique \citep{Catanzaro11,Catanzaro13}. We focused only on cool stars, those with \teff<7500\,K, since the fast rotation in hotter stars with earlier B--A types blurs the spectral lines, preventing a reliable estimate of abundances. Additionally, in these stars, compared to cool ones, the number of spectral features is also smaller.

As a first step we computed the atmospheric model for each star from their APs, previously obtained with \rotfit\ (Sect.~\ref{subsec:APS}). We produced 1D local thermodynamic equilibrium (LTE) models by means of the {\sf ATLAS9} code \citep{Kurucz1993a,Kurucz1993b} and then the corresponding synthetic spectra were generated using the radiative transfer code {\sf SYNTHE} \citep{Kurucz1981}. The next stage of our analysis consisted of comparing the target spectra (at their original resolution) with the synthetic ones (adequately broadened according to the instrumental and rotational profiles). This was carried out in 39 spectral segments of 50\,\AA\ each between 4400 and 6800\,\AA. The chemical abundances were then found by minimizing the $\chi^2$ of their differences. The final abundance of each element was estimated as the mean of the individual values from all the observed spectral lines, after applying a 2$\sigma$-clipping to remove outliers. The median absolute deviation of the individual abundances was considered a robust indicator of the uncertainty of the average value.

In this way we calculated the chemical abundances of 41 FGK-type stars belonging to the seven clusters under study. We explored 25 elements with atomic numbers up to 60, namely, C, Na, Mg, Al, Si, S, Ca, Sc, Ti, V, Cr, Mn, Fe, Co, Ni, Cu, Zn, Sr, Y, Zr, Ba, La, Ce, Pr, and Nd. We adopted the lines list from \citet{Romaniello08} for \ion{Fe}{i} and \ion{Fe}{ii} while for the rest of the elements we used those of \citet{Castelli04}, who updated the original parameters of \citet{Kurucz95}. Our abundances are always expressed taking as reference the solar composition reported by \citet{Grevesse07}. The individual stellar abundances, in terms of [X/H], are listed in Tables from \ref{tab_abb_I} to \ref{tab_abb_V}. The chemical composition of the clusters, instead, is reported in Table~\ref{tab_abb_med}. The final value for each element has been calculated as the weighted average of the abundances of the cluster members (i. e., `Y' and `Y?' stars in Table~\ref{Tab:Spectra_param}), using the individual errors as weights. Its uncertainty has been estimated as the standard deviation observed among the members. In general, this number is close or below 0.1\,dex. Only in a few occurrences we found some scatter in the abundance of 
some elements among the members of the cluster. This is the case of the Pleiades, where we have only analyzed two stars that exhibit some chemical differences for Fe, Cu and Zn.  It is worth clarifying that, for the sake of homogeneity, the [Fe/H] obtained with {\sf SYNTHE} is used throughout the discussion and chemical analysis, while that of \rotfit\ (slightly higher than the former) is displayed in Table~\ref{tab_abb_med} only for comparison. In any case, both sets of metallicities are compatible within the errors.

We find that Pleiades, ASCC\,16 and NGC\,7058 show solar metallicity while Melotte\,20, ASCC\,19, Roslund\,6 and NGC\,2232 display values about 0.1 dex lower. A spread in metallicity of almost 0.2 dex can be seen among our targets, which is more than one can expect for these clusters with such similar ages and distances. Regarding the chemical abundances we find, on average, solar composition for $\alpha$- and Fe-peak elements and a mild overabundance of about 0.2 dex for neutron-capture elements. Sulfur and zinc present some differences with respect the general trend. The former exhibits overabundant values (0.2--0.3 dex above the Sun) while the latter is slightly subsolar.

\begin{table*}
\caption{Average chemical composition ([X/H]), relative to solar abundances found by \citet{Grevesse07}, of the clusters observed in this work.}
\begin{center}\label{tab_abb_med}
\begin{tabular}{lrrrrrrr}
\hline\hline
X  &     Melotte\,20   &       Pleiades    &       ASCC\,16    &       ASCC\,19    &      NGC\,2232    &      Roslund\,6   &       NGC\,7058   \\
\hline
C  &    0.23$\pm$0.17  &    0.21$\pm$0.12 &    0.24$\pm$0.03  &    0.18$\pm$0.05  &    0.38$\pm$0.06 &    0.39$\pm$0.04  &    0.15$\pm$0.16  \\
Na &    0.01$\pm$0.09  &    0.08$\pm$0.07  &    0.05$\pm$0.04  &    0.06$\pm$0.17  &    0.03$\pm$0.06  &    0.05$\pm$0.05  &    0.17$\pm$0.04  \\  
Mg & $-$0.04$\pm$0.07  &    0.06$\pm$0.06  &    0.03$\pm$0.05  & $-$0.05$\pm$0.08  & $-$0.01$\pm$0.02  &    0.00$\pm$0.04  &    0.08$\pm$0.07  \\
Al &    0.12$\pm$0.15  &    0.16$\pm$0.08  &    0.13$\pm$0.08  &    0.17$\pm$0.05  &    0.08$\pm$0.07  &    0.10$\pm$0.09  &    0.14$\pm$0.04  \\
Si & $-$0.14$\pm$0.03  & $-$0.07$\pm$0.09  & $-$0.05$\pm$0.03  & $-$0.05$\pm$0.08  & $-$0.10$\pm$0.04  & $-$0.07$\pm$0.08  & $-$0.10$\pm$0.03  \\  
S  &    0.20$\pm$0.20  &    0.26$\pm$0.09  &    0.29$\pm$0.03  &    0.21$\pm$0.04  &    0.29$\pm$0.12 &    0.20$\pm$0.11  &    0.24$\pm$0.06  \\
Ca &    0.11$\pm$0.12  &    0.07$\pm$0.10  &    0.14$\pm$0.07  &    0.08$\pm$0.05  &    0.13$\pm$0.06  &    0.13$\pm$0.06  &    0.12$\pm$0.12  \\
Sc & $-$0.08$\pm$0.03  &    0.00$\pm$0.11  &    0.06$\pm$0.15  & $-$0.07$\pm$0.06  &    0.03$\pm$0.02  & $-$0.01$\pm$0.08  & $-$0.04$\pm$0.09  \\
Ti & $-$0.03$\pm$0.11  &    0.06$\pm$0.16  &    0.16$\pm$0.10  &    0.02$\pm$0.07  &    0.01$\pm$0.08  & $-$0.01$\pm$0.07  &    0.05$\pm$0.09  \\
V  &    0.11$\pm$0.07  &    0.18$\pm$0.01  &    0.15$\pm$0.05  &    0.17$\pm$0.10  &    0.18$\pm$0.05  &    0.11$\pm$0.04  &    0.14$\pm$0.08  \\
Cr &    0.06$\pm$0.10  &    0.13$\pm$0.13  &    0.11$\pm$0.07  &    0.03$\pm$0.12  &    0.07$\pm$0.05  &    0.10$\pm$0.06  &    0.07$\pm$0.11  \\
Mn &    0.09$\pm$0.09  &    0.09$\pm$0.16  &    0.12$\pm$0.07  & $-$0.01$\pm$0.07  &    0.11$\pm$0.03  &    0.10$\pm$0.05  &    0.09$\pm$0.09  \\
Fe & $-$0.10$\pm$0.13  &    0.03$\pm$0.21  &    0.01$\pm$0.10  & $-$0.14$\pm$0.15  & $-$0.10$\pm$0.10  & $-$0.11$\pm$0.08  & $-$0.02$\pm$0.10  \\
Co &    0.07$\pm$0.06  &    0.17$\pm$0.06  &    0.14$\pm$0.07  &    0.18$\pm$0.06  &    0.10$\pm$0.04  &    0.14$\pm$0.04  &    0.15$\pm$0.07  \\ 
Ni & $-$0.08$\pm$0.07  &    0.06$\pm$0.08  &    0.03$\pm$0.09  & $-$0.06$\pm$0.10  & $-$0.08$\pm$0.06  & $-$0.06$\pm$0.09  &    0.03$\pm$0.05  \\
Cu & $-$0.06$\pm$0.09  & $-$0.04$\pm$0.27  & $-$0.04$\pm$0.11  & $-$0.12$\pm$0.12  & $-$0.02$\pm$0.14  & $-$0.07$\pm$0.02  & $-$0.03$\pm$0.10  \\
Zn & $-$0.25$\pm$0.11  & $-$0.27$\pm$0.20  & $-$0.12$\pm$0.16  & $-$0.28$\pm$0.10  & $-$0.28$\pm$0.07  & $-$0.24$\pm$0.05  & $-$0.16$\pm$0.09  \\
Sr &    0.17$\pm$0.08  &    0.09$\pm$0.11  &    0.15$\pm$0.08  &    0.12$\pm$0.13  &    0.06$\pm$0.05  &    0.10$\pm$0.03  &    0.15$\pm$0.13  \\
Y  &    0.20$\pm$0.07  &    0.19$\pm$0.13  &    0.23$\pm$0.11  &    0.07$\pm$0.14  &    0.22$\pm$0.07  &    0.19$\pm$0.02  &    0.12$\pm$0.08  \\
Zr &    0.10$\pm$0.06  &    0.22$\pm$0.08  &    0.16$\pm$0.10  &    0.16$\pm$0.04  &    0.15$\pm$0.03  &    0.13$\pm$0.03  &    0.13$\pm$0.06  \\
Ba &    0.31$\pm$0.20  &    0.27$\pm$0.04  &    0.27$\pm$0.15  &    0.24$\pm$0.08  &    0.47$\pm$0.06  &    0.24$\pm$0.05  &    0.30$\pm$0.06  \\
La &    0.23$\pm$0.05  &    0.28$\pm$0.04  &    0.28$\pm$0.07  &    0.22$\pm$0.04  &    0.25$\pm$0.04  &    0.26$\pm$0.06  &    0.25$\pm$0.11  \\
Ce &    0.09$\pm$0.04  &    0.17$\pm$0.02  &    0.15$\pm$0.13  &    0.18$\pm$0.05  &    0.15$\pm$0.07  &    0.12$\pm$0.05  &    0.14$\pm$0.10  \\
Pr &    0.13$\pm$0.05  &    0.18$\pm$0.04  &    0.15$\pm$0.05  &    0.21$\pm$0.01  &    0.19$\pm$0.06  &         \dots~~~~ &    0.24$\pm$0.14  \\           
Nd &    0.22$\pm$0.07  &    0.25$\pm$0.05  &    0.21$\pm$0.03  &    0.16$\pm$0.03  &    0.20$\pm$0.06  &    0.23$\pm$0.07  &    0.25$\pm$0.06  \\
\hline
Fe$^*$& 0.01$\pm$0.11  &    0.08$\pm$0.13  &    0.00$\pm$0.05  & $-$0.02$\pm$0.13  &    0.03$\pm$0.06  &    0.02$\pm$0.05  &    0.05$\pm$0.08  \\
\hline
\end{tabular}
\begin{list}{}{}
 \item[$^*$] \ROTFIT\ metallicity, for comparative purpose.  
\end{list}
\end{center}
\end{table*}

\section{Chromospheric emission and lithium abundance}
\label{Sec:chrom_lithium}

\setlength{\tabcolsep}{3pt}

\begin{table*}
\caption{H$\alpha$ and \ion{Ca}{ii} H\,\&\,K equivalent widths and fluxes.}
\begin{center}
\begin{tabular}{lrccrrccc}
\hline
\noalign{\smallskip}
\multirow{2}{*}{Star}     & $EW_{\rm H\alpha}$  & $F_{\rm H\alpha}$ & \multirow{2}{*}{$\log(R'_{\rm H\alpha})$} & $EW_{\rm CaII-K}$  & $EW_{\rm CaII-H}$  & $F_{\rm CaII-K}$  & $F_{\rm CaII-H}$ & \multirow{2}{*}{$\log(R'_{\rm HK})$} \\   
      & (m\AA) & \scriptsize{($10^5$erg\,cm$^{-2}$s$^{-1}$)} & & (m\AA) & (m\AA)  & \multicolumn{2}{c}{\scriptsize{($10^5$erg\,cm$^{-2}$s$^{-1}$)}}  & \\  
\hline
\noalign{\smallskip}
Mel20-A1    &   \dots      &   \dots        &  \dots  &   20$\pm$36   &   17$\pm$30  &   1.1$\pm$2.0   &    0.9$\pm$1.7   & -5.48 \\
Mel20-A2    &	94$\pm$13  &   7.2$\pm$1.0  &  -4.96  &  328$\pm$73   &  238$\pm$73  &  20.5$\pm$5.2   &   14.9$\pm$4.9   & -4.27 \\
Mel20-A3    &	49$\pm$13  &   4.4$\pm$1.2  &  -5.25  &  207$\pm$39   &  132$\pm$24  &  19.1$\pm$4.0   &   12.2$\pm$2.4   & -4.40 \\
Mel20-S1    &  157$\pm$14  &  11.5$\pm$1.1  &  -4.74  &  408$\pm$58   &  297$\pm$51  &  23.0$\pm$4.0   &   16.7$\pm$3.3   & -4.20 \\
Mel20-S2    &  391$\pm$22  &  28.7$\pm$2.1  &  -4.34  &  560$\pm$60   &  427$\pm$52  &  31.9$\pm$4.8   &   24.3$\pm$3.9   & -4.05 \\
Mel20-S3    &  105$\pm$12  &   8.3$\pm$1.1  &  -4.91  &  347$\pm$78   &  274$\pm$73  &  23.2$\pm$6.5   &   18.3$\pm$5.8   & -4.22 \\
Mel20-S8    &  215$\pm$14  &  16.3$\pm$1.3  &  -4.60  &  441$\pm$62   &  358$\pm$62  &  26.9$\pm$5.0   &   21.8$\pm$4.6   & -4.13 \\
\noalign{\smallskip}
Plei-S2$^a$     &  102$\pm$13  &   7.0$\pm$1.1     &  -4.92         & 434$\pm$80    &  290$\pm$64  &  23.2$\pm$5.3   &   15.5$\pm$4.0   & -4.20 \\ 
Plei-S3     &   36$\pm$6   &   4.3$\pm$0.8  &  -5.40  & 149$\pm$29    &  114$\pm$28  &  25.1$\pm$5.8   &   19.1$\pm$5.3   & -4.39 \\
Plei-S5     &	73$\pm$14  &   5.3$\pm$1.0  &  -5.07  & 479$\pm$65    &  330$\pm$58  &  26.2$\pm$4.4   &   18.0$\pm$3.6   & -4.15 \\
\noalign{\smallskip}
ASCC16-S1   &  702$\pm$24  &  35.8$\pm$2.8  &  -4.09  & 1047$\pm$137  &  760$\pm$122 &  26.6$\pm$5.5   &   19.3$\pm$4.4   & -3.98 \\
ASCC16-S2   &  661$\pm$31  &  31.3$\pm$2.7  &  -4.12  & 1262$\pm$136  &  983$\pm$152 &  27.4$\pm$5.4   &   21.3$\pm$4.8   & -3.92 \\
ASCC16-S6   &  306$\pm$22  &  25.2$\pm$2.6  &  -4.45  &  692$\pm$78   &  488$\pm$65  &  49.4$\pm$11.0  &   34.8$\pm$8.12  & -3.93 \\
ASCC16-S12  &	55$\pm$10  &   5.3$\pm$1.0  &  -5.20  &  275$\pm$60   &  201$\pm$67  &  29.3$\pm$7.5   &   21.4$\pm$7.7   & -4.23 \\
\noalign{\smallskip}
ASCC19-S3   &	41$\pm$15  &   5.0$\pm$1.9  &  -5.34  &  235$\pm$72   &  153$\pm$64  &  41.6$\pm$13.7  &   27.0$\pm$11.8  & -4.21 \\
ASCC19-S4   &  145$\pm$50  &  12.8$\pm$4.5  &  -4.78  &  315$\pm$343  &   75$\pm$370 &  27.1$\pm$29.9  &    6-5$\pm$31.9  & -4.36 \\
ASCC19-S5   &	72$\pm$ 8  &   6.4$\pm$0.8  &  -5.08  &  267$\pm$33   &  189$\pm$35  &  23.6$\pm$3.7   &   16.7$\pm$3.5   & -4.29 \\
ASCC19-S6   &  314$\pm$13  &  24.9$\pm$2.0  &  -4.44  &  655$\pm$85   &  507$\pm$73  &  43.1$\pm$8.6   &   33.3$\pm$7.0   & -3.95 \\
ASCC19-S7$^a$   & 3943$\pm$190  &   217$\pm$26   &  -3.34  & 1774$\pm$316  & 1550$\pm$320 &  68.3$\pm$30.8  &   59.6$\pm$27.6  & -3.62 \\
ASCC19-S14  &	29$\pm$11  &   3.3$\pm$1.3  &  -5.49  &  196$\pm$58   &  175$\pm$62  &  29.4$\pm$9.3   &   26.2$\pm$9.8   & -4.26 \\
ASCC19-S15  &  \dots       &  \dots         &  \dots  &  229$\pm$63   &  139$\pm$49  &  56.4$\pm$16.2  &   34.2$\pm$12.3  & -4.17 \\
ASCC19-S16  & 1846$\pm$39  &  260$\pm$30    &  -3.70  &  326$\pm$39   &  270$\pm$37  &  155$\pm$34     &   128$\pm$29     & -3.67 \\
\noalign{\smallskip}
NGC2232-A2  &  440$\pm$19  &  32.2$\pm$2.0  &  -4.29  &  642$\pm$56   &  514$\pm$55  &  35.4$\pm$4.9   &   28.4$\pm$4.3   & -4.00 \\
NGC2232-A3  &  173$\pm$ 8  &  12.9$\pm$0.8  &  -4.70  &  367$\pm$43   &  298$\pm$49  &  20.9$\pm$3.6   &   17.0$\pm$3.5   & -4.23 \\
NGC2232-S3  &  146$\pm$20  &  11.6$\pm$1.8  &  -4.77  &  410$\pm$63   &  237$\pm$52  &  26.9$\pm$6.3   &   15.5$\pm$4.4   & -4.21 \\
NGC2232-S4A  &  109$\pm$17$^{\dag}$  &  9.7$\pm$1.6  &  -4.90  &  268$\pm$49   &  163$\pm$48  &  25.1$\pm$5.1   &   15.2$\pm$4.7   &  -4.28  \\
NGC2232-S4B  & 1209$\pm$247$^{\dag}$ &  16.2$\pm$3.6  &  -4.01  &  158$\pm$42  &   65$\pm$36  &  14.8$\pm$4.12  &    6.0$\pm$3.4   &   -3.90 \\
NGC2232-S9  &  138$\pm$16  &  10.6$\pm$1.3  &  -4.80  &  454$\pm$101  &  280$\pm$76  &  27.8$\pm$7.0   &   17.2$\pm$5.1   & -4.17 \\
NGC2232-S11 &	91$\pm$10  &   8.2$\pm$1.0  &  -4.98  &  318$\pm$61   &  139$\pm$41  &  30.8$\pm$6.5   &   13.4$\pm$4.1   & -4.25 \\
\noalign{\smallskip}
Ros6-S1     &  143$\pm$17  &  11.7$\pm$1.7  &  -4.78  &  326$\pm$40   &  264$\pm$46  &  23.7$\pm$6.0   &   19.2$\pm$5.4   & -4.22 \\
Ros6-S2     &  156$\pm$21  &  14.0$\pm$2.0  &  -4.75  &  296$\pm$45   &  197$\pm$38  &  25.9$\pm$4.6   &   17.2$\pm$3.6   & -4.25 \\
Ros6-S5     &  162$\pm$16  &  12.4$\pm$1.4  &  -4.73  &  399$\pm$68   &  221$\pm$66  &  25.0$\pm$5.1   &   13.9$\pm$4.4   & -4.23 \\
Ros6-S9     &	88$\pm$11  &   7.9$\pm$1.0  &  -4.99  &  437$\pm$91   &  244$\pm$56  &  39.7$\pm$9.1   &   22.1$\pm$5.5   & -4.10 \\
\noalign{\smallskip}
NGC7058-S5  &	87$\pm$24  &   8.9$\pm$2.6  &  -5.01  & 181$\pm$77    &  100$\pm$56  &  22.0$\pm$10.1  &   12.2$\pm$7.1   & -4.42 \\
NGC7058-S6  &  115$\pm$28  &  10.3$\pm$2.5  &  -4.88  & 517$\pm$146   &  200$\pm$106 &  47.2$\pm$14.0  &   18.3$\pm$9.9   & -4.07 \\
NGC7058-S7  &  116$\pm$27  &   9.2$\pm$2.3  &  -4.87  & 197$\pm$71    &  204$\pm$78  &  13.5$\pm$5.6   &   14.1$\pm$6.1   & -4.40 \\
NGC7058-S9  &  226$\pm$46  &  16.6$\pm$3.5  &  -4.58  & 303$\pm$209   &  153$\pm$217 &  17.9$\pm$12.6  &   9.1$\pm$12.9   & -4.37 \\
NGC7058-S10 &	65$\pm$26  &   4.7$\pm$1.9  &  -5.12  & 499$\pm$93    &  262$\pm$86  &  27.6$\pm$6.2   &   14.5$\pm$5.1   & -4.17 \\
NGC7058-S11 &  252$\pm$31  &  17.4$\pm$2.5  &  -4.53  & 475$\pm$156   &  271$\pm$120 &  24.2$\pm$9.0   &   13.8$\pm$6.6   & -4.19 \\
NGC7058-S12 &  218$\pm$20  &  13.2$\pm$1.9  &  -4.59  & 828$\pm$74    &  349$\pm$79  &  36.7$\pm$9.3   &   15.5$\pm$5.1   & -3.99 \\
NGC7058-S13 &  146$\pm$44  &   7.5$\pm$2.3  &  -4.77  & 597$\pm$185   &  253$\pm$144 &  16.3$\pm$5.6   &    6.9$\pm$4.1   & -4.28 \\
\hline
\end{tabular}
\end{center}
\begin{list}{}{}
\item[$^a$] Emission from both components has been integrated. 
\item[$^{\dag}$] Corrected for the contribution to the continuum flux.
\end{list}
\label{Tab:Halpha_CaII}
\end{table*}

\setlength{\tabcolsep}{5pt}

\begin{table}
\caption{\ion{Li}{i}\,$\lambda$\,6708\,\AA\ equivalent width and abundance of the cool stars observed in this work.}
\begin{center}
\begin{tabular}{lccc}
\hline
\hline
\noalign{\smallskip}
\multirow{2}{*}{Star}  & $T_{\textrm{eff}}$       &  $EW_{\textrm{Li}}$ & \multirow{2}{*}{$A$(Li)} \\
                       &       (K)       &  (m\AA)   &        \\ 
\hline
  \noalign{\smallskip}
  Mel20-A1   & 5752$\pm$69  &  51$\pm$6 & 2.33$\pm$0.10\\
  Mel20-A2   & 5843$\pm$71  & 153$\pm$6 & 2.96$\pm$0.07\\
  Mel20-A3   & 6101$\pm$70  & 115$\pm$5 & 2.97$\pm$0.07\\
  Mel20-S1   & 5774$\pm$71  & 150$\pm$9 & 2.91$\pm$0.09 \\
  Mel20-S2   & 5781$\pm$74  & 181$\pm$7 & 3.03$\pm$0.07 \\
  Mel20-S3   & 5888$\pm$110 & 146$\pm$6 & 2.97$\pm$0.09 \\
  Mel20-S8   & 5828$\pm$72  & 151$\pm$5 & 2.95$\pm$0.06 \\
  Mel20-S11A & 6119$\pm$75  &  86$\pm$5$^a$  & 2.83$\pm$0.08 \\
  Mel20-S11B & 5172$\pm$240 & 275$\pm$37$^a$ & 2.87$\pm$0.29 \\
  \noalign{\smallskip}
  Plei-S2A   & 5674$\pm$139 & 174$\pm$13$^a$ & 2.93$\pm$0.14 \\ 
  Plei-S2B   & 5371$\pm$172 & 153$\pm$14$^a$ & 2.63$\pm$.19 \\ 
  Plei-S3    & 6601$\pm$119 &  32$\pm$7 & 2.64$\pm$0.18 \\ 
  Plei-S5    & 5752$\pm$67  & 175$\pm$5 & 2.98$\pm$0.06\\ 
  Plei-S7    & 7107$\pm$12  &  22$\pm$5 & \dots \\
  Plei-S8    & 7280$\pm$269 &  20$\pm$5 & \dots \\
  Plei-S9    & 7536$\pm$308 &  12$\pm$3 & \dots \\ 
  \noalign{\smallskip}
  ASCC16-S1   & 5271$\pm$89  & 304$\pm$9  & 3.03$\pm$0.10 \\ 
  ASCC16-S2   & 5184$\pm$84  & 325$\pm$9  & 3.02$\pm$0.10 \\
  ASCC16-S6   & 5947$\pm$114 & 183$\pm$11 & 3.14$\pm$0.11 \\
  ASCC16-S12  & 6233$\pm$107 & 128$\pm$7  & 3.11$\pm$0.09 \\
  \noalign{\smallskip}
  ASCC19-S1  & 6724$\pm$135 & 108$\pm$14 & 3.30$\pm$0.14 \\
  ASCC19-S2  & 6549$\pm$170 &  25$\pm$8  & 2.49$\pm$0.25 \\
  ASCC19-S3  & 6657$\pm$119 &  52$\pm$24 & 2.90$\pm$0.30 \\ 
  ASCC19-S4  & 6062$\pm$119 & 176$\pm$30 & 3.19$\pm$0.18 \\
  ASCC19-S5  & 6084$\pm$73  & 152$\pm$5  & 3.11$\pm$0.07 \\  
  ASCC19-S6  & 5895$\pm$96  & 222$\pm$7  & 3.23$\pm$0.09 \\ 
  ASCC19-S7A & 5376$\pm$139 & 284$\pm$10$^a$ & 3.06$\pm$0.14 \\
  ASCC19-S7B & 4894$\pm$226 & 403$\pm$38$^a$ & 3.04$\pm$0.36 \\
 ASCC19-S14  & 6513$\pm$96  & 101$\pm$6  & 3.14$\pm$0.09 \\
 ASCC19-S15  & 6988$\pm$121 &  53$\pm$11 & 3.10$\pm$0.13 \\
 ASCC19-S16  & 6750$\pm$220 &  63$\pm$5  & 3.15$\pm$0.14 \\
  \noalign{\smallskip}
  NGC2231-A2  & 5778$\pm$73  & 194$\pm$7 & 3.07$\pm$0.07 \\
  NGC2231-A3  & 5803$\pm$69  & 139$\pm$6 & 2.89$\pm$0.07 \\
  NGC2232-S3  & 5895$\pm$111 & 155$\pm$9 & 3.01$\pm$0.11 \\
  NGC2232-S4A & 6070$\pm$70  & 116$\pm$7$^a$  & 2.95$\pm$0.07 \\
  NGC2232-S4B & 4130$\pm$60  & 361$\pm$81$^a$ & 1.97$\pm$0.35   \\
  NGC2232-S9  & 5849$\pm$70  & 151$\pm$10 & 2.96$\pm$0.09  \\
  NGC2232-S11 & 6099$\pm$75  & 123$\pm$7  & 3.01$\pm$0.08 \\
  \noalign{\smallskip}
  Ros6-S1    & 5942$\pm$133 & 125$\pm$5 & 2.91$\pm$0.11  \\ 
  Ros6-S2    & 6082$\pm$71  &  93$\pm$7 & 2.84$\pm$0.08  \\
  Ros6-S5    & 5844$\pm$68  & 117$\pm$7 & 2.82$\pm$0.08  \\
  Ros6-S9    & 6088$\pm$72  & 135$\pm$7 & 3.06$\pm$0.07  \\
  \noalign{\smallskip}
  NGC7058-S5 & 6324$\pm$147 &  62$\pm$19 & 2.79$\pm$0.24  \\
  NGC7058-S6 & 6087$\pm$72  & 189$\pm$22 & 3.24$\pm$0.11  \\
  NGC7058-S7 & 5901$\pm$120 & 139$\pm$15 & 2.95$\pm$0.14  \\ 
  NGC7058-S9 & 5785$\pm$88  &  88$\pm$12 & 2.63$\pm$0.13  \\
 NGC7058-S10 & 5744$\pm$83  & 190$\pm$12 & 3.03$\pm$0.10  \\
 NGC7058-S11 & 5682$\pm$114 & 205$\pm$10 & 3.04$\pm$0.10 \\
 NGC7058-S12 & 5491$\pm$157 & 202$\pm$14 & 2.90$\pm$0.16 \\
 NGC7058-S13 & 5285$\pm$90  & 132$\pm$14 & 2.49$\pm$0.13 \\
\hline
\end{tabular}
\end{center}
\begin{list}{}{}
\item[$^a$] Values corrected for the contribution to the continuum flux.
\end{list}
\label{Tab:lithium}
\end{table}

For stars belonging to young clusters both the chromospheric emission (traced by Balmer H$\alpha$ and \ion{Ca}{ii} H\&K lines in the HARPS-N spectra) and lithium absorption can be used to estimate the age \citep[see, e.g.,][and references therein]{Jeffries2014, Frasca2018}.

To this aim, the photospheric spectra produced by \ROTFIT\ and \COMPO\ with non-active, lithium-poor templates were subtracted from the observed spectra of the targets to measure the excess emission in the core of the H$\alpha$ line ($EW_{\rm H\alpha}$). The result of the subtraction of the photospheric template from the observed spectrum in the H$\alpha$ region is shown in Fig.\,\ref{fig:subtraction_halpha} 
for the single-lined objects. The residual emission in the H$\alpha$ core that has been integrated to get $EW_{\rm H\alpha}$ is shown by the hatched green areas in these plots.  
The equivalent width of the \ion{Li}{i} $\lambda$6708\,\AA\ absorption line ($EW_{\rm Li}$) was also measured in the subtracted spectra where the blends with nearby photospheric lines have been removed.
The values of $EW_{\rm H\alpha}$ and $EW_{\rm Li}$ are quoted in Tables~\ref{Tab:Halpha_CaII} and \ref{Tab:lithium}, respectively. 

\begin{figure}
\begin{center}
\hspace{-.3cm}
\includegraphics[width=9.1cm]{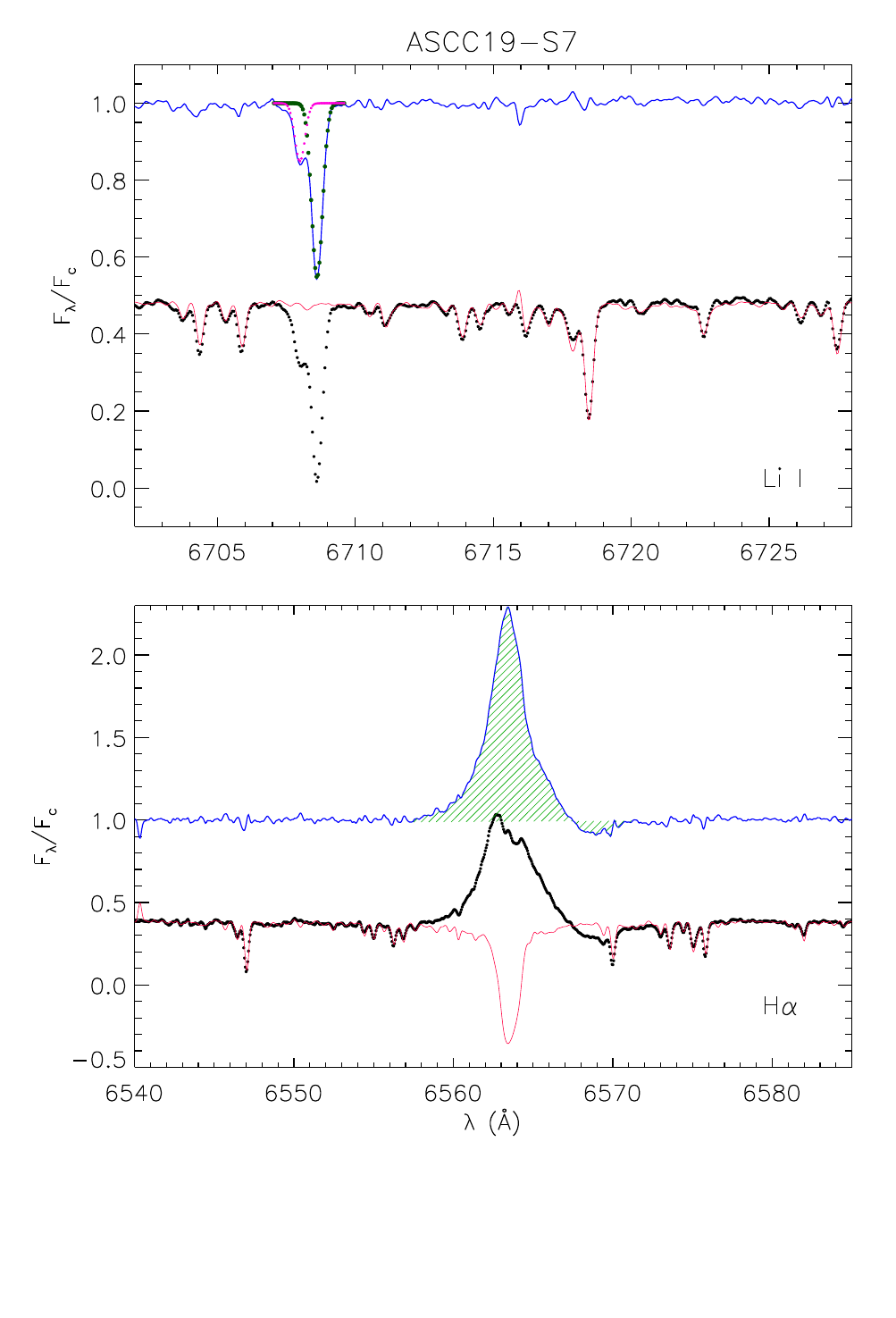}
\vspace{-1.8cm}
\caption{Subtraction of the synthetic composite spectrum (red line) from the observed spectrum of the SB2 ASCC19-S7 (black dots).
The difference washes up the photospheric absorption lines leaving only the chromospheric H$\alpha$ emission (blue line in the {\it bottom panel}) and emphasizes the \ion{Li}{i} $\lambda$6708\,\AA\ absorption lines of the two components (blue line in the {\it top panel}). 
The green and magenta dotted lines superimposed on the residual spectrum represent Gaussian fits used to deblend the \ion{Li}{i} lines of the two components.
The two components cannot be resolved in the strong and broad H$\alpha$ emission profile (also showing some redshifted extra-absorption), which has been integrated to provide a
total $W_{\rm H\alpha}^{em}$ (green hatched area in the {\it bottom panel}).
}
\label{fig:subtraction_compo2}
\end{center}
\end{figure}

For SB2 systems, spectral subtraction is strongly advisable to try to isolate chromospheric emission originating from the two components and to remove different lines of the two stars which, due to the Doppler 
shifts, can overlap the \ion{Li}{i} line of one or both components.  
The difference spectrum contains only the \ion{Li}{i} absorption lines of the two components, whose equivalent 
widths can  be measured individually, eventually by means of a two-Gaussian deblending (see Fig.~\ref{fig:subtraction_compo2} for an example). 
We note that the values of $EW_{\rm Li}$ for the components of SB2 systems, which are quoted in Table~\ref{Tab:lithium}, were 
corrected by dividing them for the contribution to the continuum of the respective component.   

Lithium is a fragile element that is burned in stellar interiors at temperature as low as 2.5$\times10^6$\,K. It is progressively depleted 
from the stellar atmosphere in a way depending on the internal structure (i.e., stellar mass) for stars with convective envelopes deep enough to 
reach temperatures at which lithium is burned. 
It can therefore be used as an age proxy for stars cooler than about 6500\,K.

We derived the lithium abundance, $A$(Li), from our values of \teff, \logg, and $EW_{\rm Li}$ by interpolating the curves of growth of \citet{Lind2009}, which span the \teff\ range 4000--8000\,K and \logg\ from 1.0 to 5.0 and include non-LTE corrections. The errors of $A$(Li) take into account both the \teff\ and $EW_{\rm Li}$ errors.
The values of $A$(Li) are listed  in Table~\ref{Tab:lithium}.

To estimate or check the ages of the clusters investigated in the present paper from the lithium content in the photospheres of its components, we used the \eagles\ code \citep[][]{Jeffries2023}. This code fits  Li-depletion isochrones to the values of \teff\ (in the range 3000-6500\,K) and $W_{\rm Li}$  of a coeval star group, such as the members of a cluster or a kinematical group or the components of a binary system. We note that \eagles\ works better for clusters with a wide \teff\ distribution of members, while in our case most of the stars are solar-like or hotter. The results of the application of this code to our data are shown in Fig.~\ref{fig:EAGLES} and summarized in Table~\ref{Tab:ages_EAGLES}, along with data from three literature sources, based on \gaia\ data and mass-derivation of ages using automated methods. We also included ASCC~123, whose age was now determined with \eagles\ for consistency. Its value is younger than the ones previously published in our papers: 100--250\,Myr \citep[][which was derived by isochrone fitting]{Frasca2019} and a Pleiades-like age \citep[][from a gyrochronological perspective]{Frasca2023a}. Our results are compatible within the errors with those obtained applying neural networks to photometric data, in a very different approach to ours, in particular with \citet{Cantat2020}. ASCC\,123 is a peculiar case, as it has been attributed a very large age by \citet{HR23} and \citet{Cavallo24} (more than 1 and 0,6 Gyr, respectively) clearly incompatible with our results, \citet{Cantat2020} and the cluster CMD \citep[see also the age of about 62 Myr derived by][]{Dias21}. In the \gaia-ESO Survey framework, \citet{Binks2021} studied the lithium depletion boundary age for NGC\,2232 and found a value of 38$\pm$3\,Myr, compatible within the errors with ours. They also highlighted the important role played by magnetic activity in young clusters.

\begin{table}[h!]
\caption{Comparison of the age (Myr) of our targets derived in this work (TW) and literature: \citet[][CG20]{Cantat2020}, \citet[][HR23]{HR23} and \citet[][C24]{Cavallo24}.}
\begin{center}
\begin{tabular}{lcccc}
\hline
\hline
\noalign{\smallskip}
Cluster  &  TW     &   CG20    & HR23   & C24\\
\hline
  \noalign{\smallskip}
Melotte~20     &  68$\pm$28       &  51  &  56$\pm$32  &  42$\pm$9   \\
Pleiades       &  75$\pm$36       &  78  & 122$\pm$71  &  87$\pm$17  \\
ASCC~16        &  14$\pm$10       &  14  &   8$\pm$2   &   5$\pm$1   \\
ASCC~19        &  $<20$           &  11  &   7$\pm$3   &   6$\pm$2   \\
NGC~2232       &  29$\pm$10       &  18  &  13$\pm$7   &  12$\pm$2   \\
Roslund~6      &  108$\pm$66      &  89  & 168$\pm$114 &  98$\pm$19  \\
NGC~7058       &  65$\pm$13       &  41  &  70$\pm$44  &  21$\pm$6   \\
ASCC~123       &  57$\pm$17       &  45  &1064$\pm$657 & 661$\pm$157 \\
\noalign{\smallskip}
\hline
\end{tabular}
\end{center}
\label{Tab:ages_EAGLES}
\end{table}

As mentioned above, another important diagnostic of chromospheric activity included in the HARPS-N spectra is represented by the emission in the cores of the \ion{Ca}{ii} H and K lines
\citep[see, e.g.,][and references therein]{Frasca2010,Frasca2011}.  
The photospheric templates in the \ion{Ca}{ii} region are made with synthetic BTSettl spectra \citep{Allard2012}, because the ELODIE templates do not contain this wavelength range. Moreover, due to the high sensitivity of 
these lines to the chromospheric activity, it is very difficult to find spectra of real stars with a pure photospheric profile clean of any chromospheric contribution in the \ion{Ca}{ii} cores and is preferable to resort to synthetic spectra. These reproduce accurately the \ion{Ca}{ii} line profiles even in their cores better than they do in the H$\alpha$ cores. The latter are indeed  severely affected by NLTE and redistribution effects that make low-activity standard stars preferable as photospheric templates to be subtracted. 
Since the signal-to-noise ratio of HARPS-N spectra of our targets is normally quite low in  the \ion{Ca}{ii} H\,\&\,K region (S/N\,$\leq$\,20), we have degraded the original resolution  (R=115,000) of the HARPS-N spectra and the synthetic spectra to R=42,000, which is the same of the ELODIE templates, improving the S/N but still keeping a resolution sufficient to identify the emission cores and measure their equivalent widths and fluxes. Furthermore, the photospheric templates have been aligned in wavelength with the target spectra by means of the cross-correlation function 
and have been rotationally broadened by the convolution with a rotational profile with the \vsini\ of the target star that is reported in Table~\ref{Tab:Spectra_param}.
The observed spectra (black lines) and the photospheric templates (red lines) are displayed in Fig.~\ref{fig:subtraction_CaII_HK} for the \ion{Ca}{ii} H\,\&\,K region.

\begin{figure*}[htb]  
\begin{center}
\includegraphics[width=8.8cm]{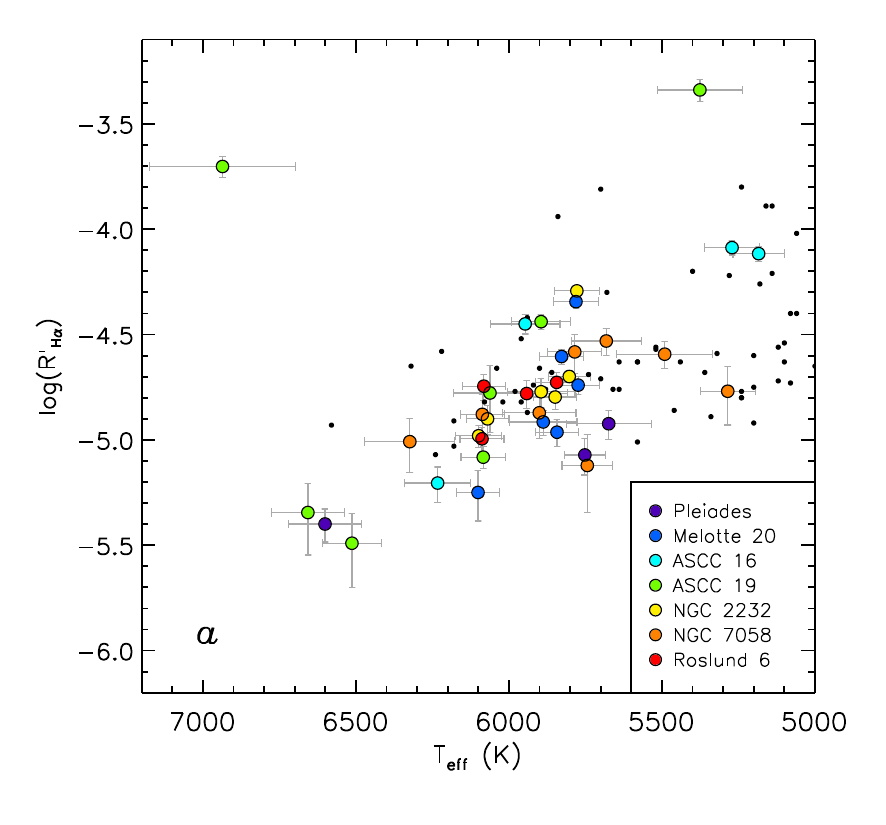}  
\vspace{-.2cm}
\includegraphics[width=8.8cm]{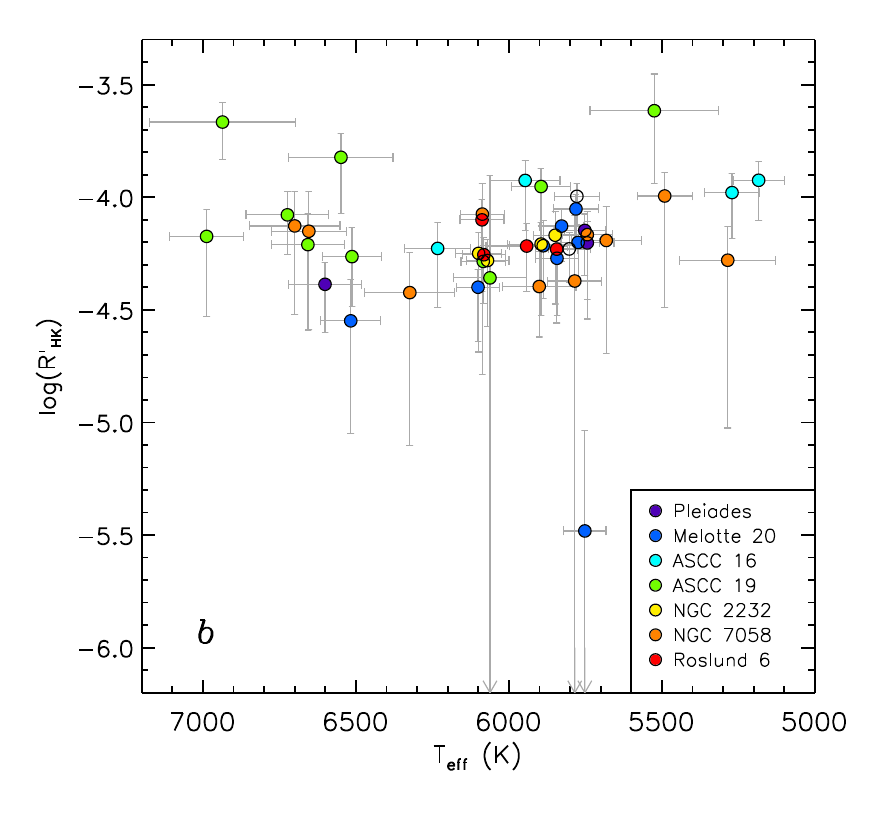}
\caption{{\it a)} The index $R'_{\rm H\alpha}$ as a function of \teff\  for the members of the seven clusters (big dots) distinguished by the colors. 
The small black dots denote the measures made by \citet{Soderblom1993a} for Pleiades stars. 
{\it b)} The index $R'_{\rm HK}$ as a function of \teff. }
\label{Fig:Flux_Teff}
\end{center}
\end{figure*}

The equivalent width of an emission chromospheric line, such as the H$\alpha$ and \ion{Ca}{ii} lines of our targets after the subtraction of the photospheric spectrum, is not the best diagnostic of the activity level, because it measures the intensity of the line with respect to the local continuum.
More effective indicators of activity for a chromospheric line are the surface flux, $F_{\rm line}$, and the ratio of line luminosity to bolometric luminosity, $R'_{\rm line}$.
For the H$\alpha$ this reads as

\begin{eqnarray}
F_{\rm H\alpha} & = & F_{6563}EW_{\rm H\alpha}, 
\end{eqnarray}
\begin{eqnarray}
R'_{\rm H\alpha}&  = & L_{\rm H\alpha}/L_{\rm bol} = F_{\rm H\alpha}/(\sigma T_{\rm eff}^4),
\end{eqnarray}
{\noindent where $F_{6563}$ is the flux at the continuum near the H$\alpha$ line per unit stellar surface area, which is evaluated from the BTSettl spectra \citep{Allard2012} at the stellar temperature and surface gravity of the target. The flux error includes both the error of $W_{\rm H\alpha}^{em}$ and the uncertainty in the continuum flux at the line centre, which is estimated propagating the errors of \teff\  
and \logg. }
Similar relations were used for the \ion{Ca}{ii} H and K lines.

The indexes $R'_{\rm H\alpha}$ and $R'_{\rm HK}$ are plotted against \teff\ in Fig.~\ref{Fig:Flux_Teff} along with the data for the Pleiades stars from \citet{Soderblom1993a}.
The chromospheric emission of the members of the young clusters investigated in the present paper is comparable with that of the Pleiades stars and roughly follows the increasing trend with the decrease of temperature, with two outliers, ASCC19-S16 and ASCC19-S7. The former source is an F2 YSO, a likely Herbig star, whose H$\alpha$ emission is more likely related to magnetospheric accretion.
The latter source is an SB2 with a strong and broad H$\alpha$ emission profile where the two components cannot be distinguished, so that it has been fully integrated providing a total flux that has been referred to the primary component. 
It is likely that the spin-orbit coupling and other effects related to the binary nature are responsible for this high flux.

\section{Discussion}\label{Sec:disc}

\subsection{Literature}

\begin{figure}
\begin{center}
\includegraphics[width=\columnwidth]{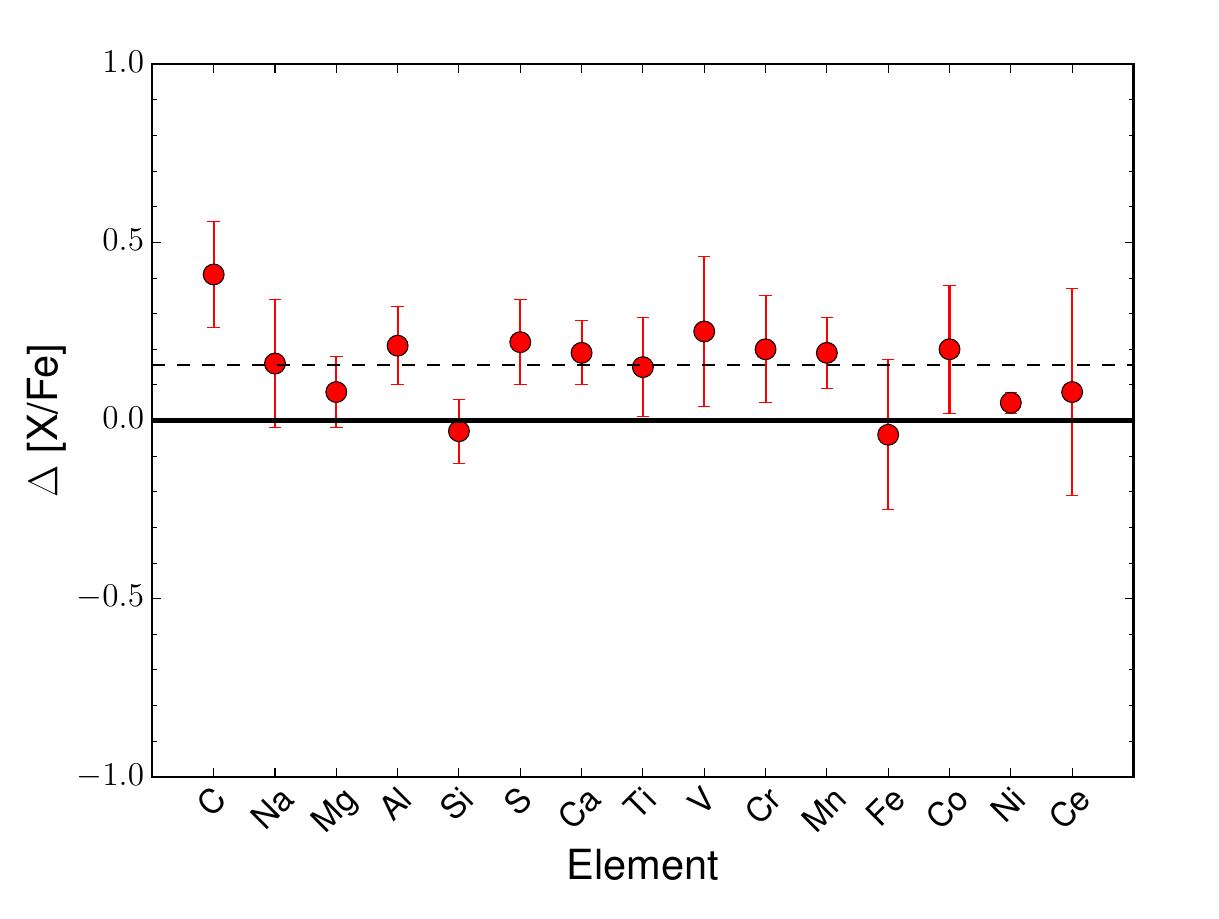}	 
\caption{Differences (ours minus theirs) in the average chemical composition derived in this work and those from APOGEE. The dashed line represents the average offset found between both data sets.}
\label{fig:comp_apo}
\end{center}
\end{figure}

We searched the literature for spectroscopic studies conducted at high resolution with which compare our results, but we did not find many references. The best one was the study performed by \citet{myers22} in the framework of the OCCAM survey. They investigated the chemical composition of 150 OCs based on the analysis of the abundances reported in the APOGEE DR17 ($R$=22\,500). 
Six out of the seven clusters studied in this work (all except Roslund\,6) were also observed by APOGEE, with which we have 15 chemical elements in common. When comparing both sets of abundances, we find that ours are, on average, 0.15\,dex higher, with C being the only element that shows a clear different value.
As stated in the APOGEE website\footnote{https://www.sdss4.org/dr17/irspec/abundances/}, the "solar scale should be close to the \citet{Grevesse07} solar values", which are the ones used as reference by us. However, they do not report these abundances, so that we cannot to put the two data sets on exactly the same scale.
This could lead to the existence of offsets between their and our data that could mitigate the observed differences. Figure\,\ref{fig:comp_apo} displays the comparison of the average chemical composition for each element. In addition, for the youngest clusters in our sample the APOGEE results might be unreliable, as its pipeline does not work optimally with stars younger than about 50\,Myr \citep{kounkel18,myers22}.

Below we discuss the papers focused on individual clusters, where possible. As previously commented, most of our targets are poorly studied, at least from a chemical point of view, and beyond the paper of \citet{myers22} there are not additional references. 

\subsubsection{NGC\,2232}

\citet{Monroe10} observed this cluster by using the multi-object spectrograph Hydra ($R$=16\,000) mounted on the 3.5-m WIYN telescope, finding a supersolar metallicity, [Fe/H]=0.49$\pm0.09$. They also derived the abundances of Al, Si and Ni,  obtaining subsolar [X/Fe] ratios for all of them: $-$0.37, $-$0.33 and $-$0.26, respectively. In that work, the authors calculated the \teff\ in two different ways (applying a photometric relationship and fitting the spectral energy distribution), obtaining different abundances as well. The values quoted here are the average of both data sets. Our results differ strongly from theirs: we find a mild subsolar metallicity ([Fe/H]=$-$0.10), solar ratios for Si (0.00) and Ni (0.02) and a slight supersolar value for Al (0.18). The cause of this disagreement is certainly due to the different methodology used and is very likely the result of the adoption of photometric temperatures by \citet{Monroe10}. In any case we consider our results as more reliable, since we used a much higher resolution than them and our findings show a much better agreement with the Galactic gradient and trends \citep[see e.g.][]{magrini23}. 

As for metallicity, in addition to \citet{Monroe10}, we also found two other references. \citet{randich22}, in the GES framework\footnote{NGC\,2232 was included in the \gaia-ESO Survey, but most of its members were only observed with GIRAFFE at intermediate resolution, and no chemical abundances beyond the metallicity were derived.}, gave a solar estimate, [Fe/H]=0.02$\pm0.05$ while the above mentioned paper by \citet{myers22} found a subsolar value, [Fe/H]=$-$0.35$\pm0.03$. Despite using high resolution, the metallicity obtained in these four works shows a large spread, of almost one dex, which is striking and calls for further studies.

\subsubsection{Pleiades}

\citet{Gebran08} studied the chemical composition of the Pleiades by observing 16 A-type dwarfs\footnote{Two of them were discovered to be chemically peculiar and will be not considered in the subsequent discussion.} with ELODIE ($R$=42\,000) and five F-type MS stars with SOPHIE ($R$=75\,000). Among their A dwarfs we found two stars that we also observed in this work, namely S7 and S8. For these objects we do not provide abundances since they are too hot for our method. However, the atmospheric parameters estimated in the present work are fully compatible with those reported by \citet{Gebran08}. As regards the chemical analysis, our results and theirs are compatible within the errors for most of the 17 elements in common, especially in the case of the F stars. They are more similar to our targets and, unlike the sample of A stars, they exhibit a smaller scatter in their chemical composition. Only the abundances of Si, V, and Ba show a clear discrepancy in both works. In the case of V and Ba the values provided by them are too high to be reliable ($\approx$\,1\,dex). However, for Si the explanation is not evident. For this element, they reported a supersolar value around $+$0.25\,dex while we found a slightly subsolar abundance ($\approx$\,$-$0.1\,dex). 

In order to determine the cluster metallicity with a great accuracy \citet{Soderblom09} conducted some observations from the Lick Observatory. They used the Hamilton echelle spectrograph ($R$=40\,000) to obtain spectra of 20 FGK stars, out of which 17 are bona fide members according to \citet{Cantat2020}. They found a solar metallicity for the Pleiades ([Fe/H]=0.03$\pm$0.02), the same value derived in this paper, in good agreement with previous works. They also calculated the chemical abundances of Na, Si, Ti and Ni, whose values are again consistent with our results.

Recently, \citet{Spina18} examined the chemical inhomogeneities in the Pleiades from five solar-like stars (none of them in common with us). They used UVES spectra ($R$=75\,000) to derive chemical abundances for 19 elements, all of which were also investigated in this work with the exception of oxygen. They performed a line-by-line differential analysis relative to the solar spectrum, a procedure somewhat different than ours that precludes a direct comparison of both sets of results. In any case, for most elements, the abundances derived in this work agree very well with those of \citet{Spina18}. Only for C, Al, Si and Co the differences are slightly greater than 0.2\,dex.

\subsection{Galactic metallicity gradient}\label{Sec:grad}

\begin{figure}
\begin{center}
\includegraphics[width=\columnwidth]{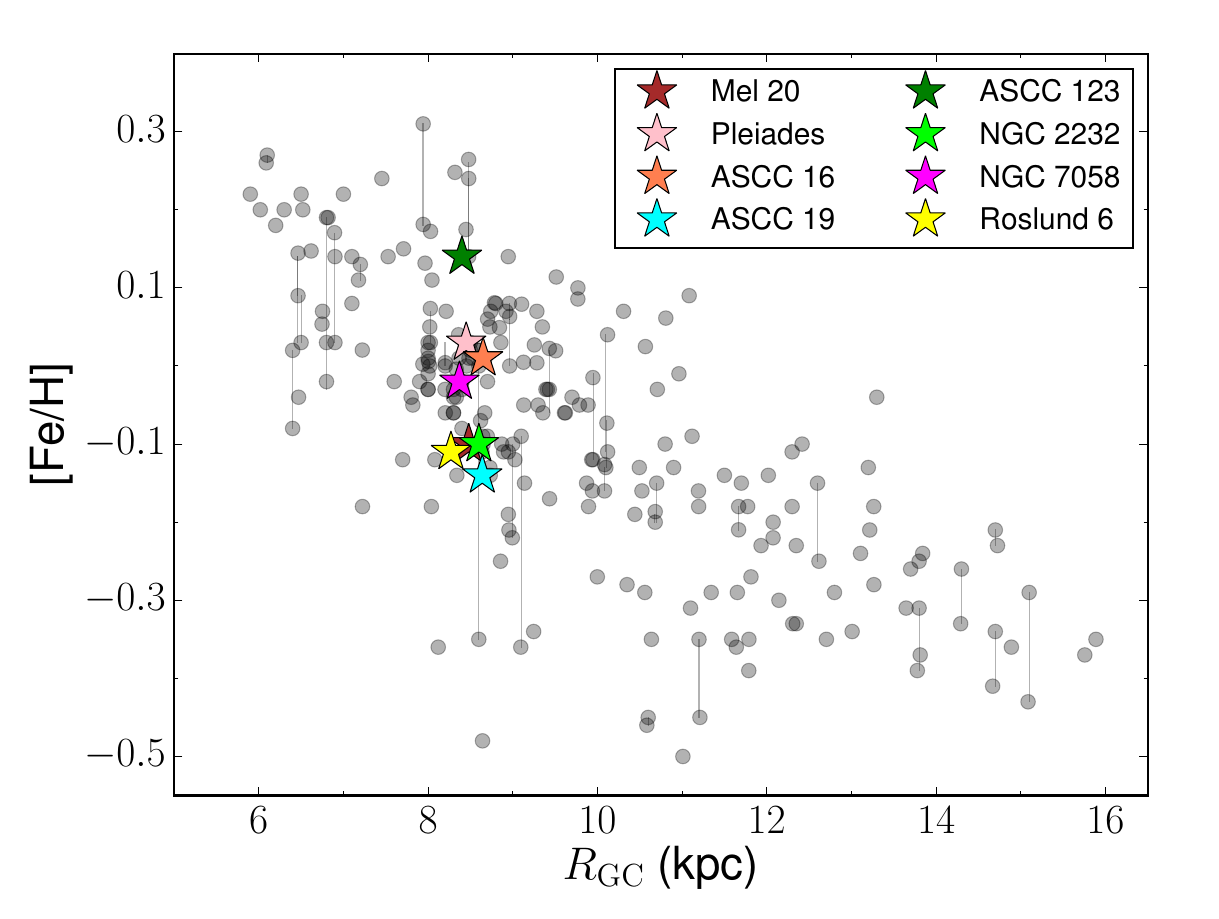}	
\caption{Radial metallicity gradient traced by literature open clusters observed at high resolution (grey circles). Vertical lines link different results for the same cluster. The objects studied in this work are highlighted with different colours.}
\label{fig:gradient}
\end{center}
\end{figure}

As done in previous papers, we compare the metallicities found in this work for our targets with those expected based on their position in the Galaxy. We used a large sample of OCs to trace the radial metallicity distribution of the Galaxy to which we contrast our results. We only collected OCs studied with high-resolution spectroscopy in the framework of surveys such as OCCAM \citep[APOGEE-DR17,][]{myers22},
GES \citep[iDR6,][]{randich22}, OCCASO \citep{occaso}, and SPA \citep{Frasca2019,DOrazi2020,Casali20,zhang21,Alonso2021}. We completed the sample with other clusters studied previously by our group \citep{6067,3105,2345,cep_oc,M39} in a similar way. In total, we gathered nearly two hundreds OCs, at Galactocentric distances ($R_{\rm GC}$) up to 16 kpc. Only a few of them were common in different datasets. With all these clusters we plot the Galactic gradient shown in Fig.~\ref{fig:gradient}. We used the \cite{Grevesse07} solar value, $A$(Fe)=7.45, to rescale the metallicities and took the values reported in \cite{Cantat2020} for the $R_{\rm GC}$. These were calculated from the analysis of the $Gaia$-DR2 astrometry, using  $R_{\odot}$=8.34 kpc as a solar reference.

Our targets are located in the vicinity of the Sun at similar distances ($R_{\rm GC}\,\approx$\,8.6 kpc). However, as discussed before, they show a spread of about 0.2 dex in metallicity, which increases up to $\approx$0.3\,dex when ASCC\,123 is also considered. In Sect.\ref{sec:rw} we will explore its possible age dependence. In any case, as seen in Fig.~\ref{fig:gradient}, their positions on the gradient are totally compatible with those of the rest of the clusters. In the figure is also visible the well-known fact that the [Fe/H] obtained at high resolution by different groups for the same objects differ by 0.1--0.2\,dex, similar to the dispersion observed in our clusters.

\subsection{Chemical composition and Galactic trends}

In an analogous way to that previously carried out with metallicity, we compare the abundances found in this work with those observed in the Galactic disc. To this aim we used the chemical trends displayed by the OCs collected in the sample described in Sect.~\ref{Sec:grad}. Additionally, we increased the number of chemical elements investigated in the GES
and SPA clusters by considering the new abundances reported in \cite{magrini23} and \cite{zhang22}, respectively.

\begin{figure*}
\begin{center}
\includegraphics[width=15cm]{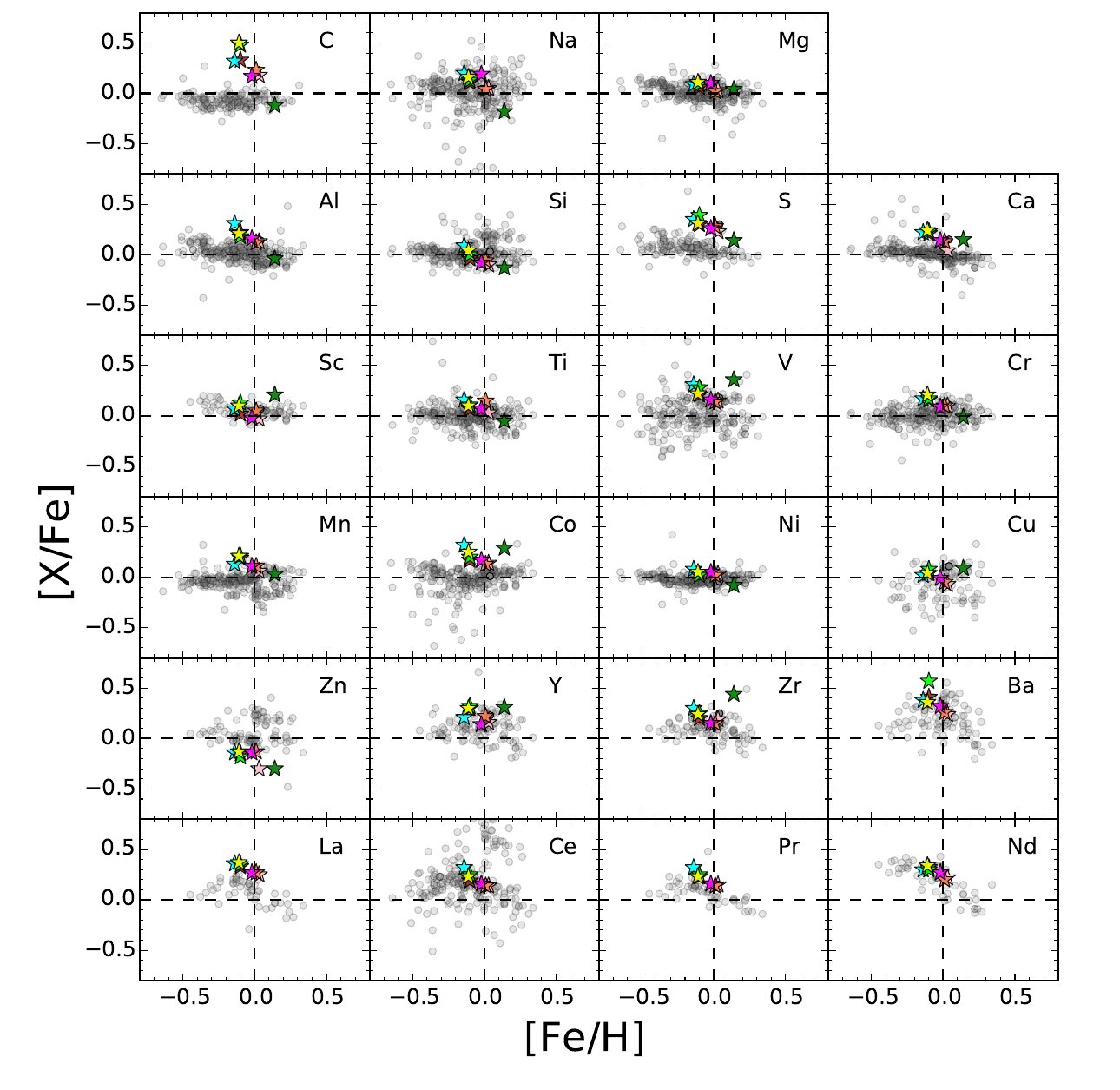}	
\caption{Abundance ratios [X/Fe] vs. [Fe/H]. Symbols and colors are the same as in Fig.~\ref{fig:gradient}. The dashed lines show the solar value.}
\label{fig:trends}
\end{center}
\end{figure*}

Figure~\ref{fig:trends} shows the Galactic trends, represented by the ratios [X/Fe] versus [Fe/H], for 23 chemical elements. With the only exception of C, the chemical composition of the clusters under study is fully compatible with the observed trends. Zinc is the only element for which our clusters show a slightly subsolar value, but close to the trend. The opposite happens with S, for which our abundances lie above the gradient, but compatible within the errors (not shown in the figure for clarity). Barium is the element that shows the largest spread among the clusters.

\section{The Radcliffe Wave}\label{sec:rw}

\begin{figure*}
\begin{center}
\includegraphics[width=18cm]{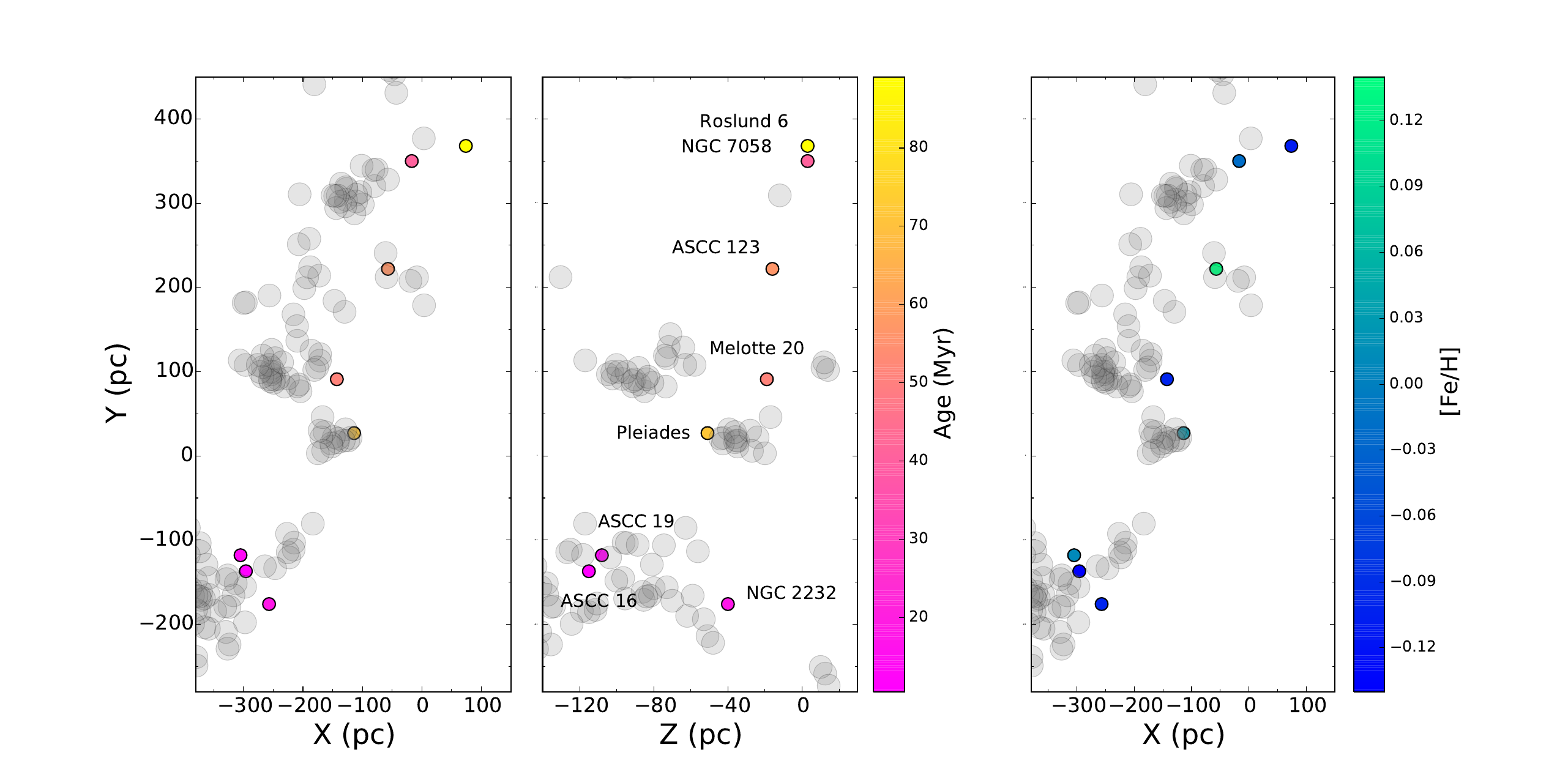}	 
\caption{Distribution of the molecular clouds forming the Radcliffe Wave in the XY and ZY planes (grey circles). Our targets are represented with different colors depending on their age (left) and metallicity (right).}
\label{fig:xyz_age}
\end{center}
\end{figure*}

In order to investigate the possible connection among the Radcliffe clusters we studied the correlation between the different variables that we have at our disposal: chemical composition, age, and position in the Galaxy. In each of these three groups we included several data sets: $R_{\rm GC}$, Cartesian Galactic coordinates (X, Y, Z), age, [Fe/H], [X/H], and [X/Fe] from the present research and the literature \citep{Cantat2020,myers22,HR23,Cavallo24}. Figure~\ref{fig:xyz_age} illustrates the distribution of our targets along the RW according to the age and metallicity obtained in this work for them. 

The degree of correlation, expressed by the Pearson's coefficient, varies according to the data sets we use. For example, we find a moderate-to-high correlation (0.65\,$<|r|<$\,0.94) between age and Galactic position when using the values derived in this work or in \citet{Cantat2020}. On the contrary, no significant correlation is found using the ages reported in \citet{HR23} and \citet{Cavallo24}. Metallicity shows the strongest correlation with age ($r\approx$\,0.8) when combining the metallicity obtained in this work (the valued derived with \rotfit\ correlates slightly better than that of {\sf SYNTHE}) and the ages of both \citet{HR23} and \citet{Cavallo24}.

Concerning the chemical composition, we find a strong correlation with age (taking as a reference the values from \citet{HR23} and \citet{Cavallo24}) of $\alpha$- ($r$=0.78), Fe-peak ($r$=0.93) and neutron-capture ($r=0.98$) elements in terms of [X/H] abundances. On the other hand, if we use the [X/Fe] ratios, the correlation of $\alpha$- ($r$=$-$0.64) and neutron-capture ($r=0.56$) elements drops and that of the Fe-peak elements disappears. As far as individual chemical elements are concerned, a correlation of at least moderate magnitude is observed for many of them. In particular,  employing the [X/H] values, we find a moderate correlation ($0.6<|r|<0.8$) for C, Na, Al, Mn, and Nd and a strong correlation ($|r|>0.8$) for Mg, Ca, Sc, V, Co, Cu, Y, and Zr. When the [X/Fe] values are used instead, the correlation is moderate for C, S, Sc, Cr, and Zn while it is strong for Na, Al, Ti, Ni, and Zr. By contrast, no correlation is found when using the ages derived in the present work or those reported in \citet{Cantat2020}.

Finally, we also used the chemical abundances ([X/Fe]) calculated by \citet{myers22}. As commented before, Roslund\,6 and ASCC\,123 were not included in their sample. They derived the abundances for the same $\alpha$ elements investigated by us,  
some Fe-peak elements (Cr, Mn, Co, Ni and Cu) and Ce as the only heavy element. Their data correlate best with those of \citet{HR23}. The dependence of [Fe/H] on age is somewhat stronger than ours, while for $\alpha$- and Fe-peak elements are similar. For the latter, on the contrary, the correlation is positive. Regarding the spatial distribution, the correlation of [Fe/H] with $R_{\rm GC}$ is slightly higher than ours, while no correlation for chemical abundances is found. Concerning the X, Y, Z coordinates a moderate level of correlation, somewhat stronger than ours, is found. 
No clear trend ($|r|$>0.6) is evident between chemical abundances and age or Galactic location. 

However, if we group the clusters by age, the correlation becomes very clear with all data sets. In this way we considered three categories. The first contained the youngest clusters of our sample: ASCC\,16, ASCC\,19 and NGC\,2232. The second group consisted of the clusters with "intermediate" age, Melotte\,20 and NGC\,7058, while the third group contained the oldest cluster, Roslund\,6. The classification of the two remaining clusters, ASCC\,123 and Pleiades, depends on the set of ages we use. Based on the ages of \citet{HR23} and \citet{Cavallo24} they should be included in the last category, whereas if we consider the ages of this work both clusters should be classified as intermediate-age clusters. Finally, according to the values of \citet{Cantat2020}, ASCC\,123 would be included in the second group and the Pleiades in the third.

\begin{figure}
\begin{center}
\includegraphics[width=\columnwidth]{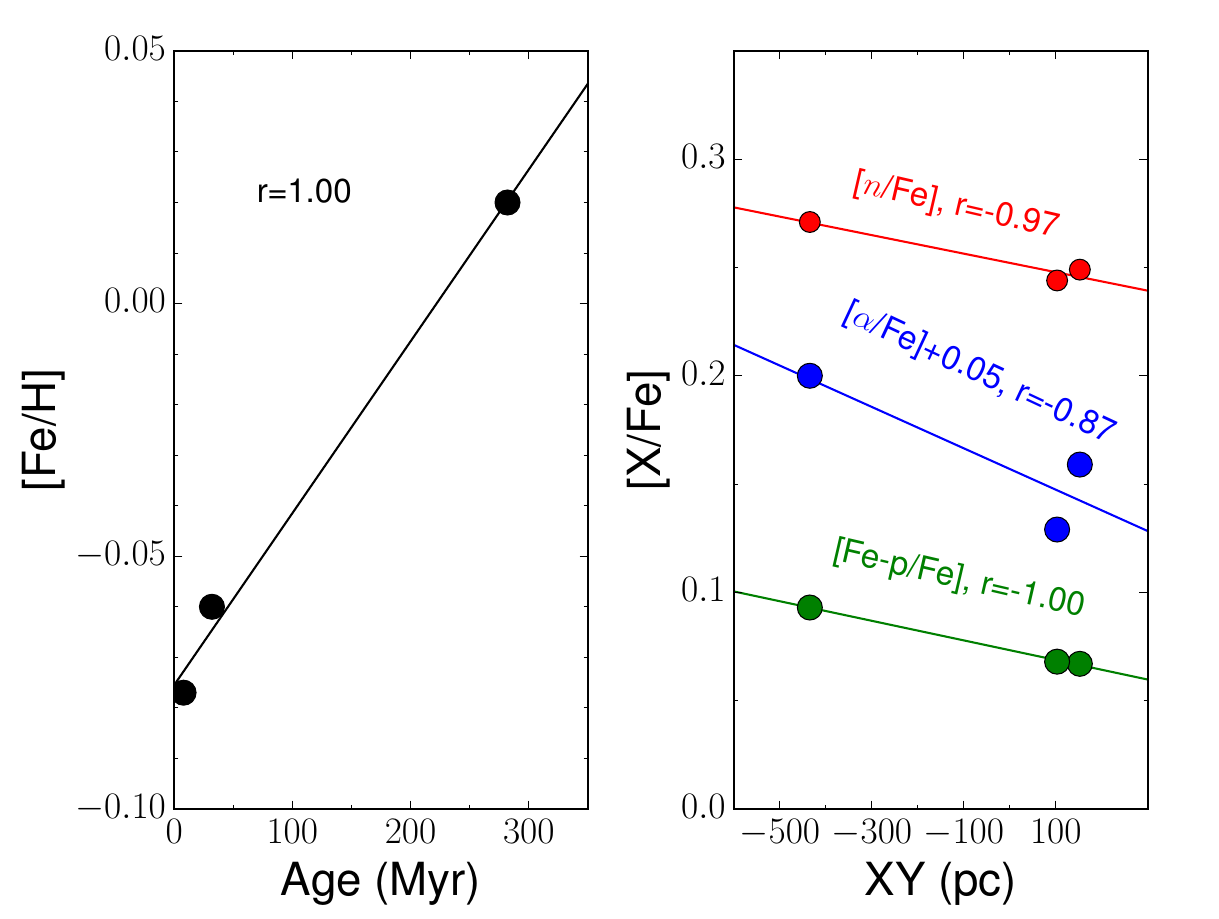}	 
\caption{Some of the correlations found in the Radcliffe clusters, grouped into age bins. Left panel: evolution of [Fe/H] over time. Right panel: distribution of chemical composition, [X/Fe], along the XZ Galactic plane. The Pearson coefficients (r) are also shown.}
\label{fig:rw_corr}
\end{center}
\end{figure}

Furthermore, to try to evaluate the significance of their position along the RW, instead of using individual coordinates as done before, we payed attention to their location in the planes XY, XZ and YZ. In this way we used the sum of the pair of coordinates that defines each plane, as a marker of the position of the cluster on it, that is, X+Y in the case of the XY plane. 
We find an age gradient across the RW, with older clusters close to the Galactic centre \citep[e. g. 0.8<$r$<0.9 in all the three planes when using the data reported in][]{Cantat2020}.
This fact confirms the main conclusion of \citet{Konietzka24}, namely the existence of a radial oscillation in the RW from the Galactic centre outwards. On the other hand,
the Pearson's coefficients between the metallicity and chemical composition derived in this work and age or position in the Wave according to \citet{HR23} and \citet{Cantat2020}, respectively, range from $\approx$0.8 to 1.0 and are listed in Table~\ref{Tab:rw_corr_groups}. As an example, some of these correlations are displayed in Fig.~\ref{fig:rw_corr}, specifically the variation of metallicity and chemical content derived in this work with respect to age and location in the XY plane reported in \citet{Cavallo24}.
Correlations are slightly stronger when [X/Fe] is used instead of [X/H],
Similar values are obtained when we used the APOGEE abundances or other ages as ours or those of \citet{HR23}. The only difference we noted is the behaviour of the Fe-peak elements (but not [Fe/H] or $\alpha$ elements): we find negative correlations between [X/Fe] and age or XYZ positions, while when considering the APOGEE abundances the correlation is positive.

From our results we infer a slight variation in the chemical composition with age and the position along the Wave, which is the first evidence of the connection between the different clusters associated to the RW. The observed metallicity trend, with younger clusters being metal-poorer compared to older ones, suggests inhomogeneous mixing or local chemical enrichment rather than a canonical triggered star formation within a large-star forming region. Additionally, it is important to note, especially in the youngest stars, the effect that strong magnetic fields can induce on the behaviour of spectral features, which could potentially alter the analysis and, therefore, the results \citep{Baratella2021}.
In any case the sample of clusters is not very large and these results have been obtained from the analysis of six to eight points (when considering the clusters separately) or only three (when we grouped them into age bins). 

\begin{table}[h!]
\caption{Pearson correlation coefficients between different variables for the Radcliffe clusters grouped by age.}
\begin{center}
\begin{tabular}{lcccc}
\hline
\hline
             & [Fe/H]   & [$\alpha$/Fe] & \scriptsize{[$Fe-peak$/Fe]} & [$n$/Fe]  \\
\hline
Age          &  ~~1.00  & $-$0.88  & $-$0.88  & $-$0.84  \\ 
R$_{\rm GC}$ & $-$0.77  &  ~~0.97  &  ~~0.97  &  ~~1.00  \\
XY           &  ~~0.81  & $-$0.93  & $-$1.00  & $-$0.97  \\
XZ           &  ~~0.76  & $-$0.94  & $-$0.99  & $-$0.98  \\
YZ           &  ~~0.87  & $-$0.88  & $-$0.73  & $-$0.94  \\
\hline
\end{tabular}
\end{center}
\label{Tab:rw_corr_groups}
\end{table}

\section{Summary}
\label{Sec:summary}

The Radcliffe Wave is a $\approx$\,3\,kpc structure in the solar neighbourhood that involves many of the nearby star-forming regions. Recently discovered, it has superseded the Gould Belt as the preferred scenario for the local star formation. In this work, for the first time, the RW is studied also from a chemical point of view with the idea of finding traces of metal content evolution throughout its history. To conduct our research we selected a sample of seven young OCs associated to the Wave namely, Melotte\,20, Pleiades, NGC\,2232, NGC\,7058, ASCC\,16, ASCC\,19 and Roslund\,6 (and added ASCC\,123, previoulsy studied by our group). Their age range varies from around 10 to 100,\,Myr and most of them are poorly studied. We present here the most complete chemical study to date of these clusters. In the case of Roslund\,6, it is also the only one.

We took high-resolution spectra with GIARPS at the TNG telescope of 53 bona fide clusters members, for which we derived radial velocities and atmospheric parameters. Most of the stars observed, 41, are slowly-rotating FGK dwarfs. For this subsample we studied the chromospheric activity and the content of Li, from which we estimated the age of the clusters. The chromospheric activity intensity traced by both the H$\alpha$ and \ion{Ca}{ii} H\&K line fluxes is consistent with the young age of these stars, but the data do not allow us to distinguish different levels or trends between the seven clusters. We find that the Li-depletion ages derived in this work agree well with the evolutionary ages reported in the literature. We find that three of the studied clusters have solar metallicity (Pleiades, ASCC\,16 and NGC\,7058) while the remaining ones (ASCC\,19, Melotte\,20, NGC\,2232 and NGC\,7058) show a slightly subsolar value, around $-$0.1\,dex. In any case, all of them follow the Galactic gradient shown by other OCs. 

In addition, we investigated their chemical composition by deriving the abundances of 25 elements, which include the main nucleosynthesis channels. Specifically we determined the chemical abundances of C, odd-Z elements (Na, Al), $\alpha$-elements (Mg, Si, S, Ca, Ti), Fe-peak elements (Sc, V, Cr, Mn, Co, Ni, Cu, Zn) and neutron-capture elements (Sr, Y, Zr, Ba, La, Ce, Pr, Nd). We find, on average, solar values for $\alpha$- and Fe-peak elements while the heaviest elements exhibit abundances $\approx$0.2\,dex higher, specially barium. Their [X/Fe] ratios are in good agreement with the Galactic trends displayed by a large sample of OCs observed at high-resolution. The abundances of C and S are slightly above these trends whereas that of Zn is just below, but in both cases, they are compatible within the errors. 

Finally, we investigated the possible link between these clusters as members of the RW. In our analysis we also included ASCC\,123, another cluster connected to the RW that was previously studied by our group with the same instrument and methodology. We calculated the Pearson correlation coefficients for different variables (age, XYZ Galactic coordinates, $R_{\rm GC}$, [Fe/H] and chemical abundances) collected from both this work and literature. We find a strong correlation between the age of the clusters and their spatial distribution confirming the radial oscillation along the RW, but no clear correlation between the chemical composition of the clusters and their age or position along the Wave emerges. However, if we do not consider individually the eight clusters, but we group them in three age classes, the correlation becomes very evident. The dependence between metallicity and chemical composition with age and position along the Radcliffe Wave supports the idea of a shared history between our targets, whose palpable result is the local enrichment found, within an episode of inhomogenous mixing. However, our sample is not very large and prevents us from going further in our conclusions. More data, such as those which will emerge from upcoming large spectroscopic surveys as WEAVE and 4MOST, are needed to confirm the first evidence claimed in this work.

\begin{acknowledgements}
We thank the referee, Dr. V.V. Bobylev, for his helpful comments and suggestions. We also acknowledge the INAF support under GO project No. 1.05.23.05.19.
This research used the facilities of the Italian Center for Astronomical Archive (IA2) operated by INAF at the Astronomical Observatory of Trieste.
This research has made use of the VizieR catalogue access tool, CDS,
Strasbourg, France \citep{10.26093/cds/vizier}. The original description 
of the VizieR service was published in \citet{vizier2000}.
This work has made use of data from the European Space Agency (ESA)
mission {\it Gaia} ({\tt https://www.cosmos.esa.int/gaia}), processed by
the {\it Gaia} Data Processing and Analysis Consortium (DPAC,
{\tt https://www.cosmos.esa.int/web/gaia/dpac/consortium}). Funding
for the DPAC has been provided by national institutions, in particular
the institutions participating in the {\it Gaia} Multilateral Agreement.
We made use of data from the 
GES Data Archive, prepared and hosted by 
the Wide Field Astronomy Unit, Institute for Astronomy, University of Edinburgh, 
which is funded by the UKScience and Technology Facilities Council.
This paper includes data collected by the \tess\  mission which are publicly available from the Mikulski Archive for Space Telescopes (MAST). 
Funding for the {\it TESS} mission is provided by the NASA's Science Mission Directorate.
AF acknowledges funding from the Large-Grant INAF YODA (YSOs Outflow, Disks and Accretion). AB  acknowledges funding from INAF Mini-Grant 2022. XF thanks the support of the National Natural Science Foundation of China (NSFC) No. 12203100 and the China Manned Space Project with No. CMS-CSST-2021-A08.


\end{acknowledgements}

\bibliographystyle{aa}
\bibliography{radclif_paper}

\newpage

\appendix

\section{Additional tables and figures}

\begin{landscape}
    
\begin{table}

\caption{$Gaia$ DR3 astrometric data and cluster membership.} 
\label{Tab:astrom}

\begin{center}
\begin{tabular}{lccccrccccc}
\hline\hline
\noalign{\smallskip}
Star         &  TIC      &  RA	    &   DEC           & $\varpi$ &  $\mu_{\alpha*}$  & $\mu_{\delta}$  & $G^a$   & $G_{\rm BP}-G_{\rm RP}^a$ & $V_{\rm broad}$  & Prob$^d$ \\ 
           &	       & (J2000)    & (J2000)     	  &  (mas)   &  (mas\,a$^{-1}$)  & (mas\,a$^{-1}$) & (mag)   &      (mag)		  &  (\kms)          &          \\ 
           
\hline
\noalign{\smallskip}
Mel20-A1    &  116991727  &  047.466998  &   +48.470950    &  4.2466  0.0191  & 19.717  0.020  & -24.251  0.018 &  10.669893  &  1.001655 &   5.6594 & \dots \\ 
Mel20-A2    &  117805849  &  049.709634  &   +49.731154    &  5.6468  0.0173  & 23.170  0.019  & -24.502  0.019 &  10.987741  &  0.892536 &  13.0912 &  1.00 \\ 
Mel20-A3    &  410705987  &  051.979241  &   +49.760330    &  5.7952  0.0158  & 22.636  0.013  & -26.262  0.015 &  10.539244  &  0.824582 &  10.8013 &  1.00 \\ 
Mel20-S1    &  431769464  &  053.574571  &   +45.730075    &  5.6585  0.0204  & 21.186  0.024  & -26.255  0.018 &  11.345673  &  0.953852 &   7.5463 &  1.00 \\ 
Mel20-S2    &  252867743  &  051.580672 &   +49.225693     &  5.7431  0.0219  & 23.253  0.020  & -24.705  0.020 &  11.409828  &  0.972621 &  14.8571 &  1.00 \\ 
Mel20-S3    &  458811423  &  050.264157  &   +58.507704    &  5.4226  0.0252  & 22.078  0.024  & -23.256  0.027 &  11.208141  &  0.963267 &  15.7313 &  1.00 \\ 
Mel20-S8    &  354180784  &  053.770928  &   +50.912543    &  5.7358  0.0184  & 22.294  0.017  & -25.745  0.018 &  11.092926  &  0.920669 &  27.2798 &  1.00 \\ 
Mel20-S11   &   65697949  &  055.273950  &   +45.793793    &  5.2512  0.0232  & 20.624  0.023  & -24.056  0.020 &  10.108564  &  0.800255 &  13.2743 &  1.00 \\ 
Plei-S2     &   61146016  &  056.657000  &   +23.787720    &  7.4719  0.0207  & 20.011  0.022  & -47.267  0.018 &  10.045166  &  0.900887 &  16.9975 &  1.00 \\ 
Plei-S3     &   61145611  &  056.659968  &   +22.919778    &  7.1526  0.0204  & 19.924  0.023  & -43.549  0.017 &   9.306278  &  0.657387 &  35.8030 &  1.00 \\ 
Plei-S5     &  114029810  &  055.143205  &   +23.682609    &  7.4986  0.0187  & 20.546  0.019  & -45.629  0.014 &  10.699779  &  0.911152 &  11.4547 &  1.00 \\ 
Plei-S7     &  405484188  &  056.830608  &   +24.139115    &  7.2693  0.0278  & 19.017  0.032  & -43.928  0.025 &   8.243902  &  0.353921 &  13.8362 &  1.00 \\ 
Plei-S8     &  405483817  &  055.930172  &   +24.374568    &  7.4128  0.0331  & 19.480  0.033  & -40.048  0.023 &   8.198288  &  0.348311 &  25.7583 & \dots \\ 
Plei-S9     &  405483707  &  056.001121  &   +24.556998    &  7.4124  0.0278  & 20.535  0.032  & -46.190  0.022 &   8.041619  &  0.272338 &  29.6633 &  1.00 \\ 
ASCC16-S1   &  459991270  &  081.073449  &   +01.037743    &  2.9002  0.0133  &  1.213  0.016  &  -0.064  0.012 &  12.495133  &  1.089946 &  27.1006 &  0.80 \\ 
ASCC16-S2   &  264460405  &  081.148573  &   +01.171941    &  2.9186  0.0129  &  1.320  0.013  &  -0.127  0.010 &  12.529669  &  1.150460 &   \dots  &  0.90 \\ 
ASCC16-S6   &  459991003  &  081.110369  &   +01.446968    &  2.8716  0.0179  &  1.519  0.020  &   0.409  0.013 &  11.047515  &  0.803285 &  40.1420 &  1.00 \\ 
ASCC16-S12  &  264484631  &  081.461492 &   +00.796933     &  2.8907  0.0153  &  1.408  0.016  &   0.156  0.012 &  10.606313  &  0.714010 &  39.1212 &  0.80 \\ 
ASCC19-S1   &   50584002  &  082.061867  &   -01.976975    &  2.8229  0.0192  &  1.484  0.018  &  -1.130  0.012 &  10.784862  &  0.610453 &   \dots  &  0.70 \\ 
ASCC19-S2   &    4350534  &  081.432092  &   -02.190045    &  2.7774  0.0245  &  1.172  0.018  &  -1.307  0.015 &  10.616993  &  0.732801 & 164.5130 &  0.80 \\ 
ASCC19-S3   &    4289678  &  081.220703  &   -02.645501    &  2.8127  0.0188  &  1.036  0.018  &  -1.550  0.014 &  10.943126  &  0.675324 & 102.5290 &  0.70 \\ 
ASCC19-S4   &   50524132  &  081.753548  &   -01.299316    &  2.8501  0.0186  &  1.118  0.018  &  -1.098  0.013 &  10.921832  &  0.769989 &  14.5996 &  1.00 \\ 
ASCC19-S5   &   50619443  &  082.328205  &   -01.310307    &  2.7580  0.0224  &  0.749  0.019  &  -1.294  0.017 &  11.291587  &  0.792629 &  22.6044 &  0.90 \\ 
ASCC19-S6   &   50619751  &  082.324391  &   -01.727577    &  2.8167  0.0237  &  1.190  0.021  &  -1.265  0.018 &  11.049373  &  0.861242 &  53.9130 &  0.90 \\ 
ASCC19-S7   &   50524284  &  081.758139  &   -01.514830    &  2.9369  0.0259  &  1.109  0.023  &  -1.257  0.017 &  11.580762  &  1.096926 &  33.7653 &  0.80 \\ 
ASCC19-S14  & 	 4293859   &  081.335941  &   -02.175241   &  2.8617  0.0374  &  1.418  0.031  &  -1.662  0.025 &  10.207284  &  0.725842 &   \dots  &  0.90 \\ 
ASCC19-S15  &   50619809  &  082.306384 &   -01.796238     &  2.9218  0.0215  &  0.804  0.018  &  -1.278  0.015 &  10.297378  &  0.499105 &  53.5383 &  0.90 \\ 
ASCC19-S16  &   50663177  &  082.577843  &   -02.032683    &  2.9615  0.0225  &  1.143  0.021  &  -1.074  0.017 &  10.352539  &  0.465540 &  63.8147 &  0.70 \\ 
ASCC19-S17  &   50524860  &  081.746516  &   -02.360650    &  2.8777  0.0519  &  1.185  0.047  &  -1.122  0.035 &   6.579845  & -0.248401 &   \dots  &  0.90 \\ 
ASCC19-S18  &   50524182  &  081.789056  &   -01.367350    &  2.7112  0.0491  &  1.203  0.046  &  -1.549  0.032 &   7.186184  & -0.212028 &   \dots  &  0.70 \\ 
ASCC19-S19  &   50529449  &  081.940690  &   -02.145474    &  2.8072  0.0447  &  1.016  0.040  &  -1.199  0.029 &   7.488637  & -0.194520 &   \dots  &  0.70 \\ 
ASCC19-S20  &    4350635  &  081.482814  &   -02.335522    &  2.8261  0.0310  &  0.929  0.031  &  -1.551  0.022 &   8.263457  & -0.088899 &   \dots  &  0.60 \\ 
ASCC19-S21  &   50720274  &  082.621456  &   -00.847423    &  2.8777  0.0249  &  1.485  0.021  &  -1.190  0.019 &   9.292233  &  0.023084 &  28.5303 &  0.80  \\ 
ASCC19-S22  &   50529597  &  081.910890  &   -01.967176    &  2.8901  0.0231  &  0.886  0.022  &  -1.315  0.016 &   9.781634  &  0.178481 &  49.0085 &  0.70 \\ 
ASCC19-S23  &   50525226  &  081.768759  &   -02.921821    &  2.8272  0.0214  &  1.180  0.019  &  -1.523  0.015 &   9.787668  &  0.144700 & 130.4069 &  0.60 \\ 
NGC2232-A2  &   42884139  &  096.948908  &   -04.675883    &  3.1433  0.0183  & -4.831  0.015  &  -1.923  0.017 &  12.104153  &  0.860097 &   \dots  &  1.00 \\ 
NGC2232-A3  &   42889061  &  097.140402  &   -04.782864    &  3.0053  0.0343  & -4.726  0.032  &  -1.663  0.031 &  12.144907  &  0.868650 &   \dots  &  1.00 \\ 
NGC2232-S3  &   42563198  &  096.615708  &   -04.556045    &  3.1514  0.0158  & -4.217  0.013  &  -2.077  0.012 &  12.126466  &  0.794903 &  26.3188 &  1.00 \\ 
NGC2232-S4  &   43152097  &  097.408543  &   -04.066444    &  2.9119  0.0152  & -4.556  0.014  &  -1.864  0.015 &  11.770985  &  0.791048 &  23.8189 &  1.00 \\ 
\noalign{\smallskip}
\hline
\end{tabular}
\end{center}
{\bf Notes.} $^a$ From \gaia~DR3 catalogue \citep{GaiaDR3}. $^b$ Membership probability according to \citet{Cantat2020}.
\end{table}
\end{landscape}

\addtocounter{table}{-1}
\begin{landscape}
\begin{table}

\caption{continued} 
\label{Tab:astrom}

\begin{center}
\begin{tabular}{lccccrccccc}
\hline\hline
\noalign{\smallskip}
NGC2232-S9  &   42563003  &  096.595911  &   -04.656253    &  3.1570  0.0166  & -4.746  0.015  &  -1.981  0.014 &  11.756686  &  0.746240 &   \dots  &  1.00 \\ 
NGC2232-S11 &   43359590  &  097.651170  &   -05.376364    &  2.7957  0.0147  & -4.900  0.015  &  -1.351  0.014 &  11.860096  &  0.722079 &  18.9535 &  0.50 \\ 
Ros6-S1     &   15065608  &  306.468650  &   +38.462216    &  2.8976  0.0097  &  5.849  0.010  &   2.079  0.011 &  12.220242  &  0.777939 &  28.9528 &  1.00 \\ 
Ros6-S2     &   14834047  &  306.454322  &   +38.791472    &  2.7869  0.0102  &  5.720  0.009  &   1.836  0.012 &  12.203323  &  0.769918 &  11.8291 &  1.00 \\ 
Ros6-S5     &   16409631  &  307.558255  &   +38.369788    &  2.8279  0.0110  &  5.978  0.010  &   2.074  0.014 &  12.389828  &  0.797299 &  12.3145 &  1.00 \\ 
Ros6-S9     &   15891271  &  307.184697  &   +40.556375    &  2.8307  0.0104  &  6.248  0.010  &   2.212  0.011 &  12.022573  &  0.738115 &   6.9740 &  1.00 \\ 
NGC7058-S1  &   63706522  &  320.451779  &   +50.835362    &  2.7443  0.0291  &  7.445  0.037  &   2.972  0.037 &   9.198004  &  0.229628 &   \dots  &  0.70 \\ 
NGC7058-S2  &   63706579  &  320.447802  &   +50.825512    &  2.7416  0.0141  &  7.424  0.018  &   2.399  0.016 &   9.662108  &  0.138198 &   \dots  &  0.80  \\ 
NGC7058-S3  &   63854407  &  320.769440  &   +50.780969    &  2.7058  0.0117  &  7.479  0.014  &   2.576  0.013 &  11.131710  &  0.588780 & 130.1522 &  1.00 \\ 
NGC7058-S4  &   63540102  &  320.274131  &   +50.821727    &  2.7069  0.0131  &  7.635  0.017  &   2.786  0.015 &  11.138639  &  0.608823 &   \dots  &  0.90 \\ 
NGC7058-S5  &   63706944  &  320.313006  &   +50.758604    &  2.7254  0.0107  &  7.323  0.012  &   2.601  0.012 &  11.577520  &  0.665600 &  36.9123 &  1.00 \\ 
NGC7058-S6  &  277385535  &  320.542333  &   +50.917663    &  2.7436  0.0088  &  7.572  0.011  &   2.869  0.010 &  12.169336  &  0.774905 &  35.6783 &  0.60 \\ 
NGC7058-S7  & 	 63706129   &  320.453566  &   +50.910386  &  2.7234  0.0092  &  7.512  0.011  &   2.636  0.010 &  12.337247  &  0.819340 &   8.6402 &  0.90 \\ 
NGC7058-S9  & 	63540039    &  320.138783  &   +50.806314  &  2.6768  0.0197  &  7.615  0.023  &   2.762  0.022 &  12.635520  &  0.889606 &   \dots  &  0.80 \\ 
NGC7058-S10 & 	63706595    &  320.450749  &   +50.822663  &  2.7190  0.0170  &  7.707  0.021  &   2.920  0.023 &  12.983939  &  0.952257 &   \dots  &  0.60 \\ 
NGC7058-S11 & 	 277384645   &  320.607971  &   +50.762039 &  2.7435  0.0101  &  7.612  0.012  &   2.764  0.012 &  13.135571  &  0.970726 &   \dots  &  1.00 \\ 
NGC7058-S12 & 	 63706150   &  320.331846  &   +50.906027  &  2.7152  0.0112  &  7.448  0.014  &   3.079  0.014 &  13.237492  &  0.987474 &   \dots  &  0.80  \\ 
NGC7058-S13 & 	  63539613   &  320.277208  &   +50.717575 &  2.7198  0.0106  &  7.411  0.012  &   2.883  0.013 &  13.291219  &  0.976221 &   \dots  &  0.90  \\ 
\noalign{\smallskip}
\hline
\end{tabular}
\end{center}
\end{table}
\end{landscape}

\begin{table*}
\caption{Chemical abundances ([X/H]), relative to solar values by \citet{Grevesse07}, for stars observed in ASCC\,16 (left) and Roslund\,6 (right).}\label{tab_abb_I}
\begin{center}
\begin{tabular}{lrrrr|rrrr}
\hline\hline
X  &          S1~~~~~  &         S2~~~~~   &         S6~~~~~   &          S12~~~~~ &         S1~~~~~   &         S2~~~~~   &         S5~~~~~   &         S9~~~~~   \\
\hline
C  &         \dots~~~~ &         \dots~~~~ &    0.26$\pm$0.12  &    0.22$\pm$0.10~  &    0.38$\pm$0.08  &    0.44$\pm$0.10  &        \dots~~~~  &    0.36$\pm$0.09  \\
Na & $-$0.01$\pm$0.11  &    0.04$\pm$0.08  &    0.07$\pm$0.10  &    0.09$\pm$0.09~  &    0.00$\pm$0.12  &    0.09$\pm$0.11  &    0.00$\pm$0.11  &    0.08$\pm$0.10  \\
Mg &    0.05$\pm$0.10  &    0.02$\pm$0.09  &    0.09$\pm$0.14  & $-$0.03$\pm$0.13~  &    0.05$\pm$0.09  &    0.00$\pm$0.11  & $-$0.04$\pm$0.09  & $-$0.03$\pm$0.10  \\
Al &    0.04$\pm$0.12  &    0.16$\pm$0.13  &    0.23$\pm$0.14  &    0.13$\pm$0.13~  &    0.00$\pm$0.09  &    0.16$\pm$0.08  &         \dots~~~~ &    0.14$\pm$0.10  \\
Si & $-$0.09$\pm$0.11  & $-$0.04$\pm$0.12  & $-$0.03$\pm$0.13  & $-$0.05$\pm$0.11~  & $-$0.12$\pm$0.15  &    0.02$\pm$0.12  & $-$0.15$\pm$0.13  & $-$0.03$\pm$0.19  \\
S  &         \dots~~~~ &         \dots~~~~ &    0.31$\pm$0.12  &    0.27$\pm$0.10~  &    0.25$\pm$0.08  &    0.06$\pm$0.10  &         \dots~~~~ &    0.25$\pm$0.10  \\
Ca &    0.13$\pm$0.11  &    0.21$\pm$0.12  &    0.18$\pm$0.14  &    0.06$\pm$0.12~  &    0.16$\pm$0.10  &    0.19$\pm$0.11  &    0.06$\pm$0.11  &    0.10$\pm$0.09  \\
Sc &    0.08$\pm$0.13  &    0.18$\pm$0.12  &    0.10$\pm$0.11  & $-$0.17$\pm$0.13~  &    0.06$\pm$0.10  &    0.03$\pm$0.09  & $-$0.07$\pm$0.10  & $-$0.11$\pm$0.12  \\
Ti &    0.12$\pm$0.13  &    0.24$\pm$0.11  &    0.19$\pm$0.12  &    0.02$\pm$0.14~  &    0.02$\pm$0.12  &    0.05$\pm$0.08  & $-$0.07$\pm$0.12  & $-$0.09$\pm$0.10  \\  
V  &    0.10$\pm$0.11  &    0.22$\pm$0.11  &    0.13$\pm$0.12  &    0.15$\pm$0.13~  &    0.15$\pm$0.11  &    0.15$\pm$0.12  &    0.07$\pm$0.11  &    0.08$\pm$0.11  \\
Cr &    0.08$\pm$0.12  &    0.19$\pm$0.12  &    0.14$\pm$0.14  &    0.02$\pm$0.14~  &    0.14$\pm$0.11  &    0.15$\pm$0.10  &    0.03$\pm$0.12  &    0.06$\pm$0.12  \\
Mn &    0.14$\pm$0.11  &    0.21$\pm$0.12  &    0.08$\pm$0.13  &    0.04$\pm$0.13~  &    0.16$\pm$0.13  &    0.11$\pm$0.12  &    0.07$\pm$0.12  &    0.05$\pm$0.13  \\
Fe & $-$0.01$\pm$0.12  &    0.09$\pm$0.11  &    0.04$\pm$0.12  & $-$0.14$\pm$0.14~  & $-$0.06$\pm$0.10  & $-$0.03$\pm$0.11  & $-$0.18$\pm$0.10  & $-$0.17$\pm$0.11  \\ 
Co &    0.11$\pm$0.12  &    0.21$\pm$0.12  &    0.17$\pm$0.14  &    0.06$\pm$0.13~  &    0.16$\pm$0.11  &    0.17$\pm$0.09  &    0.12$\pm$0.10  &    0.09$\pm$0.11  \\
Ni &    0.02$\pm$0.08  &    0.11$\pm$0.10  &    0.06$\pm$0.13  & $-$0.09$\pm$0.12~  & $-$0.01$\pm$0.12  &    0.03$\pm$0.13  & $-$0.16$\pm$0.13  & $-$0.10$\pm$0.11  \\
Cu &    0.03$\pm$0.11  &    0.02$\pm$0.12  & $-$0.04$\pm$0.13  & $-$0.21$\pm$0.14~  & $-$0.06$\pm$0.09  & $-$0.06$\pm$0.10  & $-$0.10$\pm$0.10  &         \dots~~~~ \\
Zn & $-$0.10$\pm$0.10  &    0.02$\pm$0.11  & $-$0.14$\pm$0.13  & $-$0.37$\pm$0.15~  & $-$0.29$\pm$0.11  & $-$0.25$\pm$0.10  & $-$0.19$\pm$0.11  &         \dots~~~~ \\
Sr &    0.14$\pm$0.10  &    0.18$\pm$0.09  &    0.21$\pm$0.12  &    0.03$\pm$0.12~  &    0.08$\pm$0.09  &    0.07$\pm$0.10  &    0.11$\pm$0.09  &    0.14$\pm$0.11  \\
Y  &    0.31$\pm$0.11  &    0.29$\pm$0.10  &    0.21$\pm$0.11  &    0.06$\pm$0.13~  &    0.21$\pm$0.10  &    0.20$\pm$0.10  &    0.20$\pm$0.10  &    0.16$\pm$0.11  \\
Zr &    0.16$\pm$0.12  &    0.21$\pm$0.13  &    0.26$\pm$0.15  &    0.03$\pm$0.13~  &    0.12$\pm$0.14  &    0.17$\pm$0.08  &    0.09$\pm$0.12  &    0.11$\pm$0.11  \\
Ba &    0.18$\pm$0.08  &    0.48$\pm$0.13  &         \dots~~~~ &    0.31$\pm$0.13~  &    0.23$\pm$0.12  &    0.21$\pm$0.13  &         \dots~~~~ &    0.31$\pm$0.13  \\
La &    0.26$\pm$0.10  &    0.37$\pm$0.12  &    0.29$\pm$0.14  &    0.19$\pm$0.13~  &    0.27$\pm$0.12  &    0.33$\pm$0.13  &    0.21$\pm$0.13  &    0.22$\pm$0.12  \\
Ce &    0.02$\pm$0.12  &    0.32$\pm$0.12  &    0.11$\pm$0.13  &    0.13$\pm$0.14~  &    0.04$\pm$0.09  &    0.14$\pm$0.08  &    0.16$\pm$0.10  &    0.13$\pm$0.09  \\
Pr &    0.13$\pm$0.10  &         \dots~~~~ &    0.21$\pm$0.12  &    0.13$\pm$0.12~  &         \dots~~~~ &         \dots~~~~ &         \dots~~~~ &         \dots~~~~ \\
Nd &    0.21$\pm$0.11  &    0.24$\pm$0.12  &    0.19$\pm$0.13  &    0.18$\pm$0.14~  &    0.29$\pm$0.12  &    0.29$\pm$0.11  &    0.14$\pm$0.12  &    0.19$\pm$0.11  \\
\hline
\end{tabular}
\end{center}
\end{table*}

\begin{table*}
\caption{Chemical abundances ([X/H]), relative to solar values by \citet{Grevesse07}, for stars observed in Melotte\,20.}\label{tab_abb_II}
\begin{center}
\begin{tabular}{lrrrrrrr}
\hline\hline
X  &           S1~~~~~ &          S2~~~~~  &         S3~~~~~   &          S8~~~~~  &          A1$^*$~~~~  &          A2~~~~~  &          A3~~~~~  \\
\hline
C  &         \dots~~~~ &         \dots~~~~ &    0.40$\pm$0.18  &         \dots~~~~ &         \dots~~~~ &         \dots~~~~ &    0.16$\pm$0.12  \\     
Na &    0.08$\pm$0.08  &    0.02$\pm$0.17  &    0.03$\pm$0.14  &    0.04$\pm$0.12  &    0.33$\pm$0.14  & $-$0.11$\pm$0.17  & $-$0.15$\pm$0.12  \\
Mg &    0.01$\pm$0.11  & $-$0.01$\pm$0.12  & $-$0.05$\pm$0.13  & $-$0.04$\pm$0.13  & $-$0.02$\pm$0.12  & $-$0.07$\pm$0.14  & $-$0.18$\pm$0.16  \\   
Al &    0.02$\pm$0.13  &    0.14$\pm$0.12  &    0.33$\pm$0.16  &    0.13$\pm$0.13  &    0.45$\pm$0.12  & $-$0.07$\pm$0.12  &    0.25$\pm$0.13  \\  
Si & $-$0.16$\pm$0.13  & $-$0.19$\pm$0.15  & $-$0.15$\pm$0.14  & $-$0.11$\pm$0.13  & $-$0.06$\pm$0.10  & $-$0.16$\pm$0.14  & $-$0.10$\pm$0.14  \\
S  &         \dots~~~~ &         \dots~~~~ &    0.37$\pm$0.12  &         \dots~~~~ &         \dots~~~~ &         \dots~~~~ &    0.09$\pm$0.10  \\ 
Ca &    0.16$\pm$0.11  &    0.17$\pm$0.12  &    0.16$\pm$0.09  &    0.13$\pm$0.11  & $-$0.03$\pm$0.13  &    0.02$\pm$0.12  & $-$0.14$\pm$0.16  \\
Sc & $-$0.06$\pm$0.11  & $-$0.05$\pm$0.12  & $-$0.07$\pm$0.09  & $-$0.10$\pm$0.12  &    0.06$\pm$0.14  & $-$0.09$\pm$0.12  & $-$0.12$\pm$0.13  \\  
Ti &    0.03$\pm$0.13  &    0.04$\pm$0.12  &    0.05$\pm$0.12  & $-$0.01$\pm$0.11  & $-$0.13$\pm$0.13  & $-$0.13$\pm$0.12  & $-$0.21$\pm$0.16  \\ 
V  &    0.14$\pm$0.11  &    0.16$\pm$0.11  &    0.15$\pm$0.14  &    0.12$\pm$0.10  &    0.06$\pm$0.11  &    0.04$\pm$0.14  & $-$0.01$\pm$0.13  \\   
Cr &    0.06$\pm$0.12  &    0.11$\pm$0.11  &    0.12$\pm$0.12  &    0.11$\pm$0.13  &    0.02$\pm$0.13  & $-$0.04$\pm$0.14  & $-$0.12$\pm$0.16  \\ 
Mn &    0.14$\pm$0.12  &    0.13$\pm$0.12  &    0.14$\pm$0.13  &    0.11$\pm$0.12  &    0.09$\pm$0.13  &    0.07$\pm$0.13  & $-$0.10$\pm$0.15  \\
Fe & $-$0.04$\pm$0.11  & $-$0.02$\pm$0.12  & $-$0.03$\pm$0.10  & $-$0.07$\pm$0.12  & $-$0.31$\pm$0.14  & $-$0.20$\pm$0.13  & $-$0.36$\pm$0.15  \\
Co &    0.08$\pm$0.10  &    0.09$\pm$0.10  &    0.14$\pm$0.11  &    0.08$\pm$0.11  &    0.11$\pm$0.12  &    0.07$\pm$0.12  & $-$0.04$\pm$0.11  \\
Ni & $-$0.08$\pm$0.12  & $-$0.05$\pm$0.09  & $-$0.04$\pm$0.11  & $-$0.07$\pm$0.10  & $-$0.07$\pm$0.13  & $-$0.11$\pm$0.11  & $-$0.24$\pm$0.15  \\   
Cu &    0.03$\pm$0.10  & $-$0.03$\pm$0.08  & $-$0.07$\pm$0.09  & $-$0.03$\pm$0.10  &    0.03$\pm$0.10  & $-$0.22$\pm$0.12  & $-$0.12$\pm$0.13  \\  
Zn & $-$0.24$\pm$0.08  & $-$0.21$\pm$0.10  & $-$0.22$\pm$0.11  & $-$0.22$\pm$0.09  & $-$0.20$\pm$0.15  & $-$0.25$\pm$0.12  & $-$0.50$\pm$0.14  \\
Sr &    0.16$\pm$0.12  &    0.12$\pm$0.07  &    0.09$\pm$0.08  &    0.17$\pm$0.12  &    0.23$\pm$0.15  &    0.24$\pm$0.07  &    0.30$\pm$0.12  \\
Y  &    0.22$\pm$0.13  &    0.25$\pm$0.11  &    0.24$\pm$0.13  &    0.21$\pm$0.10  &    0.12$\pm$0.13  &    0.16$\pm$0.16  &    0.07$\pm$0.14  \\
Zr &    0.11$\pm$0.12  &    0.08$\pm$0.11  &    0.16$\pm$0.13  &    0.17$\pm$0.14  &    0.04$\pm$0.13  &    0.06$\pm$0.14  &    0.02$\pm$0.12  \\ 
Mo &    0.19$\pm$0.18  &         \dots~~~~ &         \dots~~~~ &    0.25$\pm$0.17  &    0.25$\pm$0.14  &         \dots~~~~ &        \dots~~~~  \\
Ba &    0.40$\pm$0.13  &    0.22$\pm$0.12  &    0.50$\pm$0.18  &    0.35$\pm$0.13  & $-$0.21$\pm$0.17  &    0.40$\pm$0.10  & $-$0.06$\pm$0.15  \\   
La &    0.24$\pm$0.12  &    0.27$\pm$0.13  &    0.28$\pm$0.12  &    0.20$\pm$0.14  &    0.21$\pm$0.13  &    0.25$\pm$0.14  &    0.14$\pm$0.12  \\
Ce &    0.12$\pm$0.09  &    0.09$\pm$0.09  &    0.03$\pm$0.12  &    0.10$\pm$0.11  &    0.11$\pm$0.12  &    0.02$\pm$0.12  &    0.11$\pm$0.09  \\
Pr &         \dots~~~~ &    0.13$\pm$0.12  &    0.08$\pm$0.14  &         \dots~~~~ &         \dots~~~~ &         \dots~~~~ &    0.17$\pm$0.12  \\    
Nd &    0.28$\pm$0.12  &    0.27$\pm$0.13  &    0.21$\pm$0.12  &    0.24$\pm$0.13  &    0.14$\pm$0.12  &    0.16$\pm$0.15  &    0.09$\pm$0.14  \\

\hline
\end{tabular}
\end{center}
{\bf Note.} $^*$ Non member.
\end{table*}

\begin{table*}
\caption{Chemical abundances ([X/H]), relative to solar values by \citet{Grevesse07}, for stars observed in NGC\,2232 (left) and Pleiades (right).}\label{tab_abb_III}
\begin{center}
\begin{tabular}{lrrrrr|rr}
\hline\hline
X  &          S3~~~~~  &          S9~~~~~  &          S11~~~~~ &          A2~~~~~  &         A3~~~~~~   &             S3~~~~~   &         S5~~~~~   \\
\hline 
C  &         \dots~~~~ &         \dots~~~~ &    0.38$\pm$0.06  &        \dots~~~~  &         \dots~~~~~ &    0.21$\pm$0.12  &        \dots~~~~  \\
Na &    0.09$\pm$0.09  &    0.02$\pm$0.11  & $-$0.06$\pm$0.13  &    0.07$\pm$0.12  & $-$0.03$\pm$0.12~  &    0.02$\pm$0.14  &    0.12$\pm$0.12  \\
Mg & $-$0.02$\pm$0.14  & $-$0.01$\pm$0.09  & $-$0.01$\pm$0.15  &    0.01$\pm$0.13  & $-$0.03$\pm$0.07~  &    0.01$\pm$0.12  &    0.10$\pm$0.11  \\
Al &    0.16$\pm$0.15  &    0.01$\pm$0.12  &    0.07$\pm$0.10  &    0.18$\pm$0.17  &    0.07$\pm$0.18~  &    0.08$\pm$0.12  &    0.20$\pm$0.09  \\
Si & $-$0.06$\pm$0.14  & $-$0.09$\pm$0.12  & $-$0.06$\pm$0.15  & $-$0.11$\pm$0.11  & $-$0.16$\pm$0.16~  &    0.00$\pm$0.14  & $-$0.13$\pm$0.12  \\
S  &         \dots~~~~ &         \dots~~~~ &    0.29$\pm$0.12  &         \dots~~~~ &         \dots~~~~~ &    0.18$\pm$0.13  &    0.31$\pm$0.11  \\
Ca &    0.19$\pm$0.12  &    0.10$\pm$0.13  &    0.06$\pm$0.15  &    0.18$\pm$0.14  &    0.11$\pm$0.17~  &    0.02$\pm$0.09  &    0.16$\pm$0.12  \\
Sc &    0.05$\pm$0.12  &    0.03$\pm$0.13  &    0.02$\pm$0.17  &    0.02$\pm$0.14  & $-$0.01$\pm$0.16~  & $-$0.08$\pm$0.13  &    0.07$\pm$0.12  \\
Ti &    0.06$\pm$0.12  & $-$0.03$\pm$0.13  & $-$0.06$\pm$0.14  &    0.12$\pm$0.12  & $-$0.06$\pm$0.13~  & $-$0.06$\pm$0.13  &    0.16$\pm$0.12  \\
V  &    0.22$\pm$0.11  &    0.12$\pm$0.12  &    0.17$\pm$0.09  &    0.22$\pm$0.10  &    0.12$\pm$0.14~  &    0.19$\pm$0.14  &    0.17$\pm$0.12  \\
Cr &    0.12$\pm$0.11  &    0.03$\pm$0.13  &    0.00$\pm$0.14  &    0.11$\pm$0.13  &    0.04$\pm$0.13~  &    0.01$\pm$0.13  &    0.19$\pm$0.09  \\
Mn &    0.14$\pm$0.09  &    0.07$\pm$0.15  &    0.11$\pm$0.13  &    0.14$\pm$0.11  &    0.08$\pm$0.09~  &    0.00$\pm$0.08  &    0.23$\pm$0.10  \\
Fe & $-$0.01$\pm$0.10  & $-$0.18$\pm$0.11  & $-$0.21$\pm$0.14  &    0.00$\pm$0.11  & $-$0.15$\pm$0.12~  & $-$0.17$\pm$0.13  &    0.12$\pm$0.09  \\ 
Co &    0.14$\pm$0.09  &    0.07$\pm$0.10  &    0.09$\pm$0.11  &    0.14$\pm$0.10  &    0.05$\pm$0.10~  &    0.13$\pm$0.12  &    0.22$\pm$0.13  \\
Ni & $.$0.03$\pm$0.12  & $-$0.11$\pm$0.13  & $-$0.13$\pm$0.13  & $-$0.01$\pm$0.10  & $-$0.15$\pm$0.13~  & $-$0.04$\pm$0.12  &    0.08$\pm$0.06  \\ 
Cu & $-$0.09$\pm$0.12  &    0.07$\pm$0.14  & $-$0.24$\pm$0.17  &    0.00$\pm$0.14  &    0.12$\pm$0.15~  & $-$0.26$\pm$0.18  &    0.12$\pm$0.15  \\  
Zn & $-$0.24$\pm$0.07  & $-$0.27$\pm$0.06  & $-$0.37$\pm$0.07  & $-$0.17$\pm$0.10  & $-$0.27$\pm$0.13~  & $-$0.43$\pm$0.14  & $-$0.15$\pm$0.12  \\
Sr &    0.10$\pm$0.06  &    0.06$\pm$0.08  &    0.14$\pm$0.17  &    0.03$\pm$0.06  &    0.02$\pm$0.14~  &    0.00$\pm$0.15  &    0.15$\pm$0.12  \\
Y  &    0.22$\pm$0.15  &    0.24$\pm$0.16  &    0.10$\pm$0.17  &    0.28$\pm$0.13  &    0.20$\pm$0.14~  &    0.11$\pm$0.11  &    0.29$\pm$0.13  \\
Zr &    0.16$\pm$0.06  &    0.16$\pm$0.13  &    0.10$\pm$0.16  &    0.18$\pm$0.10  &    0.13$\pm$0.09~  &    0.15$\pm$0.12  &    0.27$\pm$0.11  \\
Ba &    0.50$\pm$0.06  &    0.46$\pm$0.09  &    0.35$\pm$0.14  &    0.49$\pm$0.10  &    0.45$\pm$0.08~  &    0.24$\pm$0.13  &    0.30$\pm$0.12  \\
La &    0.26$\pm$0.10  &    0.19$\pm$0.11  &    0.27$\pm$0.17  &    0.30$\pm$0.15  &    0.25$\pm$0.15~  &    0.31$\pm$0.10  &    0.26$\pm$0.07  \\
Ce &    0.23$\pm$0.14  &    0.07$\pm$0.13  &    0.09$\pm$0.13  &    0.21$\pm$0.13  &    0.15$\pm$0.13~  &    0.19$\pm$0.14  &    0.16$\pm$0.12  \\
Pr &    0.21$\pm$0.13  &    0.15$\pm$0.07  &         \dots~~~~ &    0.21$\pm$0.12  &    0.29$\pm$0.12~  &    0.15$\pm$0.14  &    0.20$\pm$0.13  \\
Nd &    0.30$\pm$0.13  &    0.16$\pm$0.11  &    0.20$\pm$0.15  &    0.18$\pm$0.12  &    0.18$\pm$0.12~  &    0.21$\pm$0.13  &    0.28$\pm$0.11  \\
\hline
\end{tabular}
\end{center}
\end{table*}

\begin{table*}
\caption{Chemical abundances ([X/H]), relative to solar values by \citet{Grevesse07}, for stars observed in ASCC\,19.}\label{tab_abb_IV}
\tiny  
\begin{center}
\begin{tabular}{lrrrrrrrrr}
\hline\hline
X  &         S1~~~~~   &         S2~~~~~   &         S3~~~~~   &          S4~~~~~  &         S5~~~~~   &         S6~~~~~   &         S14~~~~~  &         S15~~~~~  &        S16~~~~~  \\
\hline
C  &    0.11$\pm$0.14  &    0.25$\pm$0.15  &    0.20$\pm$0.16  &         \dots~~~~ &    0.16$\pm$0.12  &         \dots~~~~ &    0.24$\pm$0.12  &    0.13$\pm$0.13  &    0.16$\pm$0.11  \\ 
Na &    0.23$\pm$0.15  & $-$0.13$\pm$0.17  &    0.08$\pm$0.16  &    0.20$\pm$0.12  &    0.00$\pm$0.13  &    0.17$\pm$0.09  & $-$0.06$\pm$0.14  & $-$0.28$\pm$0.16  &    0.16$\pm$0.13  \\
Mg &    0.02$\pm$0.11  & $-$0.05$\pm$0.12  & $-$0.05$\pm$0.13  &    0.08$\pm$0.09  & $-$0.01$\pm$0.09  &    0.02$\pm$0.10  & $-$0.12$\pm$0.08  & $-$0.06$\pm$0.12  & $-$0.20$\pm$0.09  \\
Al &         \dots~~~~ &         \dots~~~~ &         \dots~~~~ &         \dots~~~  &    0.10$\pm$0.12  &    0.20$\pm$0.13  &    0.18$\pm$0.13  &         \dots~~~~ &    0.19$\pm$0.12  \\
Si &    0.06$\pm$0.10  & $-$0.09$\pm$0.13  & $-$0.07$\pm$0.12  &    0.05$\pm$0.11  & $-$0.14$\pm$0.09  & $-$0.01$\pm$0.09  & $-$0.12$\pm$0.09  & $-$0.13$\pm$0.11  &    0.05$\pm$0.10  \\
S  &    0.21$\pm$0.13  &         \dots~~~~ &         \dots~~~~ &         \dots~~~~ &    0.19$\pm$0.11  &    0.25$\pm$0.11  &    0.20$\pm$0.12  &    0.18$\pm$0.12  &    0.27$\pm$0.23  \\ 
Ca &    0.06$\pm$0.12  &    0.05$\pm$0.14  &    0.06$\pm$0.15  &    0.16$\pm$0.11  &    0.08$\pm$0.10  &    0.16$\pm$0.10  &    0.04$\pm$0.10  &    0.10$\pm$0.11  &    0.01$\pm$0.08  \\
Sc & $-$0.08$\pm$0.11  & $-$0.13$\pm$0.14  & $-$0.13$\pm$0.13  &    0.03$\pm$0.08  & $-$0.11$\pm$0.10  &    0.00$\pm$0.09  & $-$0.10$\pm$0.09  & $-$0.11$\pm$0.12  & $-$0.15$\pm$0.10  \\
Ti &    0.01$\pm$0.12  & $-$0.04$\pm$0.14  & $-$0.02$\pm$0.13  &    0.18$\pm$0.11  & $-$0.04$\pm$0.11  &    0.07$\pm$0.10  & $-$0.01$\pm$0.11  &    0.03$\pm$0.13  & $-$0.06$\pm$0.12  \\
V  &    0.23$\pm$0.12  &    0.12$\pm$0.14  &    0.00$\pm$0.14  &    0.27$\pm$0.12  &    0.10$\pm$0.11  &    0.34$\pm$0.11  &    0.15$\pm$0.10  &    0.14$\pm$0.12  &    0.10$\pm$0.11  \\
Cr &    0.03$\pm$0.12  & $-$0.12$\pm$0.15  & $-$0.01$\pm$0.13  &    0.25$\pm$0.12  & $-$0.02$\pm$0.08  &    0.21$\pm$0.11  &    0.03$\pm$0.09  & $-$0.05$\pm$0.12  & $-$0.06$\pm$0.10  \\
Mn & $-$0.09$\pm$0.13  & $-$0.08$\pm$0.13  & $-$0.05$\pm$0.12  &    0.12$\pm$0.09  & $-$0.07$\pm$0.11  &    0.06$\pm$0.10  & $-$0.04$\pm$0.10  & $-$0.06$\pm$0.12  &    0.00$\pm$0.09  \\
Fe & $-$0.17$\pm$0.13  & $-$0.29$\pm$0.15  & $-$0.22$\pm$0.14  &    0.15$\pm$0.12  & $-$0.22$\pm$0.11  &    0.06$\pm$0.10  & $-$0.23$\pm$0.10  & $-$0.20$\pm$0.13  & $-$0.24$\pm$0.11  \\ 
Co &    0.15$\pm$0.12  &    0.12$\pm$0.15  &    0.22$\pm$0.16  &    0.26$\pm$0.13  &    0.10$\pm$0.12  &    0.23$\pm$0.12  &    0.20$\pm$0.12  &    0.15$\pm$0.13  &    0.22$\pm$0.11  \\
Ni & $-$0.05$\pm$0.12  & $-$0.09$\pm$0.14  &    0.02$\pm$0.13  &    0.11$\pm$0.09  & $-$0.11$\pm$0.09  &    0.00$\pm$0.10  & $-$0.09$\pm$0.10  & $-$0.12$\pm$0.12  & $-$0.18$\pm$0.10  \\
Cu &    0.00$\pm$0.13  & $-$0.13$\pm$0.14  & $-$0.13$\pm$0.15  & $-$0.03$\pm$0.12  & $-$0.22$\pm$0.11  & $-$0.04$\pm$0.11  & $-$0.35$\pm$0.12  &         \dots~~~~ & $-$0.06$\pm$0.11  \\
Zn &         \dots~~~~ &         \dots~~~~ & $-$0.19$\pm$0.16  & $-$0.19$\pm$0.12  & $-$0.39$\pm$0.13  & $-$0.18$\pm$0.12  & $-$0.39$\pm$0.13  &         \dots~~~~ & $-$0.31$\pm$0.12  \\
Sr &    0.05$\pm$0.12  &         \dots~~~~ &         \dots~~~~ &         \dots~~~~ & $-$0.02$\pm$0.10  &    0.32$\pm$0.12  &    0.19$\pm$0.08  &    0.18$\pm$0.11  &    0.02$\pm$0.09  \\
Y  & $-$0.04$\pm$0.12  & $-$0.14$\pm$0.14  &    0.04$\pm$0.14  &    0.30$\pm$0.13  &    0.15$\pm$0.11  &    0.23$\pm$0.12  &    0.12$\pm$0.09  & $-$0.02$\pm$0.12  & $-$0.01$\pm$0.10  \\
Zr &    0.15$\pm$0.13  &         \dots~~~~ &    0.12$\pm$0.14  &         \dots~~~~ &    0.17$\pm$0.13  &    0.11$\pm$0.11  &    0.19$\pm$0.11  &    0.22$\pm$0.13  &    0.18$\pm$0.13  \\
Ba &    0.27$\pm$0.13  &    0.21$\pm$0.15  &    0.27$\pm$0.16  &         \dots~~~~ &         \dots~~~~ &    0.17$\pm$0.11  &    0.34$\pm$0.12  &    0.39$\pm$0.15  &    0.18$\pm$0.09  \\   
La &    0.21$\pm$0.11  &         \dots~~~~ &    0.16$\pm$0.13  &         \dots~~~~ &    0.23$\pm$0.09  &    0.20$\pm$0.09  &    0.27$\pm$0.10  &    0.25$\pm$0.10  &    0.20$\pm$0.09  \\
Ce &    0.21$\pm$0.13  &    0.16$\pm$0.15  &         \dots~~~~ &         \dots~~~~ &    0.10$\pm$0.11  &    0.26$\pm$0.12  &    0.17$\pm$0.11  &    0.20$\pm$0.12  &    0.14$\pm$0.12  \\
Pr &    0.20$\pm$0.11  &         \dots~~~~ &         \dots~~~~ &         \dots~~~~ &         \dots~~~~ &         \dots~~~~ &         \dots~~~~ &    0.21$\pm$0.12  &    0.22$\pm$0.11  \\
Nd &    0.19$\pm$0.14  &    0.19$\pm$0.16  &    0.16$\pm$0.08  &         \dots~~~~ &    0.13$\pm$0.12  &    0.19$\pm$0.12  &    0.18$\pm$0.13  &         \dots~~~~ &    0.11$\pm$0.12  \\
\hline
\end{tabular}
\end{center}
\end{table*}

\begin{table*}
\caption{Chemical abundances ([X/H]), relative to solar values by \citet{Grevesse07}, for stars observed in NGC\,7058.}\label{tab_abb_V}
\begin{center}
\begin{tabular}{lrrrrr}
\hline\hline
X  &          S3~~~~~  &          S4~~~~~  &          S5~~~~~  &          S6~~~~~  &          S7~~~~~  \\          
\hline
C  & $-$0.07$\pm$0.14  &    0.13$\pm$0.13  &    0.10$\pm$0.11  &    0.32$\pm$0.10  &         \dots~~~~ \\   
Na &    0.15$\pm$0.13  &    0.16$\pm$0.14  &    0.16$\pm$0.11  &    0.14$\pm$0.10  &    0.24$\pm$0.10  \\
Mg & $-$0.02$\pm$0.13  &    0.00$\pm$0.12  &    0.02$\pm$0.11  &    0.07$\pm$0.09  &    0.14$\pm$0.10  \\
Al &         \dots~~~~ &         \dots~~~~ &    0.10$\pm$0.10  &         \dots~~~~ &    0.10$\pm$0.12  \\
Si & $-$0.06$\pm$0.15  & $-$0.06$\pm$0.13  & $-$0.09$\pm$0.13  & $-$0.08$\pm$0.12  & $-$0.14$\pm$0.11  \\
S  &    0.19$\pm$0.12  &    0.28$\pm$0.12  &    0.31$\pm$0.10  &         \dots~~~~ &    0.19$\pm$0.10  \\
Ca &    0.04$\pm$0.13  &    0.00$\pm$0.14  & $-$0.09$\pm$0.11  &    0.10$\pm$0.10  &    0.24$\pm$0.10  \\
Sc & $-$0.11$\pm$0.12  & $-$0.12$\pm$0.13  & $-$0.20$\pm$0.11  & $-$0.11$\pm$0.10  &    0.03$\pm$0.12  \\
Ti & $-$0.04$\pm$0.15  & $-$0.01$\pm$0.14  &    0.01$\pm$0.12  &    0.01$\pm$0.11  &    0.11$\pm$0.12  \\
V  &    0.10$\pm$0.14  &    0.00$\pm$0.13  &    0.12$\pm$0.12  &    0.08$\pm$0.11  &    0.16$\pm$0.10  \\
Cr & $-$0.03$\pm$0.13  & $-$0.03$\pm$0.14  & $-$0.06$\pm$0.11  &    0.03$\pm$0.11  &    0.15$\pm$0.12  \\
Mn &    0.02$\pm$0.13  &    0.00$\pm$0.13  & $-$0.06$\pm$0.11  &    0.10$\pm$0.12  &    0.13$\pm$0.12  \\
Fe & $-$0.12$\pm$0.14  & $-$0.11$\pm$0.15  & $-$0.15$\pm$0.13  & $-$0.06$\pm$0.12  &    0.00$\pm$0.09  \\
Co &    0.21$\pm$0.13  &    0.25$\pm$0.13  &    0.15$\pm$0.11  &    0.16$\pm$0.12  &    0.14$\pm$0.11  \\
Ni &    0.02$\pm$0.14  & $-$0.01$\pm$0.13  & $-$0.02$\pm$0.10  & $-$0.01$\pm$0.10  &    0.09$\pm$0.11  \\
Cu &         \dots~~~~ &         \dots~~~~ & $-$0.19$\pm$0.12  & $-$0.06$\pm$0.09  & $-$0.07$\pm$0.10  \\
Zn &         \dots~~~~ & $-$0.25$\pm$0.15  & $-$0.31$\pm$0.14  & $-$0.07$\pm$0.12  & $-$0.07$\pm$0.11  \\
Sr &         \dots~~~~ &    0.05$\pm$0.15  &    0.06$\pm$0.12  &    0.35$\pm$0.13  &    0.10$\pm$0.12  \\
Y  &    0.01$\pm$0.13  &    0.11$\pm$0.13  &    0.00$\pm$0.09  &    0.09$\pm$0.10  &    0.23$\pm$0.11  \\  
Zr &    0.16$\pm$0.13  &    0.14$\pm$0.13  &    0.17$\pm$0.11  &    0.12$\pm$0.12  &    0.07$\pm$0.10  \\
Ba &    0.23$\pm$0.14  &         \dots~~~~ &    0.37$\pm$0.13  &    0.35$\pm$0.13  &         \dots~~~~ \\
La &    0.06$\pm$0.14  &    0.11$\pm$0.13  &    0.25$\pm$0.12  &    0.36$\pm$0.10  &    0.33$\pm$0.12  \\
Ce &    0.06$\pm$0.14  &    0.02$\pm$0.14  &    0.13$\pm$0.12  &    0.04$\pm$0.19  &    0.13$\pm$0.12  \\
Pr &    0.13$\pm$0.13  &    0.13$\pm$0.14  &         \dots~~~~ &         \dots~~~~ &         \dots~~~~ \\
Nd &    0.18$\pm$0.13  &    0.15$\pm$0.14  &    0.17$\pm$0.12  &    0.27$\pm$0.12  &    0.31$\pm$0.11  \\
   &                   &                   &                   &                   &                   \\  
\hline\hline
X  &          S9~~~~~  &          S10~~~~~ &          S11~~~~~ &          S12~~~~~ &          S13~~~~~ \\
\hline
C  &         \dots~~~~ &         \dots~~~~ &         \dots~~~~ &         \dots~~~~ &         \dots~~~~ \\   
Na &    0.15$\pm$0.09  &    0.20$\pm$0.08  &    0.19$\pm$0.11  &    0.16$\pm$0.10  &    0.11$\pm$0.09  \\
Mg &    0.07$\pm$0.10  &    0.08$\pm$0.11  &    0.17$\pm$0.10  &    0.17$\pm$0.09  &    0.01$\pm$0.08  \\
Al &    0.12$\pm$0.12  &    0.18$\pm$0.12  &    0.20$\pm$0.13  &         \dots~~~~ &    0.14$\pm$0.09  \\
Si & $-$0.10$\pm$0.12  & $-$0.12$\pm$0.12  & $-$0.08$\pm$0.12  & $-$0.16$\pm$0.11  & $-$0.11$\pm$0.13  \\
S  &         \dots~~~~ &         \dots~~~~ &         \dots~~~~ &         \dots~~~~ &         \dots~~~~ \\
Ca &    0.08$\pm$0.09  &    0.21$\pm$0.11  &    0.33$\pm$0.12  &    0.16$\pm$0.11  &    0.07$\pm$0.10  \\
Sc &    0.00$\pm$0.11  &    0.04$\pm$0.11  &    0.08$\pm$0.11  &    0.06$\pm$0.12  & $-$0.05$\pm$0.12  \\
Ti &    0.04$\pm$0.10  &    0.13$\pm$0.12  &    0.19$\pm$0.12  &    0.14$\pm$0.11  & $-$0.10$\pm$0.12  \\
V  &    0.10$\pm$0.12  &    0.16$\pm$0.09  &    0.30$\pm$0.10  &    0.18$\pm$0.08  &    0.03$\pm$0.11  \\
Cr &    0.04$\pm$0.10  &    0.16$\pm$0.12  &    0.27$\pm$0.12  &    0.19$\pm$0.12  & $-$0.01$\pm$0.12  \\
Mn &    0.12$\pm$0.11  &    0.13$\pm$0.10  &    0.22$\pm$0.12  &    0.20$\pm$0.11  &    0.00$\pm$0.10  \\
Fe & $-$0.06$\pm$0.10  &    0.02$\pm$0.09  &    0.19$\pm$0.12  &    0.07$\pm$0.12  & $-$0.11$\pm$0.13  \\
Co &    0.14$\pm$0.12  &    0.17$\pm$0.09  &    0.23$\pm$0.11  &    0.06$\pm$0.10  &    0.02$\pm$0.11  \\
Ni &    0.03$\pm$0.11  &    0.04$\pm$0.12  &    0.09$\pm$0.09  &    0.05$\pm$0.10  & $-$0.06$\pm$0.11  \\
Cu & $-$0.04$\pm$0.10  & $-$0.02$\pm$0.10  &    0.14$\pm$0.11  &    0.06$\pm$0.11  &         \dots~~~~ \\
Zn & $-$0.19$\pm$0.12  & $-$0.13$\pm$0.12  & $-$0.22$\pm$0.12  & $-$0.15$\pm$0.11  &         \dots~~~~ \\
Sr &         \dots~~~~ &    0.05$\pm$0.12  &    0.19$\pm$0.13  &         \dots~~~~ &    0.30$\pm$0.13  \\
Y  &    0.16$\pm$0.09  &    0.10$\pm$0.09  &    0.21$\pm$0.12  &    0.14$\pm$0.11  &    0.15$\pm$0.14  \\  
Zr &    0.17$\pm$0.11  &    0.10$\pm$0.10  &    0.23$\pm$0.12  &    0.16$\pm$0.11  &    0.02$\pm$0.12  \\
Ba &    0.26$\pm$0.11  &    0.27$\pm$0.12  &         \dots~~~~ &         \dots~~~~ &    0.33$\pm$0.11  \\
La &         \dots~~~~ &         \dots~~~~ &    0.23$\pm$0.12  &         \dots~~~~ &    0.26$\pm$0.11  \\
Ce &         \dots~~~~ &    0.27$\pm$0.10  &         \dots~~~~ &         \dots~~~~ &    0.24$\pm$0.12  \\
Pr &         \dots~~~~ &         \dots~~~~ &    0.38$\pm$0.11  &         \dots~~~~ &         \dots~~~~ \\
Nd &         \dots~~~~ &    0.24$\pm$0.12  &    0.30$\pm$0.11  &    0.30$\pm$0.12  &    0.27$\pm$0.12  \\
\hline
\end{tabular}
\end{center}
\end{table*}

\begin{figure*}
\begin{center}
\includegraphics[width=9.cm]{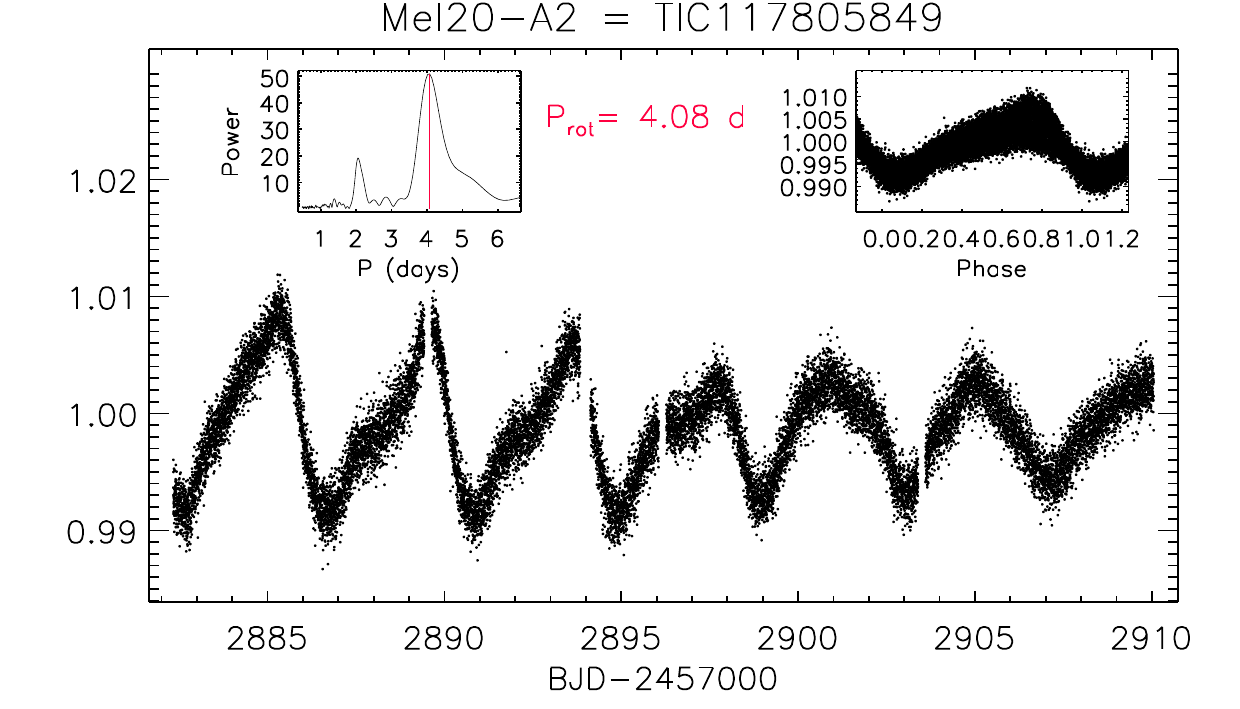}	
\includegraphics[width=9.cm]{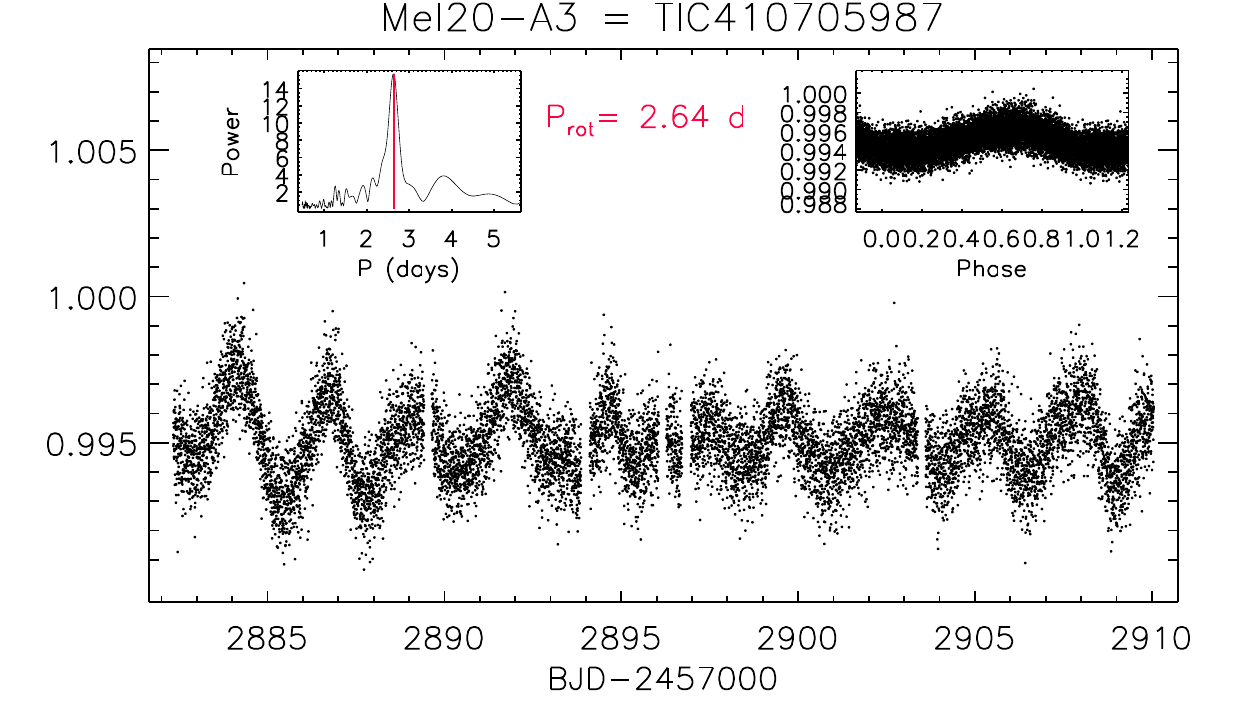}	
\includegraphics[width=9.cm]{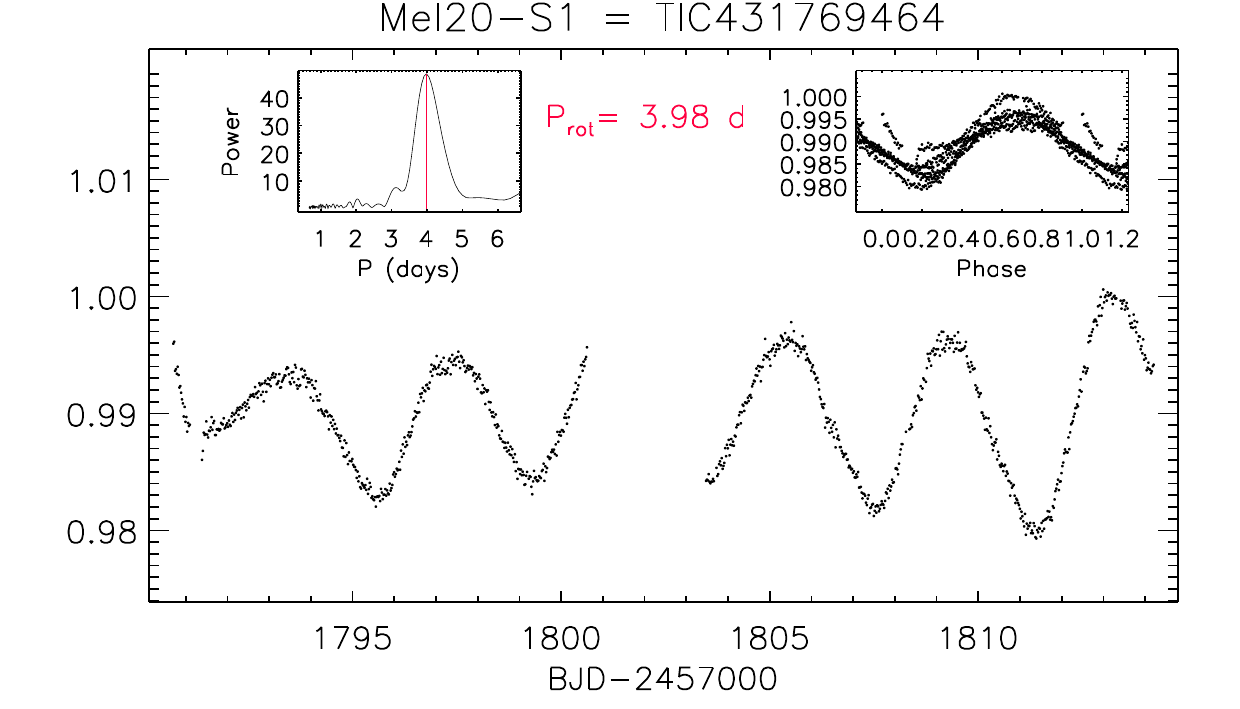}	
\includegraphics[width=9.cm]{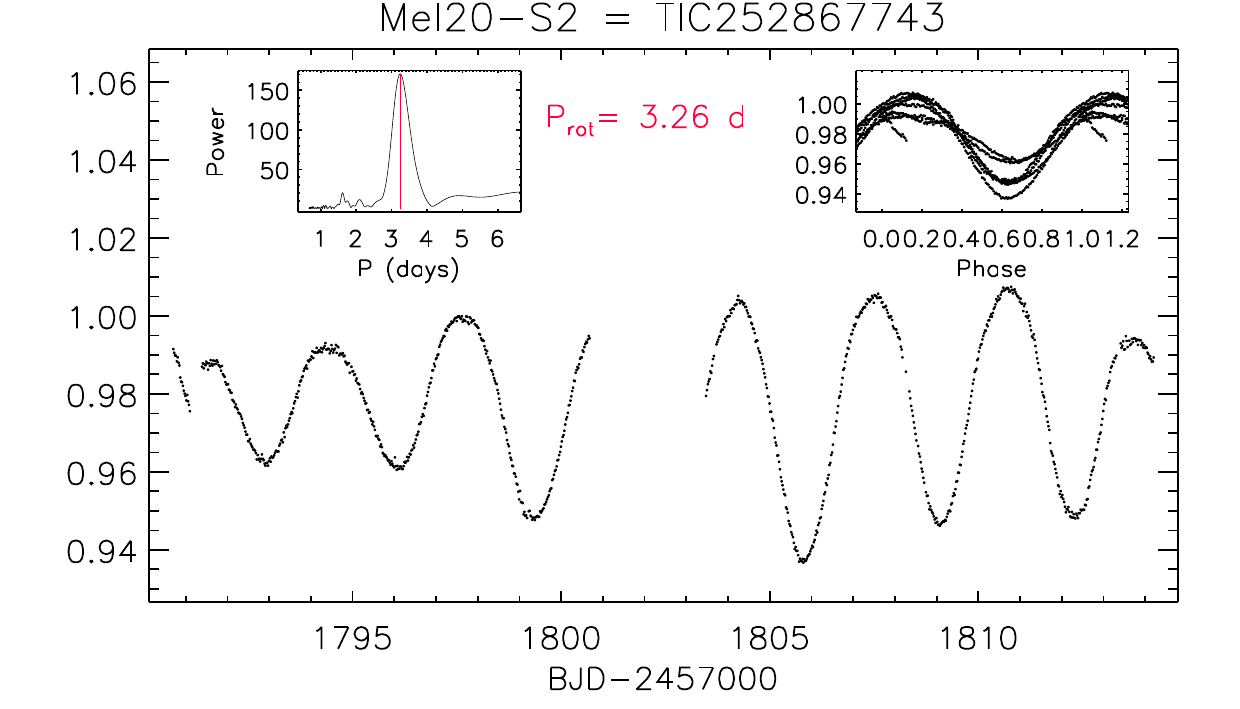}	
\includegraphics[width=9.cm]{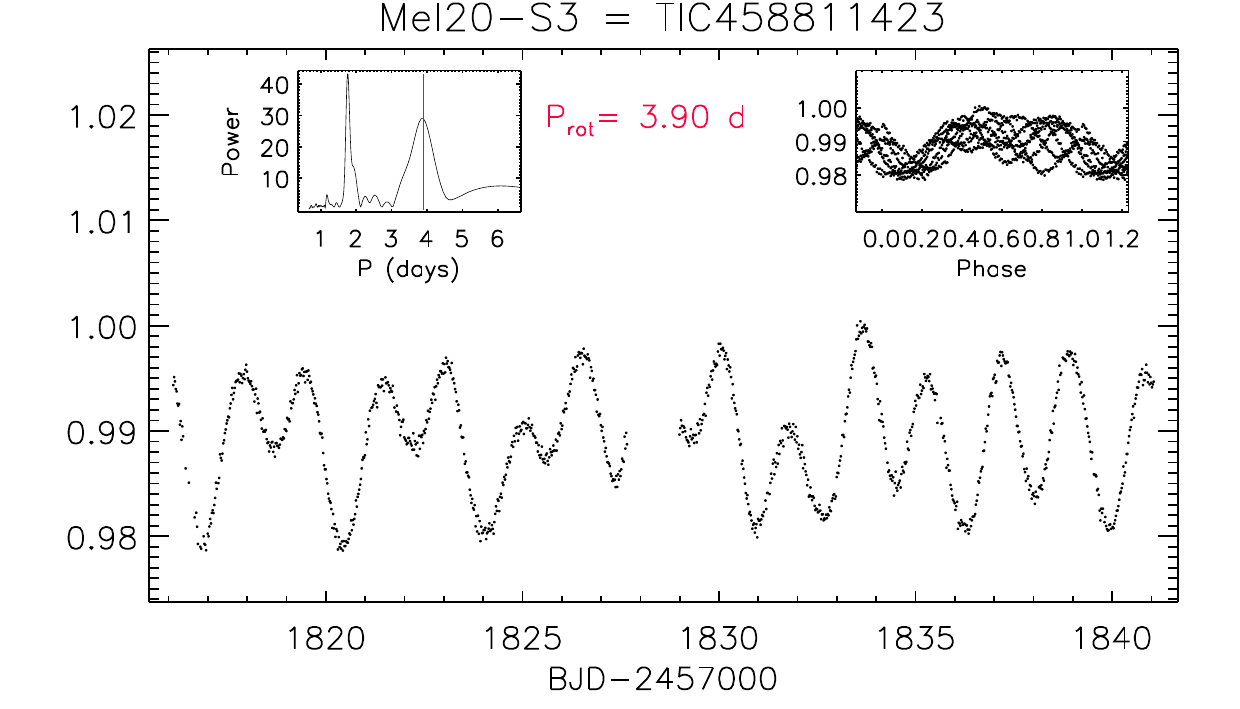}	
\includegraphics[width=9.cm]{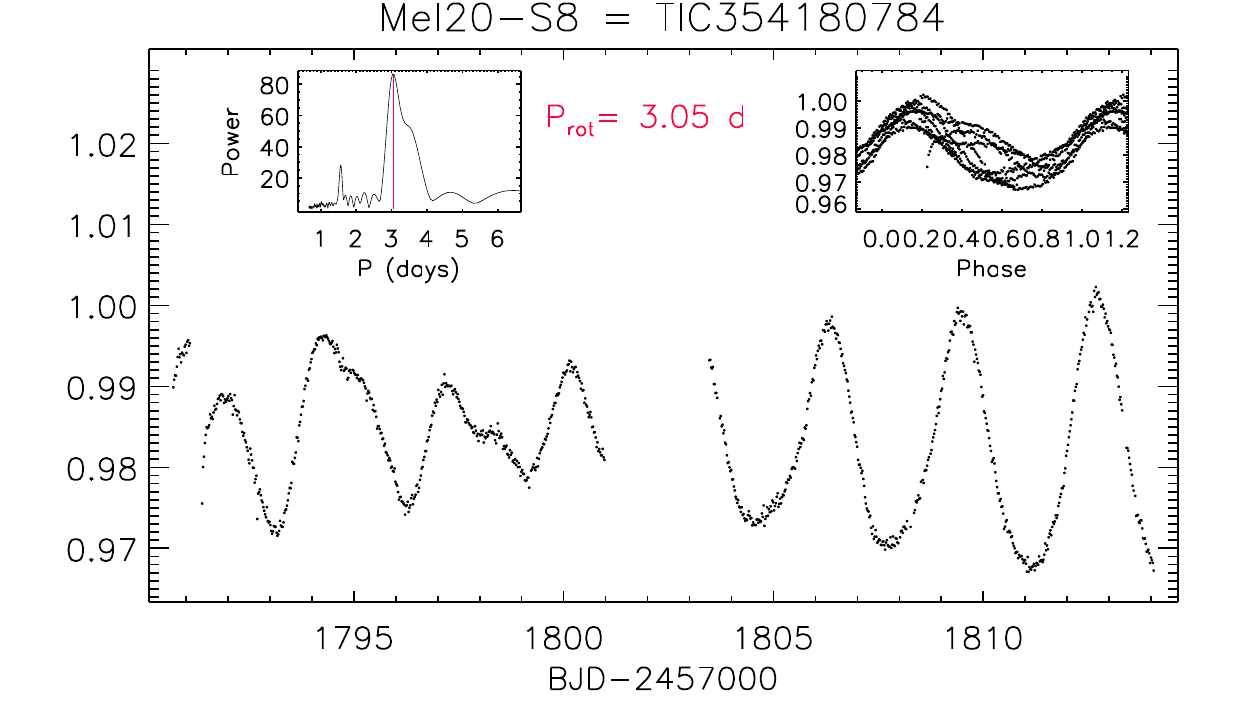}	
\includegraphics[width=9.cm]{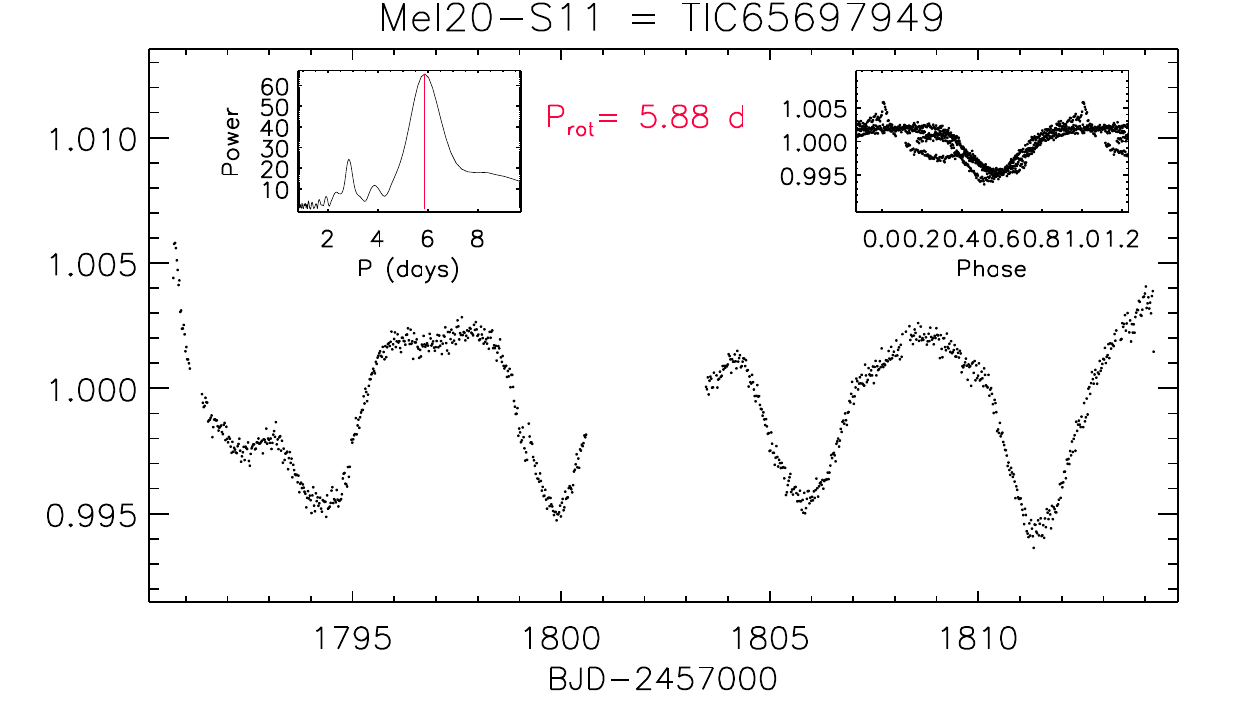}	
\vspace{0cm}
\caption{{\it TESS} light curves for the four members of Melotte~20. In each panel, the inset in the upper left corner shows the 
cleaned periodogram, with the period marked by a vertical red line and indicated with red characters. The inset in the upper right corner displays the data phased with this period.}
\label{fig:TESS_Mel20}
\end{center}
\end{figure*}

\begin{figure*}
\begin{center}
\includegraphics[width=9.cm]{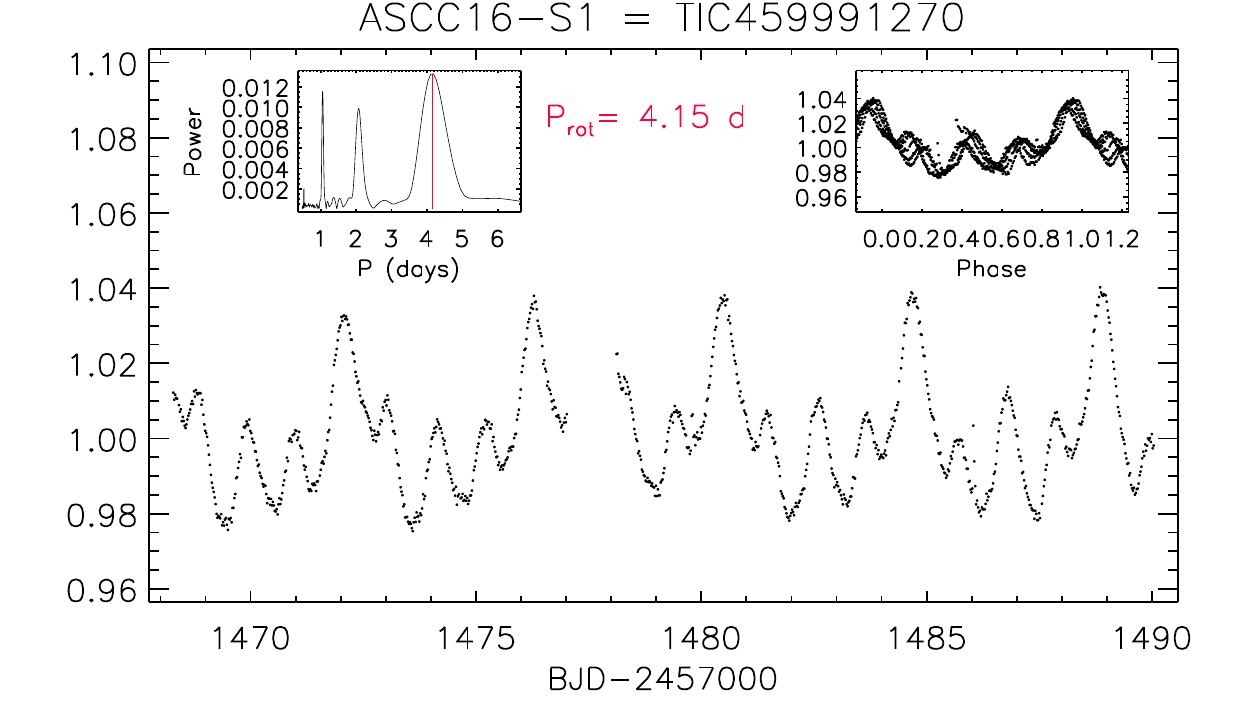}	
\includegraphics[width=9.cm]{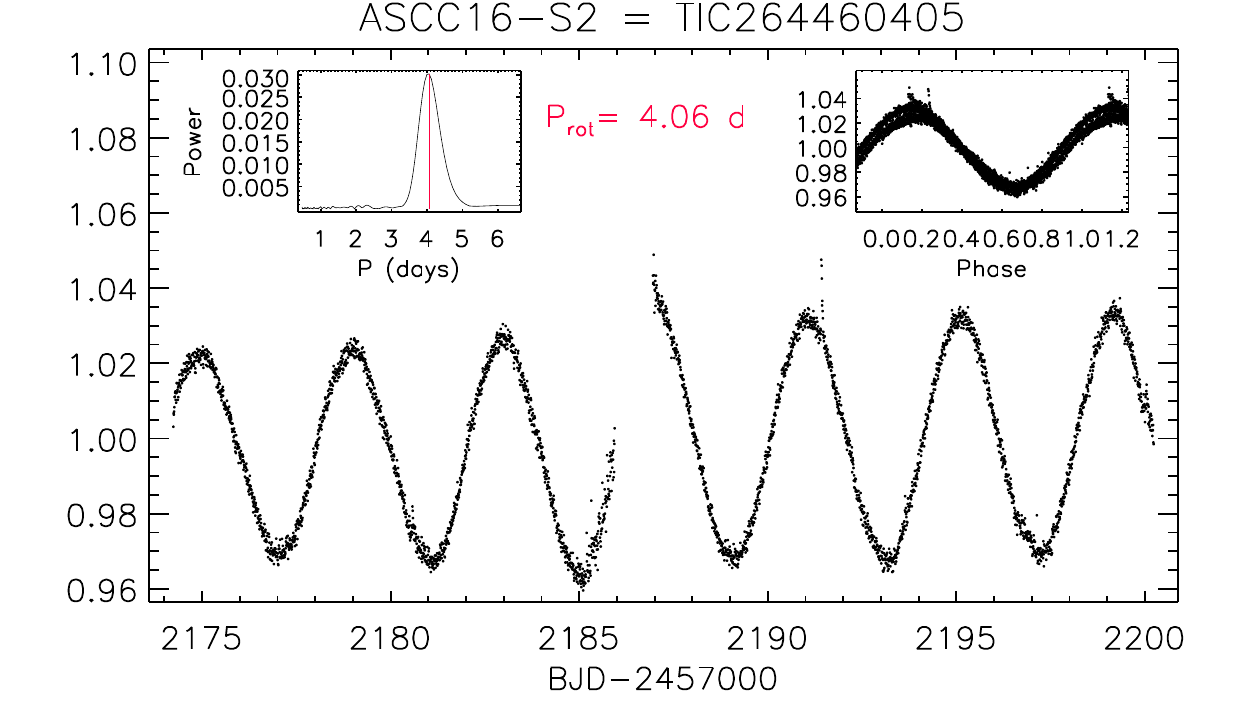}	
\includegraphics[width=9.cm]{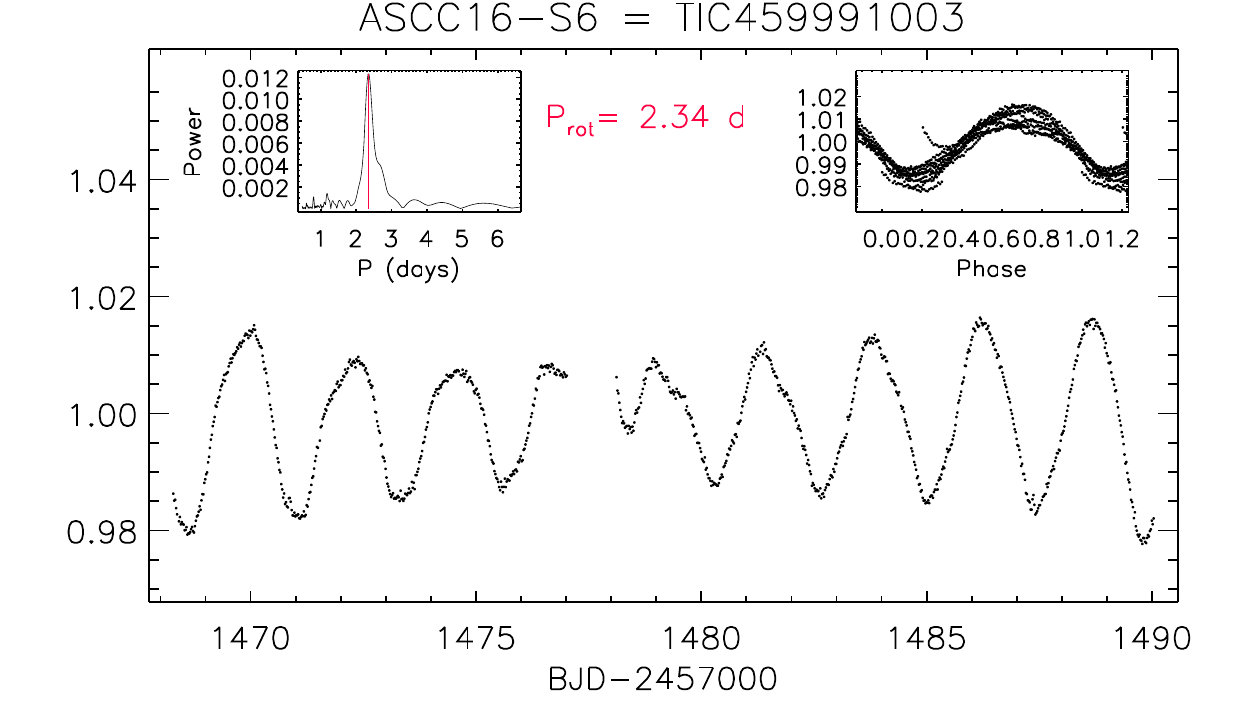}	
\includegraphics[width=9.cm]{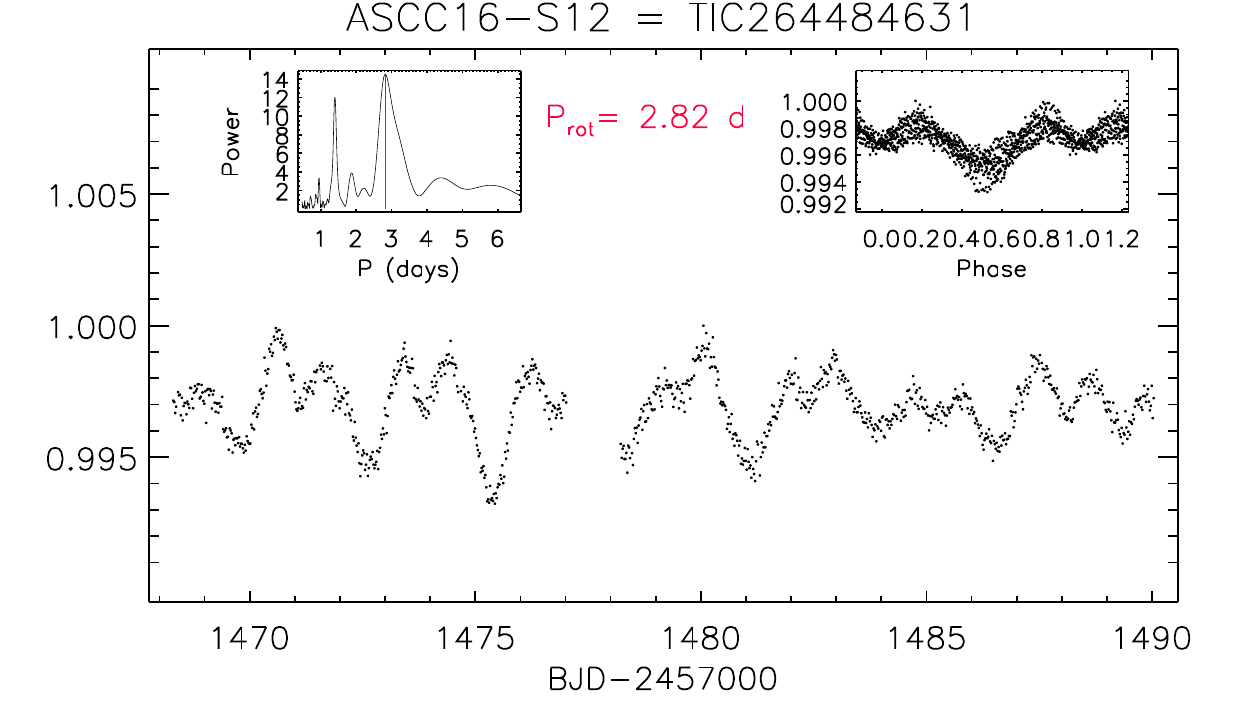}	
\vspace{0cm}
\caption{{\it TESS} light curves for the four members of ASCC~16. In each panel, the inset in the upper left corner shows the 
cleaned periodogram, with the period marked by a vertical red line and indicated with red characters. The inset in the upper right corner displays the data phased with this period.}
\label{fig:TESS_ASCC16}
\end{center}
\end{figure*}

\begin{figure*}
\begin{center}
\includegraphics[width=9.cm]{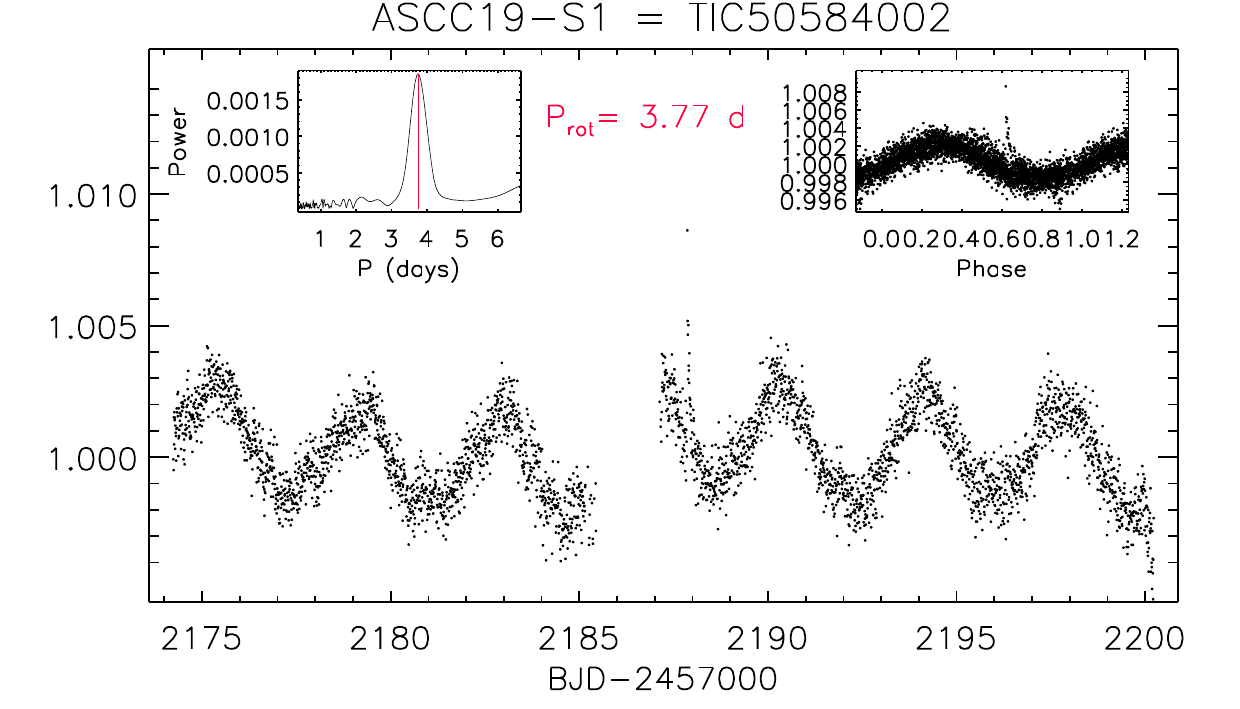}	
\includegraphics[width=9.cm]{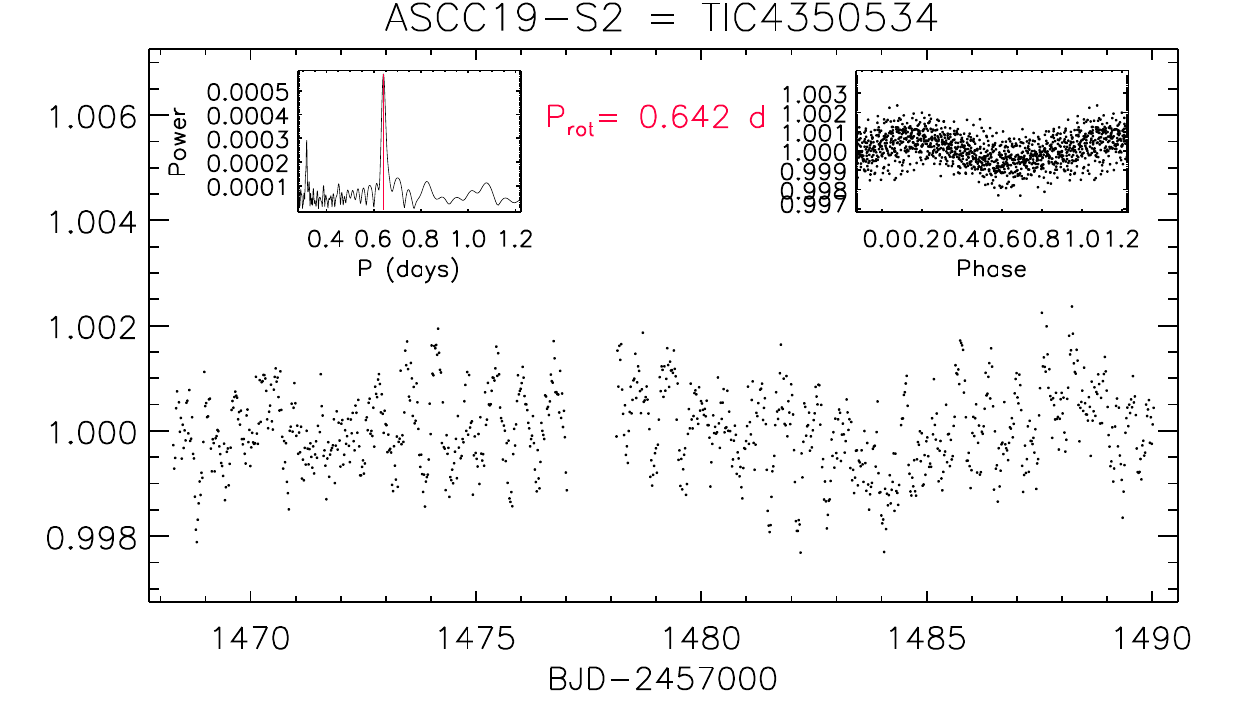}	
\includegraphics[width=9.cm]{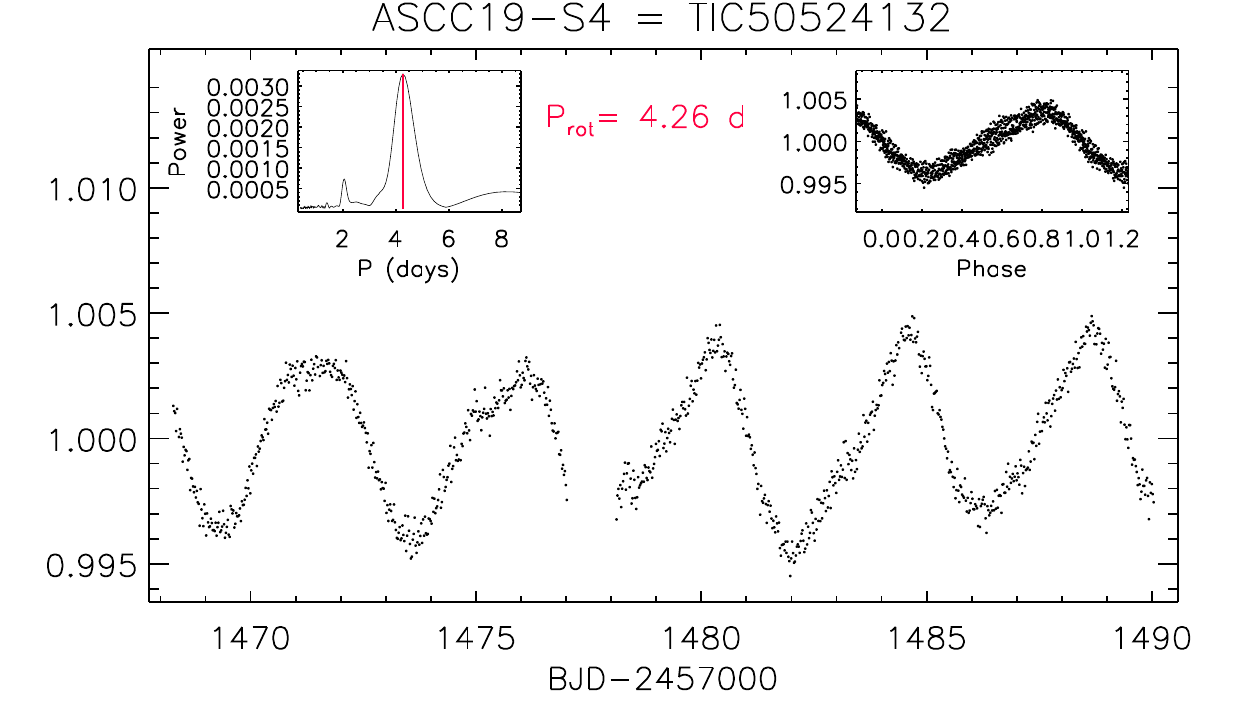}	
\includegraphics[width=9.cm]{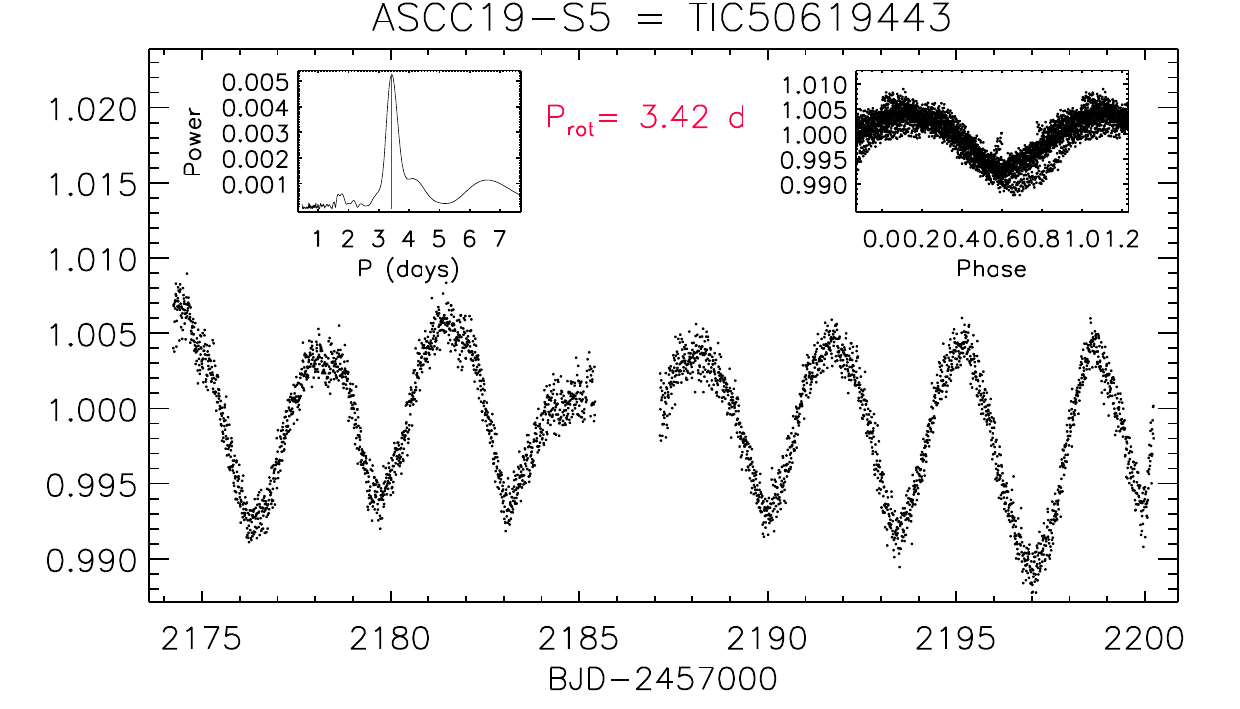}	
\includegraphics[width=9.cm]{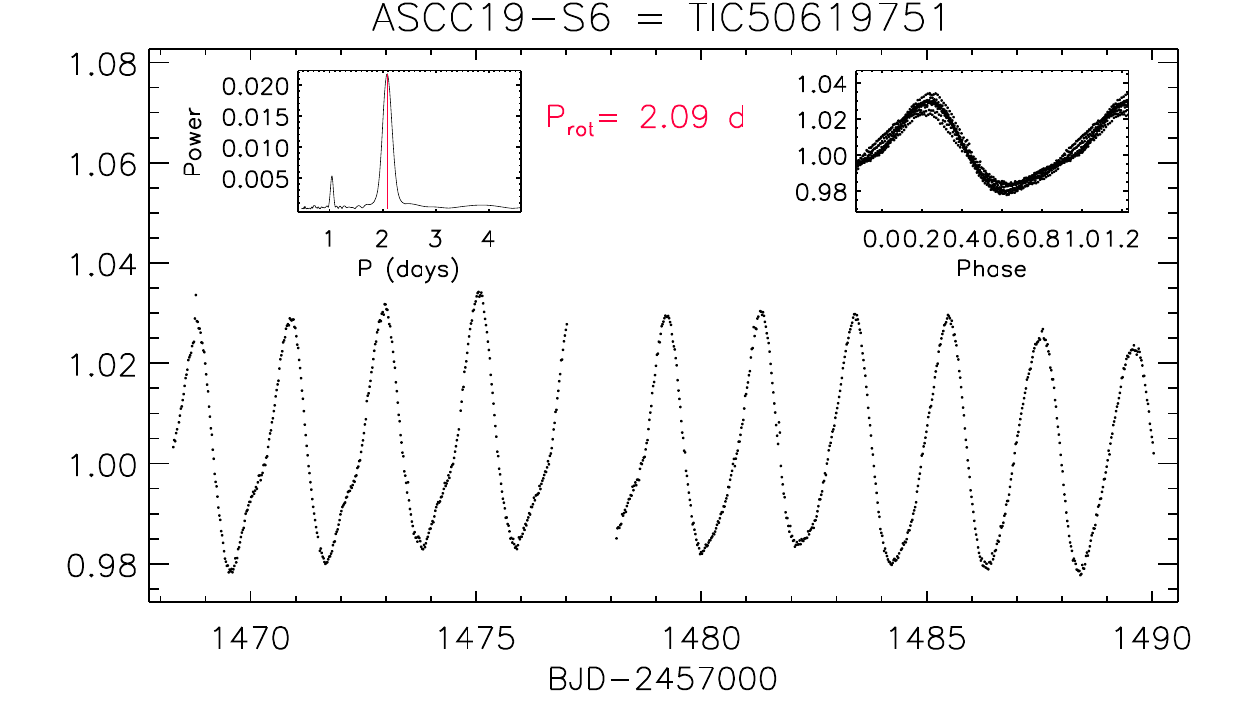}	
\includegraphics[width=9.cm]{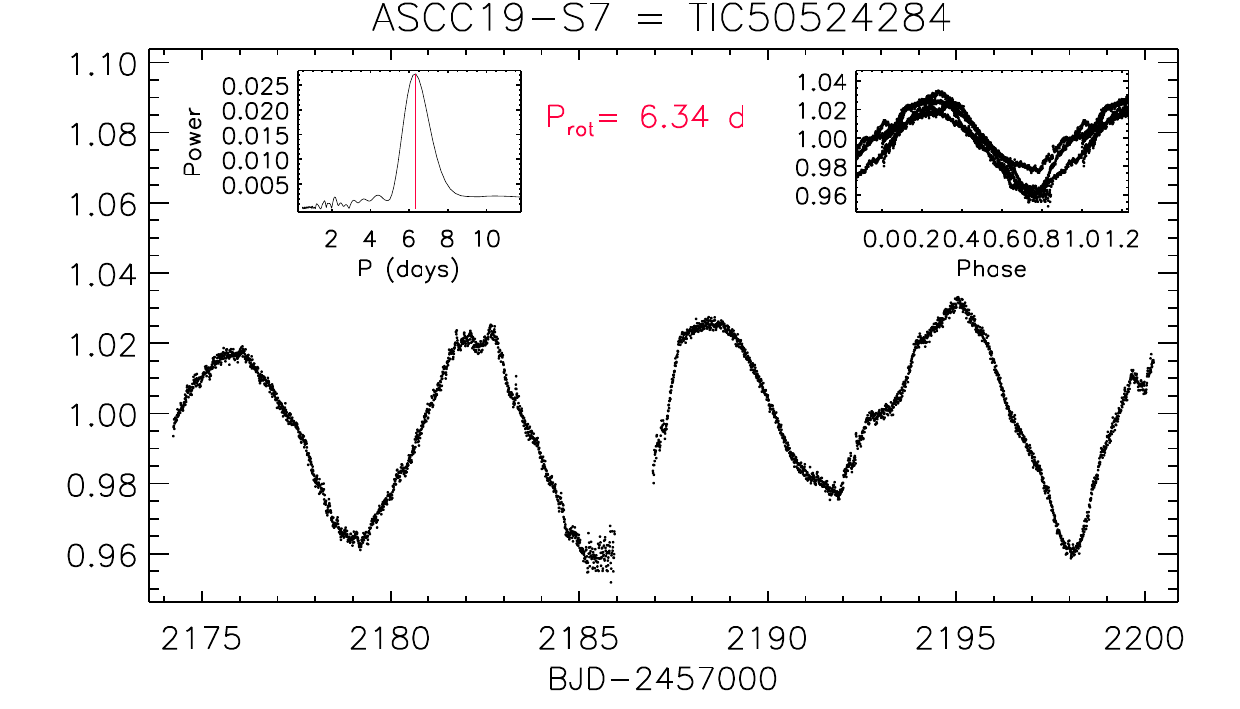}	
\includegraphics[width=9.cm]{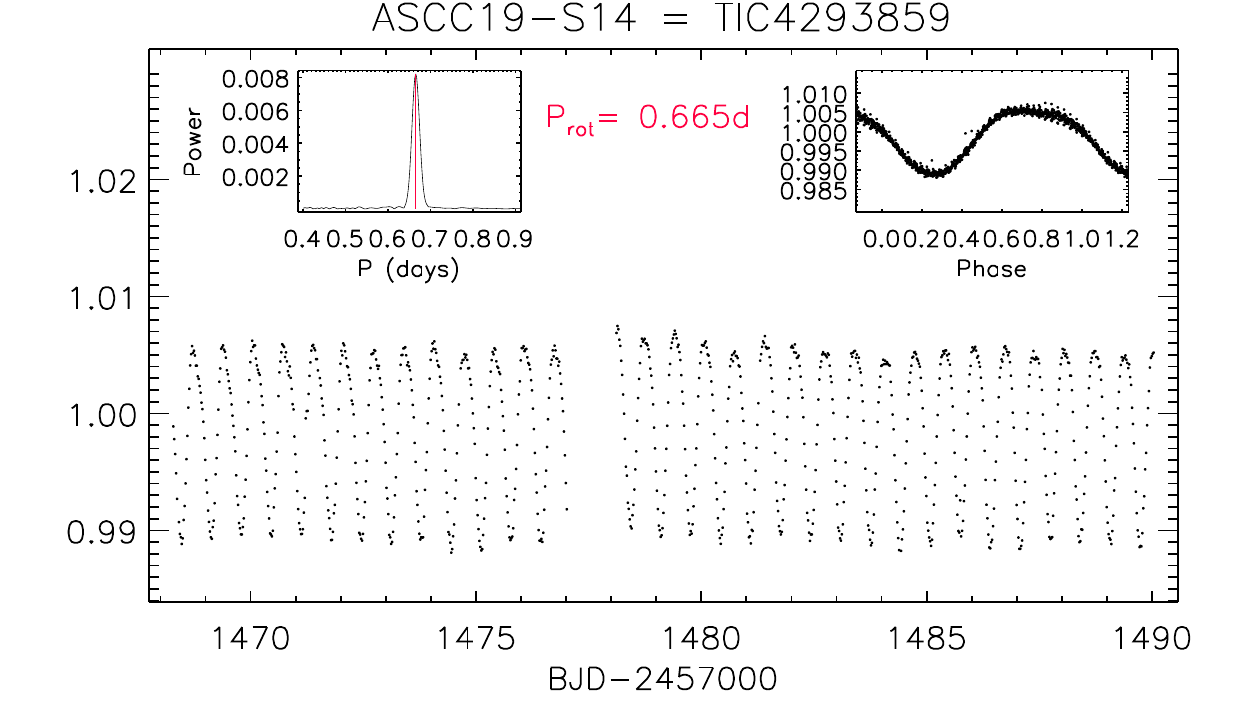}	
\includegraphics[width=9.cm]{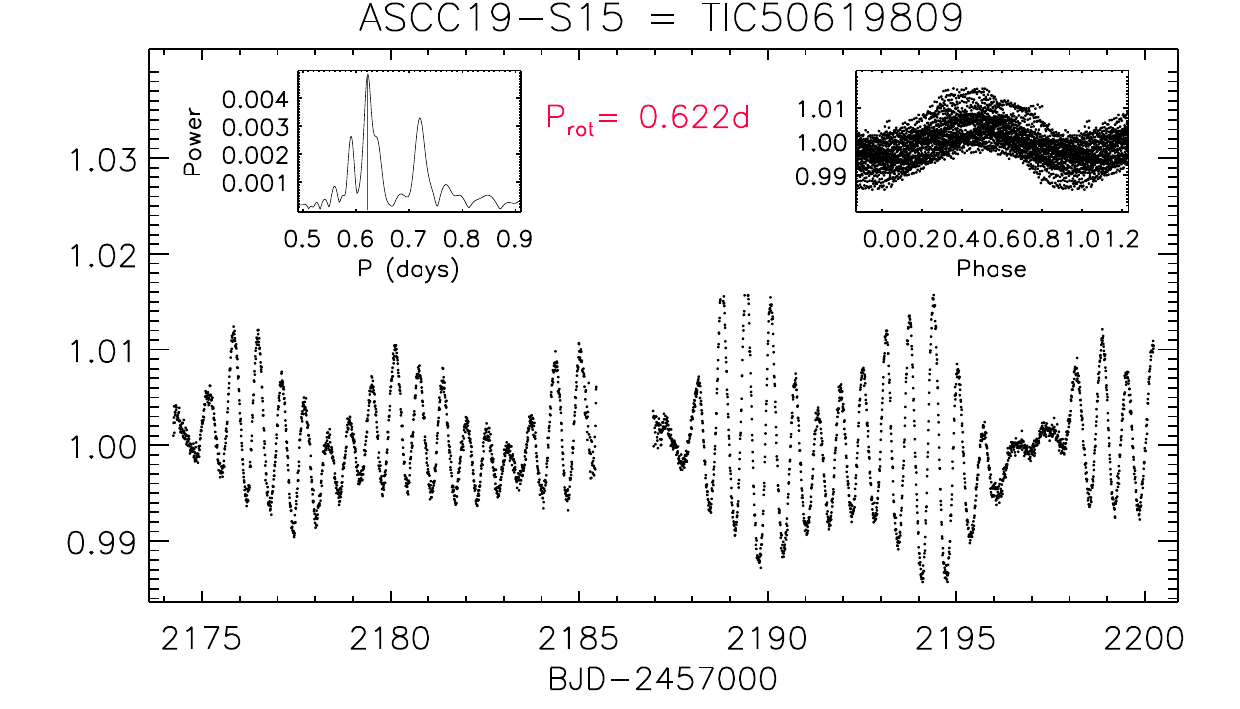}	
\vspace{0cm}
\caption{{\it TESS} light curves for eight members of ASCC~19. In each panel, the inset in the upper left corner shows the 
cleaned periodogram, with the period marked by a vertical red line and indicated with red characters. The inset in the upper right corner displays the data phased with this period.}
\label{fig:TESS_ASCC19}
\end{center}
\end{figure*}

\begin{figure*}
\begin{center}
\includegraphics[width=9.cm]{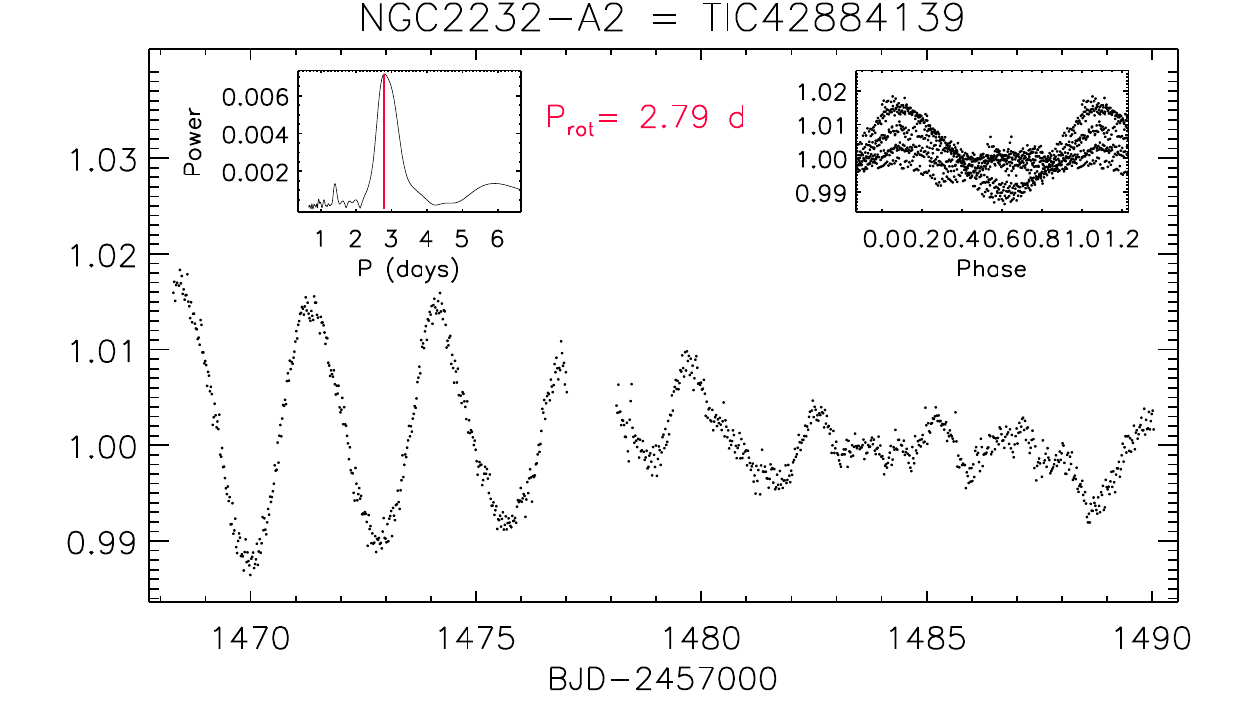}	
\includegraphics[width=9.cm]{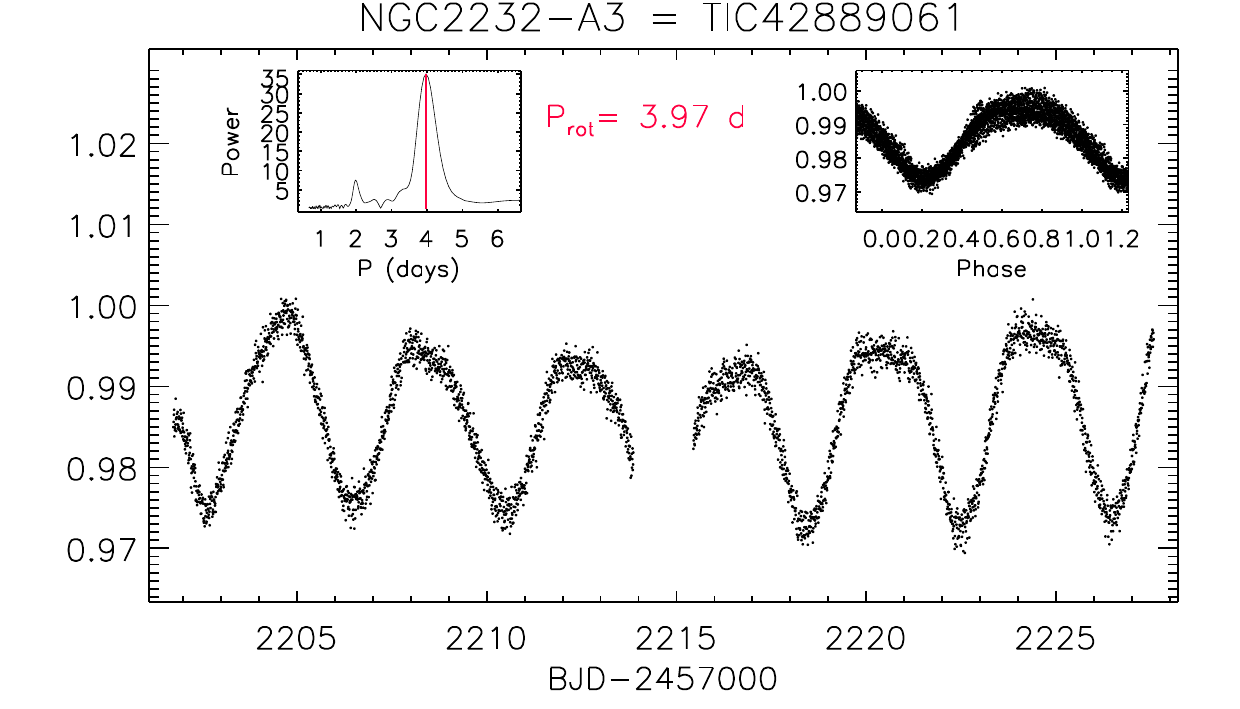}	
\includegraphics[width=9.cm]{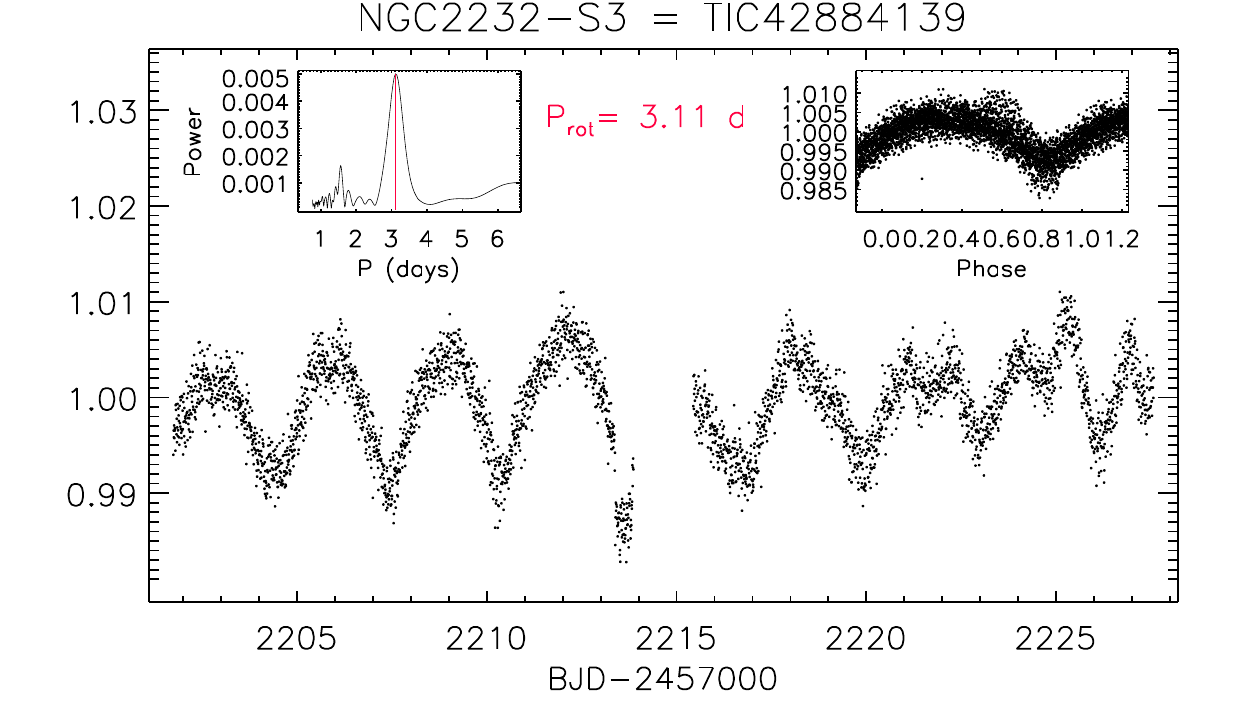}	
\includegraphics[width=9.cm]{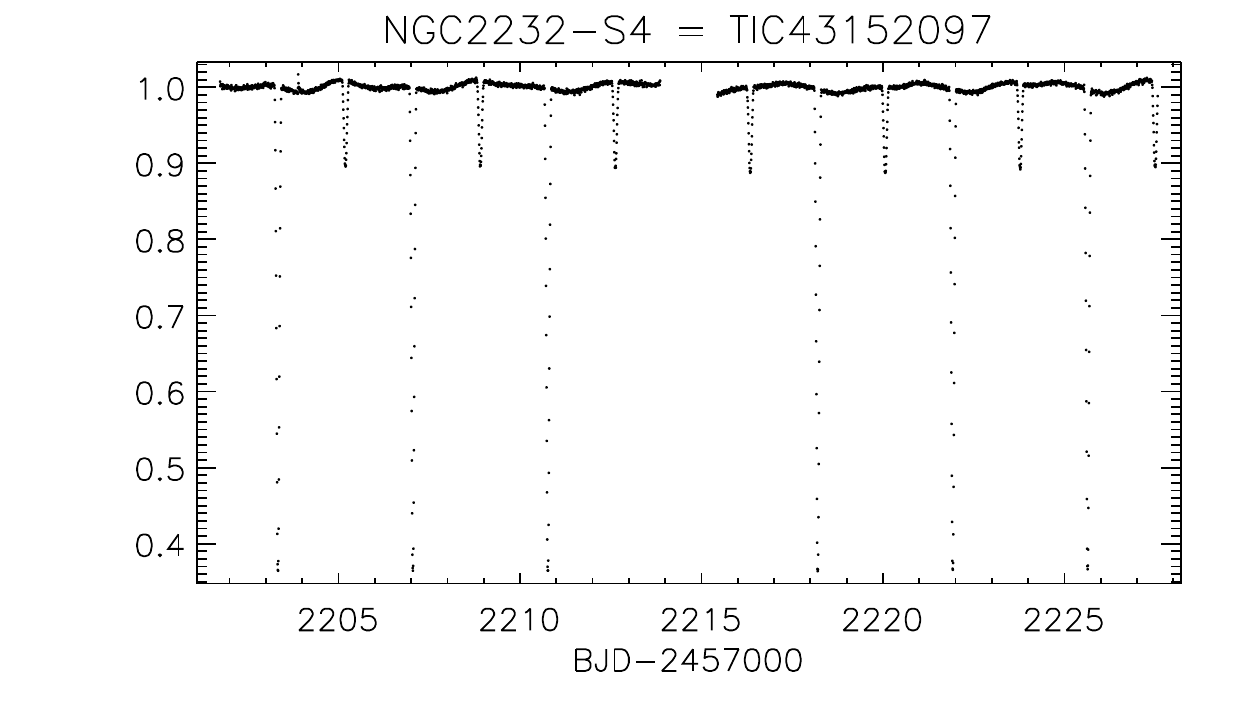}	
\includegraphics[width=9.cm]{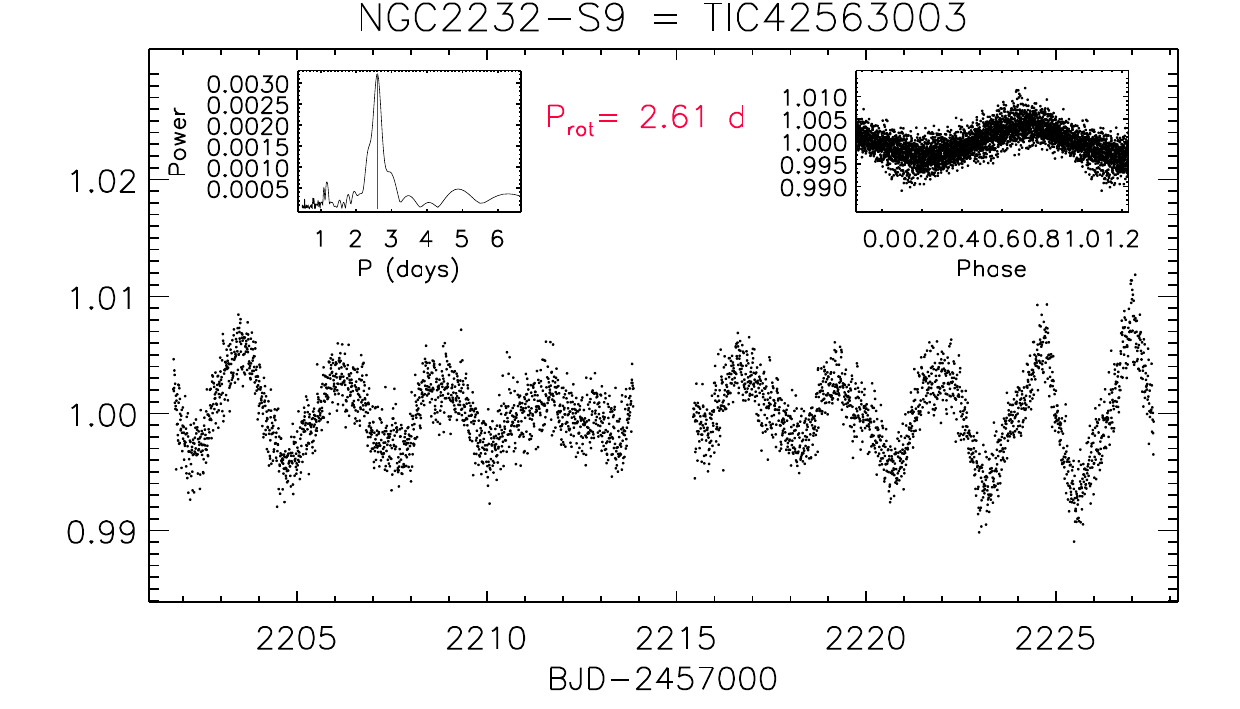}	
\includegraphics[width=9.cm]{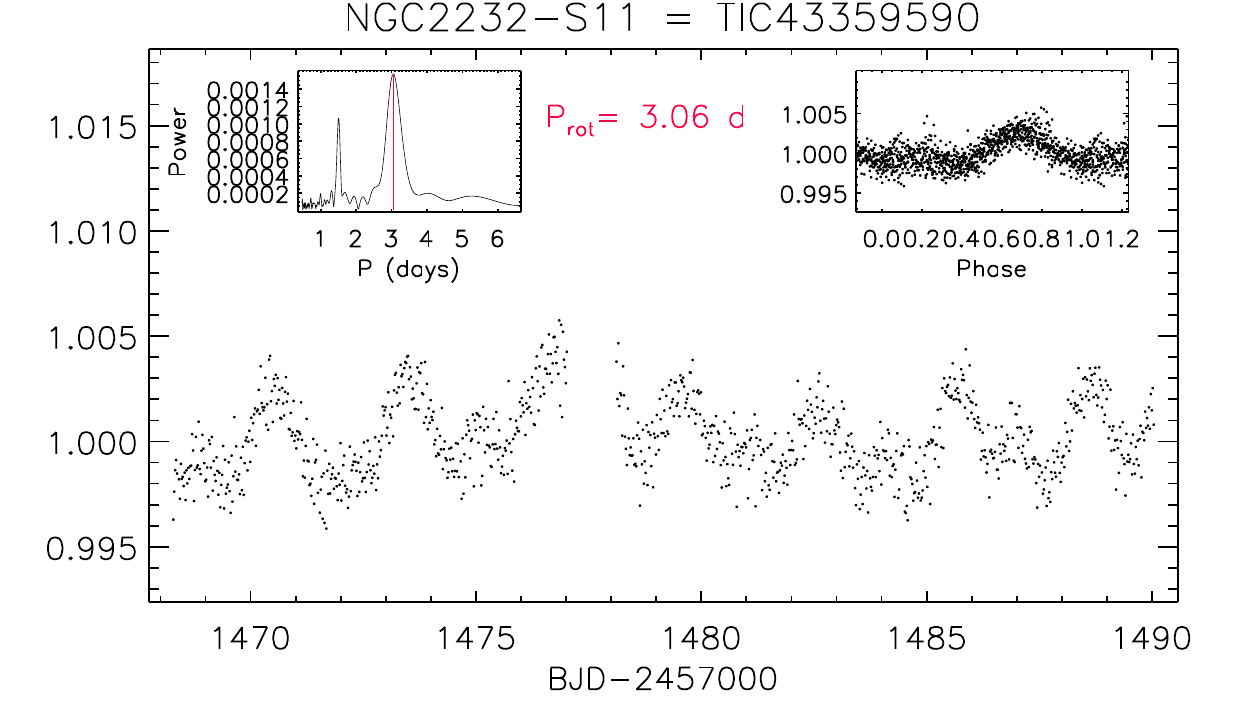}	
\vspace{0cm}
\caption{{\it TESS} light curves for six members of NGC~2232. In each panel, the inset in the upper left corner shows the 
cleaned periodogram, with the period marked by a vertical red line and indicated with red characters. The inset in the upper right corner displays the data phased with this period.}
\label{fig:TESS_NGC2232}
\end{center}
\end{figure*}

\begin{figure*}
\begin{center}
\includegraphics[width=9.cm]{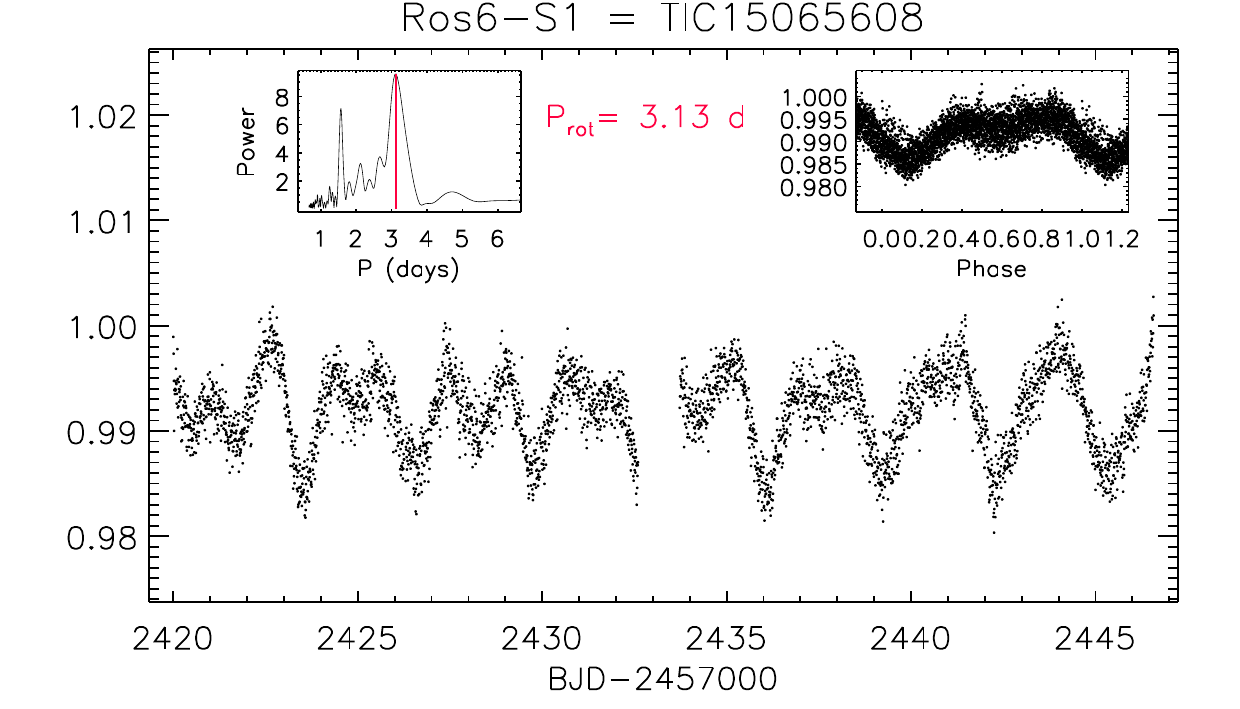}	
\includegraphics[width=9.cm]{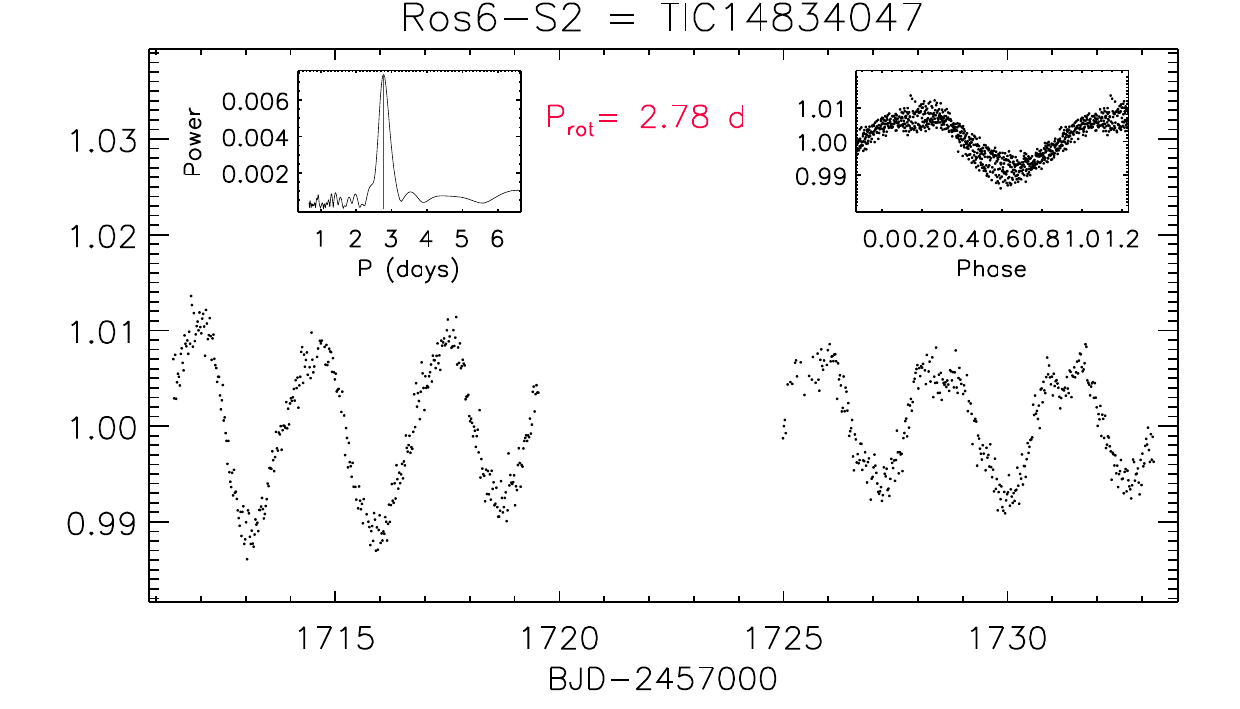}	
\includegraphics[width=9.cm]{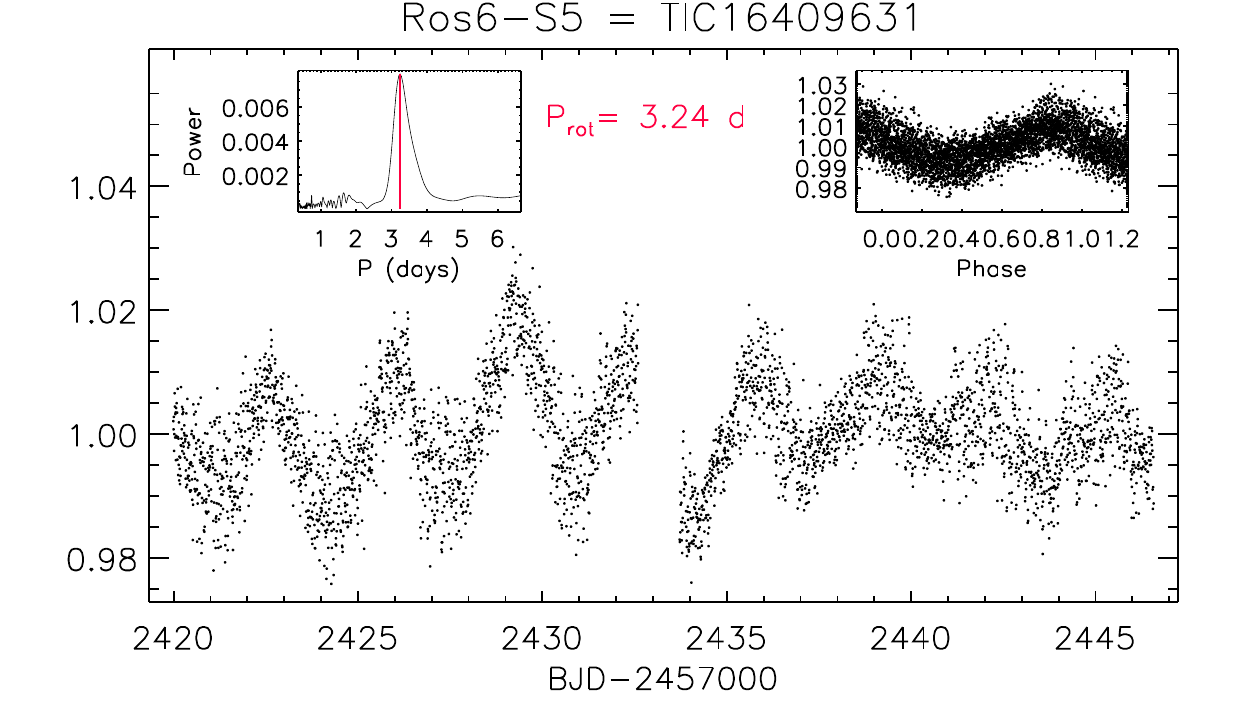}	
\includegraphics[width=9.cm]{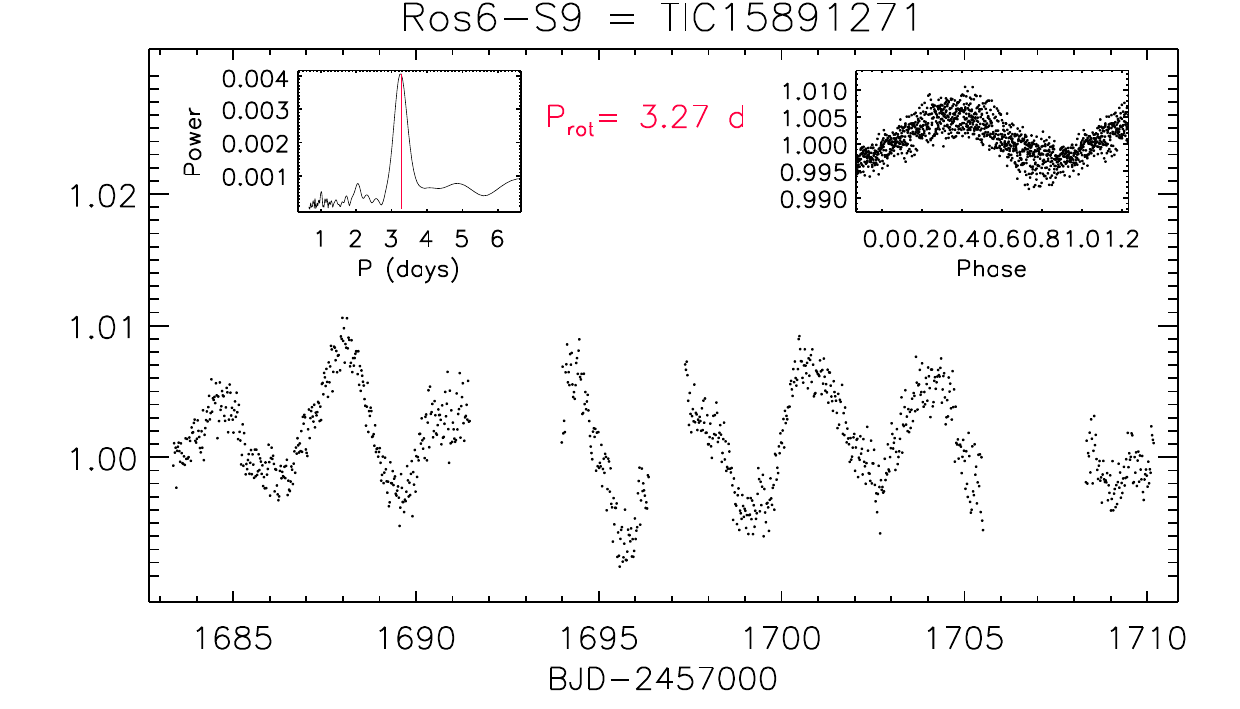}	
\vspace{0cm}
\caption{{\it TESS} light curves for the four members of Roslund~6. In each panel, the inset in the upper left corner shows the 
cleaned periodogram, with the period marked by a vertical red line and indicated with red characters. The inset in the upper right corner displays the data phased with this period.}
\label{fig:TESS_Roslund6}
\end{center}
\end{figure*}

\begin{figure*}
\begin{center}
\includegraphics[width=9.cm]{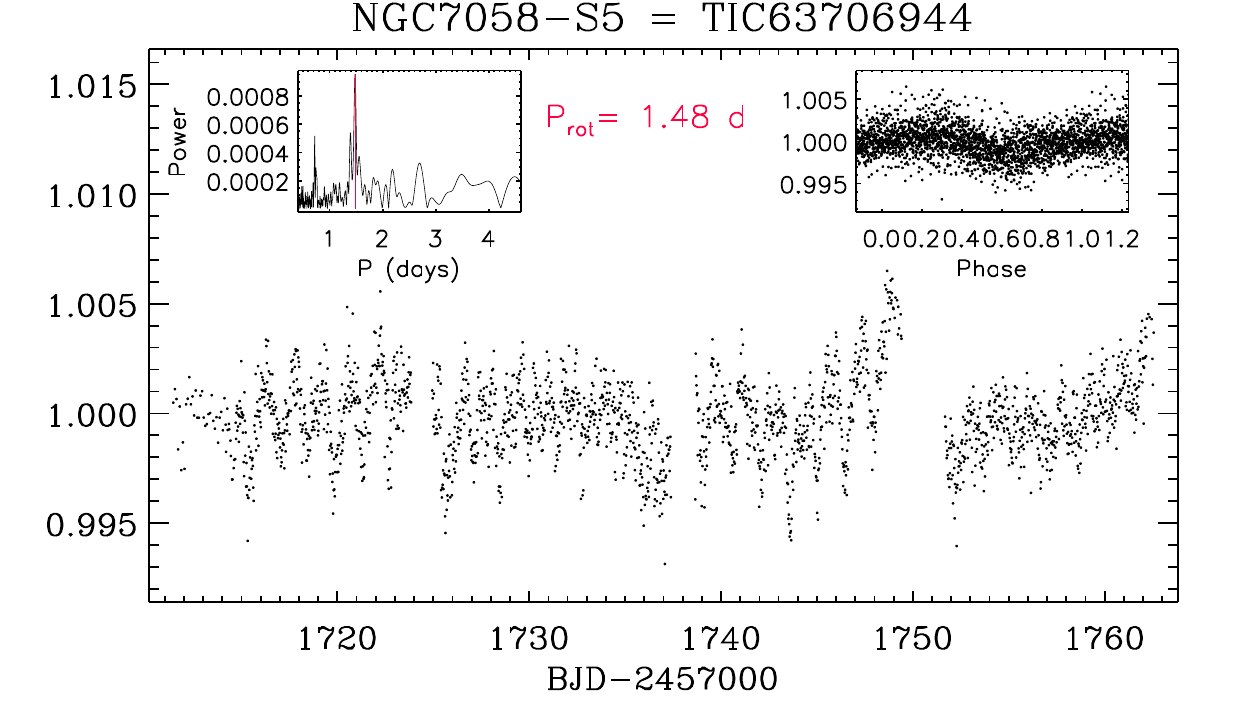}	
\includegraphics[width=9.cm]{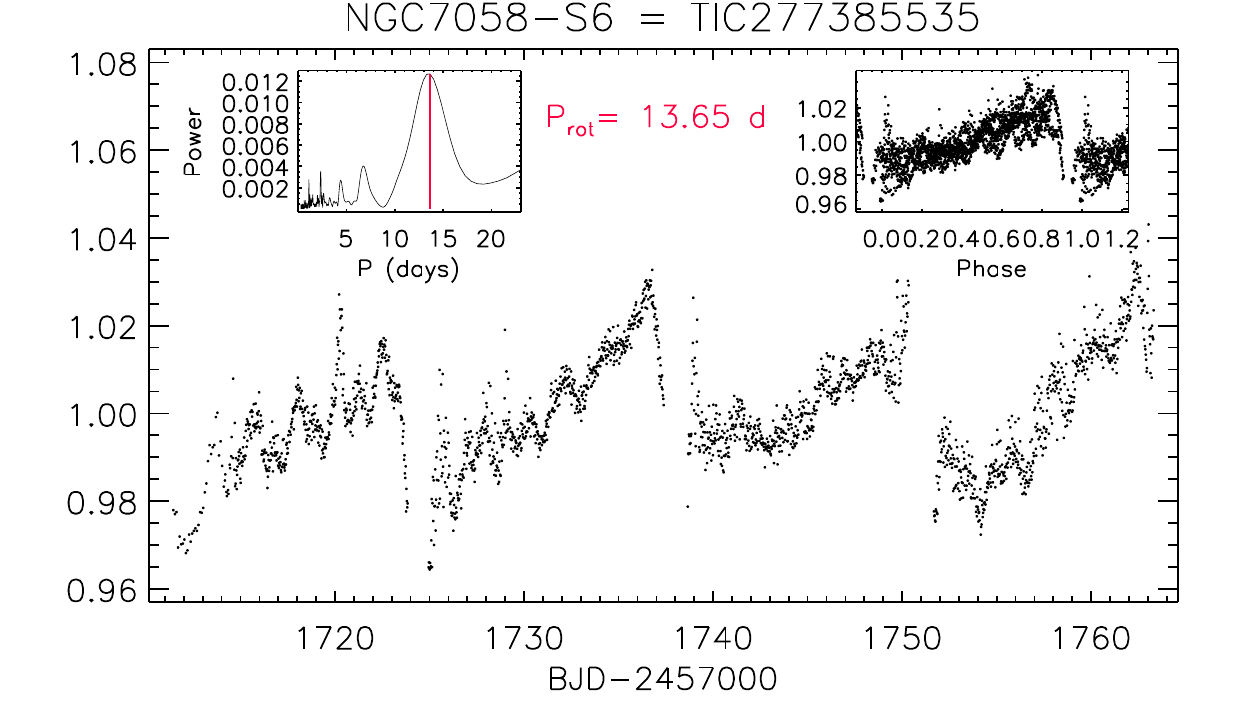}	
\includegraphics[width=9.cm]{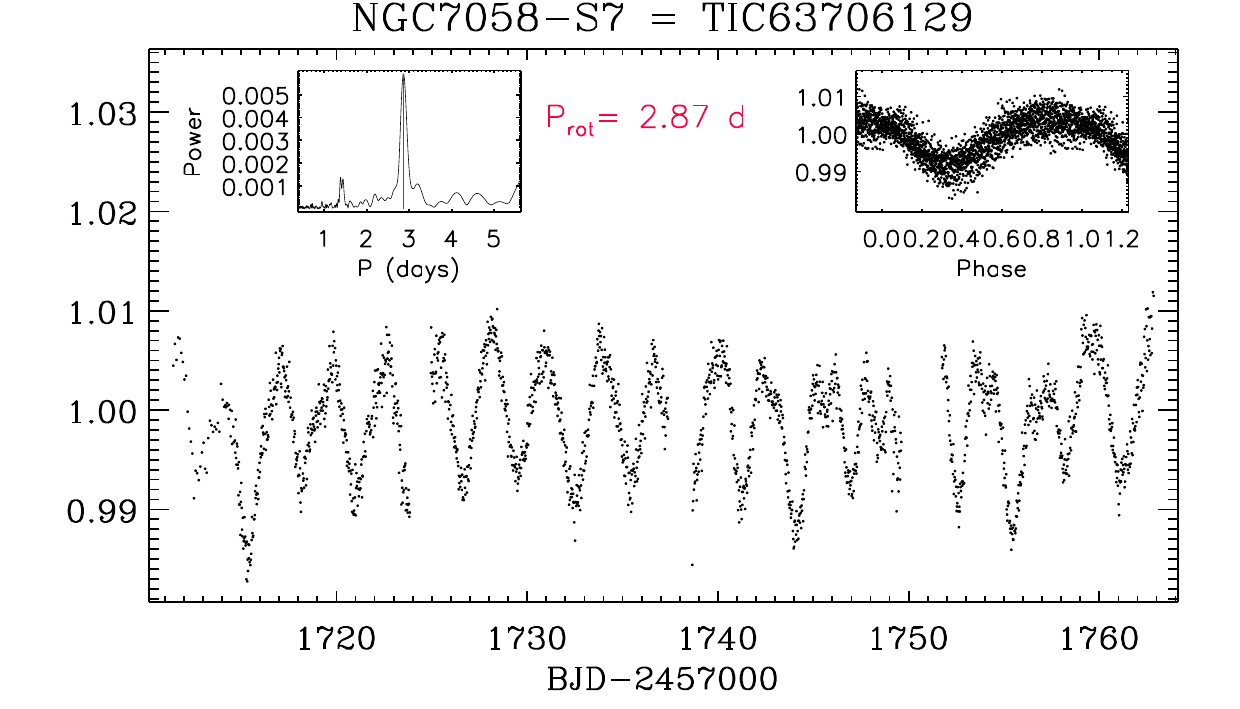}	
\includegraphics[width=9.cm]{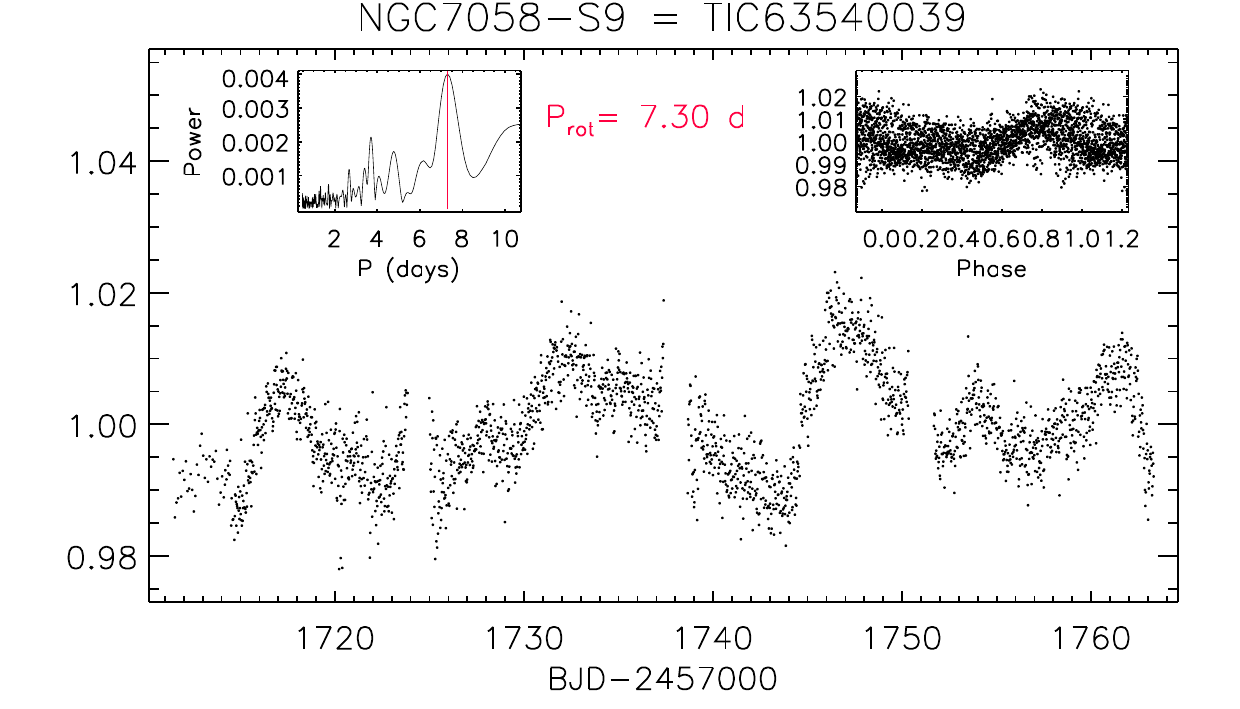}	
\includegraphics[width=9.cm]{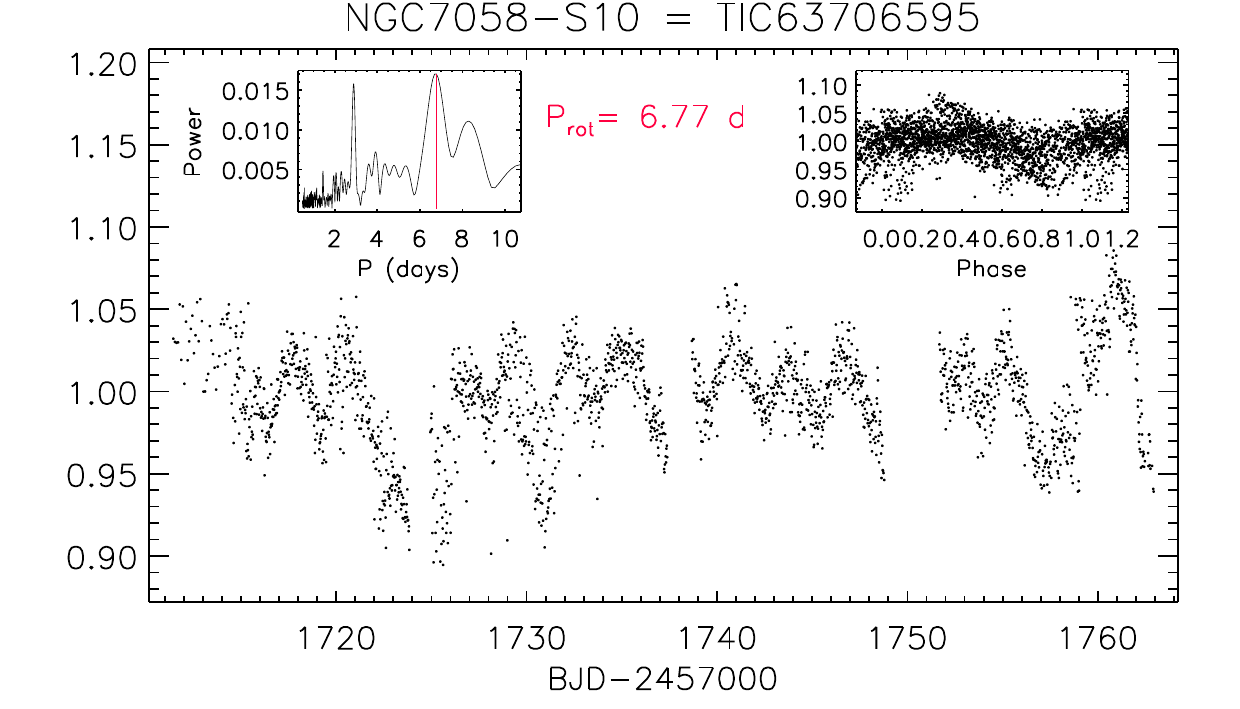}	
\includegraphics[width=9.cm]{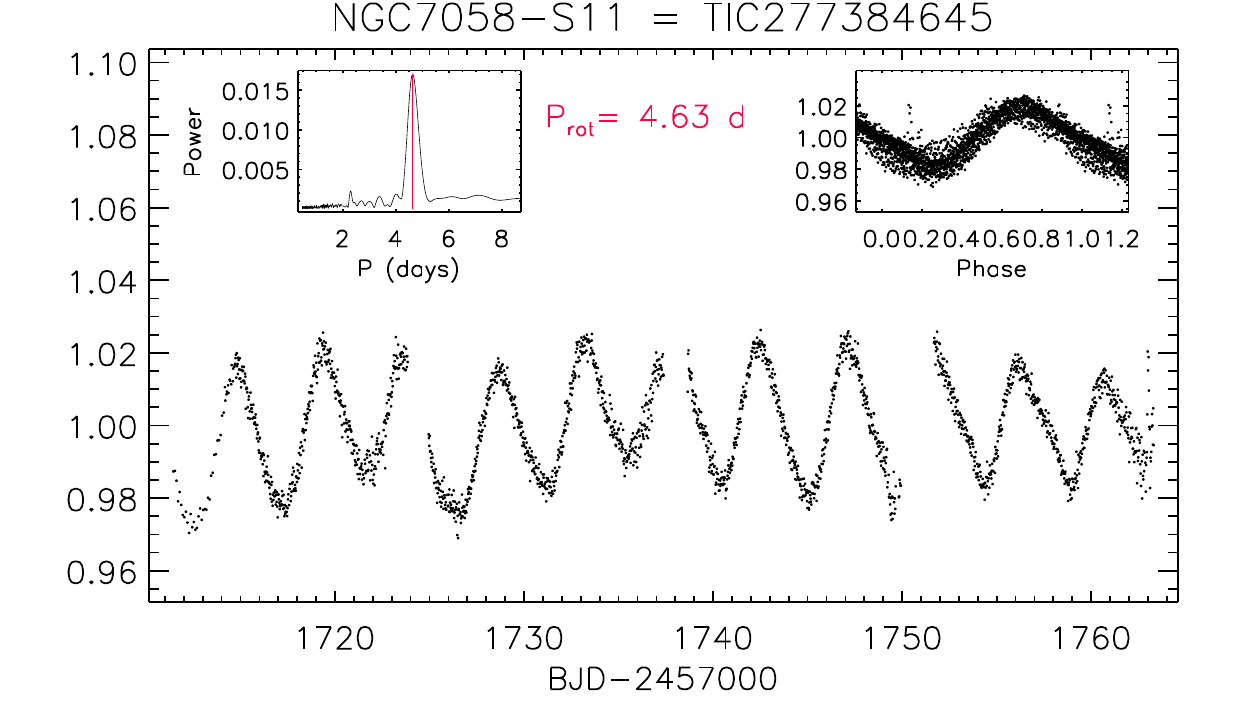}	
\includegraphics[width=9.cm]{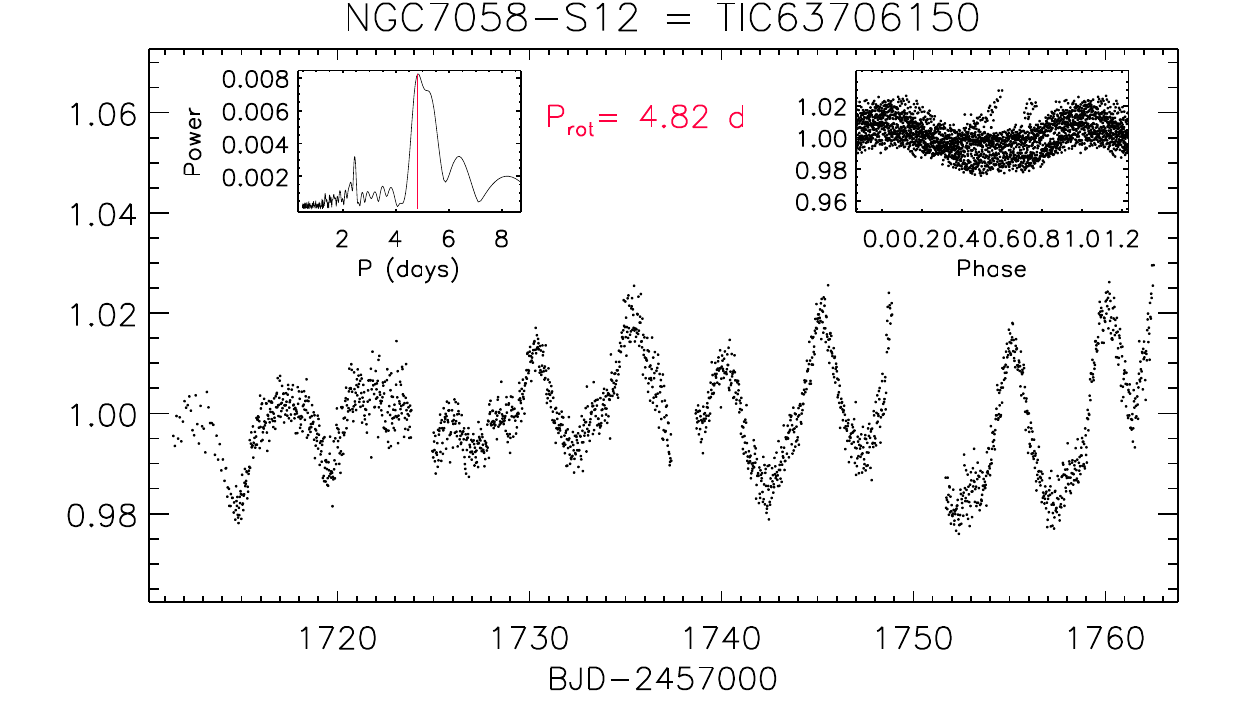}	
\includegraphics[width=9.cm]{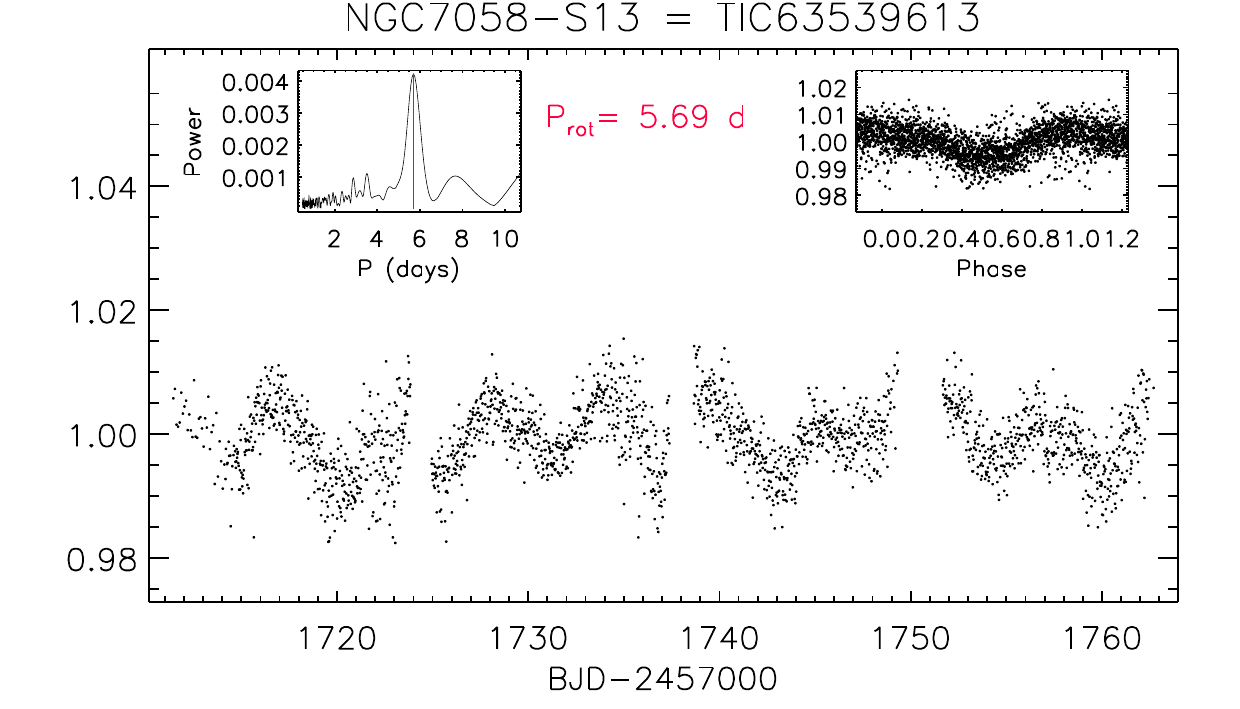}	
\vspace{0cm}
\caption{{\it TESS} light curves for eight members of NGC~7058. In each panel, the inset in the upper left corner shows the 
cleaned periodogram, with the period marked by a vertical red line and indicated with red characters. The inset in the upper right corner displays the data phased with this period.}
\label{fig:TESS_NGC7058}
\end{center}
\end{figure*}

\begin{figure*}
\begin{center}
\includegraphics[width=6cm]{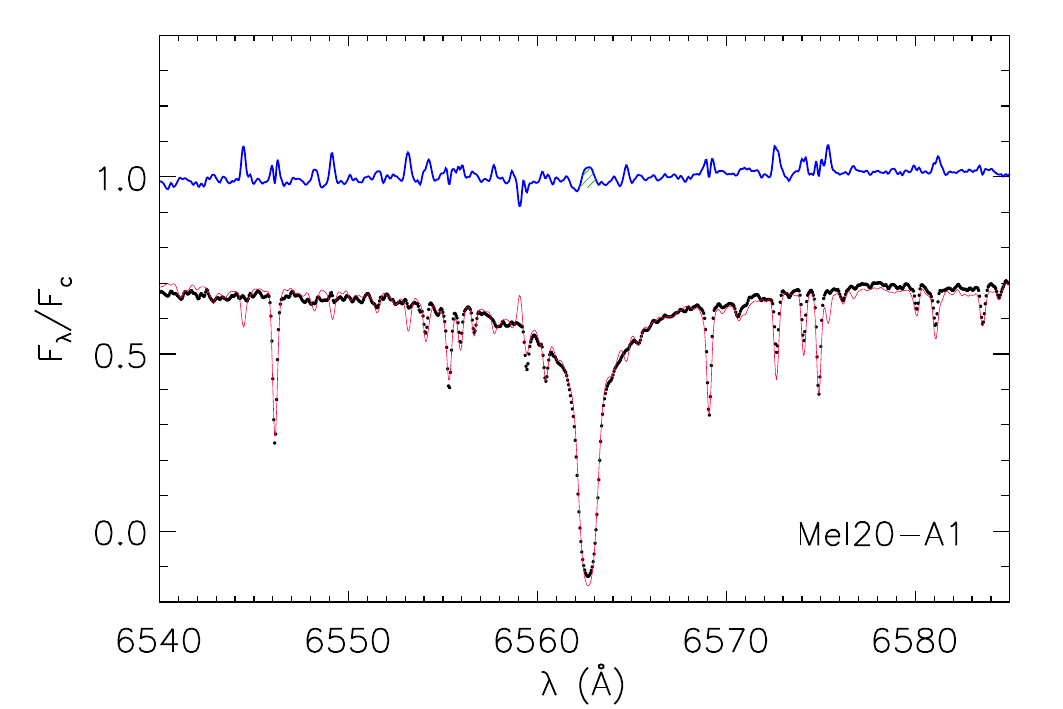} 
\includegraphics[width=6cm]{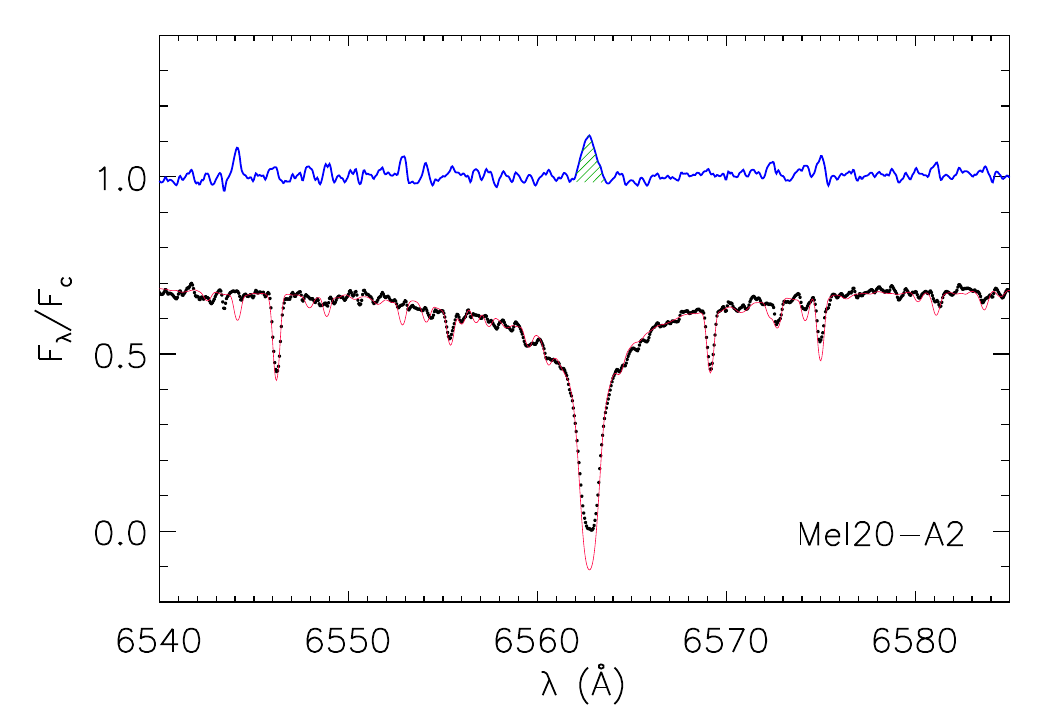}     
\includegraphics[width=6cm]{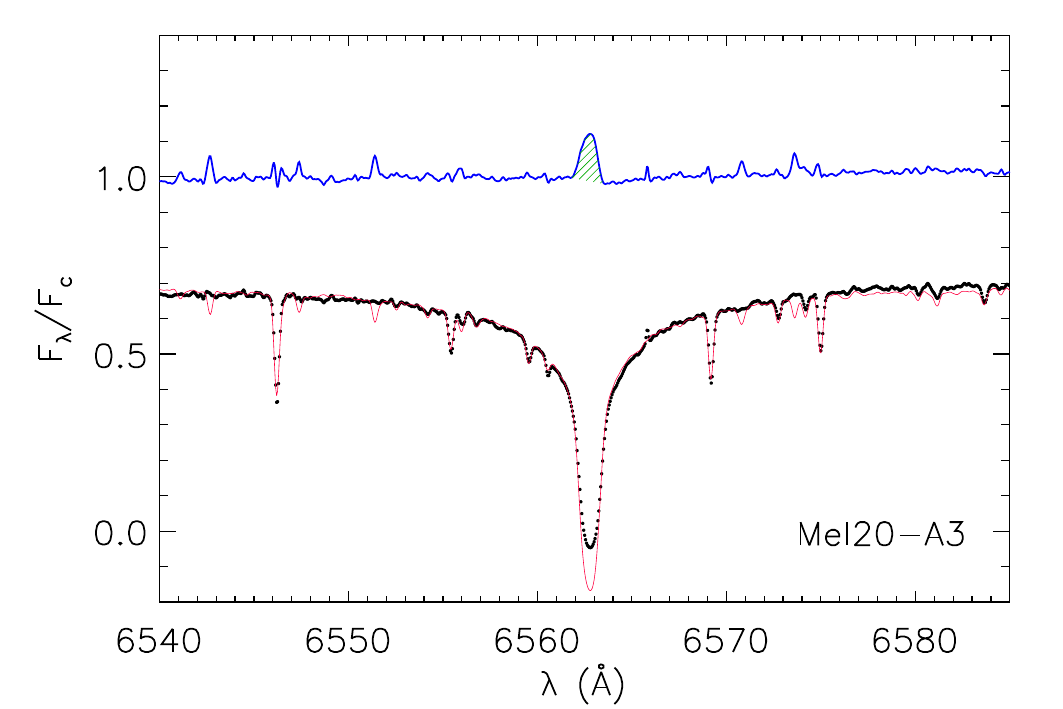}     
\includegraphics[width=6cm]{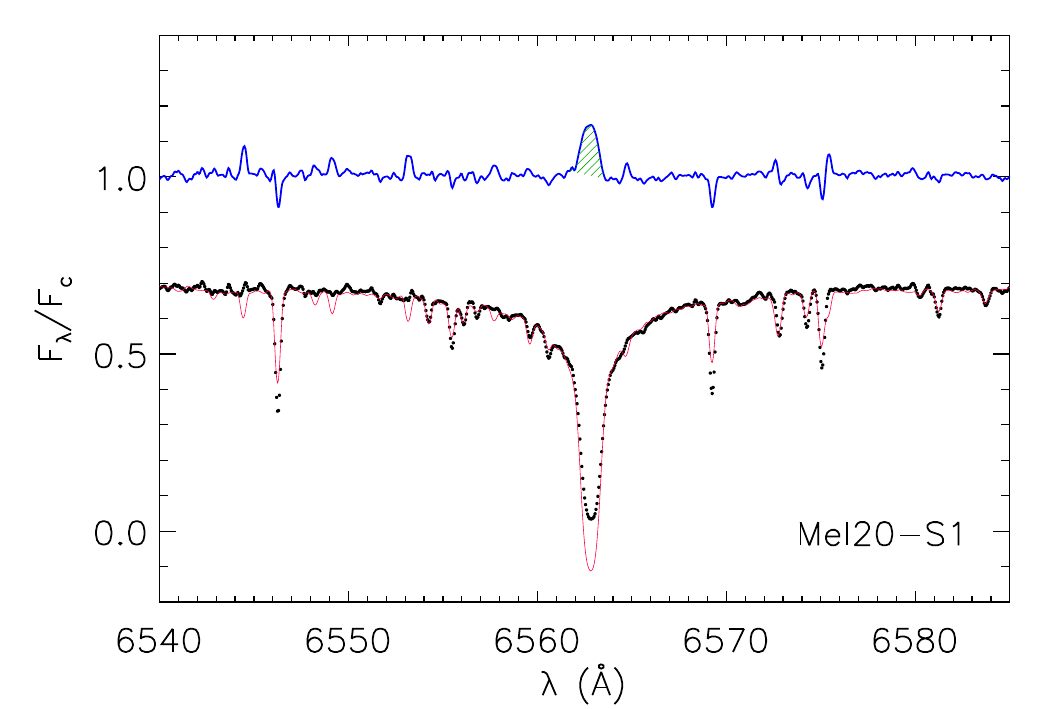}     
\includegraphics[width=6cm]{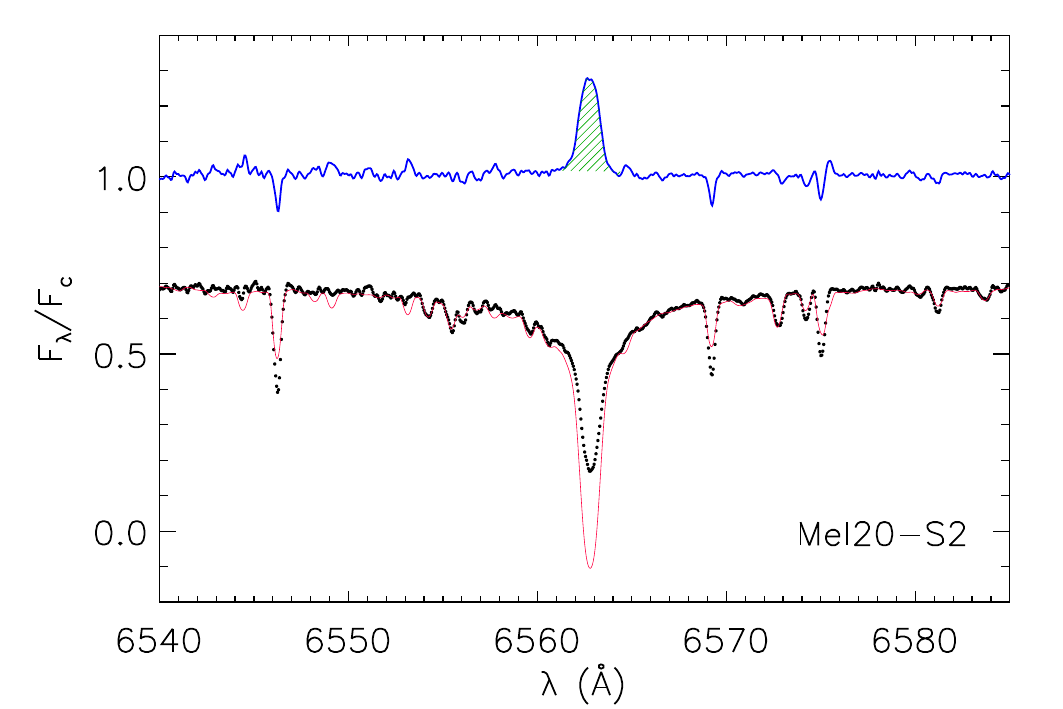}      
\includegraphics[width=6cm]{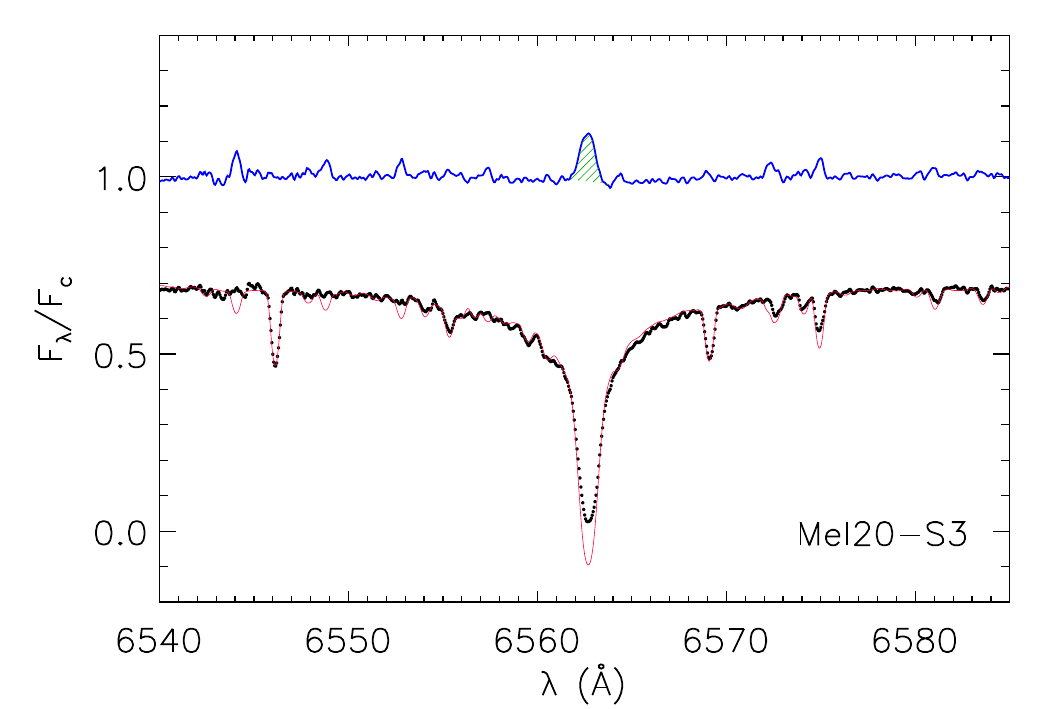} 
\includegraphics[width=6cm]{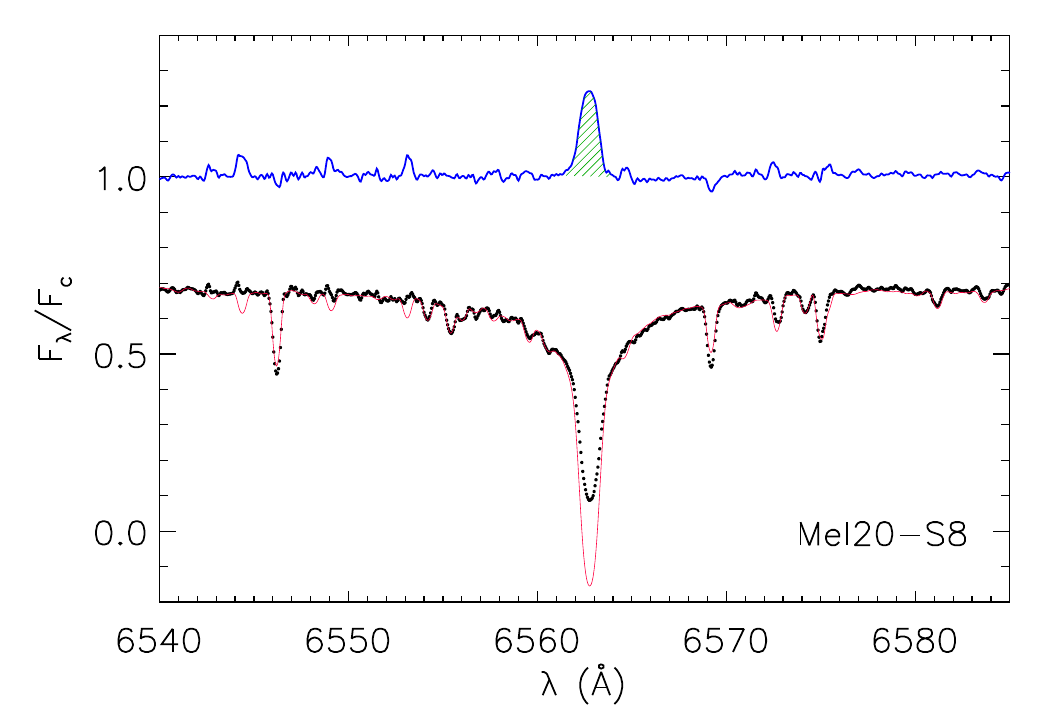} 
\includegraphics[width=6cm]{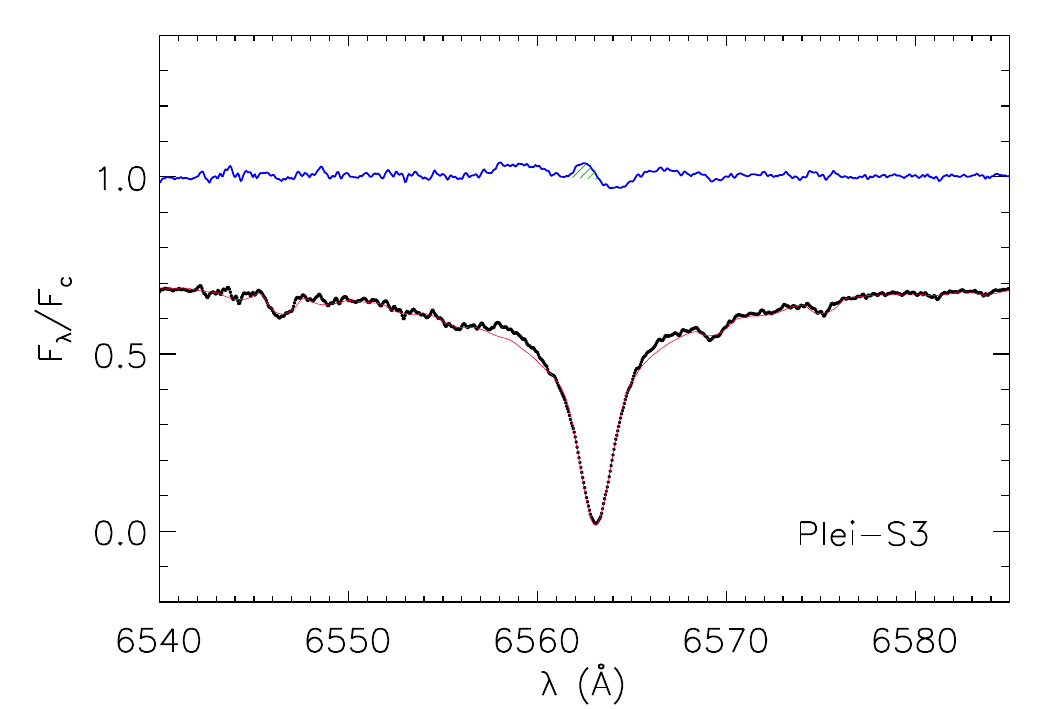} 
\includegraphics[width=6cm]{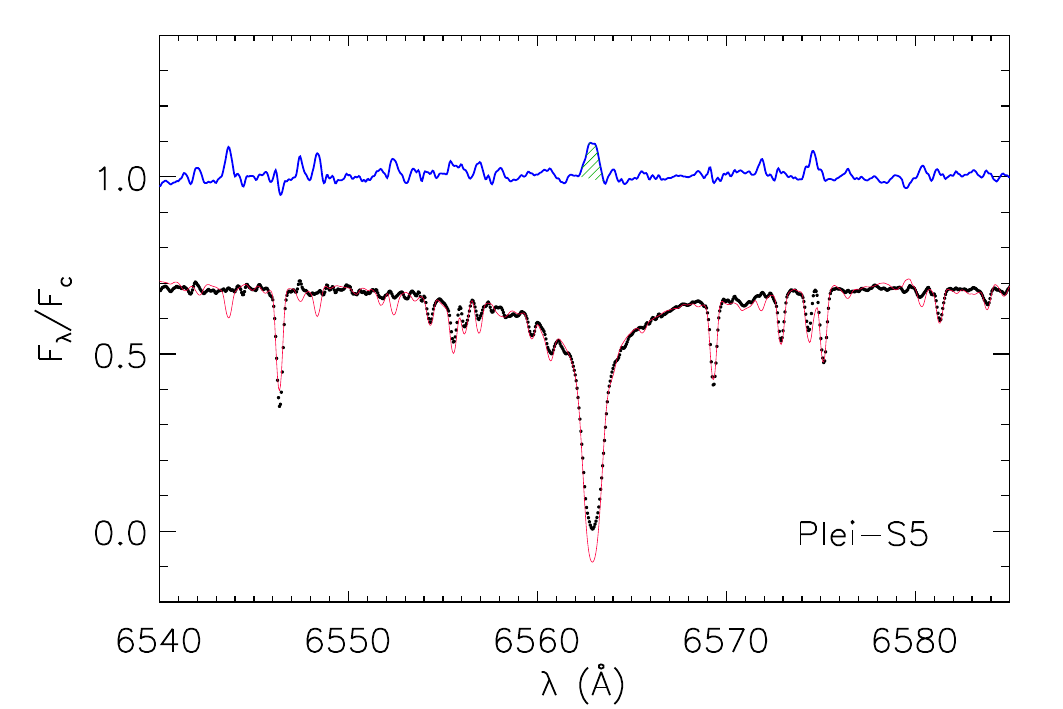} 
\includegraphics[width=6cm]{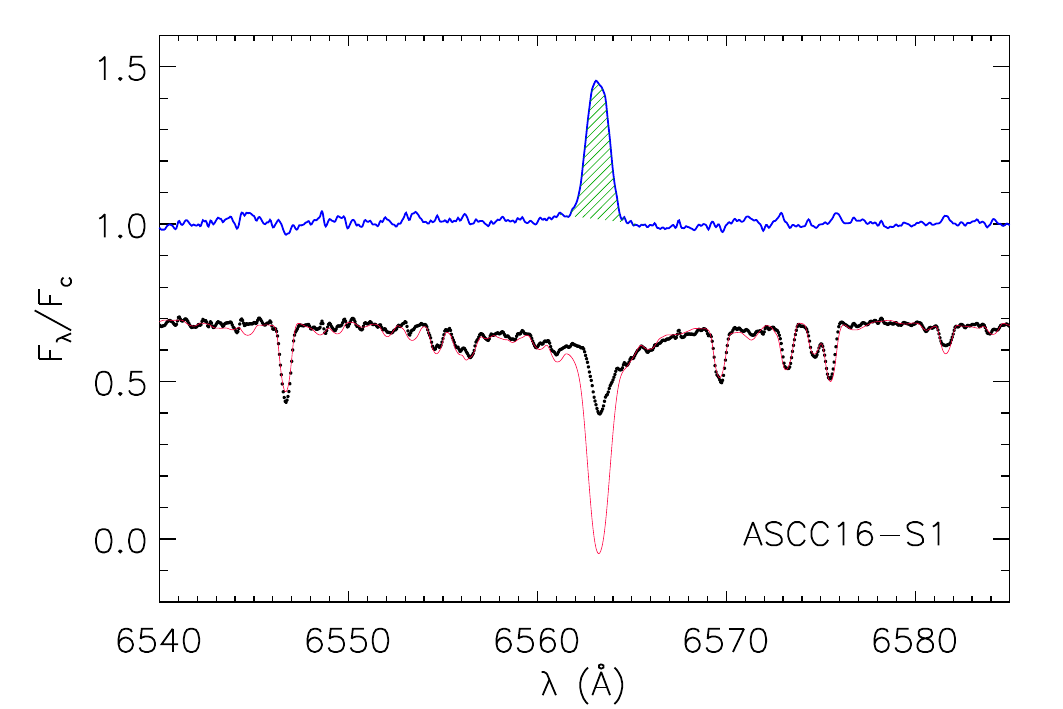} 
\includegraphics[width=6cm]{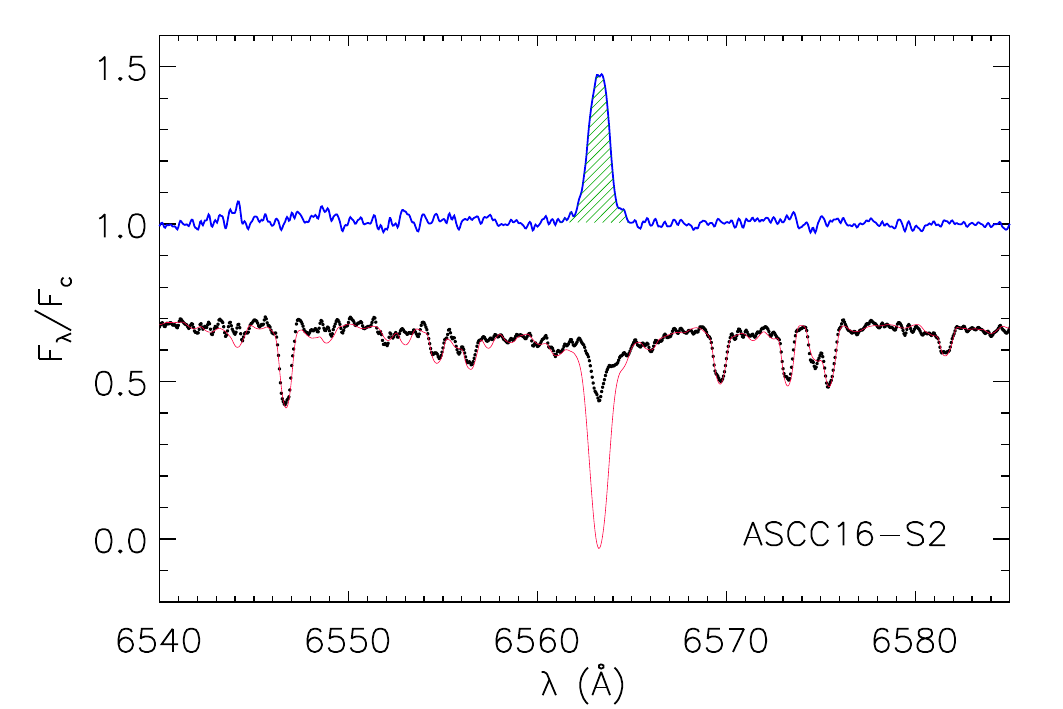} 
\includegraphics[width=6cm]{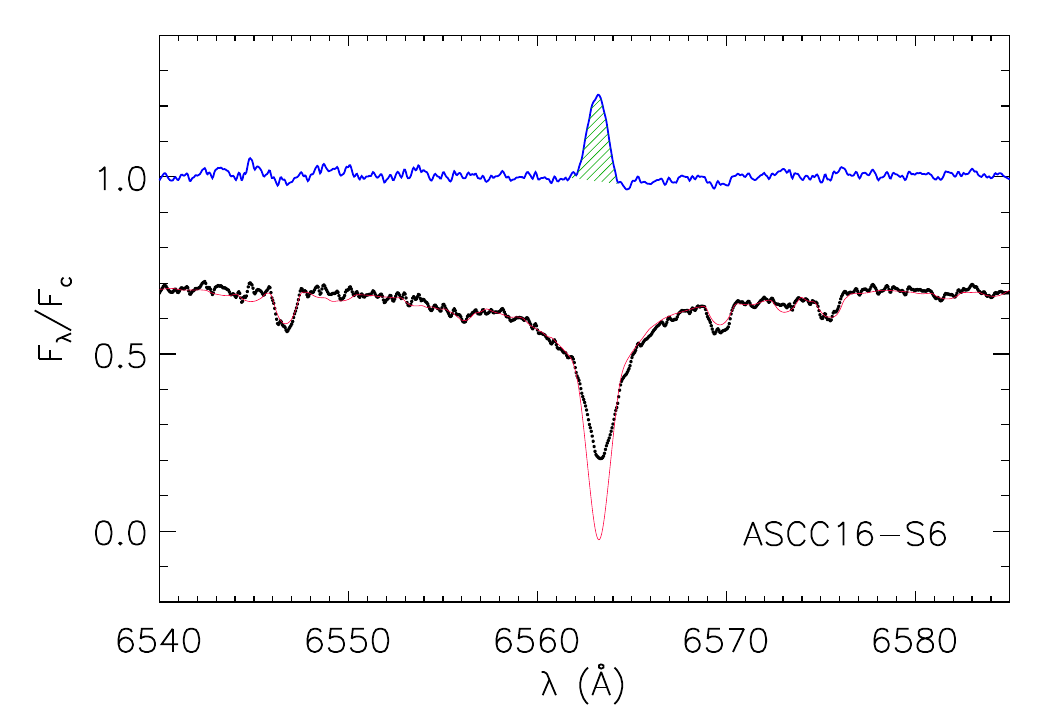} 
\includegraphics[width=6cm]{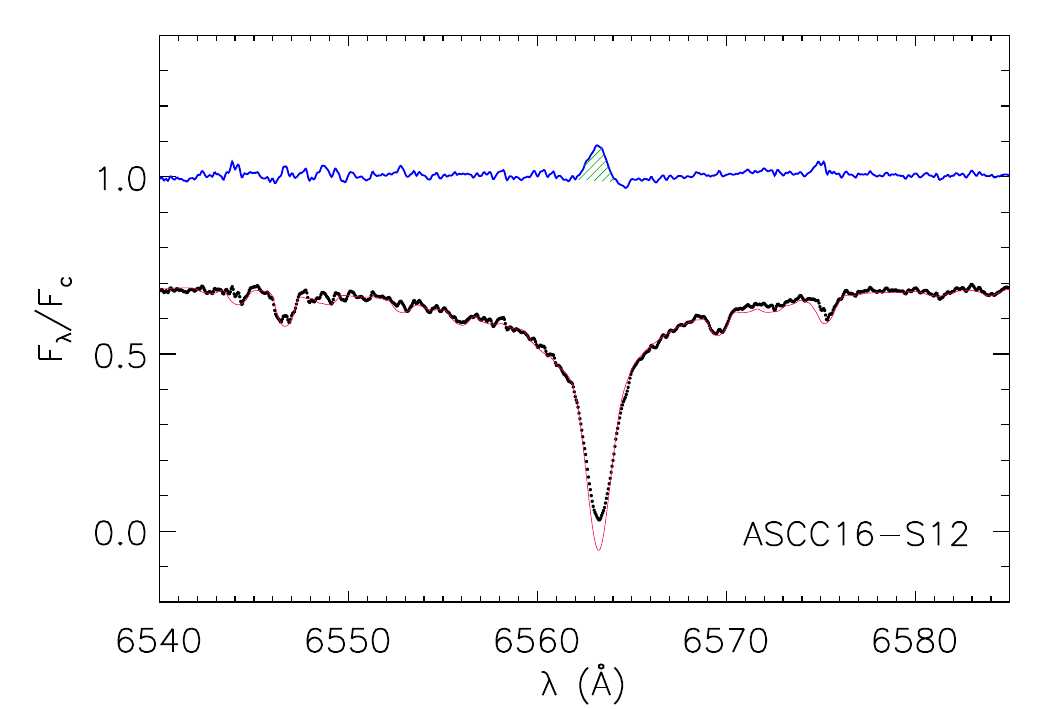} 
\includegraphics[width=6cm]{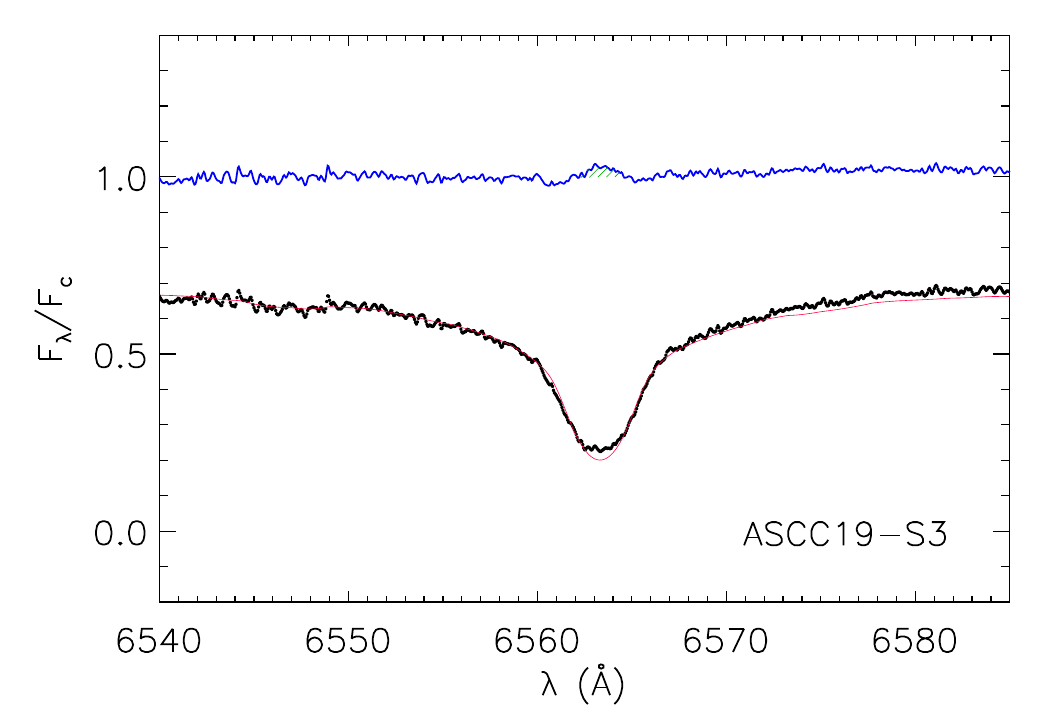} 
\includegraphics[width=6cm]{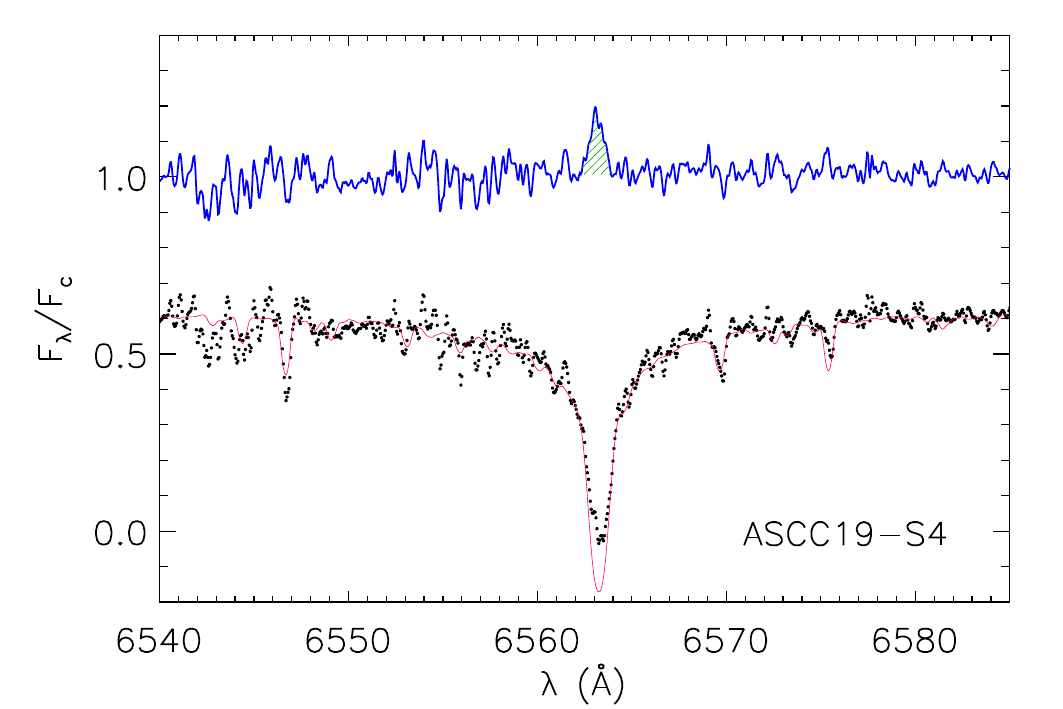} 
\includegraphics[width=6cm]{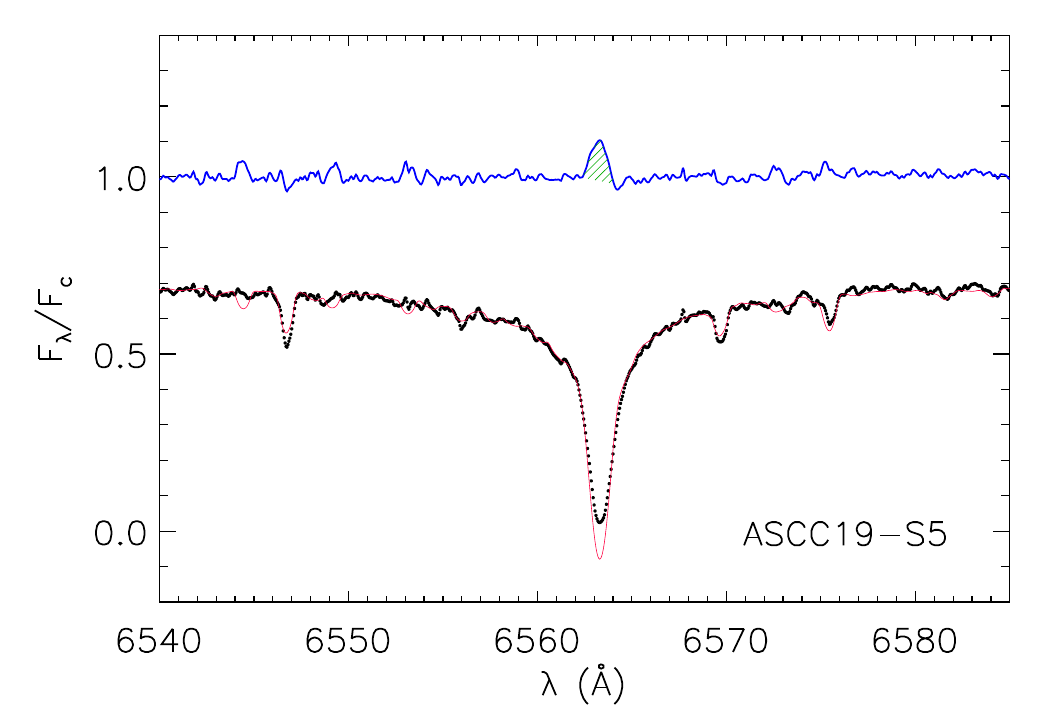} 
\includegraphics[width=6cm]{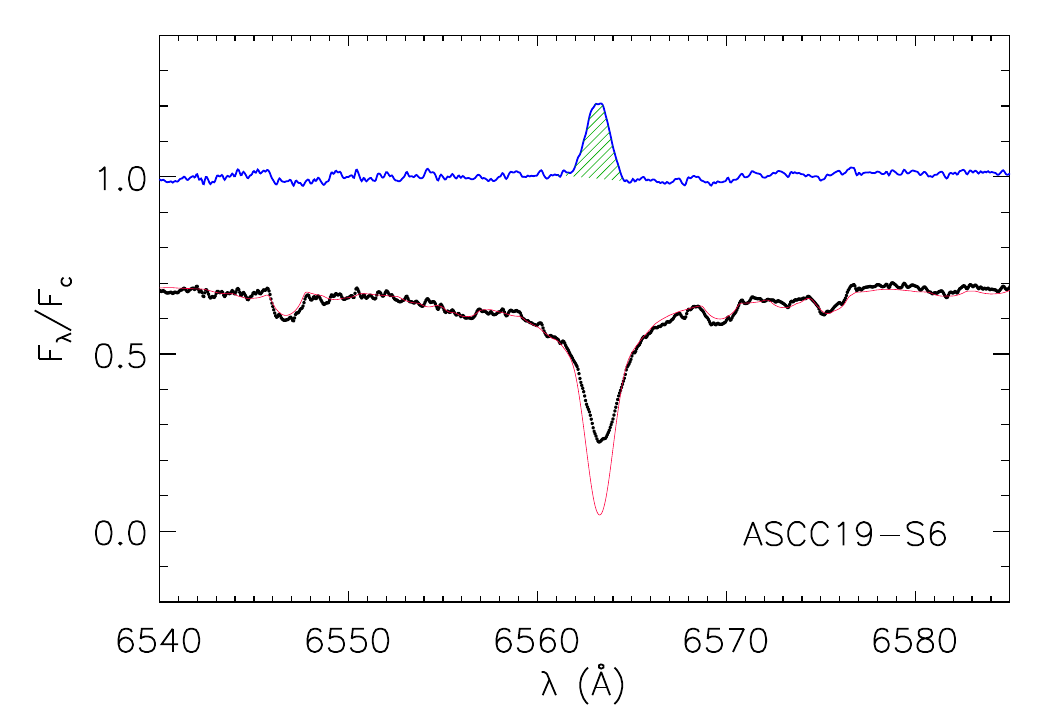} 
\includegraphics[width=6cm]{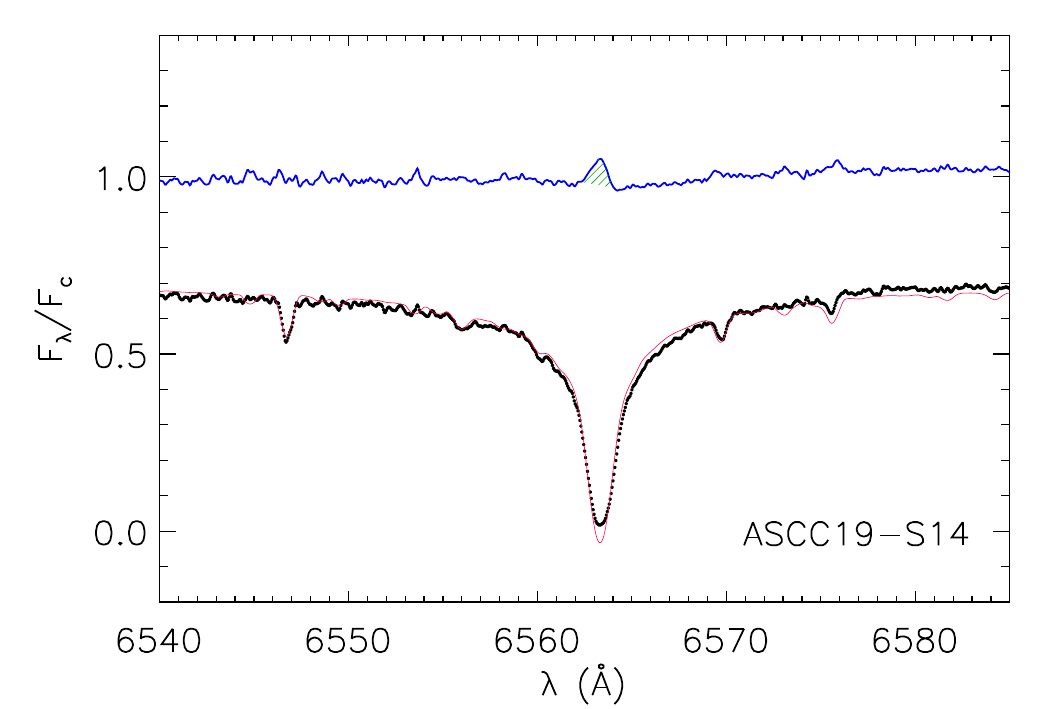} 
\vspace{0cm}
\caption{HARPS-N spectra of the investigated stars in the H$\alpha$ region. 
In each box, the non-active template (red line) is overlaid with the observed spectrum (black dots). 
The chromospheric emission which fills in the H$\alpha$ core is clearly visible 
in the subtracted spectrum (blue line in each panel). The green hatched area represents the excess H$\alpha$
emission that was integrated to obtain $W_{\rm H\alpha}$. The ID of the source is marked in the lower right corner of each box.}
\label{fig:subtraction_halpha}
\end{center}
\end{figure*}

\addtocounter{figure}{-1}

\begin{figure*}
\begin{center}
\includegraphics[width=6cm]{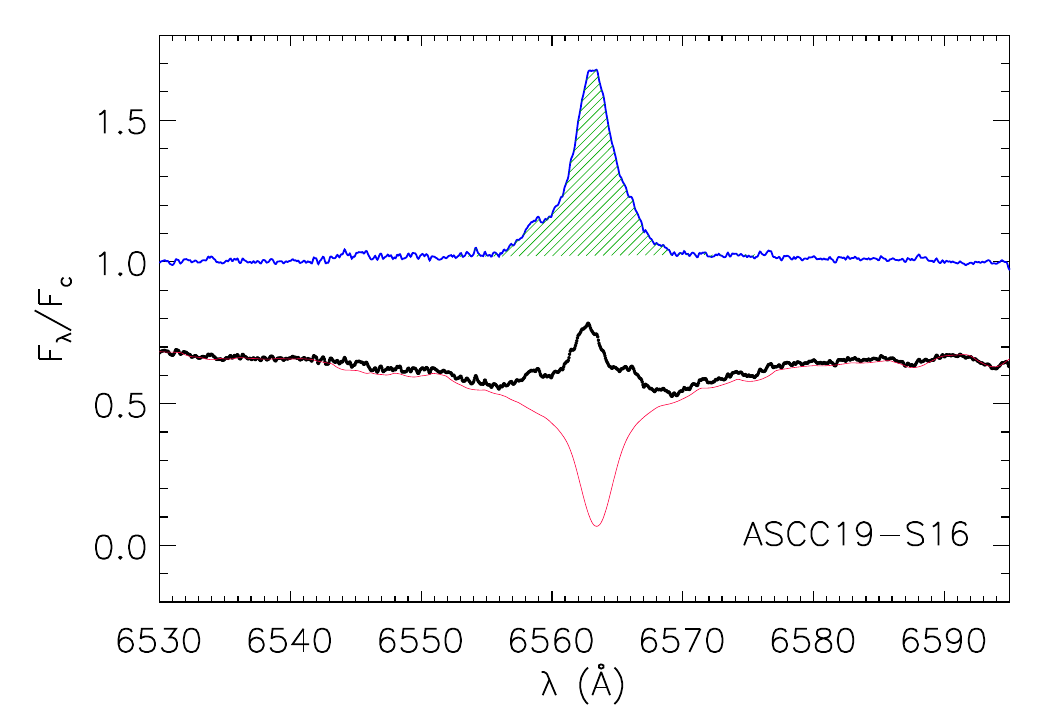} 
\includegraphics[width=6cm]{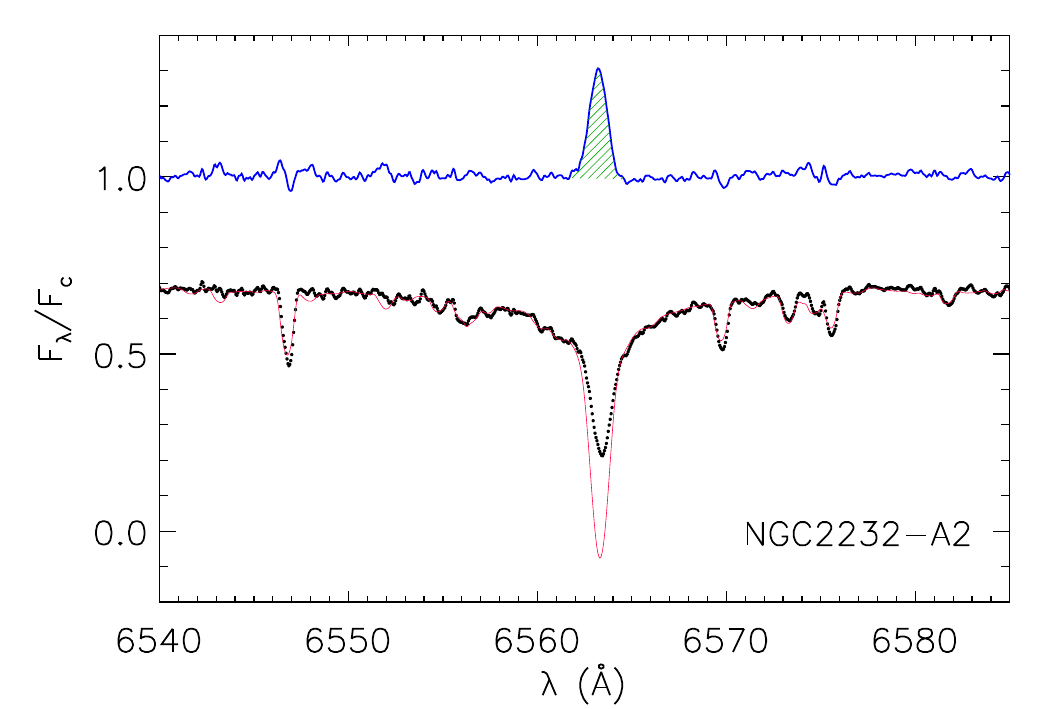} 
\includegraphics[width=6cm]{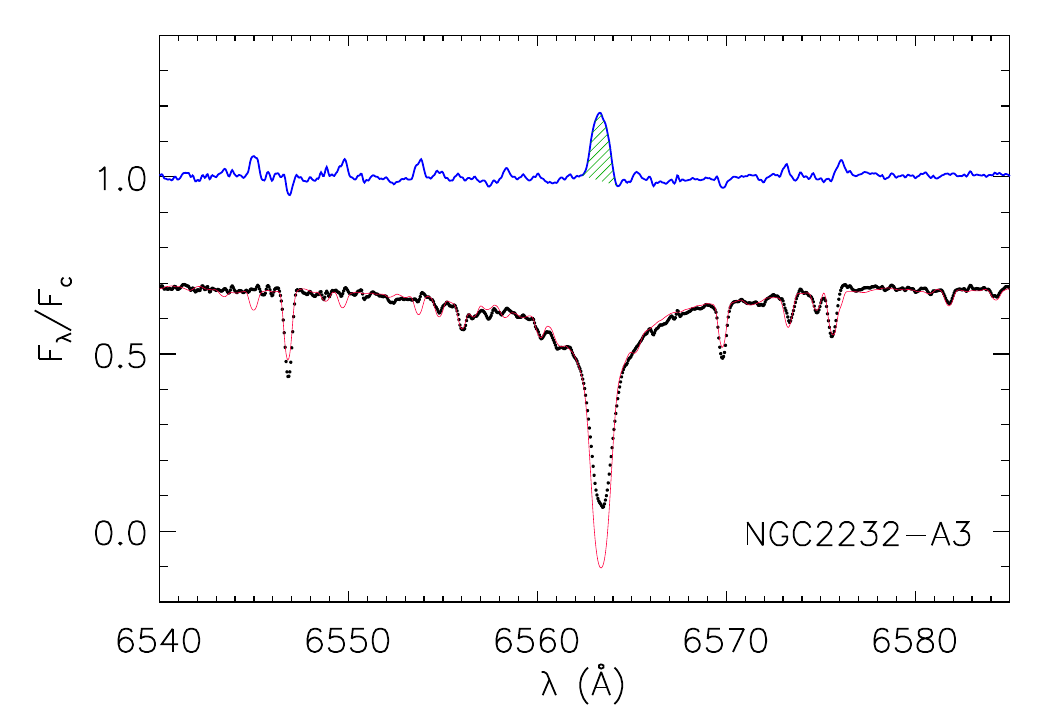} 
\includegraphics[width=6cm]{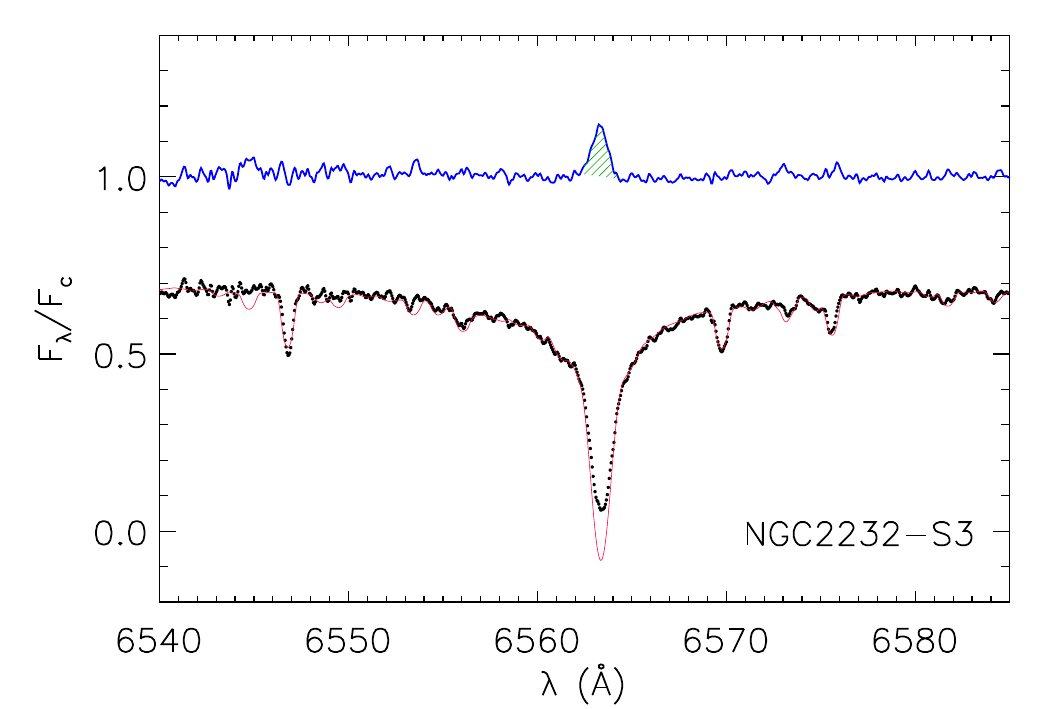} 
\includegraphics[width=6cm]{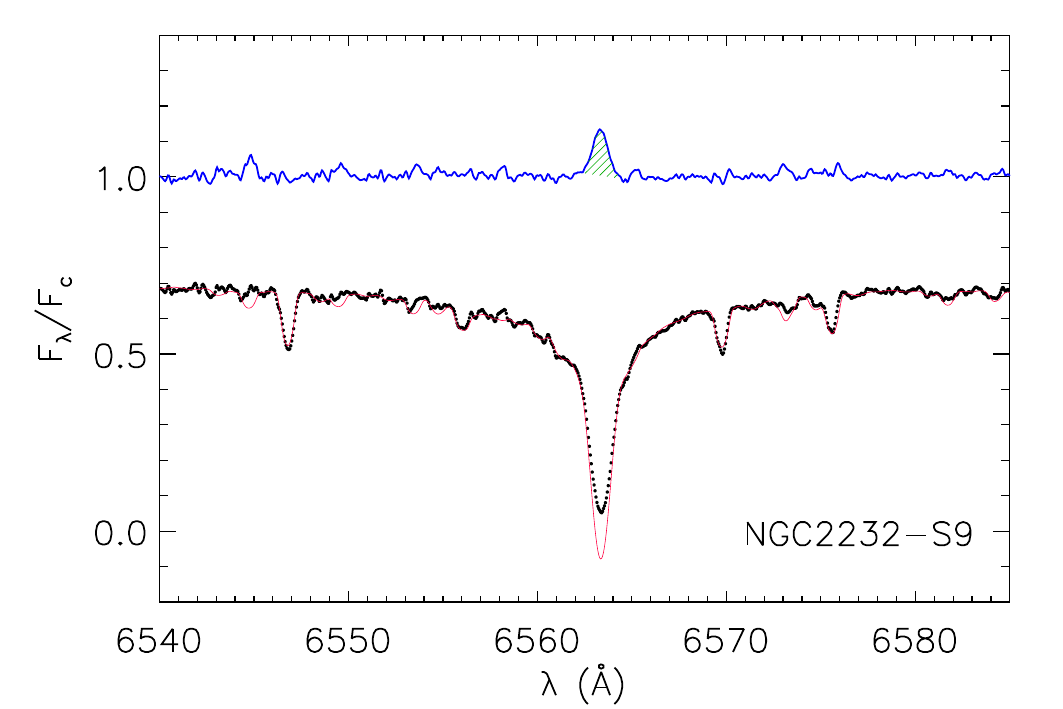} 
\includegraphics[width=6cm]{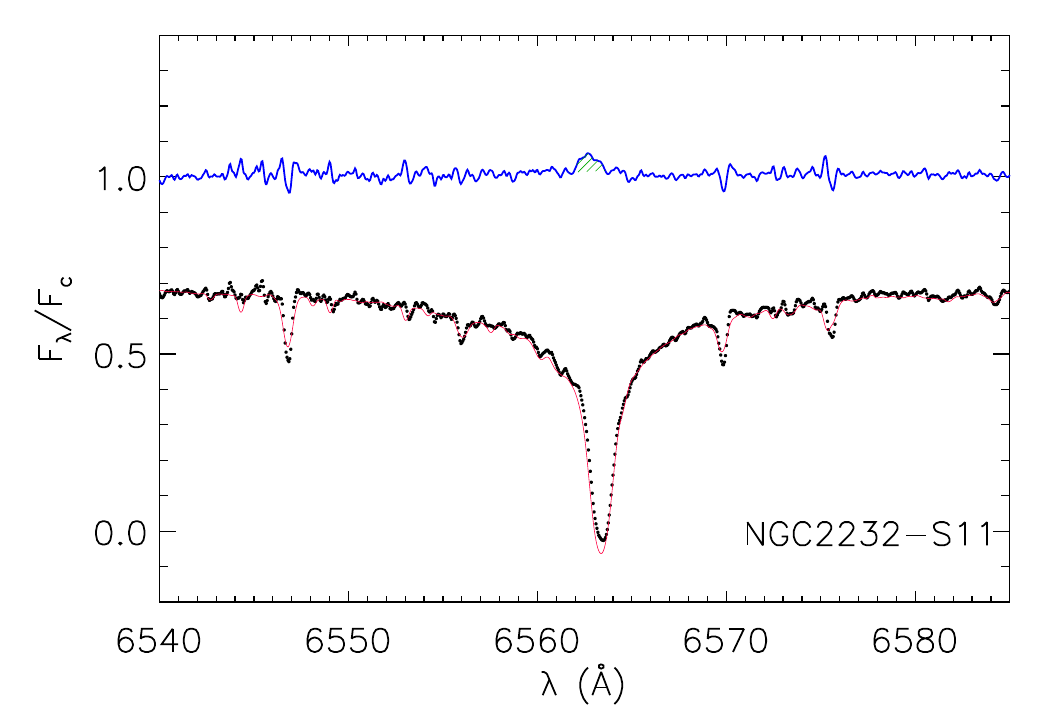} 
\includegraphics[width=6cm]{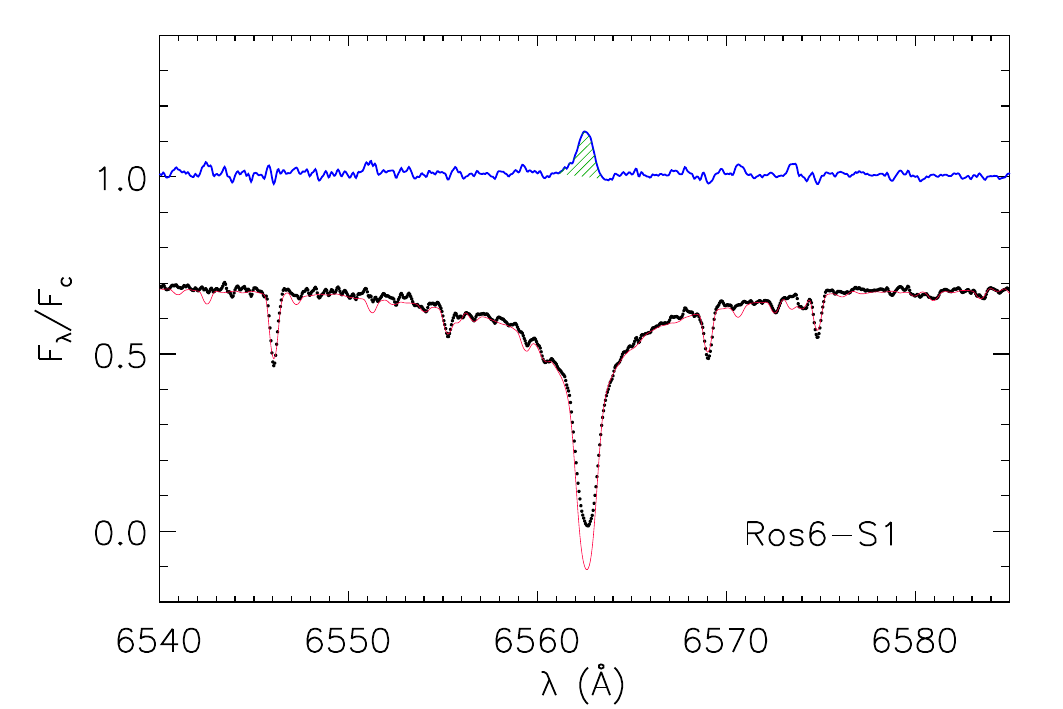} 
\includegraphics[width=6cm]{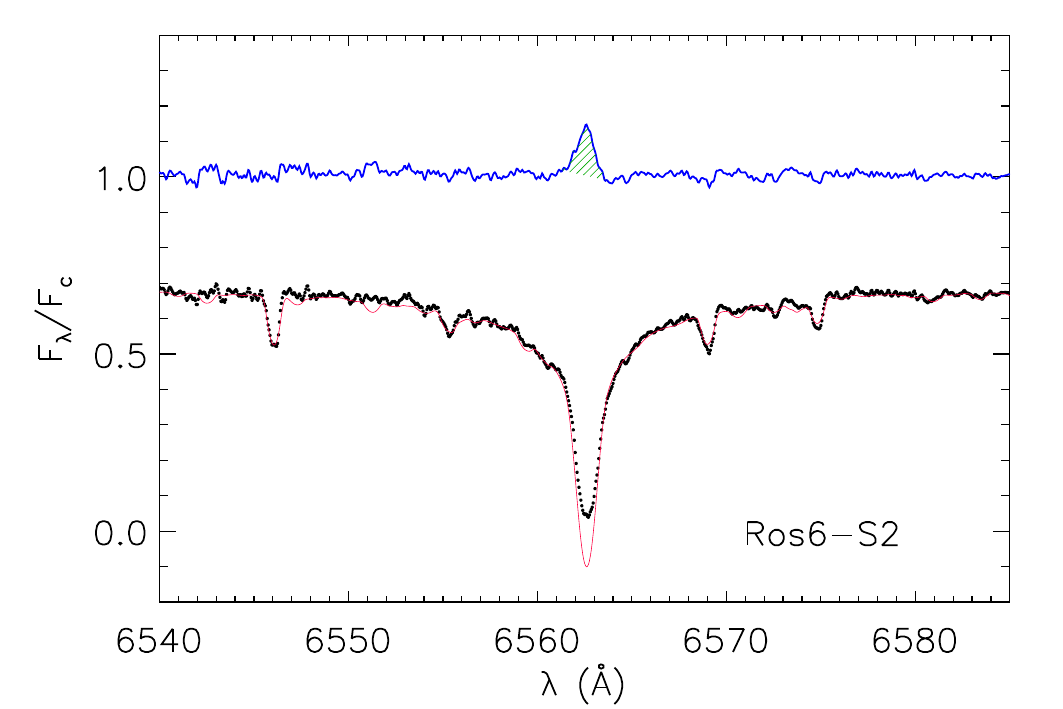} 
\includegraphics[width=6cm]{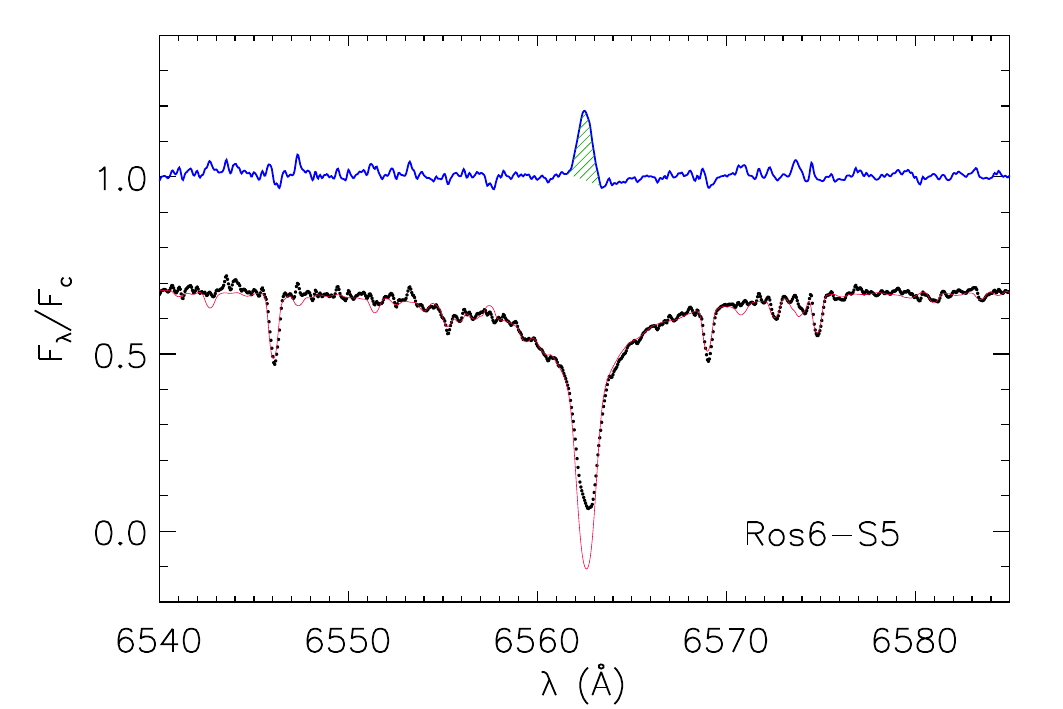} 
\includegraphics[width=6cm]{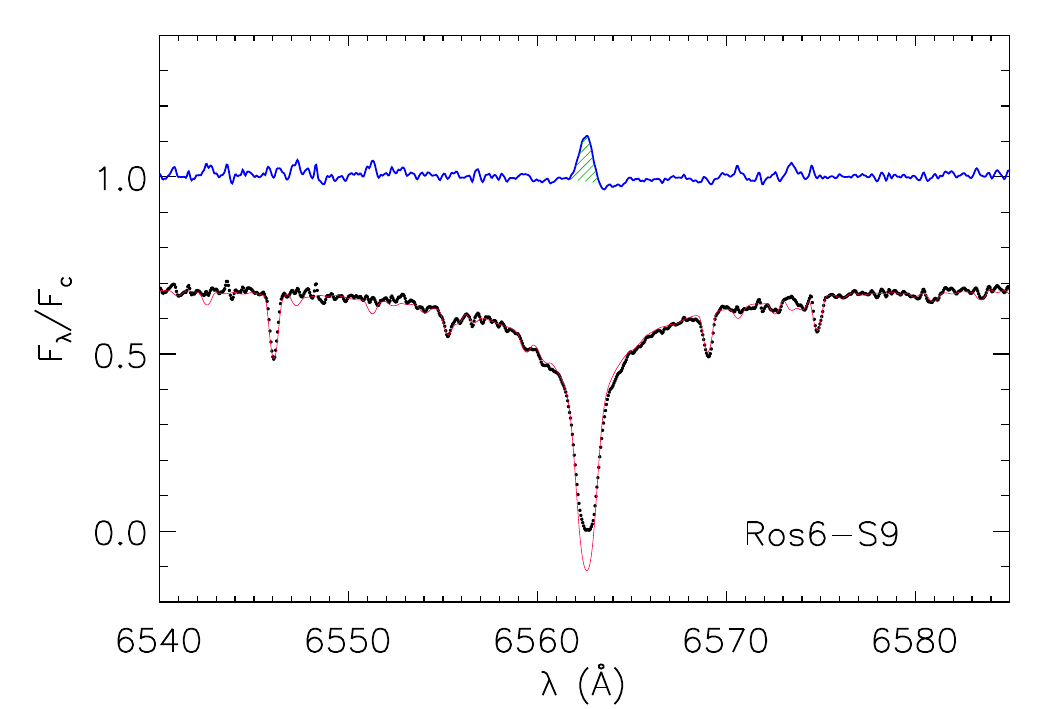} 
\includegraphics[width=6cm]{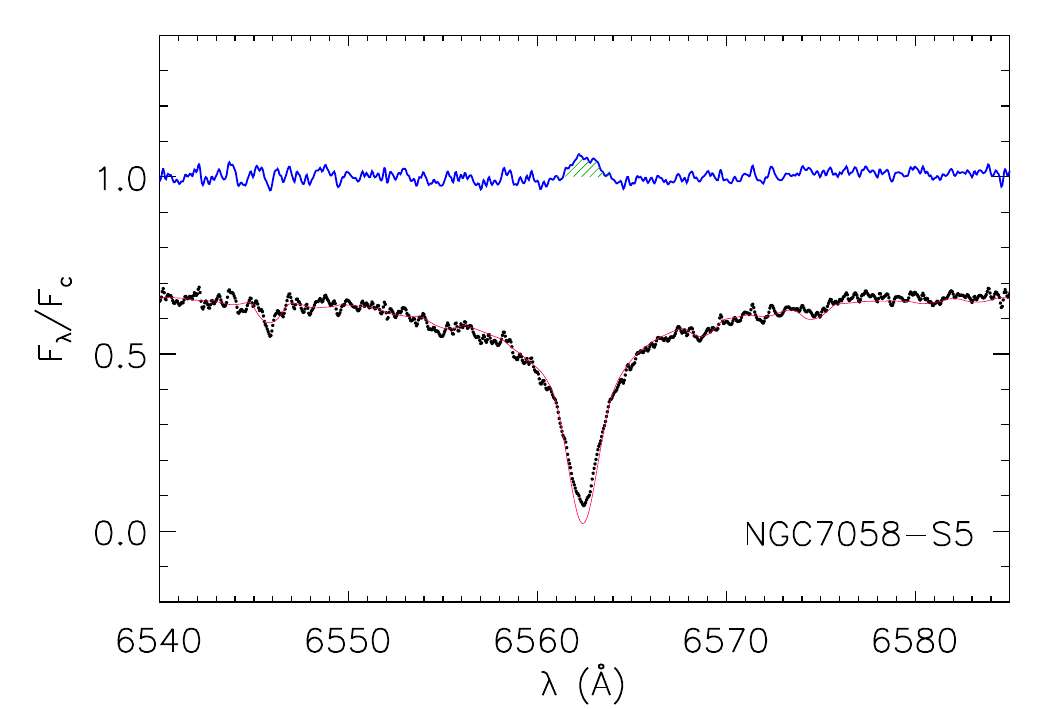} 
\includegraphics[width=6cm]{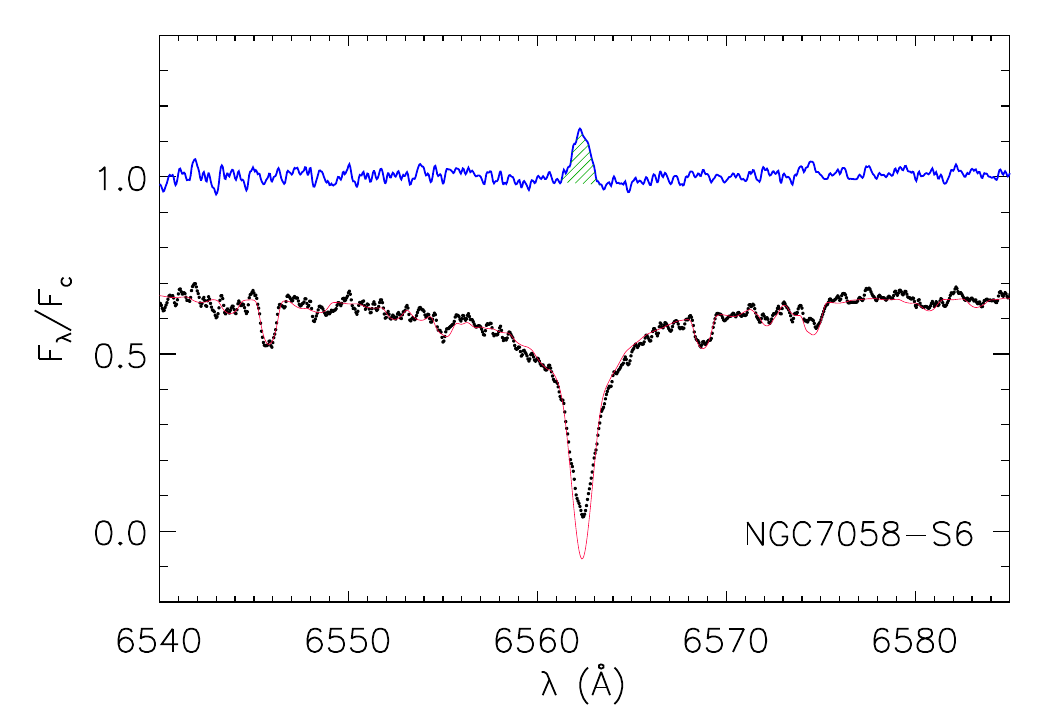} 
\includegraphics[width=6cm]{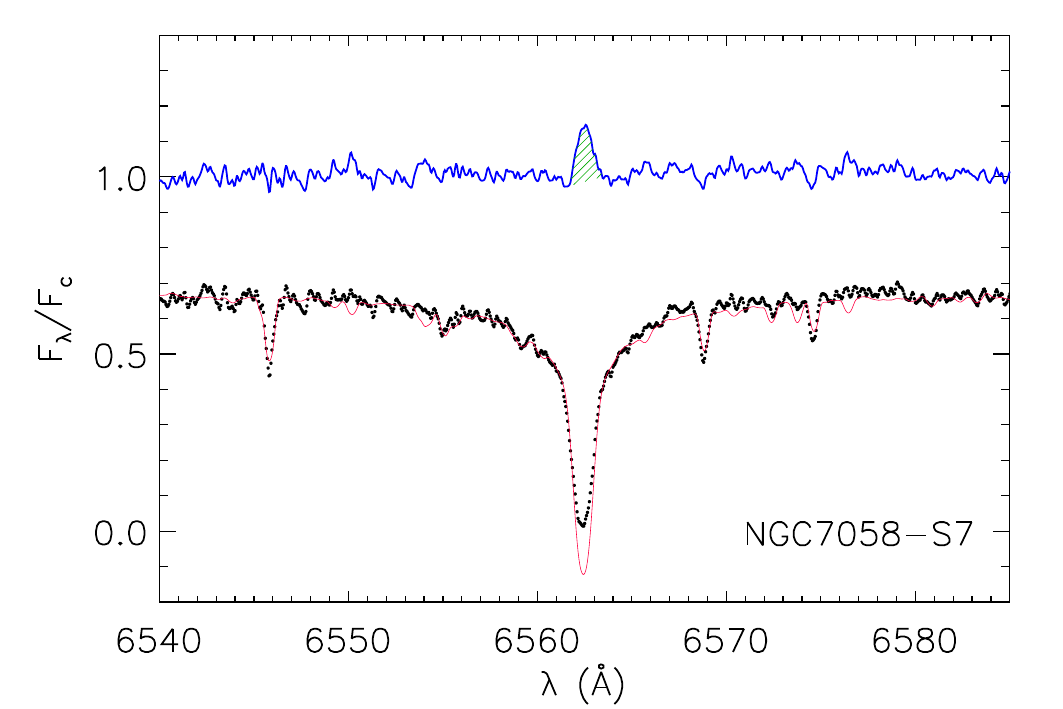} 
\includegraphics[width=6cm]{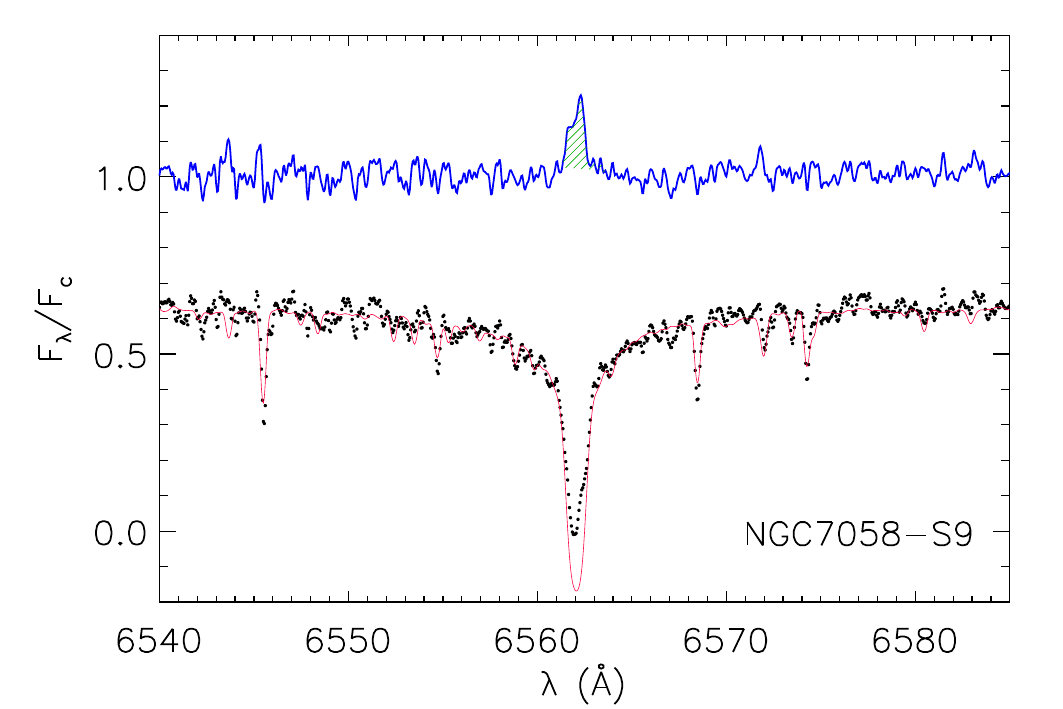} 
\includegraphics[width=6cm]{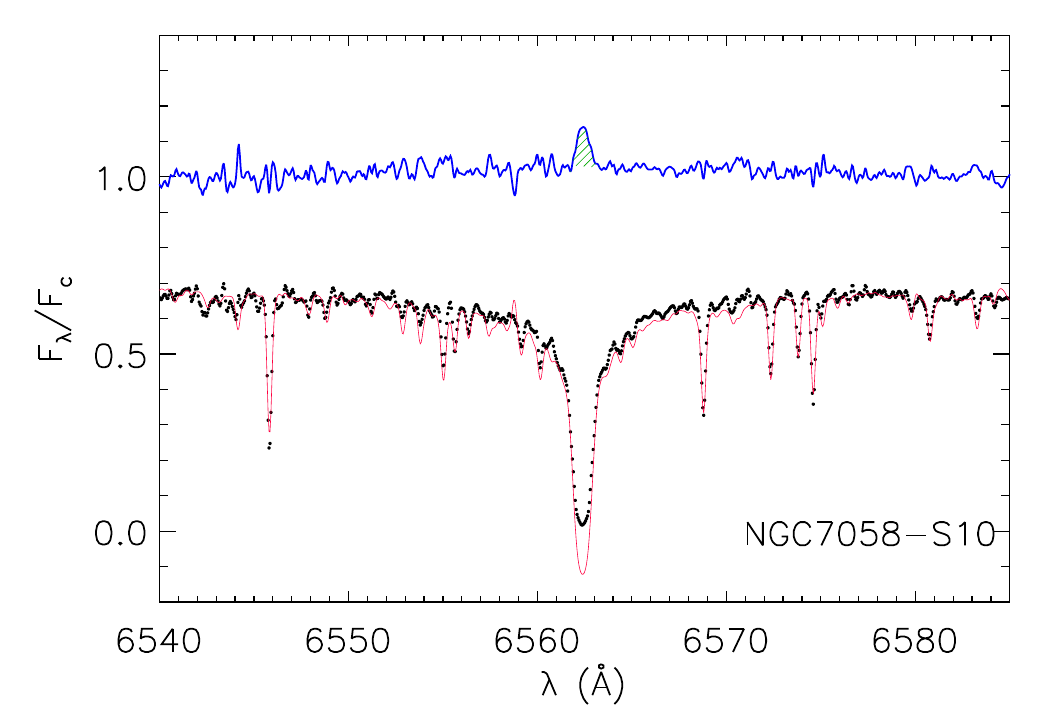} 
\includegraphics[width=6cm]{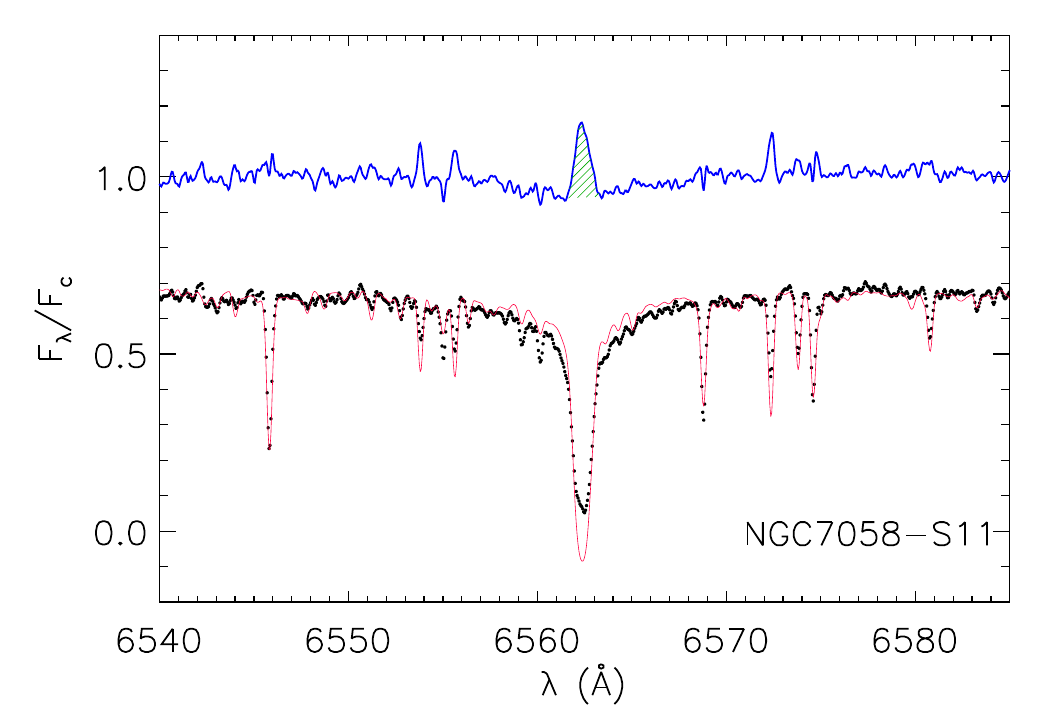} 
\includegraphics[width=6cm]{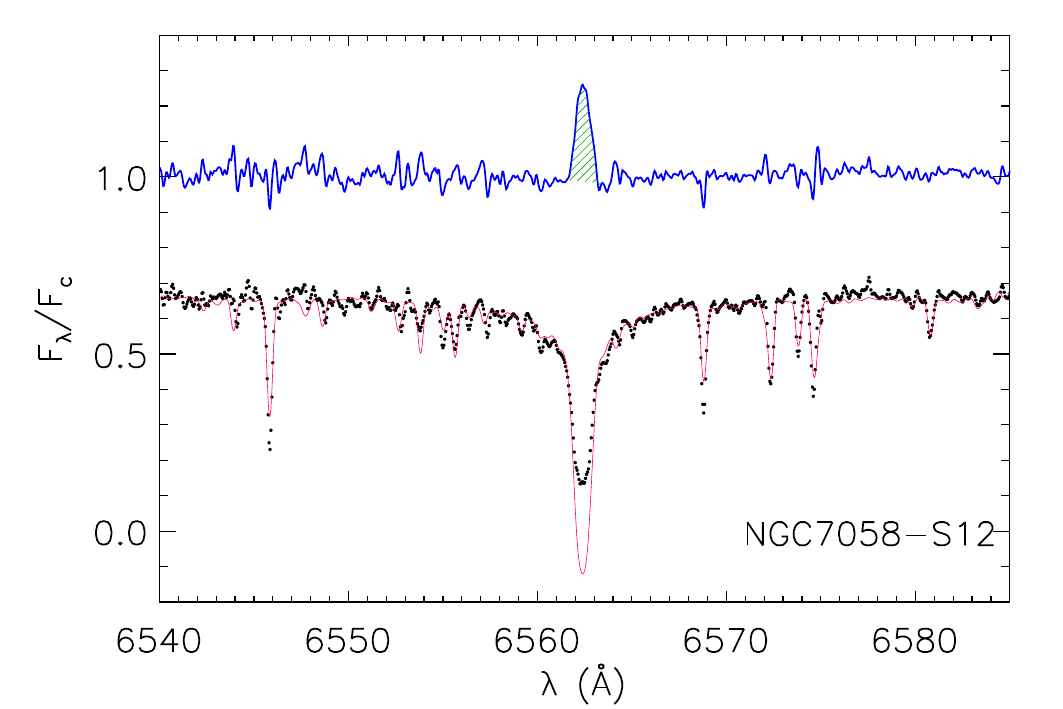} 
\includegraphics[width=6cm]{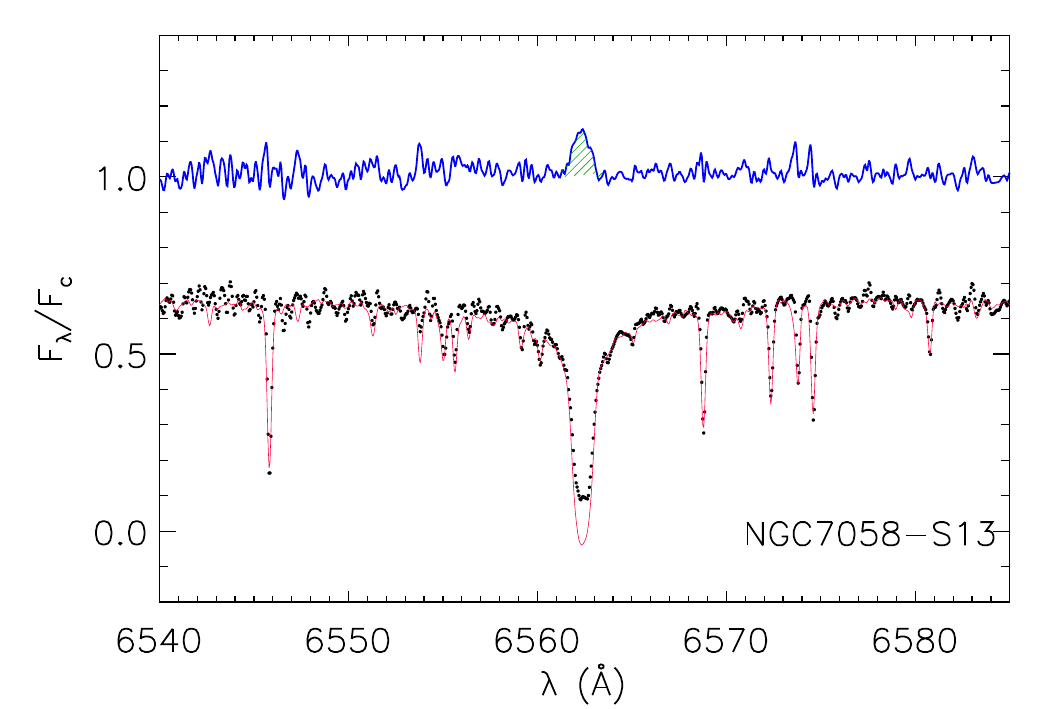} 
\vspace{0cm}
\caption{\textit{Continued}}
\label{fig:subtraction_halpha_bis}
\end{center}
\end{figure*}

\begin{figure*}
\begin{center}
\hspace{-.7cm}
\includegraphics[width=6cm]{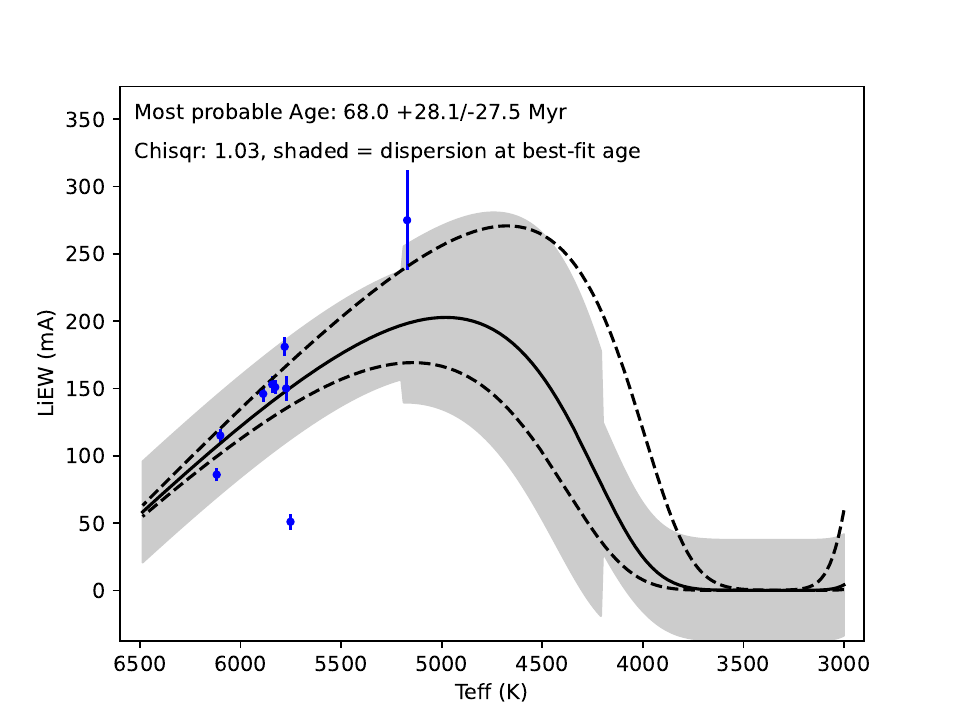}
\includegraphics[width=6cm]{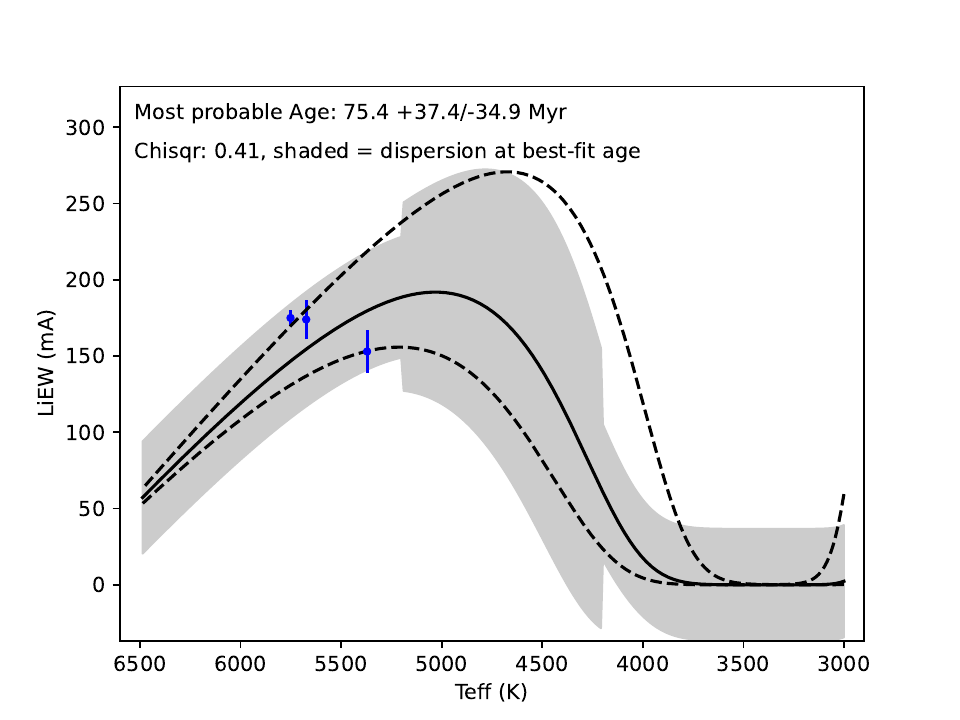}
\includegraphics[width=6cm]{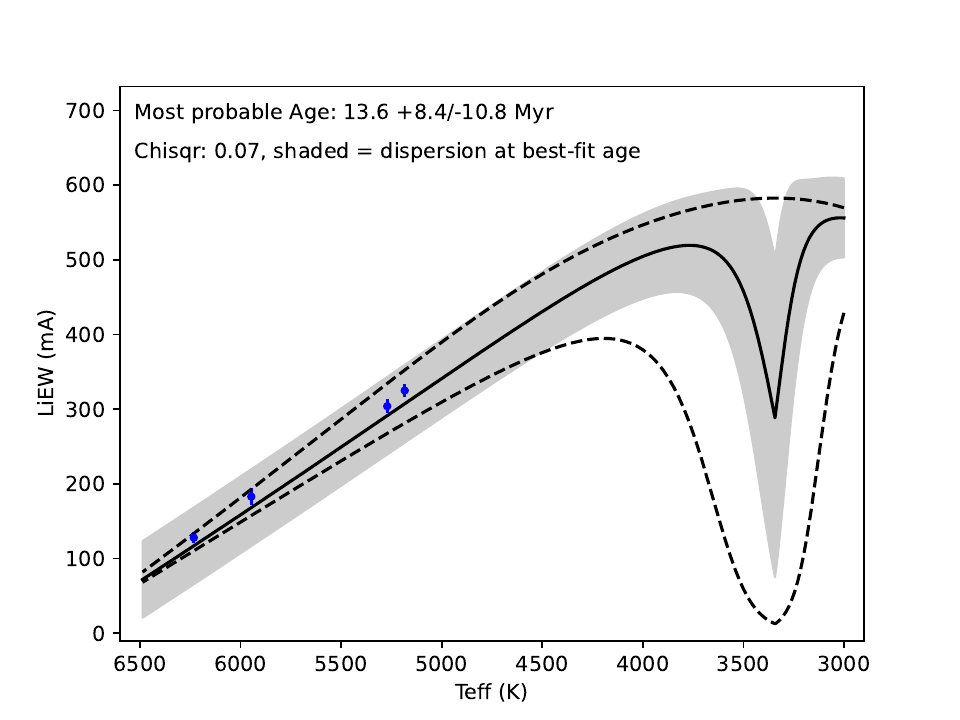}
\includegraphics[width=6cm]{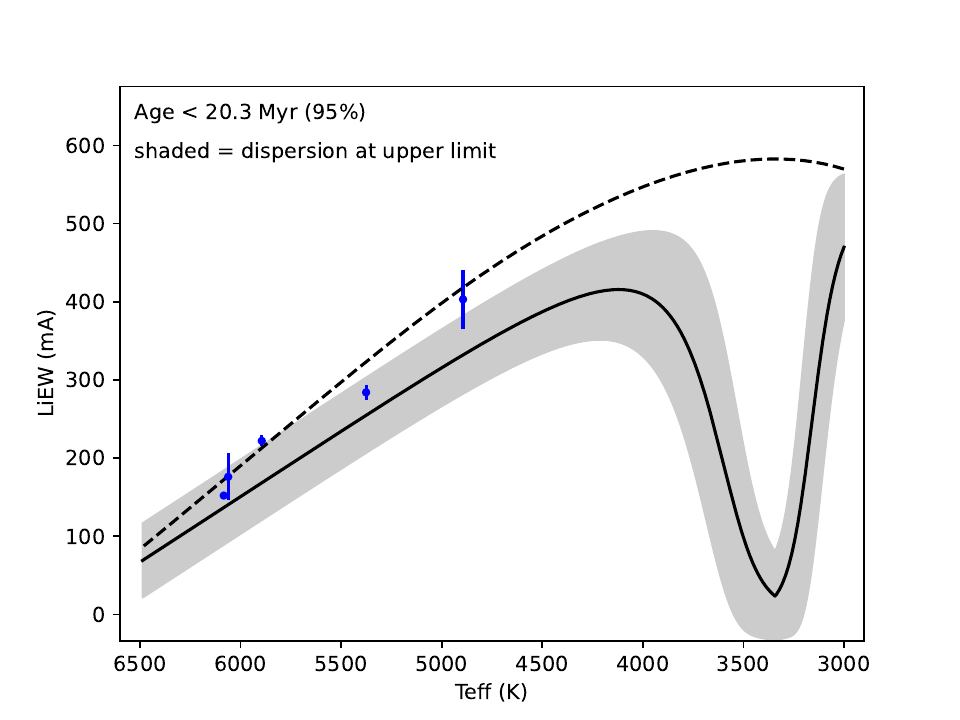}
\includegraphics[width=6cm]{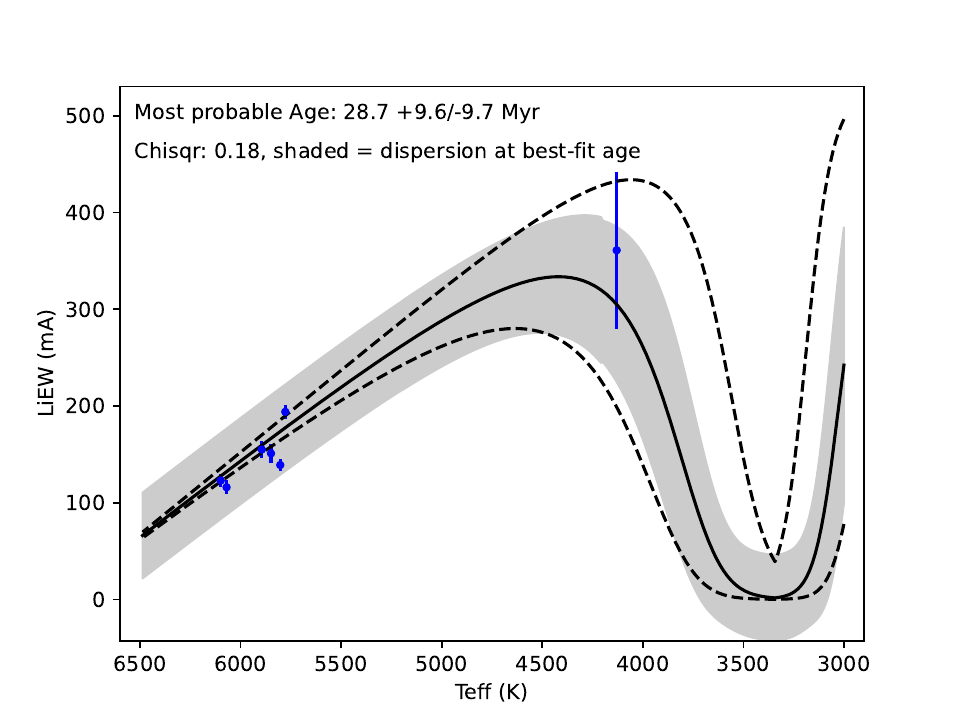}
\includegraphics[width=6cm]{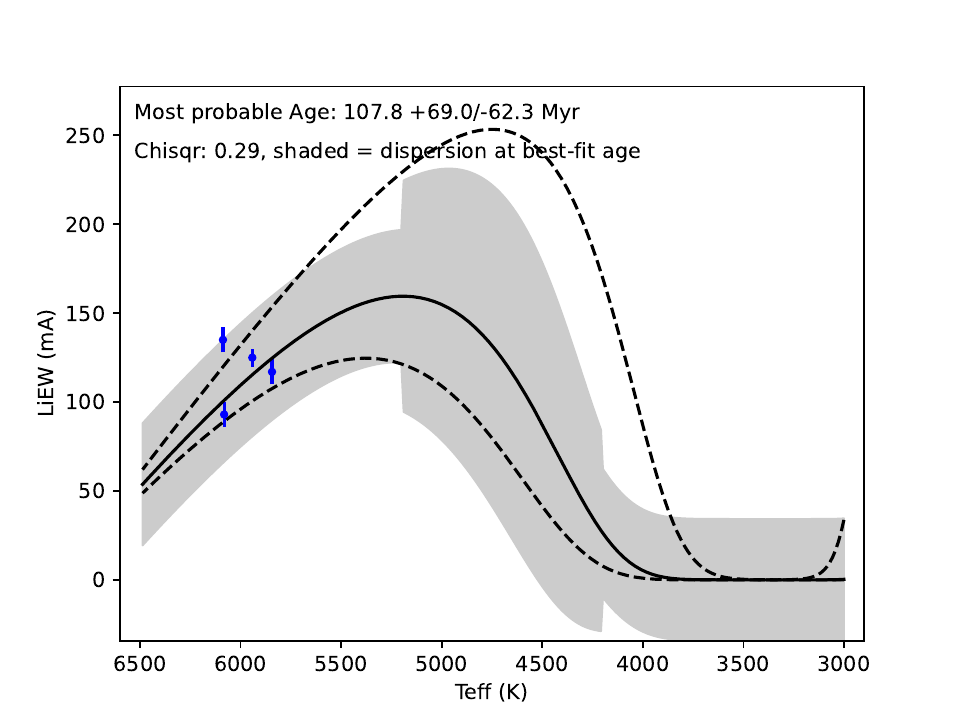}
\includegraphics[width=6cm]{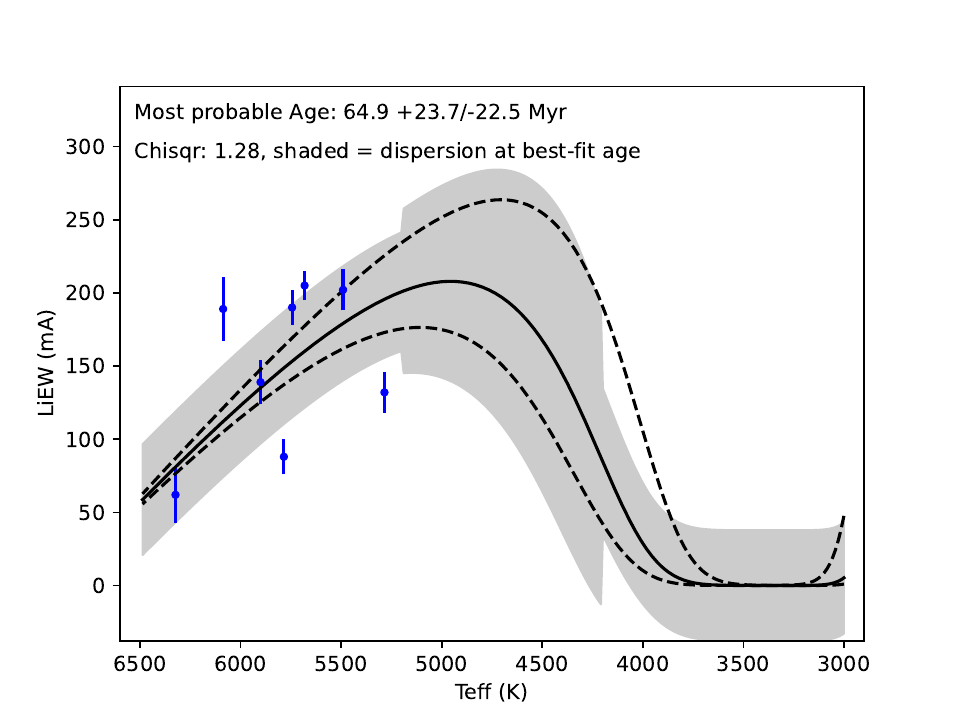}
\includegraphics[width=6cm]{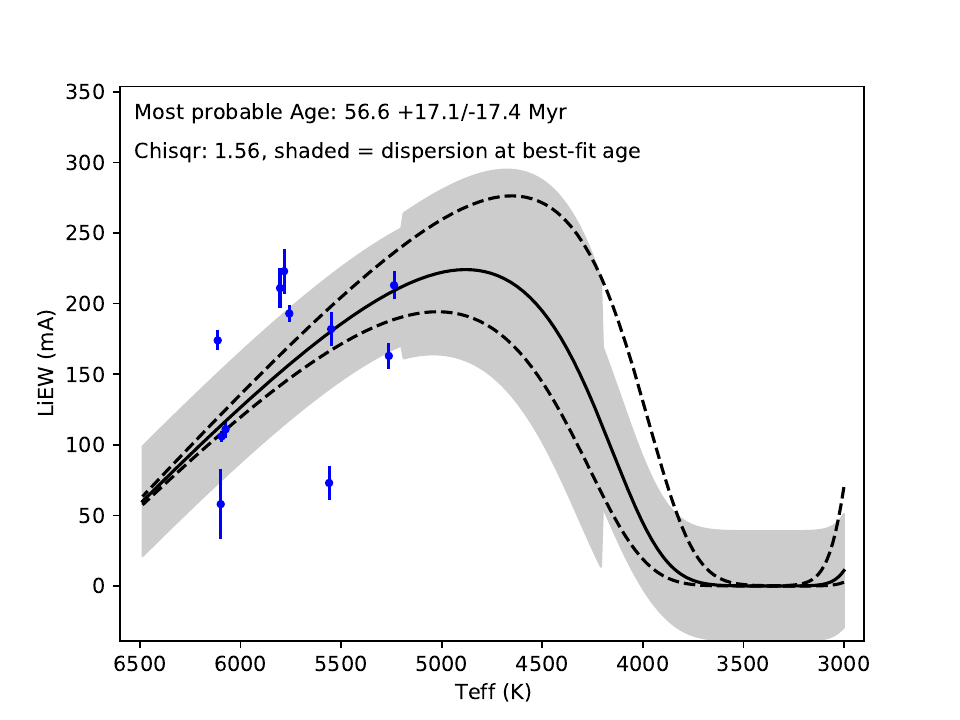}
\caption{Fit to the lithium depletion pattern of the members of the investigated clusters made with the \eagles\ code \citep{Jeffries2023}.
}
\label{fig:EAGLES}
\end{center}
\end{figure*}

\begin{figure*}
\begin{center}
\includegraphics[width=9.1cm]{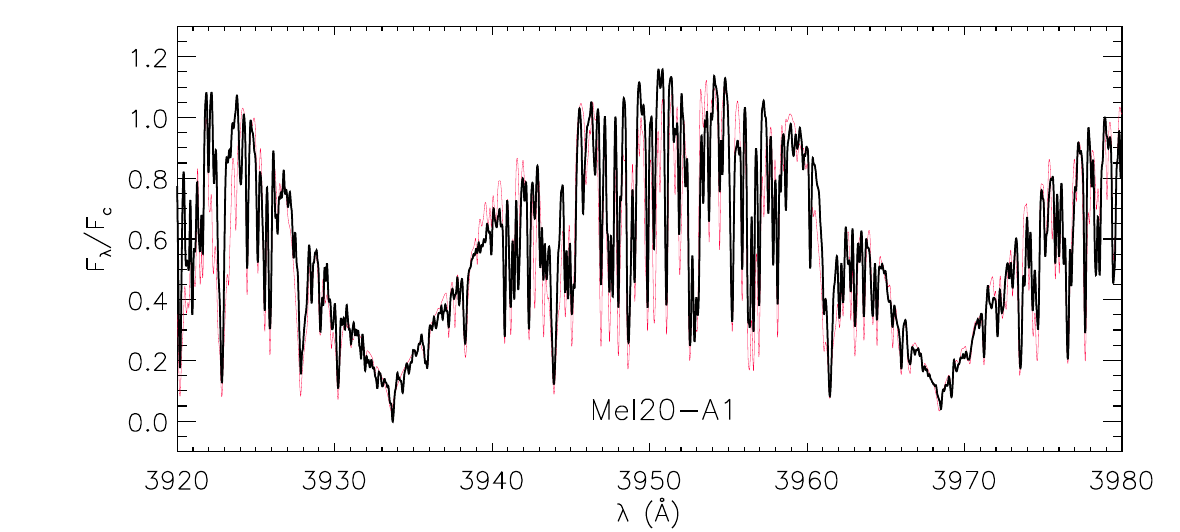} 
\includegraphics[width=9.1cm]{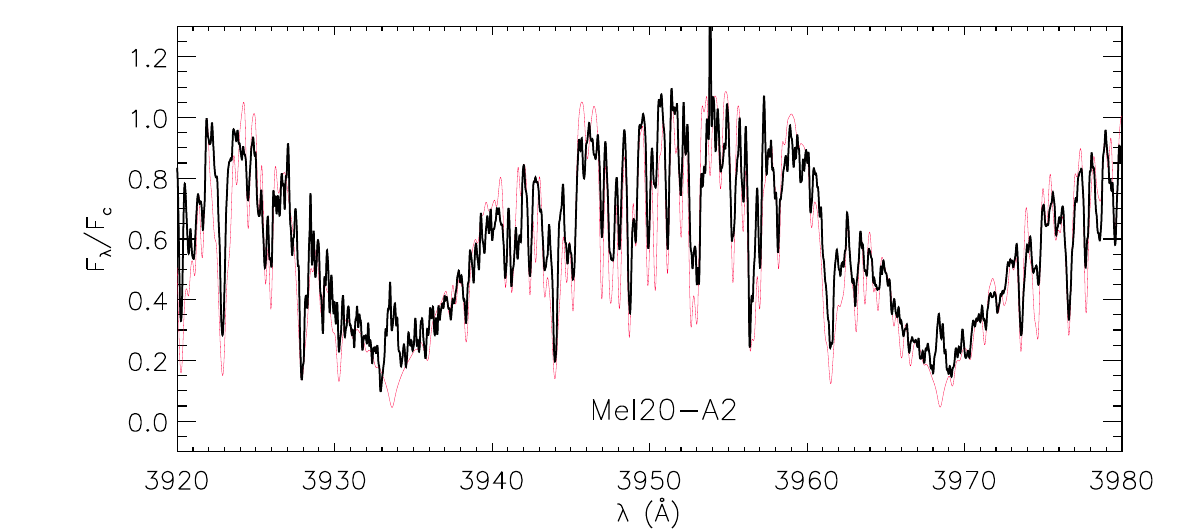}     
\includegraphics[width=9.1cm]{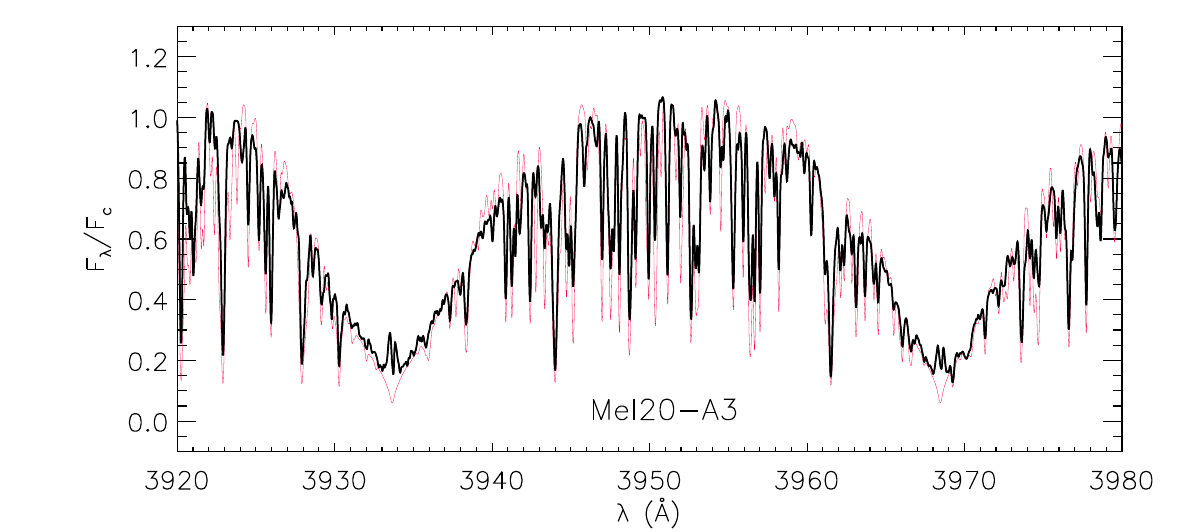}     
\includegraphics[width=9.1cm]{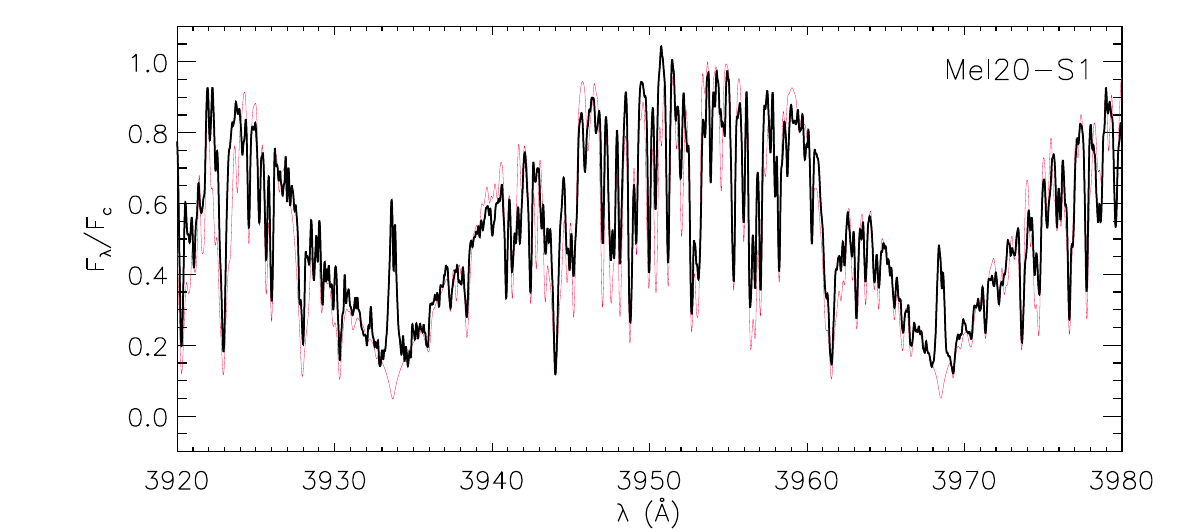}     
\includegraphics[width=9.1cm]{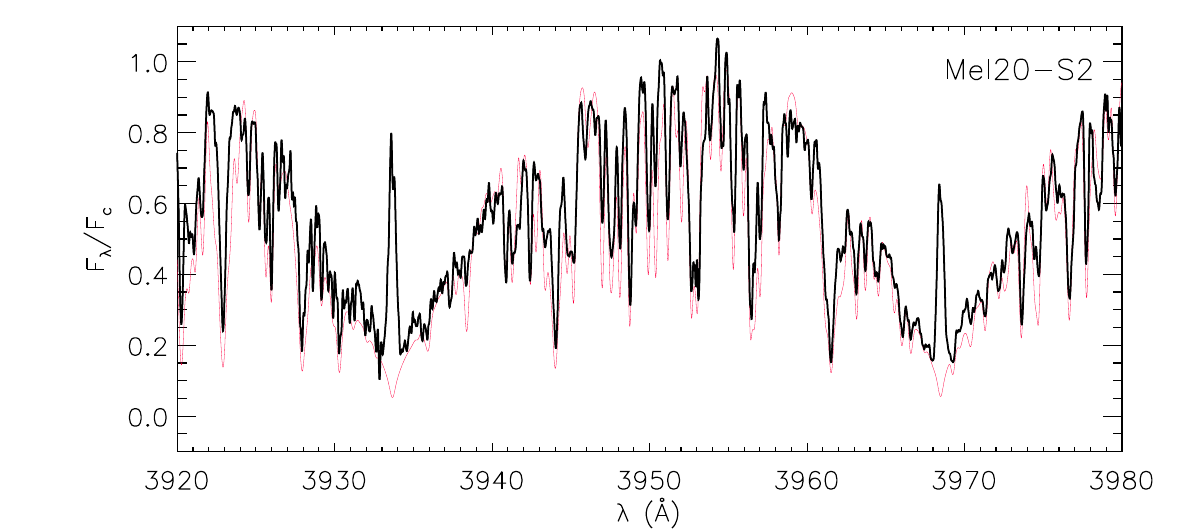}      
\includegraphics[width=9.1cm]{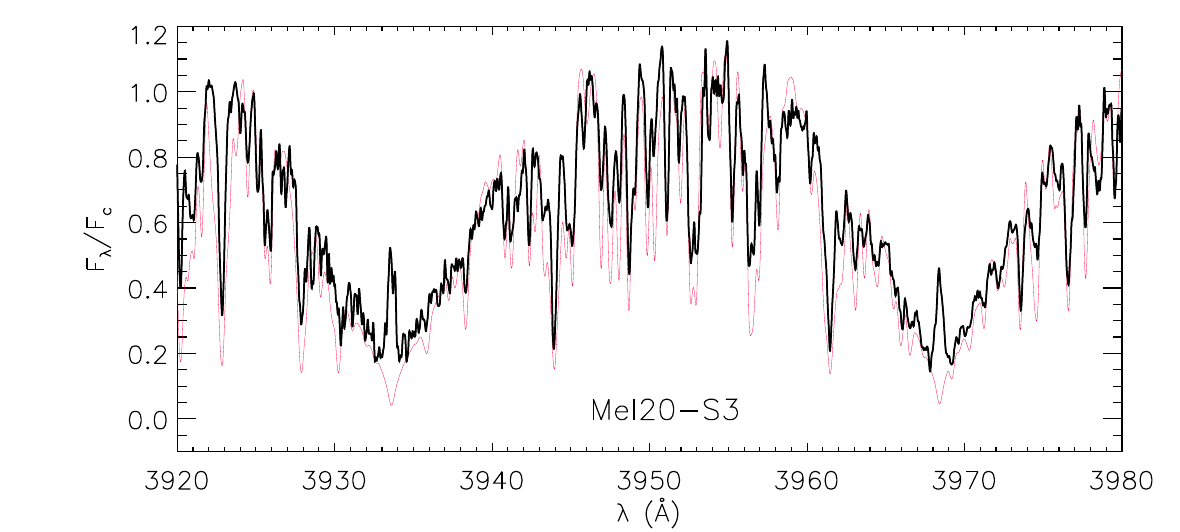} 
\includegraphics[width=9.1cm]{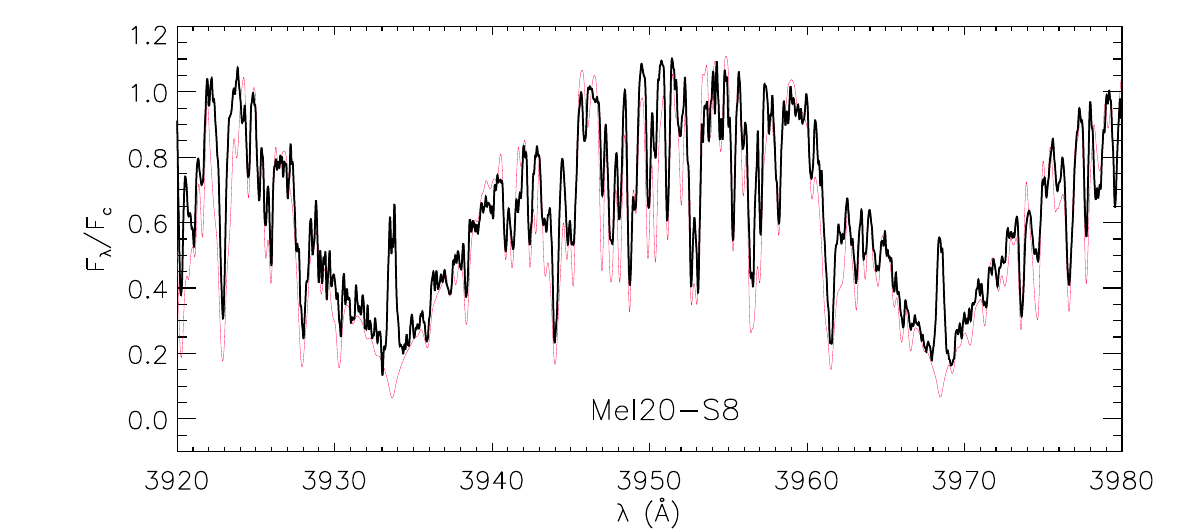} 
\includegraphics[width=9.1cm]{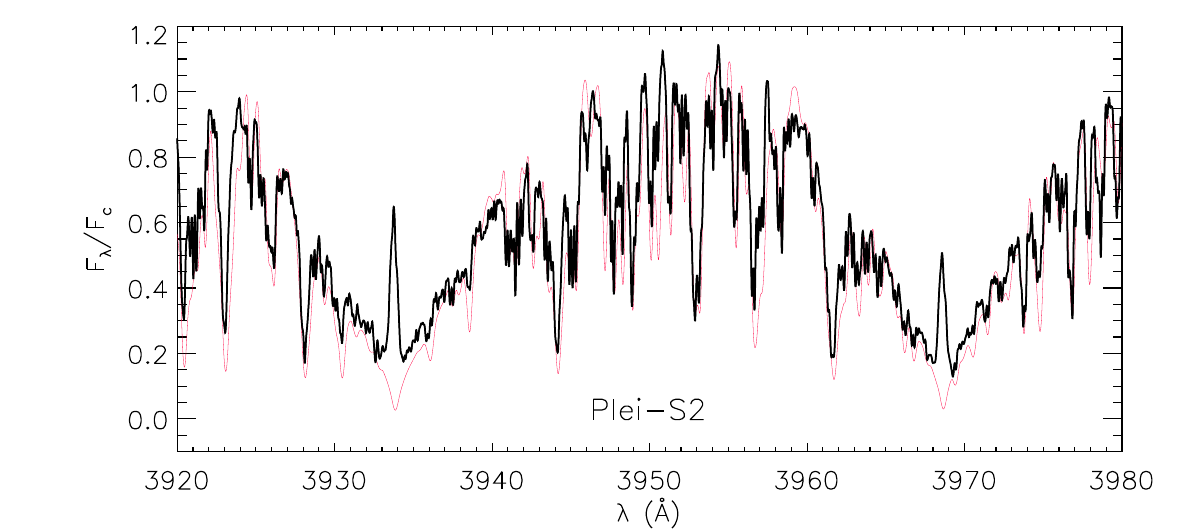} 
\includegraphics[width=9.1cm]{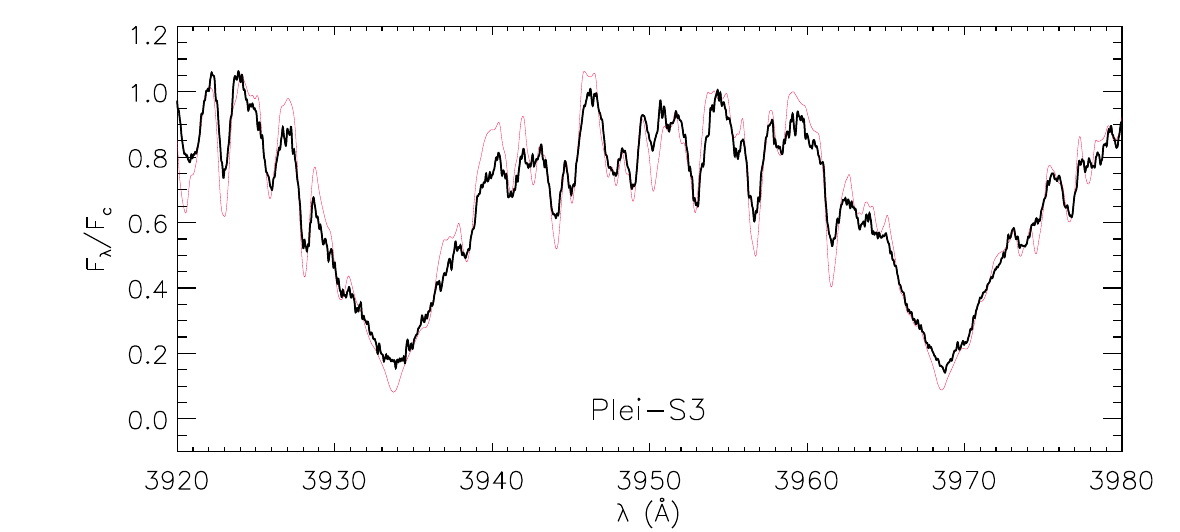} 
\includegraphics[width=9.1cm]{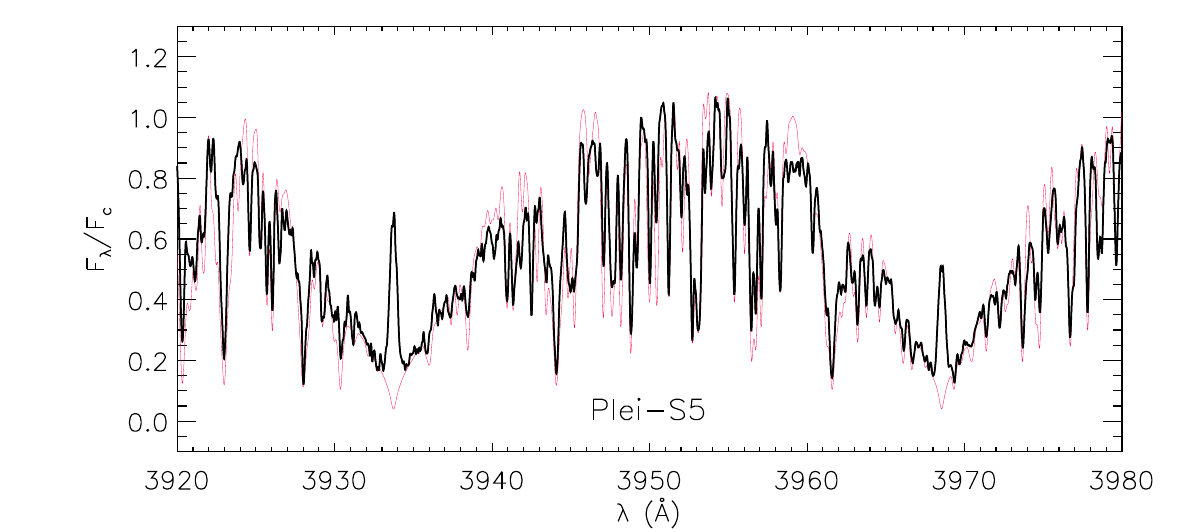} 
\vspace{0cm}
\caption{HARPS-N spectra of the investigated stars in the \ion{Ca}{ii} H\&K region. 
In each box, the non-active template (red line) is overlaid with the observed spectrum (black dots). 
The ID of the source is marked in the lower right corner of each box.}
\label{fig:subtraction_CaII_HK}
\end{center}
\end{figure*}

\addtocounter{figure}{-1}

\begin{figure*}
\begin{center}
\includegraphics[width=9.1cm]{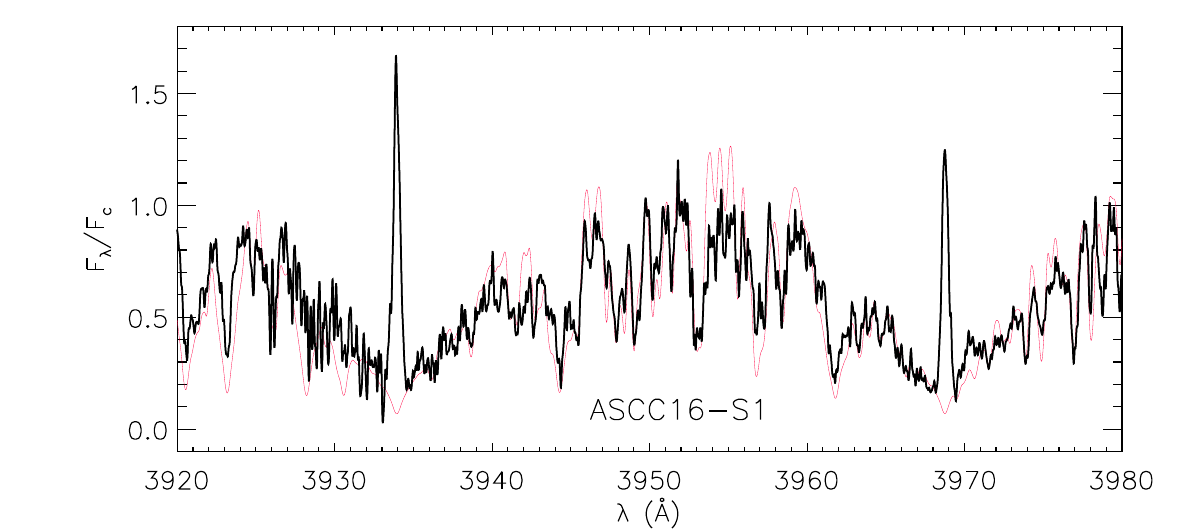} 
\includegraphics[width=9.1cm]{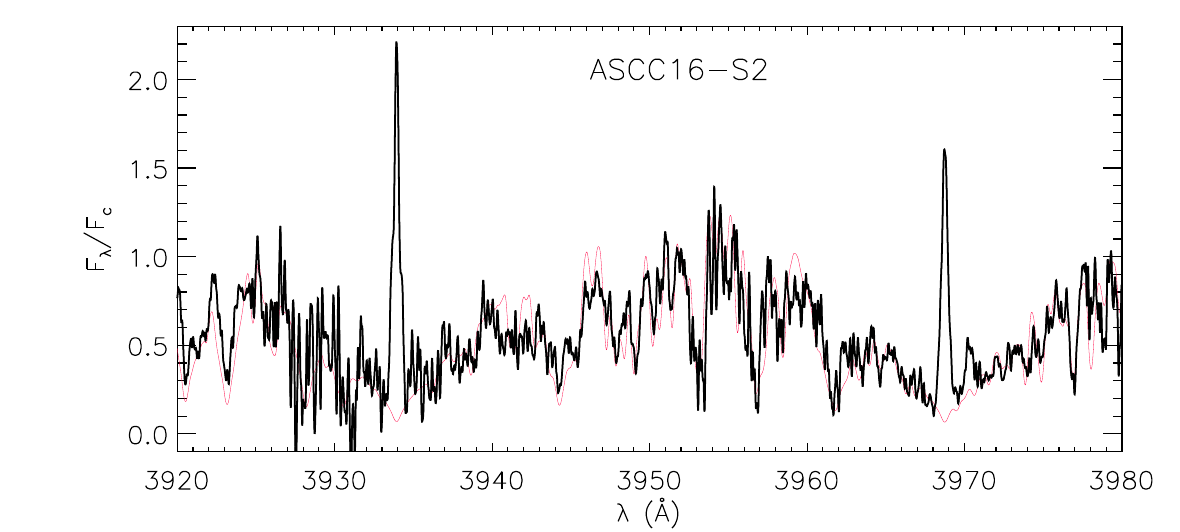} 
\includegraphics[width=9.1cm]{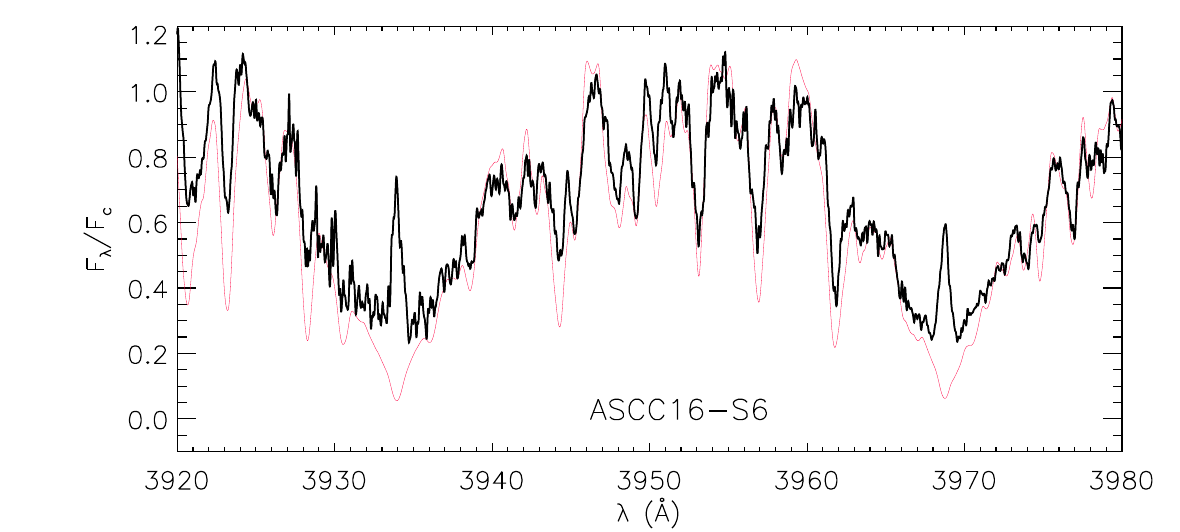} 
\includegraphics[width=9.1cm]{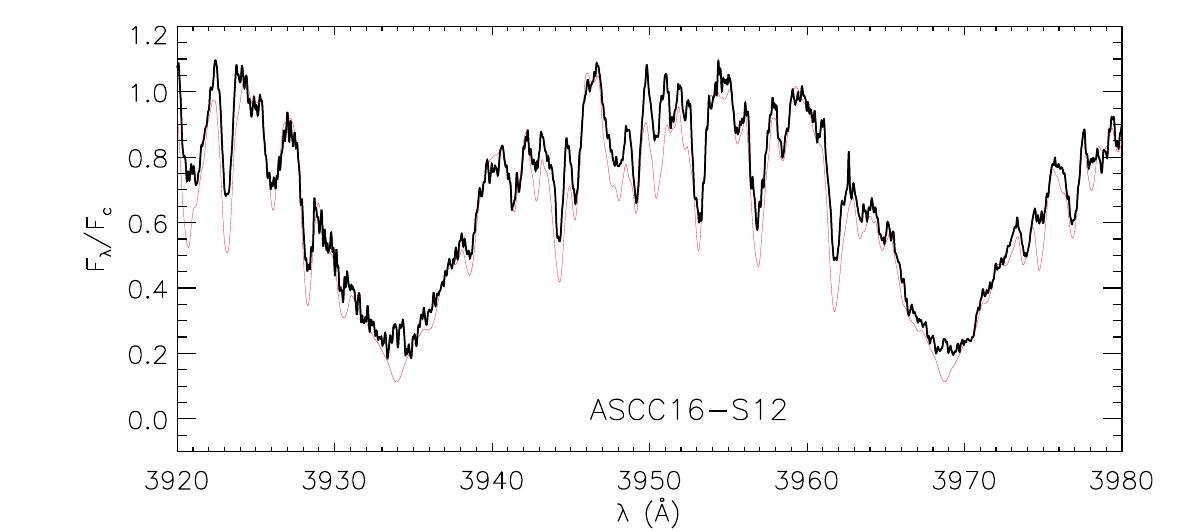} 
\includegraphics[width=9.1cm]{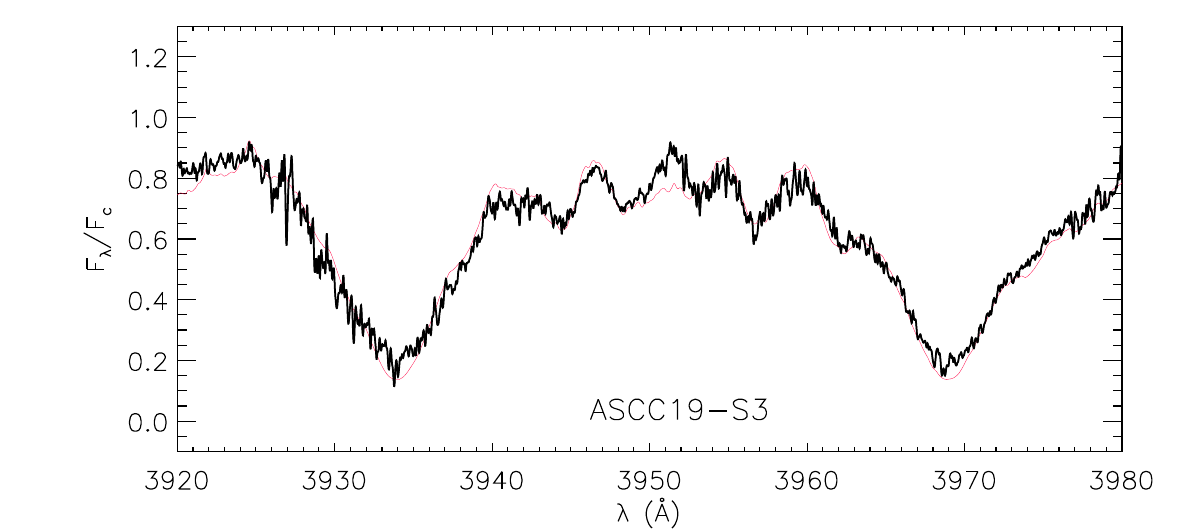} 
\includegraphics[width=9.1cm]{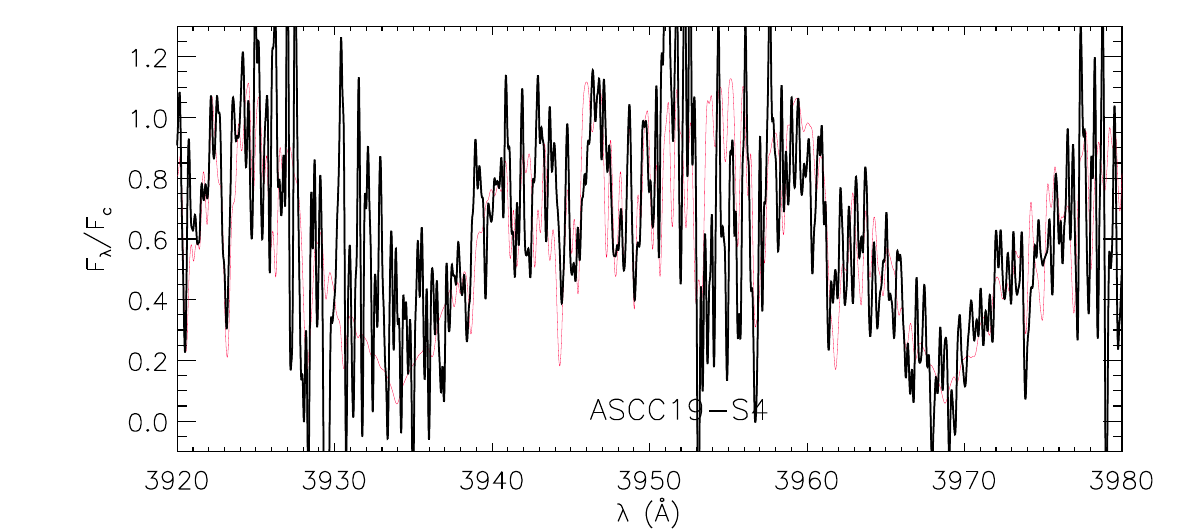} 
\includegraphics[width=9.1cm]{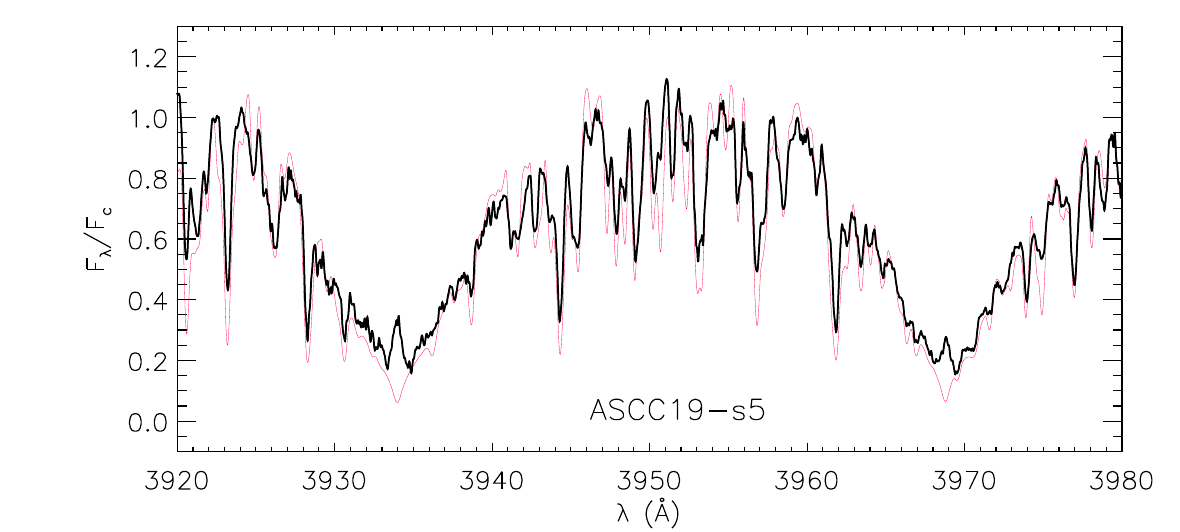} 
\includegraphics[width=9.1cm]{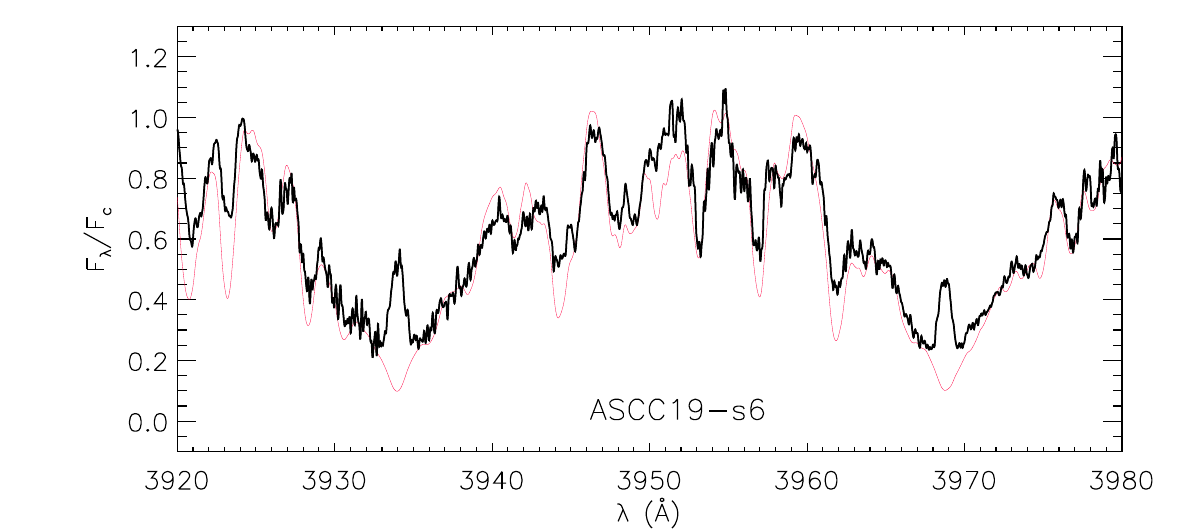} 
\includegraphics[width=9.1cm]{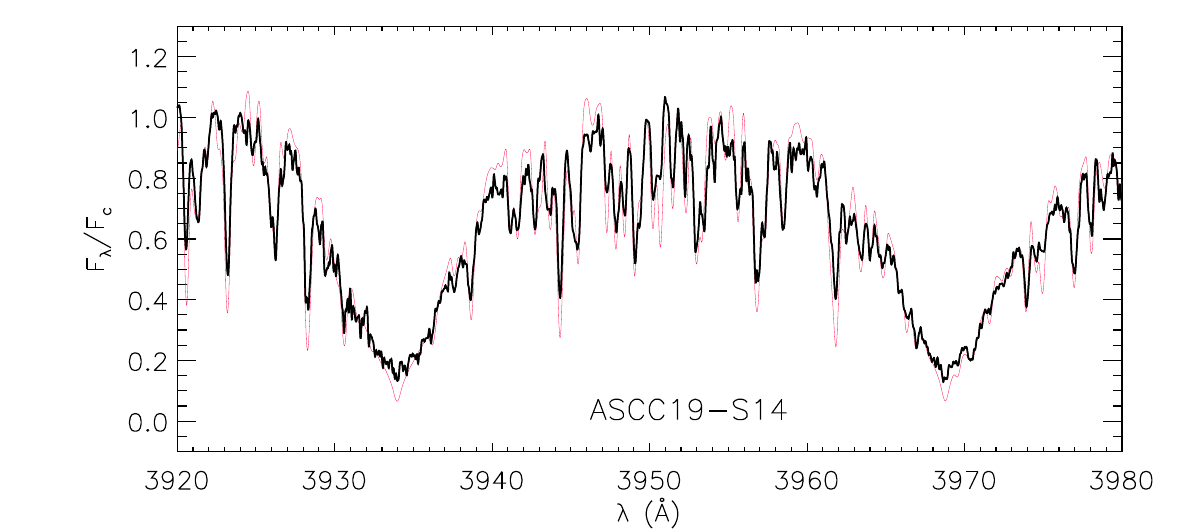} 
\includegraphics[width=9.1cm]{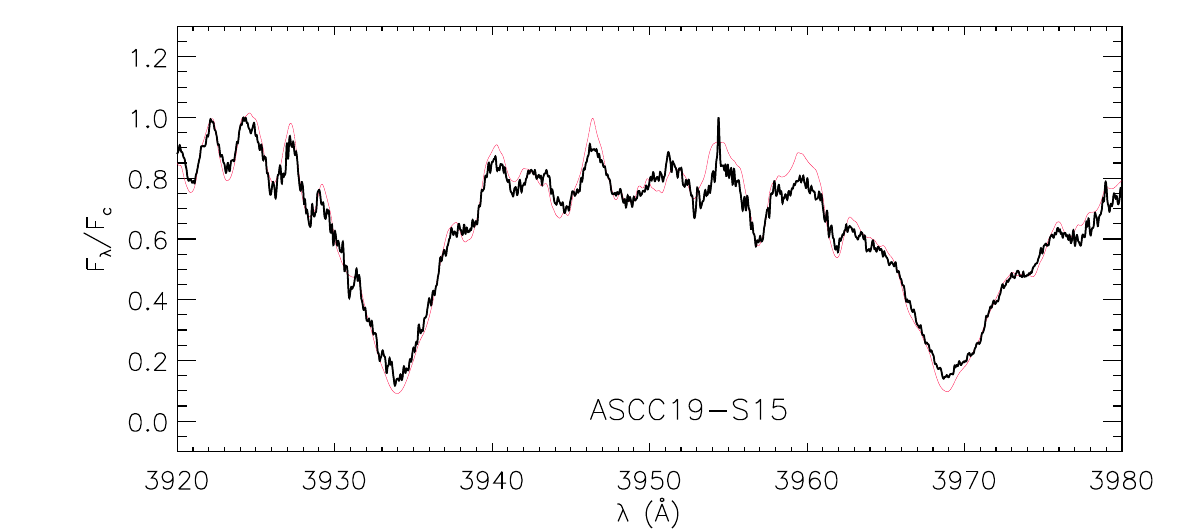} 
\vspace{0cm}
\caption{\textit{Continued}}
\label{fig:subtraction_CaII_bis}
\end{center}
\end{figure*}

\addtocounter{figure}{-1}

\begin{figure*}
\begin{center}
\includegraphics[width=9.1cm]{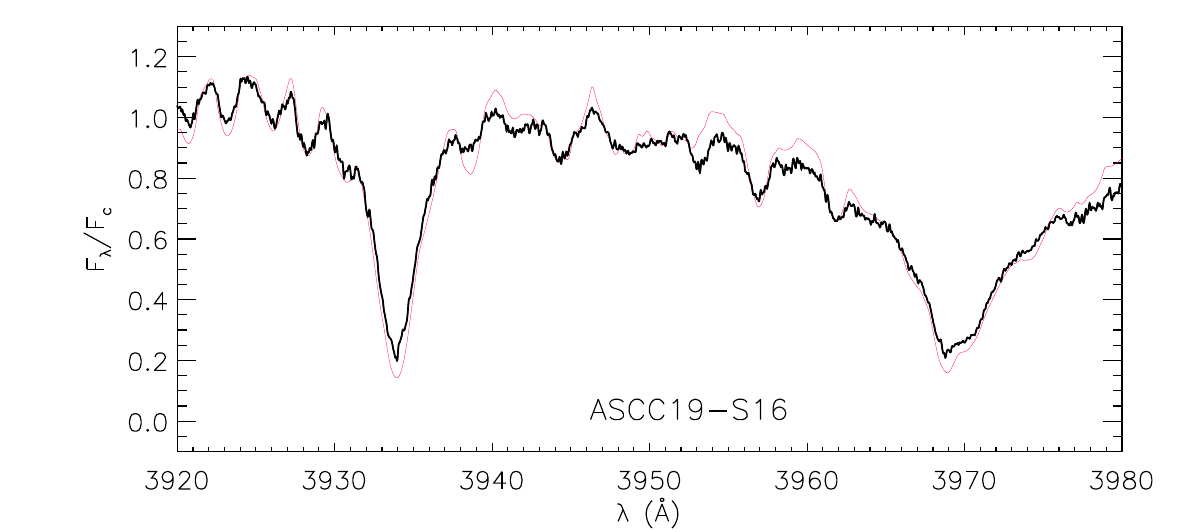} 
\includegraphics[width=9.1cm]{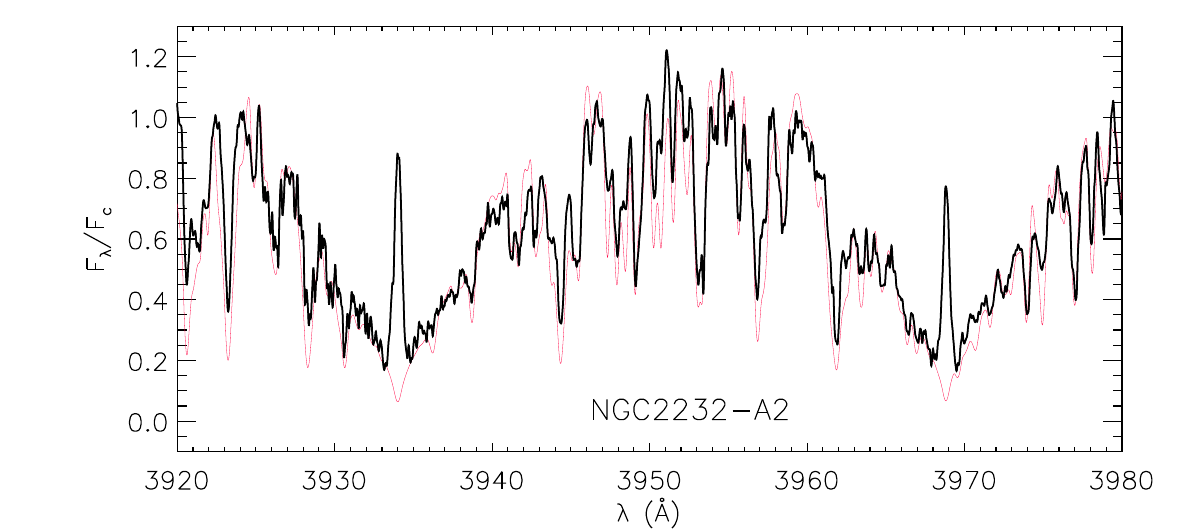} 
\includegraphics[width=9.1cm]{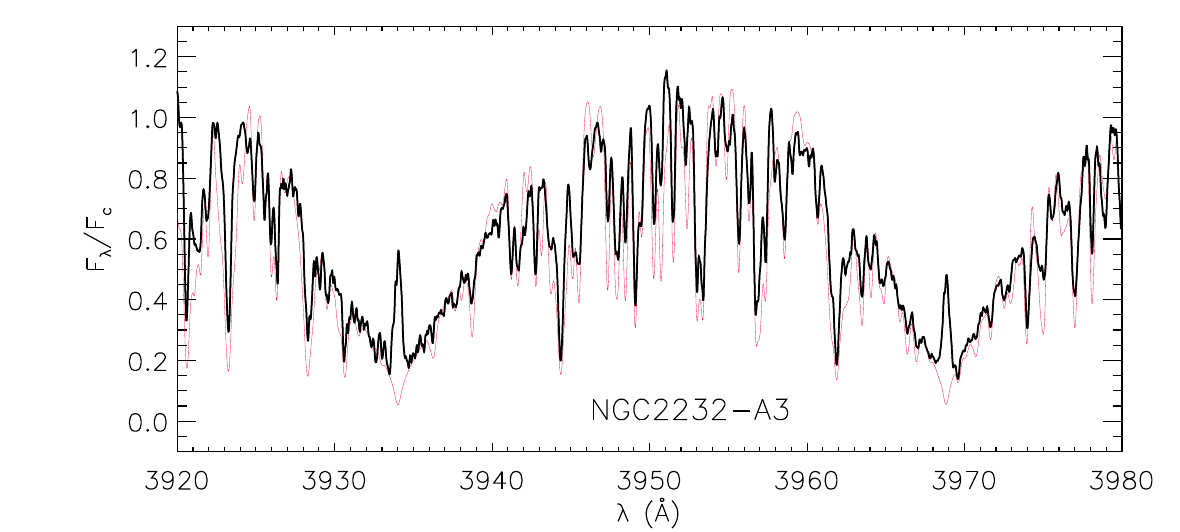} 
\includegraphics[width=9.1cm]{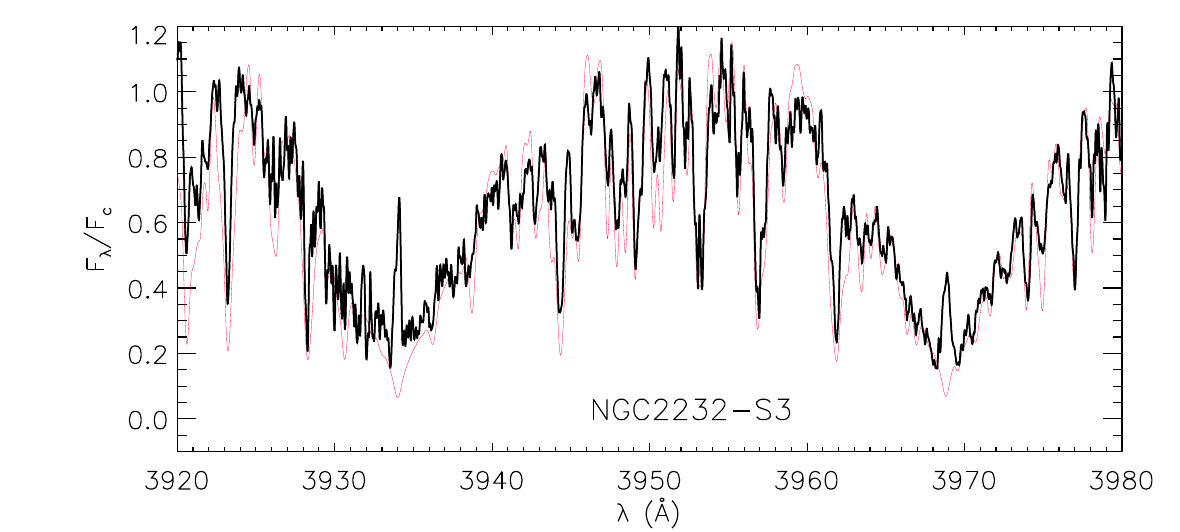} 
\includegraphics[width=9.1cm]{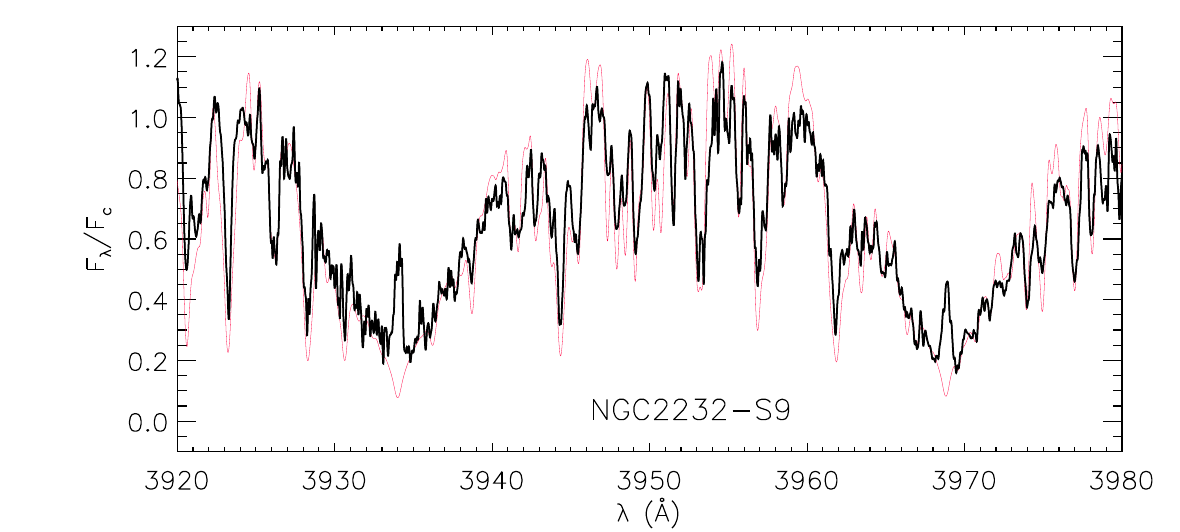} 
\includegraphics[width=9.1cm]{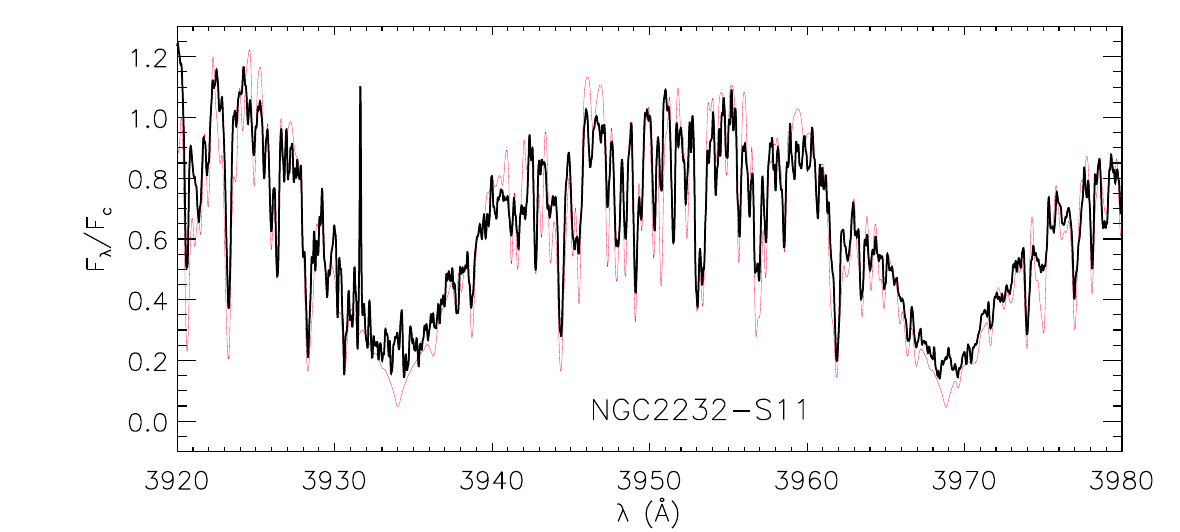} 
\includegraphics[width=9.1cm]{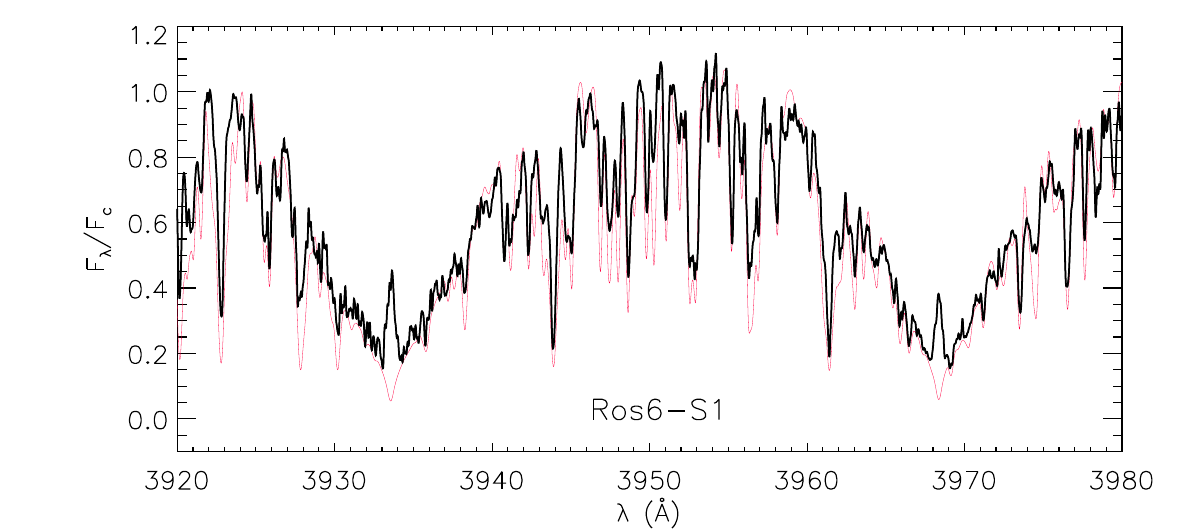} 
\includegraphics[width=9.1cm]{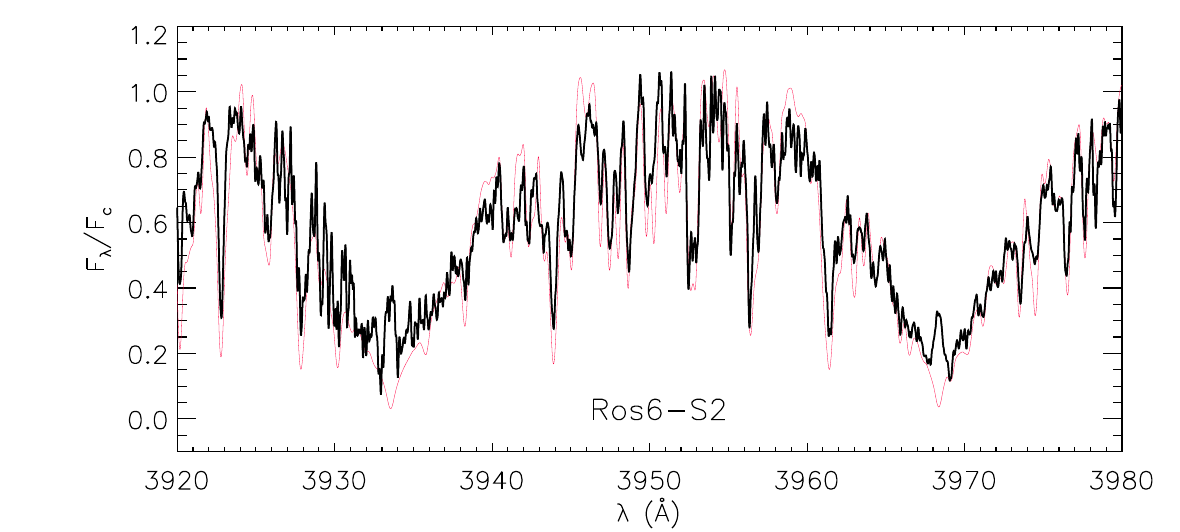} 
\includegraphics[width=9.1cm]{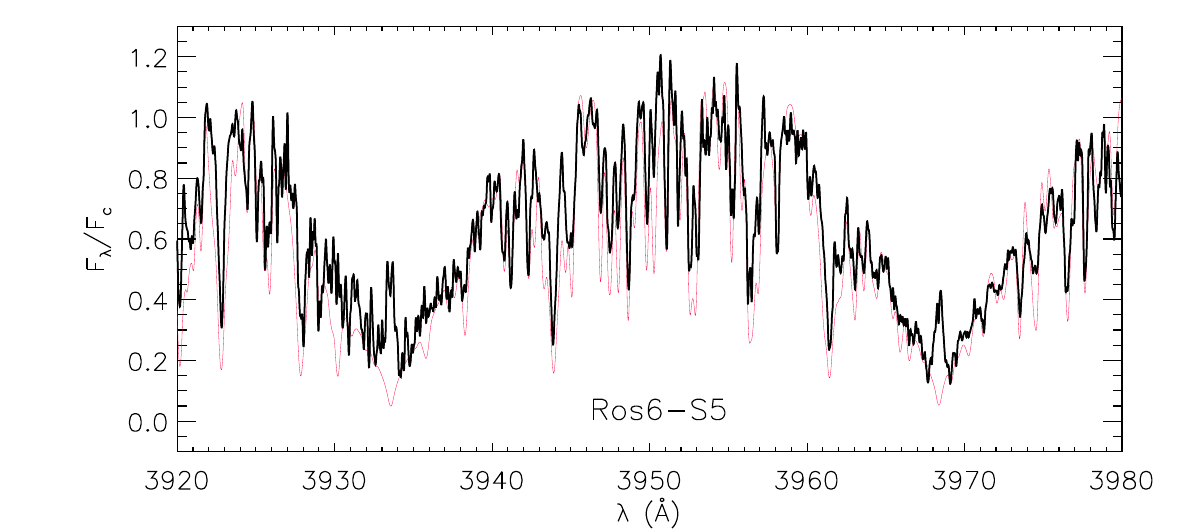} 
\includegraphics[width=9.1cm]{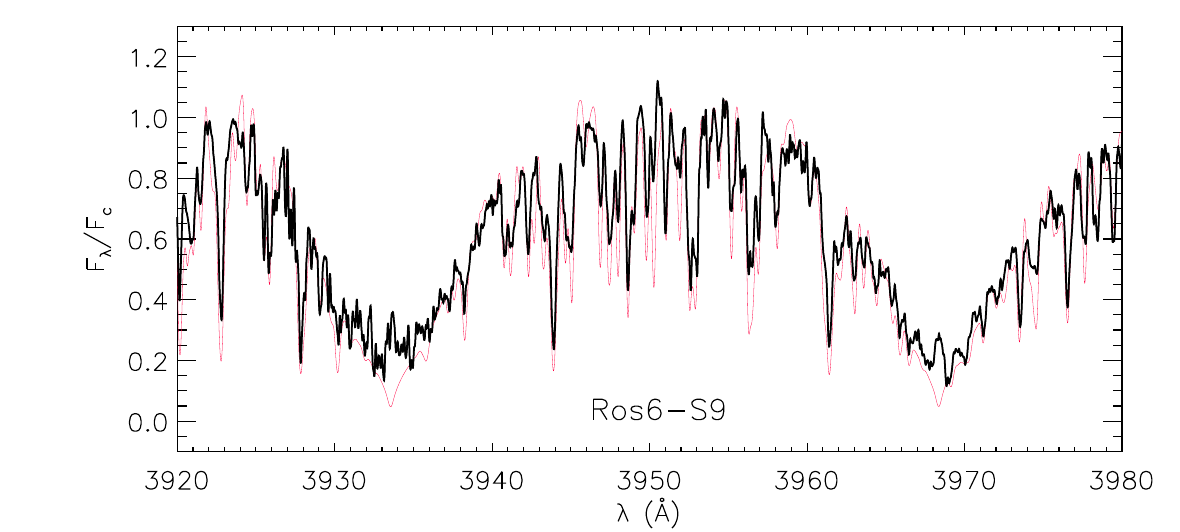} 
\vspace{0cm}
\caption{\textit{Continued}}
\label{fig:subtraction_CaII_tris}
\end{center}
\end{figure*}

\addtocounter{figure}{-1}

\begin{figure*}
\begin{center}
\includegraphics[width=9.1cm]{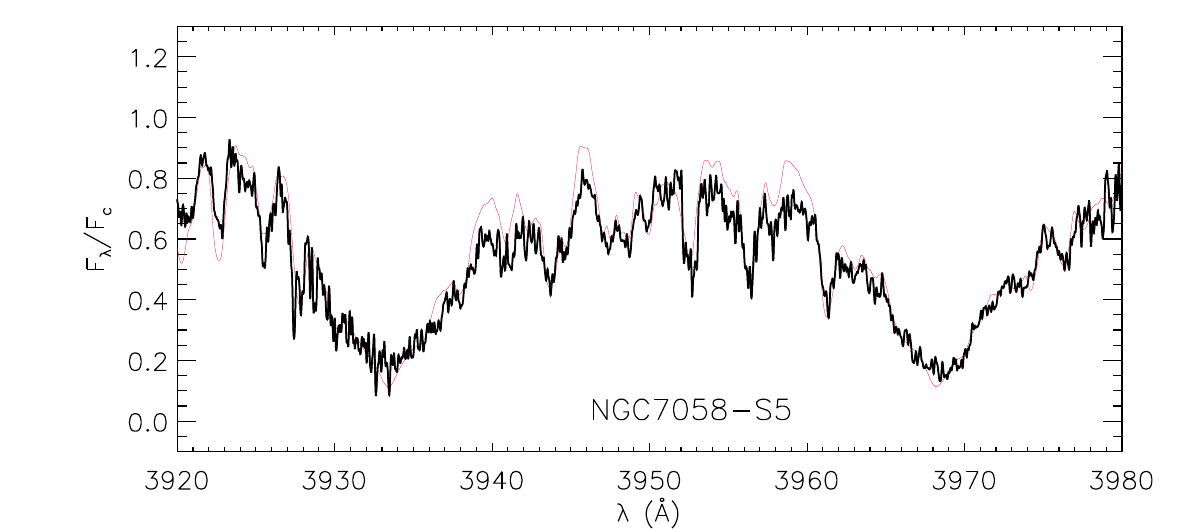} 
\includegraphics[width=9.1cm]{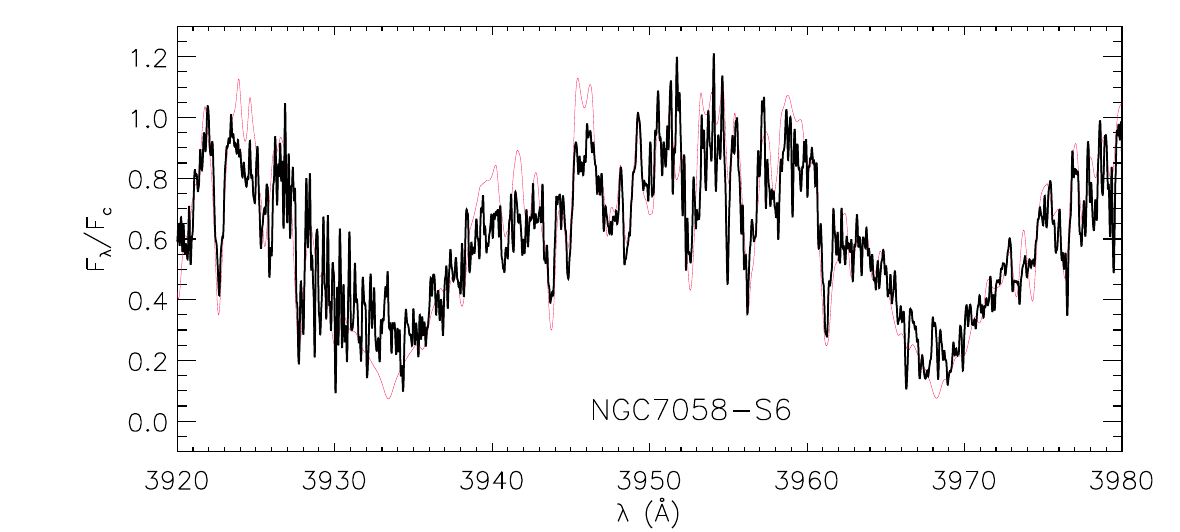} 
\includegraphics[width=9.1cm]{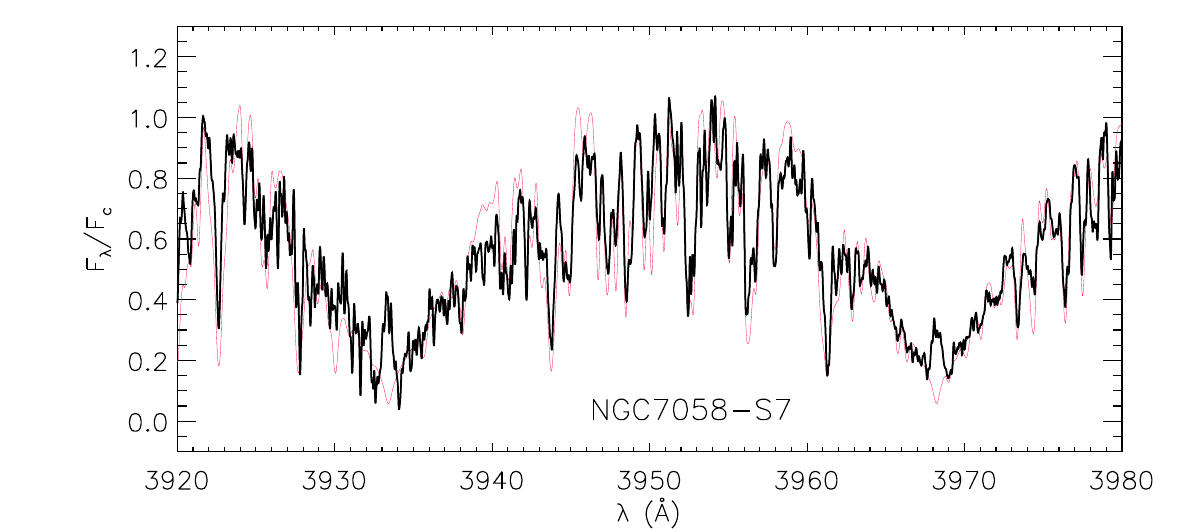} 
\includegraphics[width=9.1cm]{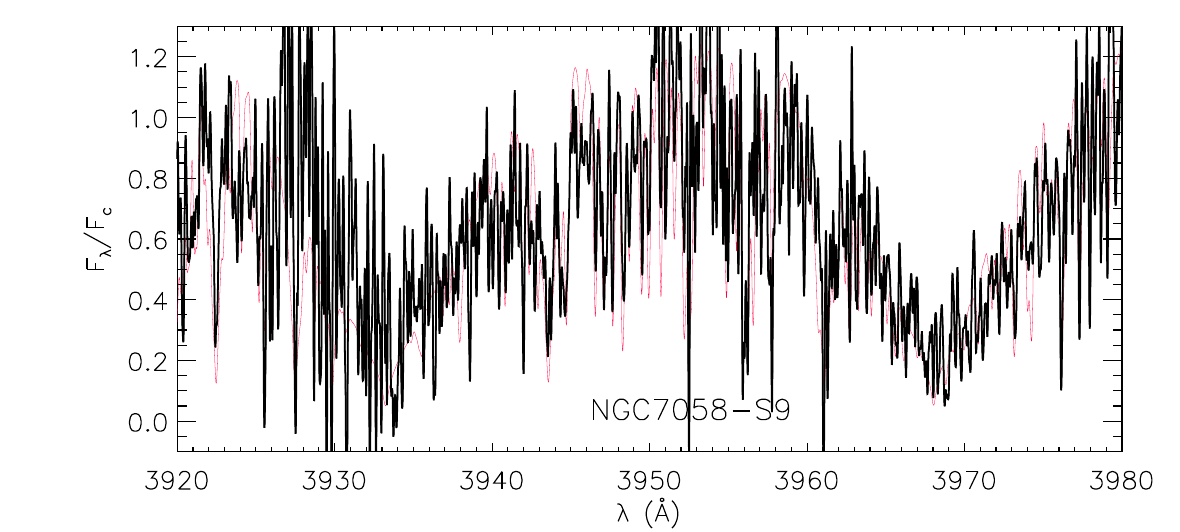} 
\includegraphics[width=9.1cm]{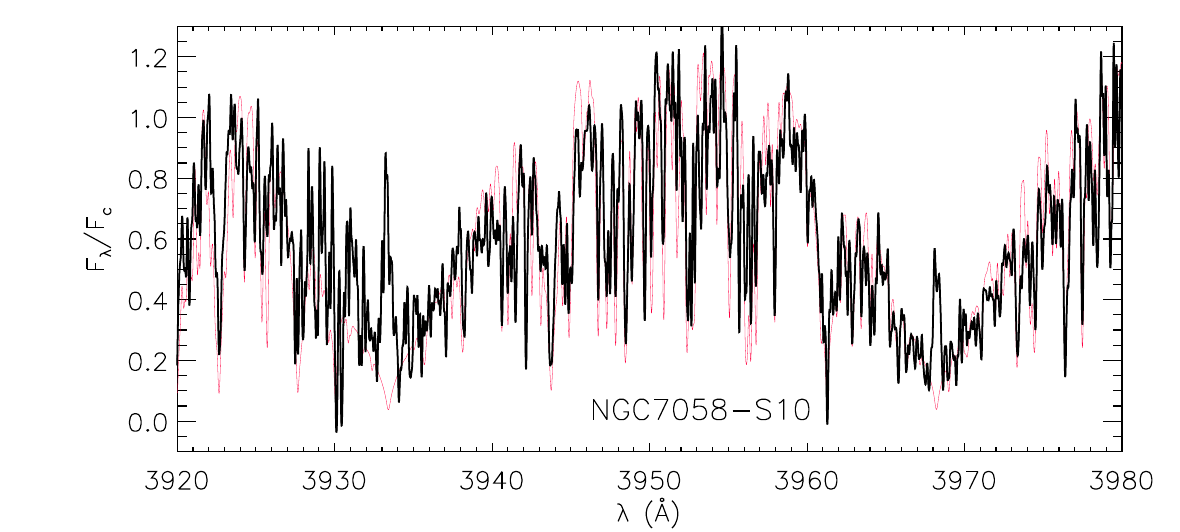} 
\includegraphics[width=9.1cm]{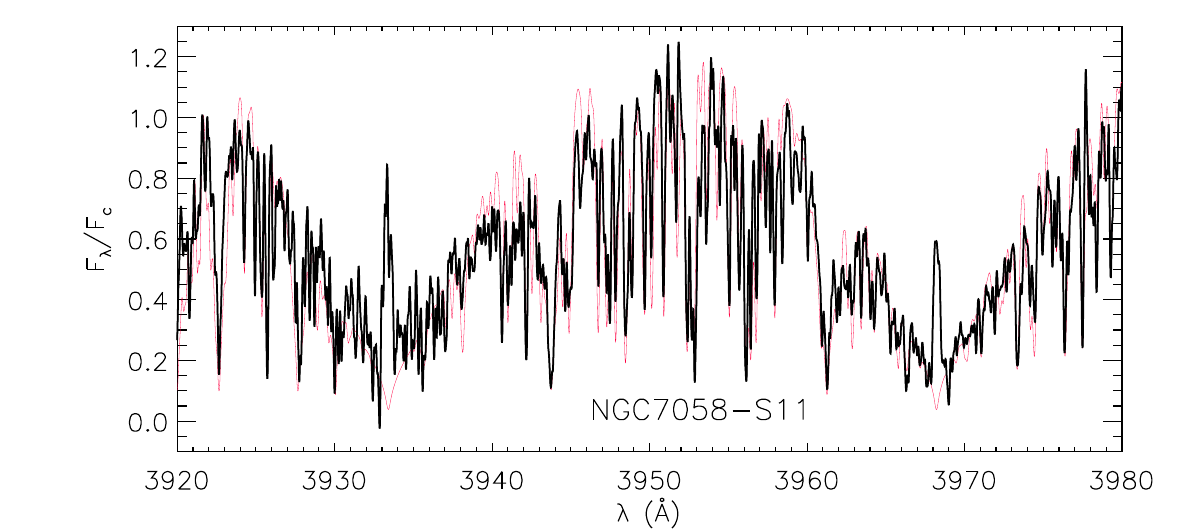} 
\includegraphics[width=9.1cm]{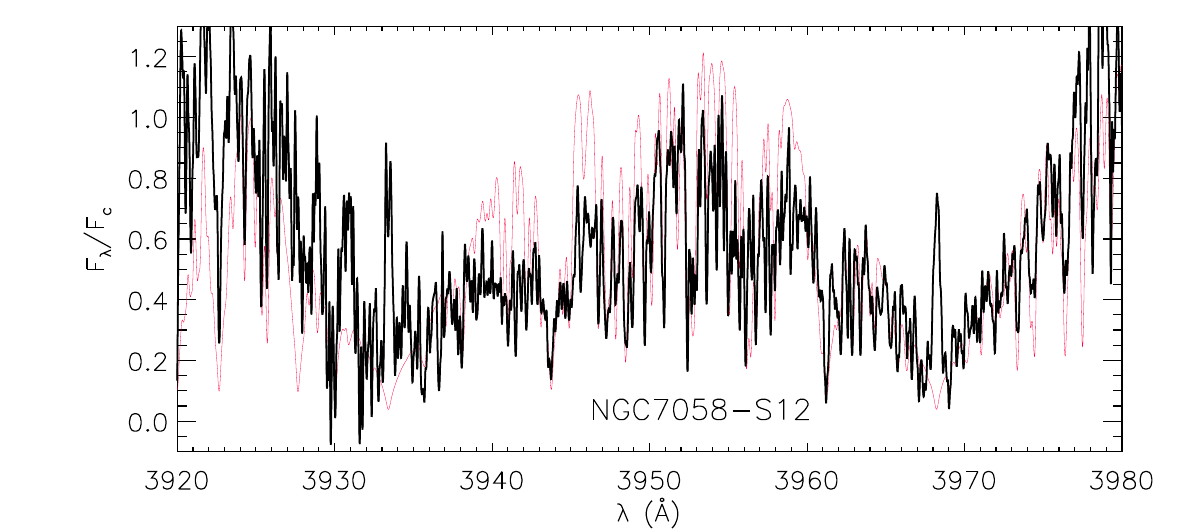} 
\includegraphics[width=9.1cm]{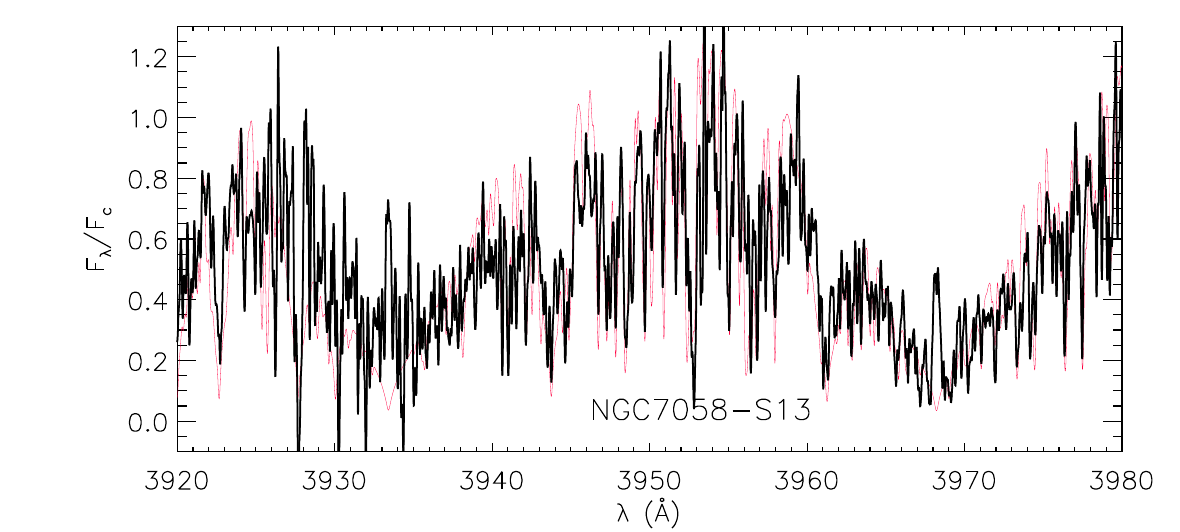}
\vspace{0cm}
\caption{\textit{Continued}}
\label{fig:subtraction_CaII_quater}
\end{center}
\end{figure*}

\end{document}